\documentclass[12pt,a4paper,final]{article}

\usepackage[margin=25mm]{geometry}
\usepackage[utf8]{inputenc}
\usepackage[OT1]{fontenc}

\usepackage[24h]{scrtime}
\usepackage{scrtime,uhrzeit,amsmath,amsfonts,amssymb,color,mathtools,mathrsfs}
\usepackage{todonotes,enumitem}

\usepackage{bm}
\let\QED\undefined 
\usepackage{bef_alex}

\newcommand{\AND}{\text{ and }}
\newcommand{\qAND}{\quad\text{and }}
\newcommand{\qANDq}{\quad\text{and}\quad}

\def\FG{\bm}
\def\ti{{\times}}

\newtheorem{exercise}{Exercise}[section]

\numberwithin{equation}{section}
\numberwithin{figure}{section}

\usepackage{psfrag,color,graphicx,tikz} 
\usepackage{pgfplots,mathrsfs}

\makeatletter
\renewcommand*\env@cases[1][1.2]{%
  \let\@ifnextchar\new@ifnextchar
  \left\lbrace
  \def\arraystretch{#1}%
  \array{@{\,}c@{\ }l@{}}%
}
\makeatother

\renewcommand{\ul}{\underline}
\newcommand{\STEP}[1]{\medskip \noindent\underline{\slshape #1}}
\newcommand{\GRAD}{\mathop{\mafo{grad}}}

\newcommand{\plF}{\pl^\rmF}

\newcommand{\DD}{\calD}
\newcommand{\UU}{u_*}
\newcommand{\AC}{\rmA\rmC}
\newcommand{\BAR}{\hspace{0.15em}\rule[-0.16em]{0.11em}{1.01em}\hspace{0.15em}} 
\newcommand{\SPE}[1]{\BAR \dot{#1} \BAR_{\DD}\hspace{-0.05em}}
\newcommand{\SPEED}[2]{\BAR \dot{#1} \BAR_{#2}\hspace{-0.05em}}
\newcommand{\SLO}[1]{\mbox{}\BAR\pl #1 \BAR_{\DD}\hspace{-0.05em}}
\newcommand{\SLOPE}[2]{\mbox{}\BAR\pl #1 \BAR_{#2}\hspace{-0.05em}}
\def\eff{\mafo{eff}}

\renewcommand{\arraystretch}{1.1}
\def\div{\mathop{\mafo{div}}}

\newcommand{\tosets}{\ensuremath\raisebox{0.05em}{${}\rightrightarrows{}$}}
\newcommand{\LB}{\lambda_{\rmB}}
\def\weak{\rightharpoonup}
\def\weaks{\overset{*}{\rightharpoonup}}
\def\Mto{\xrightarrow{\mathsf{M}}}
\def\Cto{\xrightarrow{\mafo{cc}}}

\def\Gto{\xrightarrow{\Gamma}}
\def\GEto{\xrightarrow{\Gamma_{\!\rmE}}}
\def\Glim{\mathop{\Gamma\text{-}\mafo{lim}}}
\def\GElim{\mathop{\Gamma_{\!\rmE}\text{-}\mafo{lim}}}
\def\Gwlim{\mathop{\Gamma_\rmw\text{-}\mafo{lim}}}
\def\Mlim{\mathop{\rmM\text{-}\mafo{lim}}}
\def\Gweak{\overset{\Gamma}{\rightharpoonup}}

\newcommand{\EDPto}{\overset{\mafo{EDP}}{\longrightarrow}}

\def\Rinfty{\R_\infty}

\newcommand{\Riesz}{\bbI_\rmR}

\newcommand{\dom}{\mathop{\mafo{dom}}}

\def\eff{{\mathrm{eff\,}}}%
\def\Diss{\mathop{\mathrm{Diss}}\nolimits}%

\newcommand{\scrD}{\mathscr D}
\newcommand{\Prob}{\mathrm{Prob}}
\newcommand{\sign}{\mathop{\mafo{sign}}}

\newcommand{\sfC}{\mathsf C}

\newcommand{\mfD}{\mathfrak D}

\begin{document}

\title{An introduction to the analysis\\ of gradients systems}

\author{Alexander Mielke \\[0.2em]
\normalsize WIAS Berlin and \\[-0.2em]
\normalsize Humboldt-Universität zu Berlin}

\date{7 June 2023} 
\maketitle
\vfill

\mbox{} \hfill 
\begin{minipage}{0.51\textwidth}\small\itshape
\flushright
The merit of the right gradient flow
formulation\\
 of a dissipative evolution equation is that\\
 it separates energetics and
kinetics: \\
The energetics endow the state space $M$\\ with a functional $E$, \\
the kinetics endow the state space with a \\ 
(Riemannian) geometry via the metric tensor $g$.
\\[0.5em]
\flushright Felix Otto 2001
\end{minipage}
\vfill

\centerline{\begin{minipage}{0.66\textwidth}

\centerline{\bfseries Preface}

\flushleft The present notes provide an extended version of a small lecture
course given at the Humboldt Universit\"at zu Berlin in the Winter Term 2022/23
(of 36 hours). The material starting in Section 5.4 was added
afterwards.\medskip

The aim of these notes to give an introductory overview on the analytical
approaches for gradient-flow equations in Hilbert spaces, Banach spaces, and
metric spaces and to show that on the first entry level these theories have a
lot in common. The theories and their specific setups are illustrated by
suitable examples and counterexamples.

\end{minipage}
}
\vfill

\mbox{}

\newpage
{\small
\tableofcontents
}
\newpage

\section{Introduction}
\label{s:Intro}

In this section we introduce our notions, provide a series of examples and
give motivations concerning the origins of gradients systems. 

\subsection{Gradients in the finite dimensional case}
\label{su:I.finite}

We first discuss the notion of gradient of a function $\calF:\R^d\to \R$. We
distinguish the gradient $\GRAD \calF$ and the Fr\'echet derivative $\rmD \calF$ via 
\[
\GRAD \calF(u) = \bma{c} \pl_{u_1} \calF(u)\\ \vdots\\ \pl_{u_d} \calF(u) \ema \qANDq
\rmD \calF(u) = \big( \pl_{u_1} \calF(u), \cdots , \pl_{u_d} \calF(u)\big) \in
\big(\R^d\big)^*.
\]
We will use the abbreviation ``$\GRAD$'' for general gradients and reserve the
symbol ``$\nabla$'' for PDE applications like $\Delta u = \DIV(\nabla u)$. 

For the function $\calF(u_1,u_2)=\frac12u_1^2 + \frac12 u_2^2 + \frac a4 u_2^4$ we obtain 
\[
\GRAD \calF(u) = \binom{u_1}{ u_2 {+} a u_2^3} \qANDq \rmD \calF(u) = \big(u_1,
u_2 {+} a u_2^3\big) .
\]

However we may describe the same function in polar coordinates $x= r(\cos \phi,
\sin\phi)$ giving $\wt \calF(r,\phi)= \frac12 r^2 + \frac a4 r^4(\sin\phi)^4$. 
The definition of the
gradient of $\calF$ in polar coordinates, $ \wt\nabla \wt \calF$, is no longer given by
the vector of partial derivatives but 
\[
\wt{\GRAD} \, \wt \calF(r,\phi) = \wt\bbK(r,\phi)
   \binom{\pl_r \wt \calF(r,\phi)}{\pl_\phi \wt
  \calF(r,\phi)} \quad \text{with } \wt\bbK(r,\phi)=\bma{cc} 1&0 \\ 0& 1/r^2 \ema .
\]
What is the reason for the nontrivial $\wt\bbK$?  One justification is that we
want the gradient-flow equations
\begin{equation}
  \label{eq:Euclid.Polar}
  \dot u = - \GRAD \calF(u) \qANDq \binom{\dot r}{\dot \phi} = - \wt{\GRAD}\, \wt
\calF(r,\phi)
\end{equation}
to be the same. 

However, more importantly, the right perspective is to consider the space
$\R^2$ as a manifold $M$ and $\calF:M\to \R$ as a general function. Then,
$\rmD \calF(u)$ is the differential of $\calF$ at $u$ (in differential geometry
written as $\bfd \calF(u)$). It is defined via
\[
\rmD \calF(u)[v] := \lim_{h\to 0} \frac1h \big( \calF(u{+}h v) - \calF(u)\big) 
\]
and thus we have $\rmD \calF(u) \in \mafo{Lin}(\rmT_u M;\R) = :
\rmT^*_u M$. Here we use the notion of the tangent space $\rmT_u M$ and the
co-tangent space $\rmT^*_u M$ at a point $u\in M$. We also use the duality
notation
\[
\rmD \calF(u) [v] = {}_{\rmT^*_u M}\big\langle \rmD\calF(u), v
\big\rangle_{\rmT_u M} ,
\]
where $ {}_{X^*}\langle \cdot, \cdot \rangle_X$ always means a duality pairing
between a space $X$ and its dual space $X^*$. 

However, by the definition of the gradient-flow equation $\dot u = -\GRAD
\calF(u)$ we see that the gradient has to lie in the tangent space $\rmT_u M$.
Hence, we need a mapping that maps the differential $\rmD \calF(u) \in \rmT^*_u
M$ into the vector $\GRAD \calF(u)  \in \rmT_u M$. 

This mapping is generated by a Riemannian structure $\bbG$. A pair $(M,\bbG)$
is called a Riemannian manifold, if 
\\[0.2em]
\textbullet\ $M$ is a manifold and \\ 
\textbullet\  $\bbG(u) : \rmT_u M \to \rmT^*_u M $ is symmetric and positive,
\\
\textbullet\ $g$ defined via $g(v,\wt v)_u = \langle \bbG(u) v(u),\wt
v(u)\rangle$ is a symmetric 2-tensor.\medskip

Riemannian structures are used for measuring length of curves and angles
between curves, as they define a scalar product on each $\rmT_u M$. 
For curves  $\gamma: [s_0,s_1] \to M$  one sets 
\[
\mafo{length}_\bbG(\gamma) := \int_{s_0}^{s_1} \Big( \big\langle \bbG(\gamma(s))
\gamma'(s), \gamma'(s) \big\rangle \Big)^{1/2} \dd s .
\]
When doing a transformation $u = \Phi(w)$ with $\Phi:N\to M$ 
the chain rule gives immediately
the transformation rule $\wt \bbG(w) = \rmD\Phi(w)^*\bbG(\Phi(w))
\rmD\Phi(w): \rmT_w N \to \rmT^*_w N$. 

\begin{definition}[Gradient]\label{de:Gradient}
The gradient of a function $\calF$ in a Riemannian manifold is defined via
\begin{equation}
  \label{eq:Gradient}
  \mafo{grad}_\bbG \calF(u) := \bbG(u)^{-1} \rmD\calF(u)
  = \bbK(u) \rmD\calF(u),
\end{equation}
where $\bbK(u):=\big(\bbG(u)\big)^{-1} : \rmT^*_u M \to \rmT_u M$ is called the
\emph{Onsager operator}. 
\end{definition}

For the above example in $\R^2$ we have $\bbG_\mafo{Euclid}= \binom{1\ \ 0}{0\
  \ 1}$ and the transformation $u =\Phi(r,\phi)$ into polar coordinates gives 
$\wt\bbG(r,\phi) = \binom{1 \ \ 0 \ }{0 \ \ r^2}$.  Thus we find the gradient
in polar coordinates in the following form 
\[
\wt{\GRAD{}} \wt\calF(r,\phi) = \bma{cc} 1&0\\ 0& \frac1{r^2} \ema \rmD\wt\calF(r,\phi) =
\binom{\pl_r \wt\calF(r,\phi) } {\frac1{r^2}\,\pl_\phi \wt\calF(r,\phi)} 
= \binom{r + a r^3 (\sin\phi)^4}{a r^2 (\sin\phi)^3 \cos \phi} .
\]
With this, one can indeed check that the the two ODEs in
\eqref{eq:Euclid.Polar} transform properly into each other.

\subsection{Gradient systems and their gradient-flow equations} 
 
We still stay in the framework of finite-dimensional manifolds $M$ and 
define what exactly we mean by the words ``gradient system'', ``gradient
structure'', ``gradient flow'', and ``gradient-flow equation''. 

\begin{definition}\label{de:GradSyst}
A \emph{gradient system} is a triple $(M,\calF,\bbG)$ such that $(M,\bbG)$ is a
Riemannian manifold and $\calF:M\to \R$ is a $\rmC^1$ function. 

This gradient system generates the \emph{associated gradient-flow equation}
\begin{equation}
  \label{eq:GFE.1}
    \dot u = - \mafo{grad}_\bbG\calF(u) = - \bbK(u)\rmD\calF(u) \in \rmT_u M \quad
\Longleftrightarrow \quad 0=\bbG(u)\dot u + \rmD \calF(u) \in \rmT^*_u M.
\end{equation}
We say that $u:{[0,T[}\to M$ is a solution for $(M,\calF,\bbG;u^0)$ if it
satisfies \eqref{eq:GFE.1} with $u(0)=u^0$. 
\end{definition} 

We see that $\calF$ is a Lyapunov function, i.e.\ along solutions
$u:[0,T]\to M$ the function $\calF$ is decreasing:
\begin{align*}
\frac\rmd{\rmd t} \calF(u(t)) &= \langle \rmD\calF(u), \dot u\rangle = -
\langle \rmD \calF(u),\bbK(u) \rmD \calF(u) \rangle\\
& =  - \langle \bbG(u)
\mafo{grad}_\bbG \calF(u), \mafo{grad}_\bbG \calF(u)\rangle \leq 0. 
\end{align*}

The left equation in \eqref{eq:GFE.1} will be called the rate form of the
gradient-flow equation, whereas the right equation is called the force-balance
form of the gradient-flow equation. Here $\xi=\bbG(u)\dot u$ is the viscous
force induced by the rate. We call
\[
\xi_\rmv = \bbG(u)\dot u \quad\text{or equivalently} \quad 
\dot u = \bbK(u) \xi_\rmv
\]
the \emph{kinetic relation} encoding the frictional properties of the system. 
The force $\xi= \rmD\calF(u)$ is the potential restoring force.  Of course,
kinetic relations can be more general, e.g.\ by a non-symmetric linear relation
or by nonlinear relations, see Section \ref{su:LegrNonlKinRel}. 

However, from a thermodynamical point of view the case of symmetric and
positive definite $\bbG$ or $\bbK$ is distinguished as is shown by the
fundamental work  by Lars Onsager ``\emph{Reciprocal relations in irreversible
  processes}'' \cite{Onsa31RRIP}. His ``reciprocal
relations'' were derived in the context of linearized irreversible
thermodynamics and simply mean, in modern language, the \emph{symmetry relation}
$\bbG=\bbG^*$. In fact, Onsager 
was awarded the Nobel prize for chemistry in 1968 for exactly this work, see
\begin{quote}
https://www.nobelprize.org/prizes/chemistry/1968/ceremony-speech/
\end{quote}
As Onsager and Machlup state in the follow-up work \cite[p.\,1507]{OnsMac53FIP}
[formulas slightly adapted]:
\begin{quote}
\slshape 
The tendency of the system to seek equilibrium is measured by the
\emph{thermodynamic forces} (=restoring forces) $\xi = \rmD \calS(\alpha)$ (eqn.\
(2-1)), which evidently vanish at $\alpha=0$. 

The fluxes\;(of\;matter,\;heat,\;electricity) are measured by the
time\;derivative\;$\dot \alpha$. The essential physical assumption about the
irreversible processes is that they are linear; i.e., that the fluxes depend
linearly on the forces that ``cause'' them:
\[
\bbG \dot \alpha = \xi  \qquad\text{(2-2)} \qquad\qquad \text{or} 
\qquad\qquad \bbK \xi = \dot \alpha\qquad \text{(2-3)}, 
\]
where the matrices $\bbG$ and $\bbK$ are mutual reciprocal [inverses].  

These equations express, for instance, Ohm's law for electric conduction,
Fourier's law for heat conduction, Fick's law for diffusion, and the extension
of these laws to interacting flows, e.g., anisotropic conduction (heat,
electricity), thermoelectric effects, thermal diffusion. For systems for which
microscopic reversibility holds (to which this work is confined), we have the
\emph{reciprocal relations} [symmetry relations] $\bbG = \bbG_{tr}$ (eqn.\
(2-4)), where the subscript $_{tr}$ means transpose. 
\end{quote} 
\bigskip

Under sufficient smoothness, for each $u^0\in M$ there
exists a (local or global) solution $u(t)= S_t(u^0)$ where $S_t : M \to M$ is
the \emph{gradient flow} associated with $(M,\calF,\bbG)$. Assuming that all solutions
exist globally, i.e.\ for $t\in {[0,\infty[}$ the gradient flow $(S_t)_{t\geq
  0} $ satisfies 
\begin{align*}
(1) \quad& S_0 = \mafo{id}_M \AND 
\forall \, t,r\geq 0:\ \  S_t\circ S_r = S_{t+r},
\\
(2) \quad & u(t) = S_t(u_0) \text{ is a solution for } (M,\calF,\bbG;u^0). 
\end{align*}
Property (1) is called the semigroup property of the family $(S_t)_{t\geq 0}$.

\begin{remark}[Hamiltonian systems]\label{re:HamSyst}
  The notion of gradient systems is chosen in analogy to Hamiltonian systems
  $(M,\calH,\Omega)$ (cf.\ \cite{AbrMar78FM,Arno89MMCM}) where $(M,\Omega)$ is
  a symplectic manifold with $\Omega(u):\rmT_u M\to \rmT^*_u M$ satisfying
  $\Omega(u)^*=-\Omega(u)$, $\bbJ(u)=(\Omega(u))^{-1}$ exists, and
  $\rmd \Omega \equiv 0$ (in the sense of two-forms). The associated
  Hamiltonian equations are given by
\begin{equation}
  \label{eq:HamEqn}
\dot u = \bbJ(u) \rmD\calH(u) \in \rmT_u M \quad \Longleftrightarrow \quad 
  \Omega(u)\dot u = \rmD\calH(u)  \in \rmT^*_u M .
\end{equation}
Along solutions we have $\frac{\rmd}{\rmd t} \calH(u(t)) =
\langle \rmD\calH(u), \dot u\rangle = \langle \rmD\calH(u),\bbJ(u)
\rmD\calH(u)\rangle =0$, which means energy conservation. 
\end{remark}

\begin{definition}[Gradient structure]\label{de:GradStruct}
Given a differential equation $\dot u = \bfV(u)$  on a manifold $M$ we say that
the equation has the \emph{gradient structure} $(M,\calF,\bbG)$ if
$\bfV(u)=\bbK(u)\rmD\calF(u)$ for all $u \in M$, i.e. the ODE is the
gradient-flow equation associated with $(M,\calF,\bbG)$.
\end{definition}

Note the two different perspectives:

(I) The GS $(M,\calF,\bbG)$ generates the (unique) gradient-flow 
equation $\dot u = -\bbK(u) \rmD\calF(u)$.  

(II) A given ODE can have one or many gradient structure or no at all. 
\bigskip

\begin{example}[Trivial scaling]
If $\dot u = \bfV(u)$ has the gradient structure $(M,\calF,\bbG)$, then for all
$\lambda >0$ it also has the gradient structure
$(M,\wt\calF,\wt\bbG)=(M,\lambda\calF,\lambda\bbG)$. Simply observe that
$\wt\bbK= (\lambda \bbG)^{-1}= \frac1\lambda \bbK$, such that $\lambda $
cancels. 
\end{example} 

\begin{example}[Two nontrivial structures]
Let $M=\R^2$ and $\dot u = \bfV(u)=\binom{-u_1}{-u_2-a u_2^3}$ with $a>0$. 
From above we know that we have the gradient structure 
\[
\bbG=\bbI_\mafo{Eucl} = \begin{pmatrix} 1&0\\ 0&1 
\end{pmatrix} \qANDq \calF(u)= \frac12 u_1^2 +
\frac12 u_2^2 + \frac a4 u_2^4. 
\]
However, there is another gradient structure $(\R^2,\wt\calF,\wt\bbG)$, namely 
\[
\wt\bbG= \begin{pmatrix} 1&0\\ 0&\frac1{1{+}a u_2^2}\end{pmatrix} 
\qANDq \calF(u)= \frac12 u_1^2 +\frac12 u_2^2 . 
\] 
Thus, when looking at the ODE we do not know whether the coefficient $a >0$ and
the nonlinear term $-a u_2^3$ arises because of a nonquadratic energy (as in
$\calF$) or because of a state-dependent friction law (as in $\wt\bbG$). 
\end{example}

The last example shows that different gradient structures for an ODE refer to
different physics/mechanics behind the model. The gradient structure contains
\emph{additional information} that is not contained in the ODE. 

The next example is a more recent one and relates to chemical reaction-rate
equations. 

\begin{example}[Reaction-rate equations]\label{ex:RRE} 
We consider three chemical species denoted by $X_1,\ X_2$, and $X_3$ with
densities $c_1,\ c_2$, and $c_3$, respectively. Hence, the states are
$c=(c_i)_i$ in the manifold is $M= {]0,\infty[}^3$. We consider three reactions 
\[
X_1 \rightleftharpoons X_2, \quad 
X_1+ X_2 \rightleftharpoons 2X_3, \quad 
a_1 X_1 + a_3 X_3 \rightleftharpoons b_2 X_2
\]
which follow the mass-action law, i.e.\ the reaction rates are proportional to
the corresponding monomials. The ODE reads
\[
\dot c = \bfR(c):= 
k_1(c_1{-}c_2) \bma{c}\! -1\!\\ 1\\ 0\ema + k_2(c_1c_2{-}c^2_3)\bma{c} -1\\-1\\ 2\ema 
+  k_3(c_1^{a_1} c_3^{a_3} {-}c_2^{b_2}) \bma{c} -a_1\\ b_2 \\ -a_3 \ema 
\]
It was observed in \cite{Yong08ICPD} and in a more general setting in
\cite{Miel11GSRD,MaaMie20MCRS}, that the above equation has a gradient
structure (because of the detailed-balance condition, see the references
above). If we set
\[
\calF(c) = \sum_{i=1}^3 \lambda_\rmB(c_i) \qANDq \bbK(c) =\sum_{r=1}^3 k_r
\Lambda\big( c^{\bfalpha^r},\, c^{\bfbeta^r} \big) (\bfalpha^r{-}\bfbeta^r)
\otimes (\bfalpha^r{-}\bfbeta^r) ,
\]
where $\lambda_\rmB(c) = c \log c -c +1$ is the Boltzmann function and
$\Lambda(r,\rho) = \int_0^1 r^s \rho^{1-s} \dd s = (r{-}\rho)/\log(r/\rho)$ is
the logarithmic mean of $r$ and $\rho$. The stoichiometric vectors
$\bfalpha^r, \bfbeta^r\in \N_0^3$ are given via {\small%
\renewcommand{\arraystretch}{0.9}%
\renewcommand{\arraycolsep}{0.1em}%
\[
\bfalpha^1=\bma{c} 1\\ 0 \\ 0 \ema, \
\bfbeta^1=\bma{c} 0\\ 1 \\ 0 \ema, \ 
\bfalpha^2=\bma{c} 1\\ 1 \\ 0 \ema, \ 
\bfbeta^2=\bma{c} 0\\ 0 \\ 2 \ema, \ 
\bfalpha^3=\bma{c} a_1\\ 0 \\ a_3 \ema, \ 
\bfbeta^3=\bma{c} 0\\ b_2 \\ 0 \ema.
\]}
\end{example}

One nice feature of the above model is that it nicely shows the additive
structure of $\bbK$: it is given as a sum over individual terms corresponding
to a single reaction. This additive structure will often reappear, namely
whenever there are several distinguishable dissipative processes. Their effect
will be additive on the level of $\bbK$ but not on the level of $\bbG$. Hence,
for modeling it is often more convenient to work with $\bbK$.

\subsection{Gradient structures for partial differential equations} 
\label{su:PDEexamples}

In this part we do mainly formal calculations only, and see this as a
motivation for the analysis in the following sections. Nevertheless we are
motivated by the philosophy from the smooth, finite-dimensional case discussed
in the previous section.  But now the function $\calF$ may no longer be
smooth but may attain the value $+\infty$ outside a dense set. Moreover the
operator $\bbK$ may be unbounded.

As a first example we consider the Allen-Cahn equation, which is a nonlinear
parabolic equation, sometimes called reaction-diffusion equation: 
\begin{equation}
  \label{eq:AllenCahn}
  m\dot u = \alpha\Delta u + \beta\,(u - u^3) \quad \text{in } \Omega, \qquad u=0 \quad 
  \text{on }\pl\Omega,
\end{equation}
where $\Omega$ is a bounded Lipschitz domain in $\R^d$. 

We want to show that this equation has the gradient structure $(\rmL^2(\Omega),
\calF_{\rmA\rmC} , m\Riesz )$, where $\bbI_R : H \to H^*$ denotes the Riesz
isomorphism of a Hilbert space $H$ with its dual space $H^*$. The Allen-Cahn
functional is given by 
\[
\calF_{\rmA\rmC}(u) =\left\{\ba{cl}  \int_\Omega \big( \frac\alpha2 |\nabla u|^2 +
\frac\beta4(u^2{-}1)^2\big)  \dd x & \text{for }u \in \dom (\calF_{\rmA\rmC}), \\ 
\infty& \text{for } u \in \rmL^2(\Omega) \setminus
\dom (\calF_{\rmA\rmC}),
\ea \right.
\]
where $\dom (\calF_{\rmA\rmC})= \rmH^1_0(\Omega)\cap \rmL^4(\Omega)$. Moreover
the differential $\rmD \calF_{\rmA\rmC}$ is replaced by the variational
derivative, which is defined on an even smaller set:
\[
\rmD\calF_{\rmA\rmC}(u) = - \alpha \Delta u - \beta\, (u - u^3) \quad \text{for } u \in
\dom (\rmD\calF_{\rmA\rmC}):= \rmH^2(\Omega)\cap \rmH^1_0(\Omega) \cap
\rmL^6(\Omega) . 
\]
This notion of derivative will be made rigorous in terms of the Fr\'echet
subdifferential to be introduced in Section \ref{se:GSHilbert}. Recalling our
choice $\bbG=m\Riesz$ for the Riemannian metric, we see that the
``force-balance formulation'' $\bbG \dot u = - \rmD\calF(u)$ for the given
gradient structure indeed yields the Allen-Cahn equation
\eqref{eq:AllenCahn}. 
\bigskip

Next we consider the simple linear parabolic equation
\begin{equation}
  \label{eq:ParabolicEq}
  \dot u = \Delta u \quad \text{in } \Omega, \qquad \nabla u \cdot \nu = 0
  \quad \text{on }\pl\Omega,
\end{equation}
where $\Omega \subset \R^d$ is again a bounded Lipschitz domain and
 $\nu$ is the outward unit normal on $\pl\Omega$. 
We will construct four quite different gradient structures, each of which
corresponds to a different application of this equation. 
Recall that the name for this equation is usually ``heat equation''; however,
it is sometimes also called ``diffusion equation''. 
\medskip

\underline{\bfseries Gradient Structure 1: Allen-Cahn type $\rmL^2$ gradient flow:} We
consider the GS $(\rmL^2(\Omega), \calF_\mafo{Dir}, \Riesz )$
with the Dirichlet functional defined on $\rmL^2(\Omega)$, namely
\[
\calF_\mafo{Dir}(u) = \left\{ \ba{cl} \int_\Omega \frac12|\nabla u|^2 \dd x &
  \text{for } u \in \rmH^1(\Omega), \\ 
  +\infty &\text{for } u \in \rmL^2(\Omega)\setminus\rmH^1(\Omega) \ea \right. 
\]
Here the differential can be interpreted as a convex subdifferential, which is
either empty or a singleton, namely $\rmD\calF_\mafo{Dir} (u) = -\Delta
u$. Thus, we obtain \eqref{eq:ParabolicEq} as the associated gradient-flow
equation. This will be made rigorous in Section \ref{se:GSHilbert}. \medskip

\underline{\bfseries Gradient Structure 2: $\rmH^{-1}(\Omega)$ gradient flow:}
We again consider a spatially constant Hilbert-space structure, but now in the
space $H := \big(\rmH^1(\Omega)\big){}^*=:\rmH^{-1}_0(\Omega)$ such that the
dual space is $H^*=\rmH^1(\Omega)$ and we have the Riesz isomorphism
$\Riesz : \rmH^{-1}_0(\Omega) \to \rmH^1(\Omega)$. On the formal level we
consider 
\[
\calF_{\rmL^2}(u) = \frac12\|u\|_{\rmL^2}^2 = \int_\Omega \frac12 \,u^2 \dd x
\qANDq \bbK^{(2)} \left\{ \ba{ccc}\rmH^1(\Omega)&\to& \rmH^{-1}_0(\Omega),\\
\\\xi&\mapsto & - \Delta \xi. \ea\right.
\]
Note that the $\calF_{\rmL^2}$ has domain
$\dom (\calF_{\rmL^2})=\rmL^2(\Omega) \subsetneqq
\rmH^{-1}_0(\Omega)$. Moreover, the differential $\rmD\calF_{\rmL^2}(u) = u \in
H^*=\rmH^1(\Omega)$ has the even smaller domain
$\dom\big(\rmD\calF_{\rmL^2}\big) = \rmH^1(\Omega)$. 

Again we obtain the desired gradient-flow equation 
\[
\dot u = -\bbK^{(2)}\rmD\calF_{\rmL^2}(u) = - \big({-}\Delta\big) u = \Delta u.
\]
The exact details will be made rigorous in Section \ref{se:GSHilbert}. \medskip

Two more gradient structures will be handled in the two following
subsections. They play an important role in the modeling as well as in the
initiation of a new branch of mathematics, namely optimal transport for PDEs,
see \cite{Otto01GDEE, AmGiSa05GFMS, Vill09OTON, Pele14VMEG, DanSav14LNGF,
  Sant17EMWG}. We give some more details here, because the standard parabolic
equation $\dot u= \Delta u$ is most often simply called the ``heat equation''
but sometimes also ``diffusion equation''. On the level of PDEs there is no
distinction, it is simply a parabolic equation. However, on the level of
gradient-flow equations the distinction will become apparent.

\subsection{Otto's gradient structure for diffusion}
\label{su:OttoGradStruct}

The theory of gradient systems received a major push around the
year 2000 through the seminal work of Felix Otto in
\cite{Otto96DDDE,JoKiOt98VFFP,Otto01GDEE}. It is interesting to note the title
and a citation of the latter work:
\begin{quote}
\itshape 
``The geometry of dissipative evolution equations: the porous medium equation'' 

p.\ 108: \ldots\ The merit of the right gradient flow
formulation
 of a dissipative evolution equation is that
 it separates energetics and
kinetics: 
The energetics endow the state space $M$ with a functional $E$, 
the kinetics endow the state space with a (Riemannian) geometry via the metric tensor $g$.
\end{quote} 

This work suggests the following choice of a gradient structure $(\Prob(\Omega),
\calE_{\text{Bolz}},\bbK_\text{Otto})$:
\begin{align*}
&M = \Prob(\Omega):=\bigset{u\in \rmL^1(\Omega)}{ u\geq 0 \text{ a.e., } \ts
  \int_\Omega u \dd x =1 } \subset \rmL^1(\Omega),
\\
&\calF(u)=\calE_{\text{Bolz}}(u) = \int_\Omega \LB(u(x)) \dd x \quad \text{where } 
\LB(z)=\left\{ \ba{cl} z\log z - z +1 &\text{for } z>0, \\[-0.2em]
1& \text{for } z=0, \\[-0.2em] + \infty& \text{for } z<0; \ea\right. 
\\
&\bbK_\text{Otto}(u)\xi := - \DIV\big( u \nabla \xi\big) .
\end{align*}
Of course, it was known for a century that the (relative) Boltzmann entropy
$\calE_{\text{Bolz}} $ is a good Lyapunov function for the diffusion equation
$\dot u = \Delta u$. However, introducing the (Riemannian-type) geometrical
structure $\bbK_\text{Otto}$ was the key step. In these papers, and in more
than one hundred follow-up papers, the geometry is called \emph{Wasserstein
  geometry} because calculating the corresponding geodesic distance
$d_{\bbK_{\text{Otto}}}$ one obtains the 2-Wasserstein distance $ \rmW_2$ on
  $\Prob(\Omega)$, see more on that in Section \ref{se:MetricGS}. 

On the formal level we easily see that the associated gradient-flow equation is
indeed the linear diffusion equation, if we use $\rmD\calE_{\text{Bolz}} (u)=
  \LB'(u) = \log u$ and the classical chain rule $\nabla \log u = \frac1u
  \nabla u$:  
\[
\dot u = -\bbK_{\text{Otto}} (u) \rmD\calE_{\text{Bolz}}(u) = -\big({-}\DIV\big[ u
\nabla \LB'(u)\big]\big) = \DIV( u \frac1 u \nabla u) = \DIV(\nabla u) = \Delta
u. 
\]
Of course, the works \cite{Otto96DDDE,JoKiOt98VFFP,Otto01GDEE} and the
follow-up works provide the rigorous analysis following from this choice of the
gradient structure.  Because of its big
importance in the recent developments for diffusion equation, we define the
Otto gradient (unfortunately often called Wasserstein gradient) of a general
functional $\calF(u)=\int_\Omega \big(F(u(x))- V(x)u(x)\big) \dd x$, namely
\begin{equation}
\label{eq:Def.OttoGrad}
\mafo{grad}_\text{Otto} \calF(u) := \bbK_\text{Otto}(u) \rmD \calF(u) =
-\DIV\big( u \nabla \big[F'(u){-}V]\big) .
\end{equation}

Clearly this choice is physically highly relevant (and can be justified in the
Onsager-Machlup sense \cite{OnsMac53FIP} via fluctuation theory for diffusion,
see e.g.\ \cite{DawGar87LDMV,ADPZ11LDPW,MiPeRe14RGFL}), but it leaves the range
of linear theory. The energy is nonquadratic and even enforces the positivity
of $u$. Otto's approach to diffusion applies genuinely nonlinear methods to a
linear problem, which hence opens the theory to nonlinear applications such as
the porous medium equation as in \cite{Otto01GDEE}. 
In particular, this new gradient structure has created a
whole new branch of mathematics, namely the treatment of diffusion equations
using ideas from optimal transport of probability measures, see
\cite{AmGiSa05GFMS}.

\subsection{Gradient structures for the heat equation}
\label{su:GradStructHeatEqn}

On the level of gradient systems there is a strong distinction between the heat
and the diffusion equation, which will become clear below. For diffusion a
good gradient structure is Otto's gradient structure, but it is not appropriate
for heat conduction. 

When writing the heat equation in terms of the absolute temperature $\theta>0$
we need the internal (or heat) energy $e=E(x,\theta)$ and the internal entropy $s
=S(x,\theta)$ which are related by the Gibbs relation $E'(x,\theta) = \theta
S'(x,\theta)$, where $'$ means $\pl_\theta$. One major point is that $E'$ is
called heat capacity and it must 
be positive, following the intuition that for heating up a body one has to invest
energy (e.g.\ 4.18\:Joule for heating up 1\,kg of water by 1\,Kelvin). 
By Gibbs' relation also $S'(x,\theta)>0$. 

The fundamental laws of thermodynamics say that the total energy is conserved
in a closed system while the total entropy increases. The heat equation reads
\begin{equation}
  \label{eq:HeatEqn}
  \dot e + \DIV \bfq=0 \text{ in } \Omega, \qquad \bfq\cdot \nu  =0 \text{ on } 
\pl\Omega. 
\end{equation}
Here $e(t,x)= E(x,\theta(t,x))$ and $\bfq (t,x) \in \R^d$ denotes the heat flux
that is given by Fourier's law in the form
$\bfq(t,x) = -K(x,\theta) \nabla \theta$, where $K(x,\theta)= K(x,\theta)^*>0$
is the heat conduction matrix (recall Onsager's symmetry) and $\nabla \theta$
is now the classical Euclidean gradient of the function
$\theta(t,\cdot):\Omega \to \R$. The boundary conditions $\bfq\cdot \nu =0 $
say that the body $\Omega$ is insulated such that heat cannot leave or enter
$\Omega$. Integrating over $\Omega$ we find conservation of total energy
$t \mapsto \calE(x,\theta(t))$:
\[
\frac\rmd{\rmd t } \calE(x,\theta(t))= \int_\Omega \frac\pl{\pl t }
E(x,\theta(t,x)) \dd x  
\overset{\text{\eqref{eq:HeatEqn}}} =  
\int_\Omega {-}\DIV \bfq \dd x \overset{\text{Gau\ss}}= 
- \int_{\pl\Omega} \bfq \cdot \nu \dd a = 0.
\] 

We can now try to generate the heat equation as a
(anti-) gradient-flow equation for the total entropy 
\[
 \calS(\theta) = \int_\Omega S(x,\theta(x)) \dd x \quad \text{with }
 \rmD\calS(\theta) = S'(x,\theta) ,
\] 
where ``anti'' stands for a functional that increases along solutions.

We now follow \cite{Miel11TCMG} and generalize the idea of Otto by looking for
an Onsager operator $\bbH_\text{heat}$ in the form
\[
\bbK_\text{heat}(\theta) \xi = - \frac1{E'(x,\theta)} \DIV \! \Big( \bbA(x,\theta)
\nabla\big( \frac{\xi}{ E'(x,\theta)}  \big) \Big) ,
\]
where the factor $1/E'(x,\theta)$ was introduced twice in such a way that
$\bbK^*_\text{heat}$ is still a symmetric differential operator. This prefactor
is essential to handle the term $\dot e= \pl_t( E(x,\theta(t,x)) ) 
= E'(x,\theta) \dot
\theta$ in the heat equation \eqref{eq:HeatEqn}. 

With this we calculate the anti gradient-flow equation 
\begin{align*}
\dot\theta &\ \, = + \bbK_\text{heat} (\theta) \rmD\calS(\theta) = -
\frac{1}{E'(x,\theta)}\DIV\!\Big( 
\bbA(x,\theta) \nabla \big( \frac{S'(x,\theta)}{E'(x,\theta)}\big) \Big)  
\\
&\overset{\text{Gibbs}}= -\frac{1}{E'(x,\theta)}\DIV\!\Big( 
\bbA(x,\theta) \nabla \big( \frac{1}{\theta}\big) \Big)  \ \overset*=  \ 
\frac{1}{E'(x,\theta)}\DIV\!\Big( \frac{1}{\theta^2}\
\bbA(\theta)\  \nabla \theta\  \Big),
\end{align*}
where in $\overset*=$ we used $\nabla(\frac1\theta) = - \frac1{\theta^2} \nabla
\theta$. Thus, the abstract equation leads to the heat equation 
\[
\dot e= E'(x,\theta) \dot \theta = - \DIV \bfq = \DIV( K\nabla \theta) \quad
\text{with } K(x,\theta)= \frac1{\theta^2} \bbA(x,\theta) . 
\]

This approach teaches us, just by formal arguments, that $-\nabla ({S'}/{E'})
= -\nabla (1/\theta)$ is the correct (nonlinear) term that drives heat
conduction. This is indeed important at interfaces, where the jump of
$1/\theta$ matters. 

To obtain the simple linear heat equation $\dot \theta =\Delta \theta$, we can
use $E(\theta)=\theta$ and $S(\theta)= \log (\theta/\theta_\circ)$ and have to
choose $\bbA(\theta) = \theta^2 \bbI$, i.e.\ 
\[
\bbK_\text{heat} (\theta) \xi     = - \DIV\!\big(\theta^2 \nabla \xi\big) ,
\]
which is clearly different from $\bbK_\text{Otto} $ because of the power 2 in
$\theta^2$. 

So far, the analysis for this (Riemannian) geometry has still to be developed.

\subsection{Further remarks on modeling with gradient systems}

A general approach to modeling with gradient systems is given in the expository
work \cite{Pele14VMEG}. In particular, it addresses the proper derivation of
gradient systems from microscopic stochastic models via so-called
large-deviation principles. Thus, proceeds along the path
developed in \cite{OnsMac53FIP}. 

General development of gradient structures for semiconductor models or
energy-reaction-diffusion systems, also with interfaces, can be found in
\cite{Miel11GSRD, Miel13TMER, GliMie13GSSC}.

The interplay of Hamiltonian dynamics and gradient systems can be described in
term of the framework GENERIC, which is an acronym for General Equation for
Non-Equilibrium Reversible Irreversible Coupling, see \cite{GrmOtt97DTCF12,
  Otti05BET, Grme10WG, Miel11FTDM, DuPeZi13GFVF}. This approach was also used
to couple classical thermodynamical models to quantum systems in
\cite{MitMie17EGSL, KMMR19MMSQ}, where the interaction of the quantum system
and its classical environment is modeled by a suitable Onsager operator.

\section{Gradient systems with Hilbert-space structure}
\label{se:GSHilbert}

In this section we provide a mathematical rigorous framework for gradient
systems in Hilbert spaces. By this name we do not only mean that the underlying
space is a Hilbert space $H$, but we also use the full nice properties of the
Hilbert-space geometry, i.e.\ we will always assume that $\bbG$ is independent
of the state variable $u \in H$ and equals the Riesz isomorphism $\Riesz : H
\to H^*$. Of course, this still allows us to adapt the Hilbert-space norm, if
we have an equivalent norm. For example we consider the parabolic PDE
\[
c(x) \dot u = \DIV\!\big( A(x) \nabla u\big)  - \pl_u F(x,u(x)) 
\quad \text{ in }\Omega, \qquad u=0 \quad \text{on } \pl\Omega,
\]
where $c\in \rmL^\infty(\Omega)$ with $c(x) \geq c_0>0$ a.e.\ and suitable $A$
and $F$. Then we can choose the gradient structure $(H,\calF,\bbG)$ with 
\[
H=\rmL^2(\Omega) , \quad \bbG v = \bbI_\text{Riesz} v = c v, \quad 
\calF(u)=\int_\Omega \frac12 \nabla u(x)\cdot A(x) \nabla u(x) + F(x,u(x))\big) \dd x 
\] 
for $u \in \rmH^1_0(\Omega)$ and $+\infty$ otherwise on $\rmL^2(\Omega)$. Here
$\rmH^1(\Omega)$ is the Sobolev space of functions with square integrable
gradient, and $\rmH^1_0(\Omega)$ is the closed subspace obtained by closing
$\rmC^\infty_\rmc(\Omega)$ in $\rmH^1(\Omega)$.

\subsection{Differentials and subdifferentials on Banach spaces} 
\label{su:SubDiff}

For PDEs it is essential to have a suitable notion of differential, because of
two important facts:
\\[0.2em]
\textbullet\ (even quadratic) functionals and their differentials need to be
defined on dense subsets
\\[0.2em]
\textbullet\ nonsmoothness is important in applications (contact, Coulomb
friction, plasticity, ...)
\bigskip

We are now working on general Banach spaces $X$ with dual spaces $X^*$ and dual
pairing ${}_{X^*}\langle\cdot, \cdot \rangle_X$. In particular, we avoid the
identification $H\sim H^*$ in Banach spaces. As G\^ateaux and Fr\'echet
differentials are only useful for continuous functions, we directly define
so-called subdifferentials, which are \emph{set-valued mappings}. For a mapping
$A:X \to 2^Y=\mathfrak P(Y)$ we shortly write $A:X\tosets Y$, i.e.\ for all
$u\in X$ we have $A(u) \subset Y$, where $A(u)=\emptyset$ is of course
possible.

Here we develop a theory in the spirit of Br\'ezis' foundational work, see in
particular the existence result in
\cite[Thm.\,3.6,\;p.\,72]{Brez73OMMS}. However, the approach there is
completely different, because it is based on Yosida regularizations for maximal
monotone operators whereas we use time-incremental minimization for gradient
systems. Our approach can be adapted easily to Banach spaces and metric spaces.

\begin{definition}[Subdifferentials]\label{de:Subdifferential}
Let $\calF:X \to \Rinfty:=\R\cup\{+\infty\}$ be a functional.  
The \emph{(convex) subdifferential} $\pl \calF:X\tosets X^*$ is defined via
$\pl\calF(u)=\emptyset $ for $\calF(u)=\infty$ and 
\[
\pl\calF(u) :=\bigset{\xi \in X^*}{ \forall\,w\in X:\ \calF(w)\geq
  \calF(u)+\langle \xi, w{-}u\rangle_X } \subset X^*
\]
otherwise. 
The \emph{Fr\'echet subdifferential} $\plF \calF:X\tosets X^*$ is defined
via $\plF\calF(u)=\emptyset$ for $\calF(u)=\infty$ and 
\[
\plF\calF(u) :=\bigset{\xi \in X^*}{ \calF(w)\geq
  \calF(u)+\langle \xi, w{-}u\rangle_X + o(\|w{-}u\|_X) \text{ for } w\to u } \subset X^*
\]
otherwise. The domains of $\calF$, $\pl\calF$, and $\plF\calF$ are the
subsets of $X$ defined via
\begin{align*} 
\dom(\calF)&=\bigset{u\in X}{ \calF(u)<\infty}, \quad
\dom(\pl\calF)=\bigset{u\in X}{ \pl\calF(u) \neq \emptyset}, \\
\dom(\plF\calF)&=\bigset{u\in X}{ \plF\calF(u) \neq \emptyset}.
\end{align*}
\end{definition}

By the definition, we clearly have $\pl\calF(u) \subset \plF\calF(u)$. 

\begin{exercise} Consider $X=\R$ and the following functions:
\[
\calF_1(u)= \frac14(u^2{-}1)^2, \quad  \calF_2(u)=-|u|+u^2, \quad 
\calF_3(u) = \min\{ 0, |u|{-}1, \frac12u^2-1\}.
\]
Calculate $\pl\calF$ and $\plF\calF$ for all three cases.   
\end{exercise}

\begin{exercise}  Let $\Omega$ be a smooth bounded domain in $\R^d$. 
\\[0.1em]
(A) As an example we consider the quadratic functional $\calF:\rmL^2(\Omega)\to
\Rinfty$ with 
\[
\calF(u)= \int_\Omega \frac12|\nabla u(x)|^2\dd x  \quad \text{on }
\dom(\calF)=\rmH^1(\Omega).
\]
Show that $\dom(\pl\calF) = \bigset{u \in \rmH^1(\Omega)}{ \Delta u\in
  \rmL^2(\Omega), \ \nabla u \cdot\nu=0 \text{ on } \pl \Omega}$ and
$\pl\calF(u) = \{{-} \Delta u\} \subset \rmL^2(\Omega)$. 
\\[0.1em]
(B) Consider exponents $p$ and $q$ with $1<p<q$ and let 
\[
X= \rmL^p(\Omega) \qANDq \calF(u) = \int_\Omega \frac1q |u(x)|^q \dd x.
\]
Calculate $\dom(\calF)$ and the differentials $\pl\calF$ and $\plF\calF$. 
\end{exercise}

The important property of the Fr\'echet subdifferential is that there is a
\emph{sum rule}. Similar sum rules play an important role many areas of applied
analysis: calculus of variations, optimization, abstract evolution equations,
and of course in the theory of gradient systems. 

\begin{proposition}[Sum rule for subdifferentials]
\label{pr:SumRule1}
If $\calF_1:X \to \Rinfty$ is convex and $\calF_2:X\to \R$  is
Fr\'echet differentiable (i.e.\ for all $u\in X$ we have $\calF_2(u{+}v) -
\calF_2(u)-\langle \rmD \calF_2 (u),v\rangle = o(\|v\|)$ for $v\to 0$), then 
\[
\plF \calF(u) = \rmD\calF_2(u) + \pl\calF_1(u) \subset X^*.
\]
\end{proposition}
\begin{proof} As $\calF_2(u)\in \R$ for all $u \in X$ we have
  $\dom(\calF)=\dom(\calF_1)$.

Now consider $\xi \in \pl\calF_1(u)$. Then, for $w \in X$ we have 
\begin{align*}
\calF(w) & = \calF_1(w) + \calF_2(w) \geq    \calF_1(u) + \langle
\xi,w{-}u\rangle + \rmD\calF_2(u)[w{-}u] + o(\|w{-}u\|_X) \\
&= \calF(u) + \langle \xi{+}\rmD\calF_2(u), w{-} u \rangle + o(\|w{-}u\|_X),  
\end{align*}
which means that $\rmD\calF_2(u) + \pl\calF_1(u) \subset \plF\calF(u)$. 

For the opposite inclusion we assume $\eta \in \plF\calF(u)$ and obtain 
\begin{align*}
\calF_1(w)& = \calF(w)-\calF_2(w)\\
 &\geq \calF(u) +
\langle \eta, w{-}u\rangle + o(\|w{-}u\|) -\calF_2(u)
 -\langle \rmD\calF_2(u), w{-}u\rangle + o(\|w{-}u\|)  
\\ & = \calF_1(u) + \big\langle \eta{-}\rmD\calF_2(u) ,  w{-}u\big\rangle + o(\|w{-}u\|). 
\end{align*}
By convexity of $\calF_1$ we have $\calF_1(w_\theta)\leq (1{-}\theta) \calF_1(w)
+ \theta \calF_1(u)$, where $w_\theta=(1{-}\theta) w+ \theta u$, and conclude
(by setting $w=w_\theta$ in the above estimate) 
\begin{align*}
\calF_1(w) &\geq \frac1{1{-}\theta}\big(\calF_1(w_\theta) - \theta \calF_1(u) \big) 
\\
& \overset{\text{above}}\geq \frac1{1{-}\theta}\big(\calF_1(u) +\big\langle
\eta{-}\rmD\calF_2(u) , w_\theta{-}u\big\rangle + o(\|w_\theta{-}u\|)    -
\theta \calF_1(u) \big)  
\\
&= \calF_1(u) + \big\langle
\eta{-}\rmD\calF_2(u) , w{-}u\big\rangle+ o((1{-}\theta)\|w{-}u\|) 
\\
&
\overset{\theta\to 1^-}\longrightarrow 
\calF_1(u) + \big\langle \eta{-}\rmD\calF_2(u) , w{-}u\big\rangle.
\end{align*}
Thus, we conclude $\eta {-}\rmD\calF_2(u) \in \pl\calF_1(u)$ which means
$\plF\calF(u) \subset \rmD\calF_2(u) {+}\pl\calF_1(u)$. 
\end{proof}

\begin{exercise}[Convex subdifferentials]
\label{ex:CvxSubd}
Consider a reflexive Banach space $X$ and a functional $\calF:X \to
  \Rinfty$ that is proper, lower semicontinuous, and convex.

(A) For $\xi \in X^*$ define the functional $\calG_\xi:u \mapsto \calF(u) -
\langle \xi ,u\rangle$. Show the sum rule 
$\pl\calG_\xi(u) = -\xi + \pl\calF(u)$. 

(B) Assume additionally that $\calF$ is superlinear, i.e.\
$\calF(u)/(1{+}\|u\|)\to \infty $ for $\| u \| \to \infty$. Show that the
subdifferential $\pl\calF:X \tosets X^*$ is surjective, i.e.\ for each $\xi \in
X^*$ there exists $u_\xi \in X$ such that $\xi \in \pl\calF(u_\xi)$. 
(Hint: Minimize a suitable functional.)
\end{exercise} 

\subsection{Semiconvexity and closedness of subdifferentials}
\label{su:Hilbert.ClosSubdiff}

An important class of functionals will be the following one.

\begin{definition}[Semiconvexity]
\label{de:SemiConvex}
A function $\calF:X \to \Rinfty$ is called \emph{$\lambda$-convex}, if 
\begin{equation}
  \label{eq:SemiConvex}
\begin{aligned}
  &\forall\, u_0,u_1 \in X\ \forall \, \theta\in [0,1]{:} \\ 
& \qquad 
\calF\big((1{-}\theta)u_0{+}\theta u_1\big) \leq (1{-}\theta) \calF(u_0) + \theta
\calF(u_1) - \frac\lambda 2 \theta(1{-}\theta) \|u_1{-}u_0\|_X^2. 
\end{aligned} 
\end{equation}
We simply say that $\calF $ is \emph{semiconvex} if there exists $\lambda \in
\R$ such that $\calF$ is $\lambda$-convex. 
\end{definition}

We will often use the notion of sublevels of $\calF$, namely
$S^\calF_E:=\bigset{u \in H}{ \calF(u)\leq E}$.  It is a classical fact that
$\calF$ is (weakly) lower semicontinuous if and only if for all $E\in \R$ the
sublevels $S^\calF_E$ are (weakly) closed. (For that reason, in some papers and
books, lsc functionals are simply called `closed'.)

\begin{exercise}[Convex hulls of sublevels of semiconvex functionals]
\label{exer:CvxHullSublevel}
Assume that $\calF:X\to \Rinfty$ is $\lambda$-convex. 

(A) Show that in the case $\lambda \geq 0$ the sublevels $S^\calF_E$ are
convex. 

(B) Give an example where $\calF$ is $(-1)$-convex and $S^\calF_E$ is
nonconvex for some $E\in \R$. 

(C) Consider a subset $A$ of $X$ such that $A\subset B_R(0)\cap
S^\calF_E$. Show that the convex hull $\mafo{co}(A)$ lies in $S^\calF_{\wt E}$
for a suitable $\wt E$ depending on $\lambda$ and $R$. 
\end{exercise}

Two of the fundamental properties of semiconvex functionals are
a simple global characterization of the Fr\'echet subdifferential and the
so-called closedness of the graph of $\plF \calF$. 

\begin{lemma}[Characterization of Fr\'echet subdifferential]
\label{le:CharFrechSub}
Assume that $\calF: X\to \Rinfty$ is $\lambda$-convex, then the Fr\'echet
subdifferential admits the following global representation: 
For all $u \in \dom(\calF)$ we have
\begin{equation}
  \label{eq:CharFrechet}
  \plF\calF(u)= \bigset{\xi \in X^*}{ \forall\, w\in X:\ \calF(w)\geq
    \calF(u) + \langle\xi,w{-}u\rangle + \frac\lambda2\|w{-}u\|_X^2}
\end{equation}
\end{lemma}
\begin{proof} Set $\bfA(u)$ for the right-hand side in
  \eqref{eq:CharFrechet}. As $\|w{-}u\|^2 = o(\|w{-}u\|)$ we immediately have
  $\bfA(u) \subset \plF \calF(u)$.  

For the opposite inclusion consider $\xi \in \plF\calF(u)$ and arbitrary
$x\in X$. By $\lambda$-convexity we have, with $w_\theta= (1{-}\theta) w + \theta u$,
\begin{align*}
\calF(w) &\geq \frac1{1{-}\theta}\big(\calF(w_\theta) - \theta \calF(u)
+\frac\lambda2 \theta(1{-}\theta) \|w{-}u\|_X^2 \big) 
\\
& \geq \frac1{1{-}\theta}\big(\calF(u) +\big\langle
\xi , w_\theta{-}u\big\rangle + o(\|w_\theta{-}u\|)    -
\theta \calF(u) \big)  + \frac\lambda2\, \theta \|w{-}u\|_X^2 
\\
&= \calF(u) + \langle \xi , w{-}u\rangle
 + \frac{o((1{-}\theta)\|w{-}u\|)}{1-\theta} + \frac\lambda2\, \theta
\|w{-}u\|_X^2. 
\end{align*}
Taking the limit $\theta\to 1^-$ we obtain $\xi \in \bfA(u)$ and conclude
$\plF \calF(u) \subset \bfA(u)$ as desired. 
\end{proof}

While the above lemma can be seen as a technical tool, the following
closedness property is essential for showing existence of solutions via
limiting processes. This condition parallels the important concept of
``closedness of a graph of a linear operator'' (recall the closed-graph
theorem).

\begin{definition}[Closedness of the differential]
\label{de:ClosednessDiff} 
A set-valued mapping $\bfA:X \tosets Y$ is called \emph{(strong-weak) closed}
if
\[
\left.\ba{c} u_n \to u \text{ in } X,\ y_n\weak y \text{ in } Y\\[0.2em]
            y_n \in \bfA(u_n) \ea \right\} \ \Longrightarrow \  y \in \bfA(u). 
\] 
Let $\ol\pl\calF:X\tosets X^*$ by any (sub-) differential of $\calF:X\to
\Rinfty$, then $\ol\pl\calF$ is called \emph{(strong-weak) energy closed} (in
short $E$-closed) if
\[
\left.\ba{c} u_n \to u \text{ in } X,\quad \xi_n\weak \xi \text{ in }X^*\\[0.2em]
           \sup\limits_{n\in \N} \calF(u_n) <\infty,\ \  \xi_n \in \ol\pl\calF(u_n) \ea
         \right\} \ \Longrightarrow \  \xi \in \ol\pl\calF(u).  
\] 
\end{definition}

One can also define $(\alpha,\beta)$-closedness for $\alpha,\beta\in
\{\text{weak}, \text{strong}\}$. In particular, for quadratic functionals with
$\plF\calF(u) = \{ \bbA u\}$ it may be relevant to define (weak,weak)
closedness.

\begin{exercise}[Closedness]
Consider $X=\rmL^2(\Omega) $ with $\Omega ={]0,\ell[} \subset \R^1$ 
and the functional $\calF(u)= \int_\Omega\big( \frac23|u(x)|^{3/2} + \frac12
u(x)^2\big) \dd x $. 

Show that $\calF$ is strong-weak closed but not weak-weak closed. 
\end{exercise}

Obviously, in general a subdifferential is not closed, simply consider
$\calF:\R \to \R; \ u\mapsto \frac12 u^2 - |u|$ then $\plF F(0)=\emptyset$
while $\plF \calF(u)= u - \mafo{sign}(u)$ for $u\neq 0$.  Hence
$\xi_n=1/n-1 \in \plF \calF(1/n)$ and $(u_n,\xi_n)=(\frac1n,\frac1n {-} 1)\to
(0,-1)=(u_*,\xi_*)$, but $\xi_*=-1 \not\in \plF\calF(u_*)$. 
 
However, the situation is much better for semiconvex functionals, where we can
take advantage of the global characterization of the Fr\'echet
subdifferential. 

\begin{proposition}[Closedness of $\plF\calF$]
\label{pr:ClosedFrechet}
If $\calF:X\to \Rinfty$ is proper, lower semicontinuous, and semiconvex, then 
its Fr\'echet subdifferential $\plF\calF$ is strong-weak [energy???] closed. 
\end{proposition}
\begin{proof}
We consider sequences $(u_n)_n$ in $X$ and
$(\xi_n)_n$ in $X^*$ satisfying the properties in the definition of weak-strong
closedness.  Using the global characterization of Lemma
\ref{le:CharFrechSub}  we have, for all $n\in \N$ and all $w \in X$, the estimate
\[
\calF(w) \geq \calF(u_n) + \langle \xi_n ,w{-}u_n\rangle + \frac\lambda2 \|
w{-}u_n\|^2. 
\]
In this identity we can pass to the limit $n\to \infty$ using strong lsc of
$\calF $, the weak-strong continuity of the duality product
$(v,\eta) \mapsto \langle \eta, v\rangle$, and the strong continuity of the
norm. Thus, we find
$\calF(w) \geq \calF(u) + \langle \xi ,w{-}u\rangle + \frac\lambda2 \|
w{-}u\|^2$.

As $\calF$ is proper, we conclude $\calF(u)< \infty$, and applying the global
characterization \eqref{eq:CharFrechet} gives $\xi \in \plF\calF(u)$ as
desired.
\end{proof}

Of course, this result is only one of the easy results and there are many
other possibilities for establishing closedness of subdifferentials.

\subsection{Existence via time-incremental minimization}
\label{su:Hilbert.Increm}

One of the most versatile methods of showing existence results for evolutionary
problems is that of time discretization. Fixing a time horizon $T>0$ (which
will be completely arbitrary here) we choose $N\in \N$ and define the
(constant) time step $\tau = T/N >0$. One of the main advantages of treating
gradient-flow equations is that the time-incremental problem can be formulated
as a minimization problem. Thus, we are speaking about \emph{time-incremental
  minimization} or the \emph{minimizing-movement scheme}.  

Given a gradient system $(H,\calF,\Riesz ) $ on a Hilbert space $H$ and an
initial condition $u^0\in H$, the aim is to find a solution $u \in
\rmW^{1,1}([0,T];H)$ such that 
\begin{equation}
  \label{eq:HilbertGFE}
  \Riesz  \dot u(t) \in - \plF \calF(u(t)) \quad \text{ for a.a. }t\in
[0,T], \qquad u(0)=u^0. 
\end{equation}
Note that for Hilbert spaces we have $\rmW^{1,1}([0,T];H) = \rmA\rmC([0,T];H)
\subset \rmC^0([0,T];H)$ such that posing the initial condition $u(0)=u^0$ is
well defined. 

The backward Euler time-discretization (fully implicit) is defined via 
\begin{equation}
  \label{eq:EulerTimeStep}
 0\in   \Riesz  \frac1\tau(u_k{-}u_{k-1})  + \plF \calF(u_k)  \quad
 \text{in } H^*. 
\end{equation}
Here $u_0=u^0$ is the initial condition and $u_k$ is to be found incrementally
for $k=1,\ldots,N$. Recalling that the functional 
\[
u \mapsto \frac1{2\tau} \,\| u{-}u_{k-1}\|^2 = \frac1{2\tau} \big\langle
\Riesz (u{-}u_{k-1}), u{-}u_{k-1} \big\rangle
\]
is Fr\'echet differentiable with derivative $\Riesz  \frac1\tau(u{-}u_{k-1})$
we see that \eqref{eq:EulerTimeStep} is the Euler-Lagrange equation for the
following 
\begin{equation}
  \label{eq:HilbertTIMS}
\boxed{
  \begin{aligned}
  &\text{{\bfseries time-incremental minimization scheme} for $\tau=T/N$:}\\
  &\text{set } u_0=u^0;
  \\
  &\text{ for } k=1,\ldots,N \text{ find } u_k \text{ as a minimizer of the
    functional}
  \\
  &\qquad u \ \mapsto \ \Phi^\calF_\tau(u_{k-1}; u):= \frac1{2\tau} \|
  u{-}u_{k-1}\|^2_H + \calF(u).  
\end{aligned} 
}
\end{equation}
In the case that $\calF$ is lower semicontinuous and $\lambda$-convex on $H$ we
easily see that $\Phi^\calF_\tau(w,\cdot)$ is
$(\lambda{+}\frac1\tau)$-convex. Hence, for sufficiently small $\tau>0$, the
minimizer $u_k$ is unique and minimizing $\Phi_\tau^\calF(u_{k-1}; \cdot)$ is
equivalent to solving the Euler scheme \eqref{eq:EulerTimeStep}.

Based on the discrete solution $(u_k)_{k=0,...,N}$ we are now able to define
the \emph{piecewise affine interpolant} $\wh u_\tau \in \rmC^0([0,T];H)$ and the
\emph{piecewise constant interpolant} $\ol u_\tau\in \rmL^\infty([0,T]; H)$ as
follows:
\begin{align*}
& \wh u_\tau((k{+}\theta{-}1)\tau)= (1{-}\theta) u_{k-1} + \theta u_k \quad
\text{for } k\in \{1,\ldots,N\} \text{ and } \theta \in [0,1],
\\
& \ol u_\tau(0)=u^0 \text{ and } \ol u_\tau(t)=u_k \text{ for } t\in
{](k{-}1)\tau,k\tau]}  \text{ and } k\in \{1,\ldots,N\} .
\end{align*}  
These two interpolants are constructed in such a way that the discrete equation
\eqref{eq:EulerTimeStep}  leads to the relation in the
evolutionary form
\begin{equation}
  \label{eq:DiscrEvolution}
0\ \in \   \Riesz \dot{\wh u}_\tau(t)  \ + \ \plF\calF(\ol u_\tau(t)) \quad
\text{for all } t \in [0,T]\setminus \set{k\tau}{k=0,...,N}.
\end{equation}
On each open subinterval ${](k{-}1)\tau, k\tau[}$ the terms on the right-hand
side are constant and equal the terms in \eqref{eq:EulerTimeStep}. 

The following theorem shows that in the limit $\tau\to 0^+$ we indeed obtain
convergence to a limiting function $u$ and this function indeed is a solution
of the gradient-flow equation \eqref{eq:HilbertGFE}. Thus, the following result
is not only an existence result, but it is also a convergence result for the
time-incremental minimization scheme. 

\begin{theorem}[Existence of a gradient flow for $(H,\calF,\Riesz)$]
\label{th:HilbertExist}
Consider the gradient system $(H,\calF,\Riesz)$ where $H$ is a Hilbert space
with Riesz isomorphism $\Riesz$ and $\calF: H\to \Rinfty$ is proper, lower
semicontinuous, $\lambda$-convex for some $\lambda \in \R$, and has compact
sublevels, i.e.\ for all $E\in \R$ 
the sets $S^\calF_E:=\bigset{u \in H}{ \calF(u)\leq E}$ are compact in $H$. 

Then for all $u^0\in \dom(\calF)$ the solutions $\wh u_\tau:[0,T]\to H$ obtained
from the incremental minimizing scheme converge to the unique solution $u \in
\rmW^{1,2}([0,T];H)$ of \eqref{eq:HilbertGFE}, i.e.\ 
\[
\forall\, t\in [0,T]: \quad \wh u_\tau(t) \to u(t) \text{ \ in } H.
\]
Moreover, for any two solutions $u_0$ and $u_1$ we have the
$\lambda$-contractivity estimate
\begin{equation}
  \label{eq:HilbertContract}
  \| u_1(t) -  u_0(t)\|_H \leq \ee^{-\lambda (t-s)} \,\| u_1(s) -  u_0(s) \|_H
  \quad \text{for } 0\leq s < t. 
\end{equation}
\end{theorem} 

The proof will be given in the next subsection. 

\begin{example}[Nonsmooth energy in $\R^2$]
\label{ex:NonsmoothR2}
We consider $\R^2$ equipped with the Hilbert space norm $\|v\|^2=v_1^2/a +
v_2^2/b$, i.e.\ $\bbK=\binom{a\ \ 0}{0 \ \ b}$. Moreover, we consider the
nonsmooth functional $\calF(u)=\max\{|u_1|, |u_2|\}$. Clearly, $\calF$ is convex
but nonsmooth. The subdifferential is a singleton for points not lying on the two
diagonals $u_1= \pm u_2$:
\[
\pl\calF(u) = \plF\calF(u)= \left\{ \ba{cl} 
\{\binom{\sign(u_1)}0 \} & \text{for } 0<|u_2|<|u_1|, \\ 
\{\binom0{\sign(u_2)} \} & \text{for } 0<|u_1|<|u_2|, \\ 
\bigset{\sign(u_1) \binom{\theta}{1-\theta}}{ \theta \in [0,1]} 
     & \text{for } 0\neq u_1=u_2, \\
\bigset{\sign(u_2) \binom{-\theta}{1-\theta}}{ \theta \in [0,1]} 
     & \text{for } 0\neq u_1=- u_2, \\
\bigset{\binom{\xi_1}{\xi_2} } {|\xi_1|{+}|\xi_2| \leq 1} & \text{for } 0= u_1=u_2 . 
\ea \right. 
\]
As $\calF$ is convex, we know that the GFE has exactly one solution for each
initial condition. 

We now piece together the solutions of the GFE $\dot u \in \bbK
\pl\calF(u)$. Without loss of generality we start in $u^0$ in the triangle
$u^0_1>u_2^0 >0$. As long as the solution stays in this triangle the
subdifferential $\pl\calF(u)$ is the singleton $(1,0)^\top$. Hence, we have the 
velocity $\dot u = -(a,0)^\top$, i.e.\ 
\[
u(t)= u^0 - \binom{ at}{0} \quad \text{for } t \in [0,t_1] \text{ with }
t_1:=(u^0_1{-}u^0_2)/a. 
\]
At $t=t_1$ the solution has reached the ray $u_1=u_2>0$, and it has to stay
there, i.e.\ 
\[
\dot u= \alpha \binom11 \qANDq  0 \in \alpha \binom11 + \binom{a\ \ 0}{0 \ \ b}
\binom{\theta} {1{-}\theta} .
\]
Thus, we find $\theta=b/(a{+}b)$ and $\alpha= - ab/(a{+}b)$ which gives
\[
u(t)= \big( u_2^0 - \frac{ab}{a{+}b}(t{-}t_1)\big)\binom11 \text{ for } t_1\leq
t  \leq t_2=t_1+\frac{a{+}b}{ab} u_2^0 \ \AND u(t)=0 \text{ for
}t \geq t_2.
\]
Clearly this provides a solution, and by uniqueness it is the only solution. 

We emphasize that for this system our existence theory implies the existence
of a contractive semiflow, i.e.\ uniqueness for positive times and Lipschitz
continuous dependence on the initial data. 

However, this example does not admit
any uniqueness or Lipschitz continuity backward in time. Indeed, all solutions
starting in the ball $B_R(0)$ reach $u=0$ in a finite time $t_R>0$ and then
satisfy $u(t)=0$ for $t \geq t_R$.  
\end{example}

\begin{example}[Allen-Cahn equation]
\label{ex:AllenCahn}
Here we want to show that the
result applies to the Allen-Cahn equation \eqref{eq:AllenCahn} (also called
Chafee-Infante equation) the GFE for $(\rmL^2(\Omega), \calF_{\rmA\rmC}, m \Riesz)$,
where $m$ is a positive constant. 

Let $\Omega \subset \R^d$ be a smooth bounded domain and $d\leq 3$ such that
$\rmH^1_0(\Omega) \subset \rmL^6(\Omega)$. Then, using $\alpha,\beta >0$, 
it is standard to see that the functional  
\[
\calF_{\rmA\rmC}(u) =\left\{ \ba{cl} \ds\int_\Omega \big( \frac\alpha2 |\nabla
  u|^2 + \frac\beta4(u^2{-}1)^2 \big) \dd x & \text{for } u \in \rmH^1_0(\Omega), \\
  +\infty&\text{for }u \in \rmL^2(\Omega)\setminus \rmH^1_0(\Omega) \ea \right.
\]
has domain $\dom(\calF_{\rmA\rmC})=\rmH^1_0(\Omega)$ and Fr\'echet
subdifferential $\plF\calF_{\rmA\rmC}$ given by 
\[
\plF\calF_{\rmA\rmC}(u) = \left\{ \ba{cl} \big\{ {-}\alpha \Delta u - \beta(u
  {-}u^3) \big\}& \text{for } u \in \dom(\plF\calF_{\rmA\rmC}):=
  \rmH^2(\Omega)\cap\rmH^1_0(\Omega), \\ 
 \emptyset &\text{for }u \in \rmL^2(\Omega)\setminus
 \dom(\plF\calF_{\rmA\rmC}). \ea \right. 
\] 
Moreover, we see that $u\mapsto \calF_{\rmA\rmC}(u)+ \frac {\lambda m}2\|u{-}u_0\|_H^2$
is convex if and only if $\lambda m \geq \beta $. To see this one uses that $u \mapsto
\frac\alpha2 \| \nabla u\|^2$ is quadratic and non-negative, and hence
convex. Moreover, $ z \mapsto \frac\beta4(z^2{-}1)^2 + \frac{\lambda m}2(z{-}z_0)^2$ is
convex if and only if ${\lambda m}\geq \beta$.  Thus, we conclude that
$\calF_{\rmA\rmC}$ is $(-\beta/m)$-convex. 

Thus, we conclude existence of solutions for the Allen-Cahn equation for all
initial values $u^0 \in \rmH^1(\Omega)$ and obtain Lipschitz-continuous
dependence of the solution on the initial data in the sense that 
\[
\| u_{\rmA\rmC}(t)- \wt u_{\rmA\rmC}(t)\|_{\rmL^2} \leq \ee^{\beta(t{-}s)/m}\,
\| u_{\rmA\rmC}(s)- \wt u_{\rmA\rmC}(s)\|_{\rmL^2}  \quad \text{for } 0 \leq s
< t. 
\]
\end{example}

\subsection{The first convergence proof}
\label{su:HilbertConverg}

The following proof consists of the classical steps for most constructions of
the solutions of PDEs. We give the steps in some detail to prepare for the
more advanced cases. \medskip

Step 0: construction of approximations (here via time discretization),

Step 1: a priori estimates, 

Step 2: extraction of convergent subsequences,

Step 3: identification of the equation,

Step 4: uniqueness and convergence of the full sequence.\bigskip

\noindent
In particular, we will essentially rely on the gradient structure in two
points, namely in (1) by doing energy estimates, in (3) when using the
closedness of subdifferentials, and in (4) when using semiconvexity. 
Of course, very similar steps will appear in
later sections. \bigskip

\noindent
\begin{proof}[Proof of Theorem \ref{th:HilbertExist}]
\label{proof:HilbertExist} 
We follow the above five steps.

\STEP{Step 0: Approximants via time discretization.} The time discretization
with time step $\tau = T/N$ with $N\in \N$ is described above leading to the
time-incremental minimization scheme \eqref{eq:HilbertTIMS}. We have existence
of minimizers because $\calF$ is lower semicontinuous and bounded from below by
$F_\mafo{min} = \min_{u\in H} \calF(u)$. Here we used that
$F_\mafo{min} = \inf_{u\in H} \calF(u) $ is attained by the one-sided
Weierstra\ss\ extremal principle exploiting the compactness of the sublevels of
$\calF$. Similarly $u_k $ as minimizer of $\Phi^\calF_\tau(u_{k-1}; \cdot)$
exists.

\STEP{Step 1: A priori estimates.} As $u_k$ is a minimizer of
  $\Phi^\calF_\tau(u_{k-1};\cdot)$ we have
\[
\frac1{2\tau}\| u_k{-}u _{k-1}\|^2 + \calF(u_k) =\Phi^\calF_\tau(u_{k-1};u_k)
\leq  \Phi^\calF_\tau(u_{k-1};u_{k-1}) = \calF(u_{k-1}). 
\]
From this we immediately obtain 
\begin{equation}
  \label{eq:HilbertAPriori}
\begin{aligned}
&F_\mafo{min} \leq  \calF(u_k)\leq \calF(u^0)<\infty \text{ for } 
  k\in \{0,\ldots,N\} \qANDq  
\\
&  \int_0^T \|\dot{\wh u}_\tau(t)\|^2 \dd t = \sum_{k=1}^N \tau\, 
    \big\|\frac1\tau(u_k{-}u_{k-1}) \big\|^2 \leq 2
    \big(\calF(u^0)-F_\mafo{min}\big). 
  \end{aligned}
\end{equation}
The second estimate follows by adding up the incremental
estimate for $k=1,\ldots,N$. 

\STEP{Step 2: Extraction of subsequences.} As the sequence $\wh u_\tau$
is bounded in $\rmW^{1,2}([0,T];H)$ we can extract a subsequence (not
relabeled) such that
\[
\wh u_\tau \weak u \text{ in } \rmL^2([0,T];H) \qANDq 
\dot{\wh u}_\tau \weak \dot u \text{ in } \rmL^2([0,T];H) 
\]
for a limit $u \in \rmW^{1,2}([0,T];H)$. 

Moreover, for all $t\in [0,T]$ and $\tau =T/N$ the values 
$\wh u_\tau(t)$  lie in the compact sublevel $S^\calF_{\calF(u^0)}$. Together
with the equi-continuity 
\[
\|\wh u_\tau(t){-}\wh u_\tau(s)\| \leq |t{-}s|^{1/2} \|\dot{\wh
  u}_\tau\|_{\rmL^2([0,T];H)} \leq \,|t{-}s|^{1/2} 
   \big( 2(\calF(u^0) {-} F_\mafo{min}) \big)^{1/2}
\]
we can apply the Arzel\`a-Ascoli theorem and find, after extracting a further
sequence (not relabeled), the uniform convergences
\[
\wh u_\tau  \to u  \text{ in }\rmC^0([0,T];H) \qANDq
\ol u_\tau \to u \text{ in }\rmL^\infty([0,T];H).
\]
For the second convergence we observe $ \wh u_\tau(k\tau)=\ol u_\tau(k\tau)$
and that $\ol u_\tau$ is piecewise constant. Hence we have $\|\wh u_\tau {-}\ol
u_\tau\|_{\rmL^\infty([0,T];H)} \leq \sqrt\tau \,\big(2
(\calF(u^0){-}F_\mafo{min})\big)^{1/2} \to 0$ for $\tau\to 0^+$.  

\STEP{Step 3: Identification of equation.} To show that the limit $u$ satisfies
the gradient-flow equation we define
\[
\xi_\tau= -\Riesz \dot{\wh u}_\tau \ \in \ \rmL^2([0,T];H^*) \overset\sim= \big(
\rmL^2([0,T];H)\big)^*.  
\]
By construction we have the following three properties 
\begin{equation}
  \label{eq:HilbertAss.frF}
  \begin{aligned}
&\ol u_\tau \to u \text{ in } \rmL^2([0,T];H), \quad \xi_\tau \weak \xi_*:= -\Riesz \dot u
\text{ in } \rmL^2([0,T];H^*),  \\
& \sup \calF(\ol u_\tau(t)) \leq \calF(u^0), \quad 
\xi_\tau (t) \in \plF \calF(\ol u_\tau(t)) \ \text{ a.e.\
in } [0,T]. 
\end{aligned}
\end{equation}
We now would like to apply the closedness property following from Proposition
\ref{pr:ClosedFrechet}. For this we set
\[
X= \rmL^2([0,T];H) \qANDq \mathfrak F(u(\cdot)) := \int_0^T \calF(u(t)) \dd t. 
\]
It is a simply calculation to show that $\mathfrak F$ is still proper, lower
semicontinuous, and $\lambda$-convex on
$X$ if $\calF$ is $\lambda$-convex on $H$. A deeper result is the
characterization of the Fr\'echet subdifferential of $\mathfrak F$; namely
\[
\plF\mathfrak F(u(\cdot)= \Bigset{\xi\in \rmL^2([0,T];H^*) }{ \xi(t) \in
  \plF \calF(u(t)) \text{ a.e.\ in } [0,T]}, 
\]
see Exercise \ref{exerc:EvolClosed}.  With this, we can apply Proposition
\ref{pr:ClosedFrechet} to $\mathfrak F : X \to \Rinfty$ such that
\eqref{eq:HilbertAss.frF} implies $\xi_* \in \plF\mathfrak F(u)$. Thus, using
the characterization of $\plF\mathfrak F(u)$ once again, we have
\[
\xi_*(t) = - \Riesz \dot u(t) \in \plF\calF(u(t)) \quad \text{for a.a.\ } t
\in [0,T],
\] 
which is the desired gradient-flow equation \eqref{eq:HilbertGFE} as $u(0)=u^0$
holds as well. 

\STEP{Step 4: Uniqueness and full convergence.}  For this we use that $ u \mapsto
F_\lambda(u):= \calF(u) - \frac\lambda 2\|u\|_H^2$ is convex. Clearly we have 
$\pl F_\lambda(u) = -\lambda \Riesz u + \plF\calF(u)$. Thus, for arbitrary
$u_1,u_0 \in \dom(\plF\calF)$ and $\xi_j \in \plF\calF(u_j)$ we set
$\eta_j = \xi_j -\lambda \Riesz u_j \in \pl F_\lambda(u_j)$ and obtain 
\[
\big\langle \xi_1{-}\xi_0, u_1{-}u_0 \big\rangle =  
\big\langle \eta_1{-}\eta_0, u_1{-}u_0 \big\rangle + \lambda 
\big\langle \Riesz (u_1{-}u_0), u_1{-}u_0 \big\rangle \geq 0 + \lambda
\|u_1{-}u_0\|^2,
\]
by using the monotonicity of subdifferentials of the convex function $F_\lambda$.

We now assume that we have two solutions $u_0,u_1 \in \rmW^{1,2}([0,T];H)$
which implies that $t \mapsto \| u_1(t){-}u_0(t)\|^2 $ is absolutely continuous.
Because of $-\Riesz \dot u_j \in \plF \calF(u_j(t))$ a.e., we have
\[
\frac12 \,\frac\rmd{\rmd t} \|  u_1(t){-} u_0(t)\|^2 = \langle \Riesz (\dot
u_1(t) - \dot u_0(t)) , u_1(t)-u_0(t) \rangle \leq - \lambda
\|u_1(t){-}u_0(t)\|^2.  
\]
Applying Grönwall's estimate we obtain the desired Lipschitz continuity
\eqref{eq:HilbertContract}. Assuming $u_1(0)=u_0(0)=u^0$ we obtain uniqueness of
solutions. 

Having this uniqueness we see that the choice of the subsequences does not matter
and the whole sequence $(\wh u_\tau)$ has to converge without taking any
subsequence. 
\end{proof}

We emphasize that semiconvexity was used only at two positions: (i) to show
closedness for subdifferential $\plF\calE$ and (ii) for the contraction
estimate \eqref{eq:HilbertContract}.  Thus, semiconvexity is not really
necessary for showing existence of solutions, if we obtain closedness with
other methods.

In fact, using the $\lambda$-convexity of $\calF$ it is possible to show better
even quantitative convergence rates. This is important for two reasons: first
it shows convergence without assuming the compactness of the sublevels
$S^\calF_E$, and secondly the convergence rate $ \wh u_\tau(t) {-}u(t) =
O(\tau^\alpha)$ for $\alpha =1/2$ or even $\alpha=1$ is useful in numerical
implementations.

\begin{exercise}[Evolutionary closedness]
% \cite[Prop. 16.50]{BauCom11CAMO} 
\label{exerc:EvolClosed}
  Consider a reflexive Banach space $X$ and a proper, lower semicontinuous and
  $\lambda$-convex functional $\calF: X \to \Rinfty$ and denote by
  $\plF\calF: X \tosets X^*$ its subdifferential.

(a) Define the Banach space $X := \rmL^2([0,T];X)$
with its dual $X^*= \rmL^{2}([0,T];X^*)$ and the functional 
\[
\calE:\left\{ \ba{ccc} X & \to &\Rinfty,\\ u(\cdot)& \mapsto & 
\int_0^T \calF(u(t)) \dd t. \ea \right. 
\]
Show that $\calE$ is again proper, lsc, and $\lambda$-convex. 

(b) Show that $\plF\calE$ admits the following characterization:
\[
\plF\calE(u) = \bfN(u):= \Bigset{\xi\in X^*}{ \xi(t) \in \plF
  \calF(u(t)) \text{ a.e.\ in } [0,T]}. 
\]
Hint: It is useful to know that Lebesgue points are dense, i.e.\ for a.a.\
$t\in [0,T]$ we have
$\frac1{2\delta} \int_{|t-s|<\delta } \xi(s) \dd s \to \xi(t) $ (strongly in
$X^*$) as $\delta \to 0^+$.
\end{exercise}

\begin{exercise}[Alternative proof of evolutionary closedness]
\label{exerc:EvolClosed2}\mbox{} 
\\
(A) Given a sequence $\xi_n \weak \xi \in \rmL^1([0,T];Y)$, define, for 
each $t\in [0,T]$, the accumulation set $\Xi(t) \subset Y$ via 
\[
\Xi(t):=\ol{\mafo{co}}(A_\rmw(t)) \quad \text{where } A_\rmw(t):=\bigset{y \in
  Y}{\exists\, (n_k)_k:\ n_k\to \infty \AND \xi_{n_k}(t) \weak y}.
\]
Show that $\xi(t) \in \Xi(t) $ for a.a.\ $t\in [0,T]$. (Hint: use a version
of Mazur's theorem.) 

(B) For semiconvex $\calF$ show that each $\plF\calF(u)$ is a closed and
convex set.

(C) Assume that $\calF$ is semiconvex and 
\begin{align*}
&u_n \to u  \text{ in } \rmL^1([0,T];X), \quad 
\xi_n \weak \xi \text{ in } \rmL^1([0,T];X^*), \\ 
& \sup_{n\in \N,\ t\in
  [0,T]} \calF(u_n(t)) < \infty, \quad \xi_n(t) \in \plF\calF(u_n(t)) \text{ a.e.},
\end{align*}
and conclude $\xi(t) \in \plF\calF(u(t))$, i.e.\ ``evolutionary'' closedness. 
\end{exercise}

\subsection{Completion of the Hilbert-space gradient flow via Evolutionary
  Variational Inequalities (EVI)} 
\label{su:Hilbert.EVI}

The previous existence and uniqueness theorem provides a semiflow
$(\Sigma_t)_{t\geq  0} $ on $\dom(\calF)$ defined as follows: For $u^0\in \dom(\calF)$
we set $ \Sigma_t(u^0):= u(t)$ where $u:{[0,\infty[}\to H$ is the
unique solution of the GFE \eqref{eq:HilbertGFE}. As the problem is autonomous, we
obtain the semigroup property $\Sigma_{t+r}=\Sigma_t \circ \Sigma_r$ for all
$t,r\geq 0$.  Moreover, we have a global
$\lambda$-contractivity which is completely independent of $\calF(u^0)$:
\[
\forall\, u^0,u^1\in \dom(\calF)\ \forall\, t\geq 0:\quad 
\| \Sigma_t(u^1)- \Sigma_t(u^0)\| \leq \ee^{-\lambda t} \| u^1- u^0\|.
\]
Hence, we can solve the initial-value problem for more initial data by approximating
them with elements from $\dom(\calF)$. This is possible on the closure of the
domain:
\[
\scrD := \ol{\dom(\calF)}^{H}.
\]
Note that for the Allen-Cahn equation we have $H=\rmL^2(\Omega)$ and
$\dom(\calF_{\rmA\rmC})= \rmH^1(\Omega)$. Hence, in this case we find
$\scrD=H=\rmL^2(\Omega)$ which enlarges the class of admissible initial
conditions considerably. 

For $u^0\in \scrD$ we can choose $(u^0_m)_m$ with $u^0_m\in \dom(\calF)$ and
$u_m^0 \to u^0$ in $H$. Then, there is a unique solution $u_m:{[0,\infty[}\to
H$ and we have $\| u_m(t) - u_k(t)\|_H \leq \ee^{-\lambda t} \|
u^0_m{-}u^0_k\|_H\to 0$ for $k,m\to \infty$. Hence, on each bounded interval
$[0,T]$ we have a Cauchy sequence in $\rmC^0([0,T];H)$ and we obtain a
continuous limit $u:{[0,\infty[}\to H$ with $u(0)=u^0$ and 
$\| u_m(t) - u(t)\|_H \leq \ee^{-\lambda t} \|
u^0_m{-}u^0\|_H$ for all $t\geq 0$. The question is of course, in what sense
this limit $u$ satisfies the GFE \eqref{eq:HilbertGFE}. 

The prototypical example is the Allen-Cahn gradient system $(\rmL^2(\Omega),
\calF_{\rmA\rmC}, m\Riesz) $, where $\dom(\calF_{\rmA\rmC})=\rmH^1(\Omega)$ is
dense in $H= \rmL^2(\Omega)$. The following theory will show that initial
conditions $ u^0 \in \scrD=\rmL^2(\Omega)$ can be treated. 

We summarize the result in the following theorem. Its proof is based on
Evolutionary Variational Inequalities, which are not really needed, but we can
prepare in this way to a general idea used later in metric spaces. 

\begin{theorem}[Completed gradient flow for $(H,\calF,\Riesz)$]
\label{th:CompleteGF}
Let the GS $(H,\calF, \Riesz)$ be given as in Theorem
\ref{th:HilbertExist}. Then, there exists a $\lambda$-contractive, continuous
semiflow $(S_t)_{t\geq 0}$ on $\scrD$ (``the gradient flow'' associated with
the GS $(H,\calF,\Riesz)$), i.e.\
\\[0.2em]
(S1)\quad $S_t: \scrD \to \scrD$, \ $S_0=\mafo{id}_\scrD$, \
$S_t\circ S_r = S_{t+r}$ for all $t,r\geq 0$.
\\[0.2em]
(S2)\quad For all $u^0\in \scrD$ the function
${[0,\infty[}\ni t\mapsto S_t(u^0) \in H$ is continuous,
\\[0.2em]
(S3)\quad For all $u^0,u^1 \in \scrD$ and all $t\geq 0$ we have
$\| S_t(u^1){-}S_t(u^0)\| \leq \ee^{-\lambda t} \|u^1{-}u^0\|$,
\\[0.4em]
such that for all $u^0$ the function $u:{[0,\infty[}\to H; \ t \mapsto S_t( u^0)$
is a solution of the GFE \eqref{eq:HilbertGFE}. In particular, we have
$u \in \rmW^{1,2}_\mafo{loc}({]0,\infty[} ;H)$ and
${]0,\infty[} \ni t \to \calF(u(t)) \in \R$ (finite values) is continuous,
decreasing and satisfies $\lim_{t\to 0^+} t \,\calF(u(t)) =0$.
\end{theorem}

In $\rmW^{1,2}_\mafo{loc}({]0,\infty[} ; H)$ the subscript ``$_\mafo{loc}$''
means that for all subsets $D\Subset {]0,\infty[}$ (compactly contained) we
have $\rmW^{1,2}(D; H)$

\begin{exercise}[Non-integrability of $\dot u$] Consider 
  $(H,\calF,\Riesz)$ with $\calF(u)=\frac12\langle \bbA u, u\rangle$ and
  $\bbA= \bbA^*\geq 0$ is a possibly unbounded self-adjoint operator. 

(A) Show that the gradient flow $(S_t)_{t\geq0}$ equals the classical strongly
continuous semigroup $(\ee^{-t\bbA})_{t\geq 0}$, i.e.\ $S_t(u^0)=\ee^{-t\bbA}
u^0$ for all $u^0\in \scrD=H$. 

(B) Assume further that $\bbA$ has compact resolvent and $H$ is infinite
dimensional. Show that there exists $u^0\in \scrD=H$ such that for $t\mapsto
u(t)=\ee^{-t\bbA} u^0$ we have $\dot u\not \in \rmL^1({]0,\tau[};H)$ for any
$\tau>0$.  
\end{exercise} 
\medskip

Before starting the proof we develop a preliminary theory for EVI for
Hilbert spaces. The full theory was developed in \cite{AmGiSa05GFMS,
  Sava07GFDS, DanSav14LNGF, MurSav20GFEV} and will be studied further in
Section \ref{se:MetricGS}. The major advantage of the EVI formulation is that
it is a weak form in the classical sense: all solutions constructed above also
solve (EVI)$_\lambda$.  Moreover, it does not need any time derivative $\dot u$
nor any subdifferential $\plF \calF$. Thus, taking limits in EVI will be
especially simple.

For the solutions $u \in \rmW^{1,2}([0,T];H)$ constructed above and arbitrary
$w\in H$, we have 
\begin{align*} 
\frac12\frac\rmd{\rmd t} \| u(t){-}w\|^2 &= \langle \Riesz \dot u(t),
u(t){-}w\rangle  =  \langle \xi(t) , w{-} u(t) \rangle \quad \text{for }
  \xi(t) \in \plF\calF(u(t))
\\
&\overset{\calF\;\lambda\text{-cvx}}\leq 
\calF(w)- \calF(u(t)) - \frac\lambda2 \,\| u(t){-}w\|^2. 
\end{align*}
This is already the Differential form of the Evolutionary Variational
Inequality 
\begin{equation}
  \label{eq:DEVI.la}
 \text{(DEVI)}_\lambda \quad
\left\{ \quad \begin{aligned}
  &\forall\, w\in \dom(\calF)\ \forall_\mafo{a.a.} t
  \in {[0,\infty[}: \\
& \frac12\,\frac\rmd{\rmd t} \|u(t){-}w\|^2 +
  \frac\lambda 2  \|u(t){-}w\|^2 \leq \calF(w) - \calF(u(t)). 
\end{aligned}\right.
\end{equation}
We see that DEVI is much weaker, because we do not need to impose the
existence of $\dot u$. Instead we only need to impose absolute continuity of $t
\mapsto \| u(t){-}w\|^2$. 

We can simplify further by applying a Gr\"onwall estimate and using that $t
\mapsto \calF(u(t))$ is decreasing. Then, no derivative is needed any more and
we can impose conditions for all $s\geq 0$ and all $t>0$. This leads to the
final Evolutionary Variational Inequality: 
\begin{equation}
  \label{eq:EVI.la}
 \text{(EVI)}_\lambda \quad
\left\{ \quad \begin{aligned}
  &\forall\, w\in \dom(\calF)\ \forall\, s\geq0\ \forall\, t>s:\\
& \frac12\, \|u(t){-}w\|^2 \leq  \frac12\,\ee^{-\lambda(t{-}s)} \| u(s){-}
w\|^2 + M_\lambda(t{-}s)\, \big( \calF(w) - \calF(u(t)) \big). 
\end{aligned}\right.
\end{equation}
where $M_\lambda(\tau)= \int_0^\tau \ee^{-\lambda(\tau-s)} \dd s$. We emphasize
that we
need $\calF(u(t)) < \infty$ for $t>0$ but not for $t=0$. 

Indeed, starting from (DEVI)$_\lambda$ we define $\rho (t)= \frac12
\ee^{\lambda t} \|u(t){-}w\|^2$ and obtain $\dot \rho \leq \ee^{\lambda
  t}(\calF(w){-}\calF(u(t))$. Integration over $[s,t]$ we find 
\[
\rho(t) \leq \rho(s) + \int_s^t \ee^{\lambda r}(\calF(w){-}\calF(u(r))\dd r
\leq \rho(s) + \int_s^t \ee^{ \lambda r} \dd r \:\big( \calF(w)
{-}\calF(u(t))\big). 
\]
Now, inserting the definition of $\rho$ and multiplying with $\ee^{-\lambda t}$ 
gives (EVI)$_\lambda$.

The main observation for completing the proof of Theorem \ref{th:CompleteGF} is
that all functions $t \mapsto S_t(u^0)$ with $u^0\in \scrD$ satisfy
(EVI)$_\lambda$. In Section \ref{se:MetricGS} we will show that (EVI)$_\lambda$
already characterizes the solutions uniquely, thus (EVI)$_\lambda$
characterizes the gradient flow $(S_t)_{t\geq 0}$ on $\scrD= \ol{\dom(\calF)}
\subset H$ completely.

\begin{proposition}[Gradient flow and EVI]
\label{pr:GradFlowEVI}
Let the GS $(H,\calF,\Riesz)$ and the gradient flow $S_t:\scrD\to \scrD$ be
given as in Theorem \ref{th:CompleteGF}. Then, for all $u^0\in \scrD$ the
functions $u:{[0,\infty[} \mapsto  S_t(u^0) \in \scrD\subset H$ satisfy
(EVI)$_\lambda$. Moreover, we have $\calF(u(t))< \infty$ for $t>0$ and $\limsup_{t\to
  0^+}\, t\, \calF(u(t)) = 0$. 
\end{proposition}
\begin{proof}
By construction, we already know that for all $u^0\in \dom(\calF)$ the
solutions $u(t)=S_t(u^0)$ satisfy (EVI)$_\lambda$. For all other initial
conditions $u^0 \in \scrD\setminus \dom(\calF)$ we can choose a sequence
$u^0_m$ with $u^0_m\in \dom(\calF)$ and $u^0_m \to u^0$ in $H$. The
corresponding solutions $u_m=S_t(u^0_m)$ satisfy 
\[
\frac12\|u_m(t){-}w\|^2 \leq \frac12\ee^{-\lambda (t-s)} \|u_m(s){-}w\|^2 +
M_\lambda(t{-}s) \big( \calF(w) - \calF(u_m(t))\big)
\]
for $t > s \geq 0$ and $w \in \dom(\calF)$. By $\lambda$-contractivity we have
uniform (strong) convergence of $u_m$ to $ u: t \mapsto S_t(u^0) $ on all
compact subsets of ${[0,\infty[}$. Thus, we can pass to the limit in the first
two terms. In the last term we can use the lower semicontinuity
$\calF(u(t)) \leq \liminf_{m\to \infty} \calF(u_m(t))$ and
$M_\lambda(t{-}s)\geq 0$.

This limit passage would even be allowed if $\calF(u(t)) =\infty$, however
(EVI)$_\lambda$ gives, for all $w \in \dom(\calF)$ and $s=0$ the upper bound
\[
\calF(u(t)) \leq \calF(w) + \frac{\ee^{-\lambda t}}{2 M_\lambda(t)} \,
  \|u^0{-}w\|^2. 
\]
As $\calF$ is proper, we have shown $\calF(u(t))<t$ for $t>0$. Moreover,
multiplying by $t>0$ we can take the limsup for $t\to 0^+$ and find
\[
\limsup_{t\to 0^+} \;t\, \calF(u(t)) \ \leq \ \frac12 \| u^0{-}w\|^2 \quad
\text{for all } w \in \dom(\calF).
\]
As $\dom(\calF)$ is dense in $\scrD$ we find the desired result $\limsup_{t\to
  0^+} t\,\calF(u(t))=0$. 
\end{proof}

To appreciate the last relation concerning the boundedness of $t \calF(u(t)$, we consider the
example $(\rmL^2(\Omega),\calF_\mafo{Dir}, \Riesz)$ with
$\calF_\mafo{Dir}(u) = \frac12 \|\nabla u\|_{\rmL^2}^2$ on
$\dom(\calF_\mafo{Dir})=\rmH^1_0(\Omega)$. From linear PDE theory we know the
explicit estimate $\| u(t)\|_{\rmH^1} \leq Ct^{-1/2} \| u(0)\|_{\rmL^2} $ which
corresponds to the statement
$t\,\calF_\mafo{Dir}(u(t)) \leq \frac12C^2\|u(0)\|_{\rmL^2}^2$. Hence, our
general and abstract theory recovers a very similar behavior which is optimal
in the sense that the power $\alpha=1$ cannot be decreased without losing 
boundedness of $t^{\alpha} \calF(u(t)$. \medskip

We are now in the position to study the remaining properties of the
completion of the semiflow.\medskip

\noindent
\begin{proof}[Proof of Theorem \ref{th:CompleteGF}]
It remains to show that $u(t)= S_t(u^0) $ is differentiable a.e. and that GFE
holds.

We observe that $\wt u:{[0,\infty[} \mapsto u(t{+}t_*)$ is a solution of the
GFE \eqref{eq:HilbertGFE} with $\wt u(0)=u(t_*)$. For this, first note that
$\calF(\wt u(0)) = \calF(u(t_*))<\infty$. Hence, there is a unique solution
$\wh u \in \rmW^{1,1}_\text{loc}({[0,\infty[};H)$ with $\wh
u(0)=u(t_*)$. Moreover, by the semigroup property and $\wt u(t)=u(t{+}t_*)$ we
have
$\wt u(t) = u(t{+}t_*) =S_{t+t_*}(u^0) = S_t(S_{t_*}(u^0) ) = S_t(u(t_*))= \wh
u(t)$. Thus, we conclude that the solutions obtained in the limit
$u_m(\cdot) \to u(\cdot)$ are differentiable, as desired.
\end{proof}

It is possible to establish many more properties of the gradient flows in
Hilbert spaces. E.g.\ in \cite[Thm.\,3.1(5+6)]{Brez73OMMS} it is shown that for
convex functionals $\calF$ all solutions $u$ of the GFE \eqref{eq:HilbertGFE} 
have the property that the one-sided derivatives 
\[
\dot u^+(t) = \lim_{h\to 0^+} \frac1h \big( u(t{+}h) - u(t)\big) 
\]
exist and can be identified with the ``norm-minimal selection'' in the
subdifferential of $\calF$, i.e.\ for all $t>0$ one has
\[
-\Riesz \dot u^+(t) =  \pl^0 \calF(u(t)) := \mafo{arg\!\;min}\bigset{\| \xi\|_{H^*} 
}{ \xi \in  \pl\calF(u(t)) }.
\]
Moreover, the mapping $ {]0,\infty[} \ni t \mapsto \| \dot u^+(t)\|$ is
decreasing. 

Indeed the latter property is not so surprising if we use the
$\lambda$-contractivity \eqref{eq:HilbertContract} for the two solutions
$t \mapsto u(t)$ and $t \mapsto u(t{+}h)$. After dividing by $h>0$ we obtain 
\[
\big\| \frac 1h \big(u(t{+}h) - u(t)\big) \big\| \ \leq \ 
\ee^{-\lambda(t-s)} \big\| \frac1h\big(
u(s{+}h) - u(s) \big) \big\| .
\]
Thus, for $\lambda\geq 0$ we obtain that the norms of the difference quotients
are decreasing. Clearly, if the limits exist in the strong sense, then they are
still decreasing.

\section{Generalized gradient systems in Banach spaces} 
\label{se:GSBanach}

In this section we generalize the theory in a twofold way. First we go from
Hilbert spaces to Banach spaces and second we generalize the linear kinetic
relation $\xi = \bbG(u)\dot u$ or $\dot u = \bbK(u)\xi$ to \emph{nonlinear
  kinetic relations}, which allows a much larger set of applications.

\subsection{Legendre duality and nonlinear kinetic relations}
\label{su:LegrNonlKinRel}

Now the kinetic relation $ X\ni \dot u = v \leftrightarrow \xi \in X^*$ 
cannot be given by a simple linear map such as the Riesz isomorphism $\Riesz: H \to
H^*$, because $X$ and $X^*$ are not isomorphic in general.

The typical replacements in general (separable, reflexive) Banach spaces are 
maximal monotone operators $\bfA: X\tosets X^*$ and their inverse 
\[
\bfA^{-1} :X^*\tosets X; \ \bfA^{-1}(\xi):=\bigset{v \in X}{ \xi\in \bfA(v)},
\]
which is again a maximal monotone operator, because the notion of (maximal)
monotonicity is symmetric:
\[
\bfA \text{ monotone } \Longleftrightarrow \ \forall\, v_1,v_0\in X\ \forall \, 
\xi_1\in \bfA(v_1), \:\xi_0 \in \bfA(v_0):\ \langle \xi_1{-}\xi_0,
v_1{-}v_0\rangle \geq 0. 
\]
The corresponding evolution equations are then called \emph{doubly nonlinear
  equations} (cf.\ \cite{ColVis90CDNE, Coll92DNEE}): $0 \in \bfA(\dot u) +
\plF\calF(u(t)) - \ell(t)$, where $\ell \in \rmL^1([0,T]; X^*)$ is a general
external forcing. 

To obtain a theory of generalized gradient systems we consider a subclass of
these kinetic relations that encode a nonlinear version of the Onsager
symmetry. This class is given by \emph{subdifferentials of convex potentials}.

To motivate this, we first consider the quadratic functional $\Psi:H\to \R;\ v
\mapsto \frac12\langle \bbA v , v\rangle$. Then the differential reads 
\[
\rmD \Psi(v)= \bbG v \text{ with } \bbG = \frac12(\bbA{+}\bbA^*) ,
\]
such that $\bbG$ automatically enjoys the Onsager symmetry $\bbG^*=\bbG$ for
$\bbA \in \mafo{Lin}(H;H^*)$. Moreover, 
$\bbG\geq 0$ is equivalent to $\Psi(v)\geq 0$ for all $v\in H$.  However, it is
easy to see that $\bfA(v)=\{ \bbA v\}$ defines a (maximal) monotone operator
$\bfA:H\tosets H^*$ if and only if $\frac12(\bbA{+}\bbA^*) \geq 0$, i.e.\ the
skew-symmetric part $\frac12(\bbA{-} \bbA^*)$ is completely arbitrary.

Secondly, we consider time-incremental minimization schemes in the form 
\[
u_k \text{ \ minimizes } \ u \mapsto \tau\Psi\big(\frac1\tau(u{-}u_{k-1})\big)
+ \calF(u)
\]
for a convex function $\Psi:X\to \Rinfty$. 
Assuming the sum rule, the Euler-Lagrange equation reads $0 \in \pl\Psi\big(
\frac1\tau(u_k{-}u_{k-1})\big) + \plF\calF(u_k)$ which will be interpreted as
the backward-Euler (fully implicit) discretization of the evolutionary inclusion
$0 \in \pl\Psi(\dot u(t)) + \plF\calF(u(t))$. 

\begin{definition}[Dissipation potential]
\label{de:DissPot} 
A function $\Psi:X\to \Rinfty$ is called a \emph{dissipation potential on $X$},
if $\Psi$ is lower semicontinuous, convex and satisfies $\Psi(v)\geq
\Psi(0)=0$.  

We call the Legendre-Fenchel dual (conjugate)
$\Psi^*=\mathfrak L \Psi:X^* \to \Rinfty$ the \emph{dual dissipation potential
  for $\Psi$}. It is defined by 
\[
\Psi^*(\xi)=(\mathfrak L\Psi)(\xi):= \sup\bigset{\langle \xi,v\rangle -
  \Psi(v)}{ v \in X} .
\]
\end{definition}

To justify the above name ``dual dissipation potential'', note $\Psi^*=\mathfrak
L\Psi$ is automatically convex and lsc. Moreover, $\Psi(0)\geq 0$
implies $\Psi^*(\xi)\geq 0$, whereas $\Psi(v)\geq 0$ for all $v\in X$ implies
$\Psi^*(0)=0$.

On a Hilbert space $H$ we have 
\[
\Psi(v) = \frac12\langle \bbG v, v\rangle \quad \Longleftrightarrow \quad 
\Psi^*(\xi)=\frac12\langle \xi, \bbK\xi\rangle \text{ with } \bbK=\bbG^{-1}. 
\]
If $p\in {]1,\infty[}$ and $(X,\|\cdot\|_X)$ is a Banach space with dual space
$(X^*,\|\cdot\|_{X^*})$, then 
\[
\Psi(v) = \frac1{p}\, \| v\|_X^p  \quad \Longleftrightarrow \quad 
\Psi^*(\xi)= \frac1{p^*}\,\|\xi\|_{X^*}^{p^*} \text{ with } p^* = \frac p{p{-}1}.
\]

A trivial but nevertheless important consequence of the definition of
$\Psi^*=\mathfrak L\Psi$ is the
\begin{equation}
  \label{eq:FenchYoungI}
  \begin{aligned}
&\text{\emph{Fenchel-Young inequality}:} \qquad 
\forall\, (v,\xi)\in X\ti X^*:\quad \Psi(v)+ \Psi^*(\xi)\geq \langle \xi,
v\rangle.  
  \end{aligned}
\end{equation}
We refer to \cite{Fenc49CCF} for the first occurrence in $X=\R^n$ and
\cite[Prop.\,13.15]{BauCom17CAMO} for a general theory (in Hilbert spaces). 

The following important relation is the basis of the term ``\emph{duality
  theory}''. 

\begin{lemma}[Legendre transform is an involution]
\label{le:LegeInvol}
Assume that $X$ is reflexive, i.e.\ $X^{**}= (X^*)^*=X$. Then, $\mathfrak L$
maps proper, lsc, convex functions on $X$ onto  proper, lsc, convex functions
on $X^*$ and vice versa. Moreover, $\Psi^{**}=(\Psi^*)^*=\mathfrak L (\mathfrak
L(\Psi)) = \Psi$, i.e.\ $\mathfrak L$ is an involution. 
\end{lemma}
\begin{proof} The definition of $\mathfrak L$ immediately shows that $\Psi^*$
  is again proper, lsc, and convex. 

Using the Fenchel-Young inequality \eqref{eq:FenchYoungI} we easily obtain, for
all $v\in X$,  
\begin{align*}
(\Psi^*)^*(v) & = \sup\bigset{\langle \xi,v\rangle -\Psi^*(\xi)}{ \xi\in X^*}
% \\ &
\leq  \sup\bigset{ \Psi(v) }{ \xi\in X^* } = \Psi(v).
\end{align*}
To show $\Psi(v) \leq \Psi^{**}(v)$ we use that $\Psi$ is convex and lsc. For
fixed $v_0\in X$ and $a_0<\Psi(v_0)$ there exists $\xi_0\in X^*$ such that
$\Psi(v) \geq a_0 + \langle \xi_0,v{-}v_0\rangle$. This implies $\Psi^*(\xi_0)
\leq -a_0+\langle \xi_0,v_0\rangle$ and hence $\Psi^{**}(v_0)\geq a_0$. As $a_0
<\Psi(v_0)$ was arbitrary, we conclude $\Psi(v_0)\leq \Psi^{**}(v_0)$.
\end{proof}

With this, the pair $(\Psi,\Psi^*)$ is called a \emph{conjugate pair} as
$\Psi^*=\mathfrak L\Psi$ and $\Psi=\mathfrak L(\Psi^*)$. We will use the word
`primal dissipation potential' for $\Psi$ and `dual dissipation potential' for
$\Psi^*$, but of course, the notion of `primal' and `dual' can be interchanged
in the case of reflexive spaces $X$ and $X^*$.    

For our theory the most important duality result are the so-called Fenchel
equivalences which we formulate explicitly here as a theorem, even though in
some textbooks they are considered simple lemmas or exercises. See also
\cite[Thm.\,16.29]{BauCom17CAMO} for a proof in the Hilbert space setting.

\begin{theorem}[Fenchel equivalences \cite{Fenc49CCF}] 
\label{th:FenchelEquiv}
Consider a reflexive Banach space $X$ with dual $X^*$ and consider a conjugate
pair $(\Psi,\Psi^*)$ of proper, lsc, and convex functions. Then, for all
$(v_0,\xi_0) \in X\ti X^*$ the following
five statements are equivalent:
\begin{enumerate}
[label={\upshape(\roman*)}, leftmargin=4em, rightmargin=2em]

\item $v_0$ minimizes  the functional $ v \mapsto \Psi(v) - \langle \xi_0,
  v\rangle$ \hfill (optimality of $v\in X$);\hspace*{2em}
\item $\xi_0 \in \pl\Psi(v_0)$ \hfill (subdifferential inclusion in $X^*$);
\item \label{Cond:FenchOptim}
  $\Psi(v_0) + \Psi^*(\xi_0) \leq \langle \xi_0,v_0\rangle$ \hfill (optimality
  condition in $\R$); 
\item $v_0 \in \pl\Psi^*(\xi_0)$ \hfill (subdifferential inclusion in $X$);
\item $\xi_0$ maximizes the functional $\xi \mapsto \langle
  \xi,v_0\rangle - \Psi^*(\xi)$  \hfill (optimality of $\xi \in X^*$).
\end{enumerate}
Here in \ref{Cond:FenchOptim} we can either write ``$\,\leq\,$'' or ``$\,=\,$'',
because  the Fenchel-Young inequality \eqref{eq:FenchYoungI} always gives
``$\,\geq\, $''. 
\end{theorem}
\begin{proof} Obviously, the equivalences (i) $\Leftrightarrow$ (ii)  and (iv)
  $\Leftrightarrow$ (v) easily follow by the Euler-Lagrange equation for the
  convex functionals. 

Thus, it remains to show (ii) $\Leftrightarrow$ (iii) and (iii)
$\Leftrightarrow$ (iv). By duality the two equivalences can be proved in
the same way, so we concentrate on the first. 

``(ii) $\Rightarrow$ (iii)'' \ Starting from (ii) gives $\Psi(v) \geq
\Psi(v_0) + \langle \xi_0,v{-}v_0\rangle$. Inserting this into the definition
of $\Psi^*(\xi_0)$ we immediately obtain $\Psi^*(\xi_0)\leq \langle
\xi_0,v_0\rangle - \Psi(v_0)$, which is (iii). 

``(iii) $\Rightarrow$ (ii)'' \ (iii) is the upper bound $\Psi^*(\xi_0) \leq
\langle \xi_0,v_0\rangle - \Psi(v_0)$, which using $\Psi=\Psi^{**}$ implies the
lower bound 
\[
\Psi(v)=\sup\bigset{\langle \xi,v\rangle - \Psi^*(\xi)}{ \xi \in X^*} \geq
\langle \xi_0,v\rangle - \Psi^*(\xi_0) \geq \langle \xi_0,v{-}v_0\rangle +
\Psi(v_0). 
\]
Hence, we have $\xi_0 \in \pl\Psi(v_0)$ which is (ii). 
\end{proof}

We emphasize that the equivalence (ii) $\Longleftrightarrow$ (iv) shows that
the set-valued mapping $X\ni v \tosets \pl\Psi(v) \subset X^*$ is exactly the
inverse mapping (in the sense of set-valued monotone operators) of the
set-valued mapping $X^*\ni \xi \tosets \pl\Psi^*(\xi) \subset X$.

\begin{example}[Viscoplasticity]
\label{ex:ViscoPlast}
As a nontrivial and mechanically important example we treat viscoplasticity,
where $p\in X=\rmL^2(\Omega;\R^{d\ti d}_\mafo{sym})$ denotes the plastic
distortion. The viscoplastic dissipation potential depends on the plastic rate
$\pi = \dot p$ and takes the form  
\[
\Psi( \pi) = \int_\Omega \psi(\pi(x)) \dd x \ \text{ with } \psi(\pi)=
\sigma_\mafo{yield}|\pi| + \frac\mu2|\pi|^2  
 \quad \text{where } |\pi|^2 = \sum_{i,j=1}^d \pi_{ij}^2 .
\]
The dual variables are the plastic (back-) stresses $\Sigma_\rmp\in
\rmL^2(\Omega;\R^{d\ti d}_\mafo{sym})$. 
Clearly, we have (cf.\ Exercise \ref{exerc:EvolClosed}) 
\[
\pl\Psi(\pi) = \Bigset{\Sigma_\rmp \in \rmL^2(\Omega;\R^{d\ti d}_\mafo{sym}) }{
  \Sigma_\rmp (x) \in \pl \psi(\pi(x)) \text{ a.e.\ in }\Omega }.
\]
where the pointwise subdifferential $\pl\psi$ is given by  
\[
\pl\psi(\pi) = \left\{ \ba{cl} \sigma_\mafo{yield}\ol{B_1(0)} & \text{for } \pi
  =0, \\ \big\{\,\frac{ \sigma_\mafo{yield}}{|\pi|} \,\pi + \mu \pi \big\} &
  \text{for } \pi  \neq 0. \ea \right.
\]
The relation $\Sigma_\rmp \in \pl \psi(\pi)$ can be inverted explicitly, giving
\[
\pi = \rmD \psi^*( \Sigma_\rmp) \quad \text{with } \psi^*(\Sigma_\rmp)=
\frac1{2\mu} \big( \max\{0, |\Sigma_\rmp|{-}\sigma_\mafo{yield}\}\big)^2. 
\]
Note that this relation shows that $\pi=\dot p =0$ whenever $|\Sigma_\rmp| \leq
\sigma_\mafo{yield}$, i.e.\ if the stress does not reach the threshold
$\sigma_\mafo{yield}$ for yielding.

In particular, we find the dual dissipation potential depending on the plastic stress: 
\[
\Psi^*(\Sigma_\rmp) = \int_\Omega \psi^*(\Sigma_\rmp (x) ) \dd x .
\]
\end{example}

\begin{exercise}[Dissipation functions]
\label{ex:DissFunction}
For a differentiable dissipation potential $\Psi:X\to [0,\infty] $ we  define the 
\[
\text{\emph{dissipation function} } \ \mafo{Diss}_\Psi(v)= \langle \rmD
\Psi(v), v\rangle. 
\]
(a) Show that $\mafo{Diss}_\Psi(v) \geq \Psi(v)$ and give an example where
$\mafo{Diss}_\Psi$ is non convex. 

(b) Discuss the equality $\Diss_\Psi(v) = \Psi(v)$.

(c) Assume that $\Psi$ is differentiable and positively $p$-homogeneous, i.e.\
$\Psi(\lambda v) = 
\lambda^p\Psi(v)$ for all $\lambda>0$ and $v \in X$, and show $\Diss_\Psi(v) =
p\Psi(v)$. 

(d) Assume now that $\Psi$ is only radially differentiable, i.e.\ for all $v\in X$
the function ${]0,\infty[} \ni \lambda  \to 
\Psi(\lambda v)$ is differentiable. Show that for each $v\in X$ the values of
$\langle \xi, v\rangle $ are constant for all $\xi \in \pl\Psi(v)$.   Conclude
\[
\Diss_\Psi(v) = \Psi(v) + \Psi^*(\xi) \quad \text{for all }\xi \in \pl\Psi(v). 
\]
{\small Hint: Consider $g(\lambda)= \Psi(\lambda v) + \Psi^*(\xi) - \langle
  \xi,\lambda v\rangle$. }

The function $v \mapsto \mafo{Diss}_\Psi(v)$ is called (primal) 
dissipation function, and similarly $\xi \to
\mafo{Diss}_{\Psi^*}(\xi)$ is called dual dissipation function. 
These functions are often used
in modeling, especially when their subdifferentials are single-valued. 
However, they have to be clearly distinguished from the
dissipation potentials. They can be used in the energy-dissipation balances
below, see e.g.\ \eqref{eq:EDB.ODE}, but have weaker properties. 
\end{exercise}

\begin{exercise}[Duality of properties]
\label{exerc:DualProps}
On a reflexive Banach space $X$ consider a pair of Legendre dual functions
$\Psi:X\to \Rinfty$ and $\Psi^*:X^*\to \Rinfty$. 

For a general lsc, convex functional $\Phi:Y\to \Rinfty$ consider the properties:

(P1) $\Phi(0)\leq 0$;

(P2) $\Phi(y)\geq 0$ for all $y$;

(P3) $\Phi(y)\geq c\|y\| -C $ for all $y$;

(P4) $\Phi$ is superlinear, i.e.\ $\Psi(u)/\|u\|\to \infty$ for $\|u\|\to
\infty$;

(P5) $\Phi(y) \leq M$ for all $y \in B_R(0)\subset Y$;

(P6) $\Phi$ takes only finite values;
 
(P7) $v\mapsto \Phi(v)-\langle \eta,v\rangle $ has a unique minimizer;

(P8) $\pl \Phi(w)$ is single-valued.

\noindent
Try to find implications or equivalences like ``if $\Psi$ satisfies (P$n$) then
$\Psi^*$ satisfies (P$k$). 
\end{exercise}

\subsection{Generalized gradient systems and the gradient-flow equations}
\label{su:GenGS}

Above we have always used $v \in X$ as a placeholder for the rate $\dot u$. Of
course, in general systems we may have \emph{state-dependent kinetic
relations}. We start again with the manifold setting $M$ where now the
state-dependent (primal) dissipation potential $\calR$ is defined on the
tangent bundle:
\[
\calR{:}\: \rmT M \to [0,\infty] \text{ such that } \forall\, u \in M : \
\calR(u,\cdot){:}\, \rmT_u M \to [0,\infty] \text{ is a dissipation potential}. 
\]
The dual dissipation potential $\calR^*:\rmT^*M\to [0,\infty]$ 
is obtained at fixed $u\in M$, i.e.
\[
\calR^*(u,\xi) := \big( \mathfrak L (\calR(u,\cdot)) \big) (\xi)
\]
When we write subdifferentials of $R$ or $\calR^*$ we always mean
subdifferentials with respect to the second variable in the linear space
$\rmT_u M$ or $\rmT_u^* M$:
\begin{align*}
\pl \calR(u,v)&:= \pl_v \calR(u,v) = \pl \big( \calR(u,\cdot)\big)(v)
\subset  \rmT_u^* M \qANDq
\\ 
\pl \calR^*(u,\xi)&:= \pl_\xi \calR^*(u,\xi) = \pl \big(
\calR^*(u,\cdot)\big)(\xi) \subset \rmT_u M.
\end{align*}

\begin{definition}[Generalized gradient system: ODE case]
\label{de:GenGS}
A triple $(M,\calF,\calR)$ (or equivalently $(M,\calF,\calR^*)$) is called a
\emph{generalized gradient system}, if $M$ is a manifold, $\calF:M \to
\R$ is a differentiable function and $\calR: \rmT M\to [0,\infty]$ (or
equivalently $\calR^*:\rmT^* M \to [0,\infty]$) is a (state-dependent)
dissipation potential. 

The associated gradient-flow equation is given by 
\[
0 \in \pl\calR(u,\dot u) + \rmD\calF(u) \subset \rmT_u^* M \quad
\Longleftrightarrow\quad 
\dot u \in \pl\calR^*(u, {-}\rmD\calF(u)) \subset \rmT_u M.
\]
\end{definition}

By the Fenchel equivalences, we can also reformulate the gradient-flow equation
by the optimality condition, which is a power identity:
\begin{equation}
  \label{eq:Banach.PowerId}
  \calR(u,\dot u) + \calR^*(u,{-}\rmD\calF(u)) = - \langle \rmD\calF(u), \dot
u\rangle = - \frac\rmd{\rmd t} \calF(u(t)). 
\end{equation}
This equation we can integrate and obtain the \emph{energy-dissipation balance}
\begin{equation}
  \label{eq:EDB.ODE}
  \forall\, 0<s<t: \qquad \calF(u(t)) + \int_s^t \big( \calR(u,\dot u) {+}
  \calR^*(u,{-}\rmD\calF(u)) \big) \dd r = \calF(u(s)),
\end{equation}
which simply states that the energy $  \calF(u(t))$ at the later time $t$ plus
the dissipated energy in the time interval $[s,t]$ give exactly the energy
$\calF(u(s))$ at the earlier time $s$. 

Of course, by the results from Exercise \ref{ex:DissFunction} we can write the
energy-dissipation balance also in one of the following simpler forms
\[
\calF(u(t)) + \int_s^t\! \mafo{Diss}_{\calR(u,\cdot)}(\dot u) \dd r = \calF(u(s))
\ \text{ or } \  
\calF(u(t)) + \int_s^t \!\mafo{Diss}_{\calR^*(u,\cdot)}(-\rmD\calF(u)) \dd r = \calF(u(s)).
\]
But there is a major difference between \eqref{eq:EDB.ODE} and the latter two
forms. The formulation involving ``$\calR{\oplus}\calR^*$'' is derived from the
optimality condition, and thus we will be able to show that the EDB
\eqref{eq:EDB.ODE} is still equivalent to the full gradient-flow equations. The
same is not true for the latter two formulations, which hold along all
solutions, but do not characterize the solutions.  Thus, to emphasize this
fact, we will sometimes insist that (EDB) is always assumed to be in
``\:$\calR{\oplus}\calR^*$ form''.  \bigskip

We now turn to the case of infinite dimensional evolution equations on a
reflexive Banach space $X$. Of course, for PDEs or abstract evolutionary equations one
typically needs several Banach spaces, so here $X$ denotes the space in which
rates $\dot u$ typically are located; other spaces associated with the energy
will be implicitly defined by $\dom(\calF)$ or $\dom(\plF \calF)$. For
simplicity, we will again use the Fr\'echet subdifferential $\plF \calF: X
\tosets X^*$ but other choices might be possible. 

\begin{definition}[Generalized gradient systems: PDE case]
A triple $(X,\calF,\calR)$ (or equivalently $(X,\calF,\calR^*)$) 
is called a \emph{generalized gradient system} on
the Banach space $X$, if $\calF: X \to \R_\infty$ is a lower semicontinuous
functional and $\calR$ is a primal dissipation potential meaning that 
$\calR(u,\cdot):X\to [0,\infty]$ is a dissipation potential for all $u\in X$
(or equivalently $\calR^*(u,\cdot):X^*\to [0,\infty]$). The associated
gradient flow equation is given by 
\begin{equation}
  \label{eq:Banach.GFE}
  0 \in \pl\calR(u(t),\dot u(t)) + \plF\calF(u(t)) \text{ a.e.\ in } [0,T]
\ \  \Longleftrightarrow \ \ 
\left\{ \ba{c} \dot u(t) \in \pl \calR^*(u,{-}\xi(t)) \\
   \text{and } \xi(t) \in
  \plF\calF(u(t))\\
   \text{ for a.a. } t \in [0,T]. \ea \right.
\end{equation}
Subsequently, we will often use the pair $(u,\xi)$ to denote the solutions. 
\end{definition}

\begin{example}[Doubly nonlinear diffusion equation] For $p,q\in {]1,\infty[}$
    we consider the space $X=\rmL^q(\Omega)$, the energy $\calF$ with
    $\calF(u)=\int_\Omega \frac1p |\nabla u|^p\dd x$ for $u \in
    \rmW^{1,p}(\Omega)$ and $+\infty$ otherwise in $X$. Moreover, we consider
    the dissipation potential $\calR(u,v)= \int_\Omega\frac1q(2{+}\cos u)
    |v|^q\dd x$. Both differentials $\plF\calF$ and $\pl\calR$ are
    single-valued and we obtain the doubly nonlinear diffusion equation
\[
 (2{+}\cos u)|\dot u|^{q-2}\dot u = \Delta_p u := \DIV\big( |\nabla u|^{p-2}
 \nabla u\big) \text{ \ in } \Omega, \qquad \nabla u \cdot \nu = 0 \text{ on
 }\pl\Omega,
\]
as the associated gradient-flow equation. 
\end{example} 

For course, as in the ODE case we can replace the two formulations in
\eqref{eq:Banach.GFE} by the optimality condition
\begin{equation}
  \label{eq:Banach:D.Optim}
  \calR(u(t),\dot u(t)) + \calR^*(u(t),{-}\xi(t))= - \langle \xi(t),\dot
  u(t)\rangle \AND \xi(t) \in \plF \calF(u(t)) \ \text{ a.e.\ in } [0,T]. 
\end{equation}
However, integration of this relation is no longer trivial for two reasons:
First we cannot simply assume that ``$\calR{\oplus}\calR^*$'' is integrable for
solutions $(u,\xi)$, and secondly, the application of the chain rule to
$\langle \xi, \dot u\rangle$ may be not valid. These two points will be
discussed in the following subsection. 

However, at this stage we can see already the main impact of the gradient
system on the gradient flow equation. The gradient structure provides an easy
way for setting up a 
\begin{equation}
  \label{eq:Banach:TIM}
 \boxed{
  \begin{aligned}
 &\text{time-incremental minimization scheme for time step $\tau = T/N>0$:}\\
 & u^\tau_0 = u^0 \in X \quad \text{(given initial value)}\\
 & \text{for } k=1,\ldots, N \text{ find } u^\tau_k \text{ as minimizer of the
   functional }\\
&\qquad \Phi^{\calF,\calR}_{\tau} (u_{ k-1 }; \,\cdot\,): X\ni u\mapsto \tau
\calR\big(u_{k-1}, \frac1\tau(u{-}u_{k-1})\big) + \calF(u). 
  \end{aligned}
 }
\end{equation}

As in the Hilbert-space case the minimizers $u_k$ (for notational convenience
we drop the superscript for the upcoming calculation) satisfy the Euler-Lagrange equation
\[
0 \in \pl\calR \big( u_{k-1}, \frac1\tau(u_k{-}u_{k-1})\big) +
\plF \calF(u_k) 
\]
or equivalently
\begin{equation}
   \label{eq:Banach:IncrGFE}
   \exists\, \xi_k\in X^*: \quad \xi_k \in \plF\calF(u_k) \qANDq {-} \xi_k \in
\pl\calR \big( u_{k-1}, \frac1\tau(u_k{-}u_{k-1})\big) . 
\end{equation}
The last relation can be used to apply the Fenchel equivalences giving 
\[
\calR \big( u_{k-1}, \frac1\tau(u_k{-}u_{k-1})\big) + \calR^*( u_{k-1},
{-}\xi_k) = - \big\langle \xi_k, \frac1\tau(u_k{-}u_{k-1}) \big\rangle .
\]
Assuming further that $\calF$ is $\lambda$-convex we can estimate the
right-hand side by using $\calF(u_{k-1}) \geq \calF(u_k)+ \langle \xi_k,
u_{k-1}{-}u_k\rangle + \frac\lambda2 \|u_k{-}u_{k-1}\|^2$. After
multiplying by $\tau>0$ we arrive at a discrete type of energy-dissipation inequality:
\begin{equation}
  \label{eq:Ban.DiscEDB.la}
  \tau\Big(\calR \big( u_{k-1}, \frac1\tau(u_k{-}u_{k-1})\big) + \calR^* 
    ( u_{k-1}, {-}\xi_k) \Big) \leq  \calF(u_{k-1}) - \calF(u_k) 
  - \frac\lambda2 \|u_k {-} u_{k-1} \|^2. 
\end{equation}

Reintroducing the superscript $\tau$ in $u^\tau_k$ and $\xi^\tau_k$ again, 
We can now define the four interpolants $\wh u_\tau,\ \ol u_\tau$, and $\ul
u_\tau$ from $[0,T]\to X$ and $\ol\xi_\tau:[0,T]\to X^*$ as follows:
\begin{equation}
  \label{eq:BanachInterpol}
\begin{aligned}
& \wh u_\tau((k{+}\theta{-}1)\tau)= (1{-}\theta) u^\tau_{k-1} + \theta u^\tau_k \quad
\text{for } k\in \{1,\ldots,N\} \text{ and } \theta \in [0,1];
\\
& \ol u_\tau(0)=u^\tau_0 \AND\ol u_\tau(t)=u^\tau_k \text{ for } t\in
{]k\tau{-}\tau,k\tau]}  \text{ and } k\in \{1,\ldots,N\} ;
\\
& \ul u_\tau(t)=u^\tau_k \text{ for } t\in
{[k\tau,k\tau{-}\tau]}  \AND k\in \{0,\ldots,N{-}1\} \AND \ul u_\tau(T)=u^\tau_N;
\\
& \ol\xi_\tau(0)=0 \in X^* \AND \ol\xi_\tau(t)=\xi^\tau_k \text{ for } t\in
{](k{-}1)\tau,k\tau]}  \AND k\in \{1,\ldots,N\}.
\end{aligned}  
\end{equation}
Here $\ol u_\tau$ and $\ol\xi_\tau$ are the left-continuous, piecewise constant
interpolants, $\ul u_\tau$ is the right-continuous, piecewise constant
interpolant, and $\wh u_\tau$ is the continuous piecewise affine interpolant
which has the piecewise constant derivative 
\[
\dot{\wh u}_\tau(t) = \frac1\tau \big( u^\tau_k - u^\tau_{k-1}\big) \quad \text{for
} k\in \{1,\ldots,N\} \AND t\in {]k\tau{-}\tau, k\tau[}.
\]

With these definitions we can rewrite the incremental Euler-Lagrange equation 
\eqref{eq:Banach:IncrGFE} as an approximate equation on $[0,T]$ as follows
\begin{equation}
   \label{eq:Banach:ApproxGFE}
   \ol\xi_\tau(t) \in \plF\calF(\ol u_\tau(t)) \qANDq 0\in 
\pl\calR \big( \ul u_\tau(t), \dot{\wh u}_\tau(t)\big)  + \ol\xi_\tau(t)
  \quad \text{for a.a. } t \in [0,T].  
\end{equation}
Note that all four interpolants are needed because our scheme is
``semi-implicit'', namely implicit in the functional $\calF$ and explicit in
the state-dependence of $\calR$. 

We may also consider the discrete energy dissipation inequality
\eqref{eq:Ban.DiscEDB.la}. After summation over $k=1,\ldots,N$ we find 
\begin{equation}
  \label{eq:Banach.DiscrEDI}
  \calF(\ol u_\tau(T)) + \int_0^T \!\!\Big(\calR \big( \ul u_\tau, \dot{\wh
    u}_\tau\big) {+}  \calR^*\big( \ul u_\tau, {-} \ol\xi_\tau\big)  
\Big) \dd t \leq \calF(u^0) - \frac{\tau\lambda}2\!
  \int_0^T  \!\|\dot{\wh u}_\tau(t)\|^2 \dd t.  
\end{equation}
In the following we will show that we can pass to the limit $\tau\to 0^+$ in
this discrete energy-dissipation inequality and thus find solutions.

\subsection{The energy-dissipation principle} 
\label{su:BanachEDP}

We have already seen in the previous subsection that in the ODE case we obtain
the energy-dissipation balance \eqref{eq:EDB.ODE}, i.e.\  for all
solutions $u$ of the gradient-flow equation we have the Energy-Dissipation
Inequality
\[
\text{\rm (EDI)} \qquad \calF(u(T)) + \int_0^T\big( \calR(u,\dot u)
{+}\calR^*(u,{-} \rmD\calF(u))\big) \dd t \leq \calF(u(0)).
\]
In fact, \eqref{eq:EDB.ODE} gives the balance with ``$=$'' instead of ``$\leq
$'', but we want to make the point that even the estimate is 
 \emph{equivalent} to solving the GFE 
 \begin{equation}
   \label{eq:Banach.GFE2}
    0 \in \pl\calR(u,\dot u) + \rmD \calF(u) \qquad
       \text{a.e.\ in } [0,T].
 \end{equation}
The argument involves only the Fenchel theory and the chain rule. Indeed, by
the chain rule, (EDB) can be rewritten as
\[
 \int_0^T\big(  \calR(u,\dot u)
{+}\calR^*(u,{-} \rmD\calF(u)) + \langle \rmD\calF(u),\dot u\rangle \big) \dd t
\leq 0.  
\]  
However, by the Fenchel-Young estimate we know that the integrand is
nonnegative. Thus, we conclude that it must be $0$ a.e.\ in $[0,T]$. But this
implies the power identity \eqref{eq:Banach.PowerId}. But by the Fenchel
equivalence this implies the GFE \eqref{eq:Banach.GFE2}.\medskip 

To make this argument also rigorous for the nonsmooth setting in
infinite-dimensional Banach spaces,  we need a corresponding \emph{abstract
  chain rule}. At this point we simply give a definition that exactly provides
what we need, and in Section \ref{su:ChainRule}  we then show that this
condition can be obtained in the Banach-space setting under suitable conditions
such as semiconvexity of $\calF$.

\begin{definition}[Abstract chain rule condition]
\label{de:AbsChainRuleCond}
We say that a GS $(X,\calF,\calR)$  satisfies the (abstract) chain rule, if the
following holds:
\begin{equation}
  \label{eq:ACRcond}
\left.
\begin{aligned}
&\text{If\/ } u\in \rmW^{1,1}([0,T];X)  \AND \xi\in \rmL^1([0,T];X^*) 
 \text{ satisfies } \ \sup_{[0,T]} \big| \calF(u(t)) \big| < \infty, 
\\ 
&\xi(t) \in \plF\calF(u(t)) \text{ a.e.\ in } [0,T] \qANDq 
   \int_0^T \!\!\big( \calR(u,\dot u) {+} \calR^*(u,{-}\xi)\big)
\dd t < \infty, \ \ 
\\
&\text{then }t \mapsto \calF(u(t)) \text{ is absolutely continuous, } 
\big(t\mapsto \langle \xi(t),\dot u(t)\rangle\big) \in \rmL^1([0,T]),
\\
&\AND \frac\rmd{\rmd t } \calF(u(t)) = \langle \xi(t),\dot u(t)\rangle \quad
\text{ a.e.\ in } [0,T].
\end{aligned}
\right\}
\end{equation}
\end{definition}

With this we are ready to state a precise version of the so-called
energy-dissipation principle, which concerns roughly that solving the
gradient-flow equation is equivalent to finding a function satisfying the
energy-dissipation inequality (EDI). However, we warn the reader that sometimes
the gradient-flow equation as a PDE may have solutions that do not have finite
energy (cf.\ \cite[Rem.\,2.8]{ScSeZe12AUBE})
and such solutions are not covered by this principle. 

Several versions of the  \emph{Energy-Dissipation Principle} were used
previously, see e.g.\ \cite[Thm.\,3.3.1]{Miel16EGCG}.
The following precise, but still very general version is due
to Riccarda Rossi and Artur Stephan, see \cite{MiRoSt22?SSAG}. 

\begin{theorem}[The Energy-Dissipation Princple (EDP)]
\label{th:Banach.EDP}
Consider the generalized gradient system $(X,\calF,\calR)$ on a separable Banach
space $X$ that satisfies the abstract chain-rule condition \eqref{eq:ACRcond}.
Then, for all pairs $(u,\xi) \in \rmW^{1,1}([0,T];X) \ti \rmL^1([0,T];X^*)$
with $\xi(t) \in \plF \calF(u(t))$ a.e.\ in $[0,T]$ 
the following two statements are equivalent:\medskip

(A) $(u,\xi)$ satisfies \ $\sup_{t\in [0,T]} \calF(u(t)) < \infty$ \ and
\begin{equation}
  \label{eq:Banach.EDI}
  \text{\rm (EDI)} \qquad \calF(u(T)) + \int_0^T\big( \calR(u,\dot u)
{+}\calR^*(u,{-} \xi)\big) \dd t \leq  \calF(u(0)) <\infty. \medskip 
\end{equation}

(B) $(u,\xi)$ satisfies the gradient-flow equation 
\begin{equation}
  \label{eq:Banach.GFE3}
  0 \in \xi(t) + \pl\calR(u(t),\dot u(t)) \qANDq \xi(t) \in \plF\calF(u(t))
  \quad \text{ for a.a. } t \in [0,T]
\end{equation}
 and  the energy-dissipation balance in $\calR\calR^*$ form:
\begin{equation}
  \label{eq:Banach.EDB}
  \text{\rm (EDB)} \qquad \calF(u(t)) + \int_s^t \big( \calR(u,\dot u)
{+}\calR^*(u,{-} \xi )\big) \dd r =  \calF(u(s)) <\infty 
\end{equation}
for $0\leq s < t\leq T$.
\end{theorem}
\begin{proof} \STEP{(B) $ \Longrightarrow $ (A).} This direction is trivial.

\STEP{(A) $ \Longrightarrow $ (B).} We proceed exactly as in the ODE
case. We start from (A). Because of $\calF(u(0))<\infty$ and $\calF(u(T))> -
\infty$ we conclude that the finiteness of the dissipation integral in the
$\calR\calR^*$ form in Assumption \eqref{eq:ACRcond} is satisfied. Hence,
we can apply the assumed abstract chain rule and rewrite 
$\calF(u(T))-\calF(u(0))$ in the form $\int_0^T \langle \xi(t),\dot u(t)\rangle
\dd t$. Combining this with (EDI) we find
\[
 \int_0^T \Big(\calR(u,\dot u)
{+}\calR^*(u,{-} \xi ) + \langle \xi,\dot u\rangle  \Big) \dd t =-\delta \leq 0,  
\]
where $\delta\geq 0$ is the gap (RHS minus LHS) in (EDI).  
By the Young-Fenchel inequality the integrand is nonnegative, hence we conclude
 $\delta=0$ which means that (EDI) is in fact (EDB)$_{[0,T]}$ given in
 \eqref{eq:Banach.EDB}. Moreover, the nonnegative integrand must be $0$ a.e.\
 in $[0,T]$, which implies the identity $\calR(u,\dot u)
{+}\calR^*(u,{-} \xi ) =- \langle \xi,\dot u\rangle$. By the Fenchel
equivalences this is equivalent to the GFE \eqref{eq:Banach.GFE3}. 

By the abstract chain rule we can also integrate the identity $\calR(u,\dot u)
{+}\calR^*(u,{-} \xi ) =- \langle \xi,\dot u\rangle$ on the subinterval
$[s,t]\subset [0,T]$ and thus obtain (EDB)$_{[s,t]}$. 
\end{proof}

To illustrate the EDP we look at a very simple example, namely the Hilbert-space
GS $(\rmL^2(\Omega),\calF_\mafo{Dir},\Riesz)$ with $\calF_\mafo{Dir}(u) =
\frac\alpha2 \| \nabla u\|_{\rmL^2}^2 $ and $\dom(\calF_\mafo{Dir})=
\rmH^1_0(\Omega)$. Then we have $\xi=\plF\calF_\mafo{Dir}(u)=- \alpha \Delta u$ and
$\calR^*({-}\xi)=\frac12 \|\alpha \Delta u\|_{\rmL^2}^2$. Thus, (EDI) can be
written in the form 
\begin{align*}
0&\geq \frac\alpha2 \|u(T)\|_{\rmL^2}^2 + \int_0^T \big( \frac12\|\dot
u\|_{\rmL^2}^2 + \frac12 \|\alpha \Delta u\|_{\rmL^2}^2 \big) \dd t 
 -\frac\alpha2 \|u(0)\|_{\rmL^2}^2 
\\
&=  \int_0^T \big( \frac12\|\dot
u\|_{\rmL^2}^2 + \frac12 \|\alpha \Delta u\|_{\rmL^2}^2  +\langle{-}\alpha
\Delta u , \dot u\rangle \big) \dd t \ = \ \int_0^T  \frac12\big\|\dot
u - \alpha \Delta u \big\|_{\rmL^2}^2  \dd t
\end{align*}

The major importance is that all terms in the left-hand side of (EDI) have good
lower semicontinuity properties when passing to limits of approximating
sequences. Hence starting from the discrete energy-dissipation inequality
\eqref{eq:Banach.DiscrEDI} it is reasonable to end up with (EDI) if suitable
technical conditions hold, see Section \ref{su:BanachExist}. To finalize the
proof we will then use the abstract chain rule to invoke the energy-dissipation
principle to obtain solutions.

\subsection{The abstract chain rule}
\label{su:ChainRule}

We want $\frac\rmd{\rmd t} \calF(u(t)) = \langle \xi(t) , \dot u(t)\rangle$
under as general as possible conditions. Can it work for nonsmooth energies?

\begin{example}[Chain rule for nonsmooth $\calF$]
\label{ex:CR.nonsmooth}
We consider  $X=\R^2$ and the nonsmooth, but convex functional 
$\calF(u_1,u_2)=\max\{|u_1|, |u_2|\}$. 
For $u=(y,y)$ with $y>0$ we obtain the set-valued subdifferential
$\pl\calF(u)=\bigset{(\theta,1{-}\theta) \in \R^2}{ \theta \in [0,1]}$.  

Thus, for the curve $u(t)=(y(t),y(t))$ with $y(t)>0$ we obtain elements in the
subdifferential $\xi(t)=(\theta(t),1{-}\theta(t))$, where $\theta\in [0,1]$ is
completely arbitrary. 

Moreover, we have $f(t)=\calF(u(t))=y(t)$ which implies 
\[
\dot y(t) =\dot f(t) \overset{\text{CR}}= \langle \xi(t), \dot u(t)\rangle =
\langle \tbinom{\theta}{1-\theta}, \tbinom{\dot y}{\dot y} \rangle = \dot y.
\]
Indeed the chain rule holds, although $\xi \in \plF \calF(u(t))$ is not unique. 
\end{example}

\begin{example}[Classical Gelfand evolutionary triple] 
For solving parabolic equation one often considers a so-called Gelfand triple
$V \overset\rmd\subset H \overset\sim= H^* \overset\rmd\subset V^*$. 

By approximation the solutions $u$ are constructed with 
$u \in \rmW^{1,2}([0,T];V^*) \cap \rmL^2([0,T];V)$. A major step is then to
show that this implies $ u \in \rmC^0([0,T];H)$. 

Sometimes one even shows that 
the mapping $t \mapsto \|u(t)\|_H^2$ is absolutely continuous with
\[
\frac12\,\frac\rmd{\rmd t} \|u(t)\|_H^2 = {}_{V^*}\langle \dot u(t), u(t)\rangle_V.
\]
This is typically used when solving the diffusion equation $\dot u= \Delta u$
with $V=\rmH^1_0(\Omega)$, $H=\rmL^2(\Omega)$, and $V^*=\rmH^{-1}(\Omega)$. Then 
\[
\frac12\,\frac\rmd{\rmd t} \|u\|_{\rmL^2}^2 = {}_{\rmH^{-1}}\langle \dot u, 
u\rangle_{\rmH^1_0} = {}_{\rmH^{-1}}\langle \Delta u, u\rangle_{\rmH^1_0} =
-\int_\Omega |\nabla u|^2\dd x . 
\]
\end{example}

A more general chain rule was established in \cite[Lem.\,3.3,
p.\,73]{Brez73OMMS} for general convex functionals on a Hilbert space (literal
interpretation with ``$A$'' replaced by ``$\pl\calF$'') :
\begin{quote}
  \underline{LEMMA 3.3}. Let $u \in \rmW^{1,2}(0,T;H)$ be such that 
   $u(t)\in \dom(\pl\calF)$ a.e.\ in ${]0,T[}$. Suppose there exists 
  $g\in \rmL^2(0,T;H)$ such that $g(t)\in \pl\calF(u(t))$ a.e.\ in
  ${]0,T[}$.  Then the function  $t \mapsto \calF(u(t))$ is absolutely
  continuous. 

Denote by $\mathscr T$ the set of points $t\in {]0,T[}$ such that $u(t) \in
\dom(\pl\calF)$ and that $u$ and $\calF\circ u$ are differentiable. Then, for
all $t \in \mathscr T$ we have
 
\mbox{}\qquad  $\frac\rmd{\rmd t} \calF(u(t)) = \langle h,\dot u(t)\rangle$ 
\qquad for all $h \in \pl\calF(u(t))$.
\end{quote}

We will generalize such a result to $\lambda$-convex functionals on a Banach
space. Our result is based on the theory developed in \cite{MiRoSa13NADN,
  MieRos23BVSI} which relies on ideas in \cite[Thm.\,1.2.5]{AmGiSa05GFMS}.  
Of course, the result in \cite[Prop.\,A.1]{MieRos23BVSI} is much more general,
in particular the condition of $\lambda$-convexity is weakened significantly. 

The following result will use the \emph{quantitative Young estimate} for the
dissipation potential $\calR$:
\begin{equation}
  \label{eq:QYE}
\exists\, c_\rmY, C_\rmY>0\  \forall\, u, v \in X \ \forall\, \xi \in  X^*:
\quad
\calR(u,v) + \calR^*(u,\xi)\geq c_\rmY \|v\|_X\|\xi\|_{X^*} - C_\rmY. 
\end{equation}
If $\calR$ only depends on $v$ through its norm, i.e.\ $\calR(u,v)=\rho(u,
\|v\|)$, then one has $c_\rmY=1$ and $C_\rmY=0$, see the discussion in
\cite{MieRos23BVSI}.  Another case where \eqref{eq:QYE} holds is given when
$\calR$ has uniform upper and lower $p$-growth, namely  $c \|v\|^p - C
\calR(u,v) \leq C\|v\|^p +C$. Then, $\calR^*(u,\xi) \geq \wt c\|\xi\|^{p^*} -\wt
C$ and \eqref{eq:QYE} follows. 

In the following result the main chain-rule property is totally independent of
the dissipation potential $\calR$, i.e.\ it is a property of $(X,\calF)$
alone. The connection to the gradient systems $(X,\calF,\calR)$ is coming when
we want to establish the integrability condition $\int_0^T \| \dot
u\|_X\,\|\xi(t)\|_{X^*} \dd t < \infty  $ via the quantitative Young estimate
\eqref{eq:QYE}.  

\begin{theorem}[Chain rule in Banach spaces]
\label{thm:BanachChainRule} On a reflexive Banach space $X$ consider a 
 proper, lsc, and semiconvex functional $\calF:X\to \Rinfty$. Then, the
following chain rule holds:
\begin{quote}
If $u\in \rmW^{1,1}([0,T];X)$ and $\xi\in \rmL^1([0,T];X^*)$ satisfies 
\ $\sup_{[0,T]} \big|\calF(u(t)) \big| <\infty$, 
\[
\xi(t) \in \plF\calF(u(t)) \text{ a.e.\ in } [0,T], \qANDq 
 \int_0^T \| \dot u\|_X\,\|\xi(t)\|_{X^*} \dd t < \infty,
\]
then $t \mapsto \calF(u(t))$ is absolutely continuous and $\frac\rmd{\rmd t }
\calF(u(t)) = \langle \xi(t),\dot u\rangle$ a.e.
\end{quote} 
In particular, if  $\calR$ satisfies the
quantitative Young estimate \eqref{eq:QYE}, then the abstract chain rule
\eqref{eq:ACRcond} holds.  
\end{theorem}
\begin{proof} To shorten the presentation we abbreviate $f(t):=\calF(u(t))$ and
  set $\Sigma:=\bigset{t \in [0,T]}{\plF\calF(u(t)) \neq \emptyset}$. By
  assumption $\Sigma$ is a set of full measure.  
 
\STEP{Step 1: Absolute integrability of $\wt f$ under arc-length parametrization.} 
We first consider the case that $\|\dot u(t)\|=1$ a.e.\ in $[0,T]$. Then, we
immediately have $\| u(t_1){-}u(t_0)\| \leq |t_1{-}t_0|$. 

Choosing arbitrary $t_{j-1} <t_j$ in $\Sigma$ we use $\lambda$-convexity of $\calF$ to
obtain 
\begin{align*} 
f(t_j)-f(t_{j-1}) &\geq \langle \xi(t_{j-1}), u(t_j){-}u(t_{j-1})\rangle +
\frac\lambda2 \|  u(t_j){-}u(t_{j-1})\|^2\\
& \geq - \|
\xi(t_{j-1})\|\,(t_j{-}t_{j-1}) + \frac\lambda2 |t_j{-}t_{j-1}|^2  \qANDq
\\
f(t_{j-1})-f(t_j) & \geq \ - \|
\xi(t_j)\|\ (t_j{-}t_{j-1}) \ + \frac\lambda2 |t_j{-}t_{j-1}|^2,
\end{align*}
where the second inequality follows from the first by interchanging $t_{j-1}$
and $t_j$.  

For an arbitrary interval $[s,t]$ with $s,t\in \Sigma$ we choose partitions $s= t_0<t_1 < 
\cdots < t_N = t$ with $t_j \in \Sigma$ and add up the inequalities which leads to 
\begin{align}
\label{eq:Ba.LowUppEst}
\sum_{j=1}^N\! \big( {-}\|\xi(t_{j-1})\|(t_j{-}t_{j-1})
             + \frac\lambda2 |t_j{-}t_{j-1}|^2\big) &\leq 
f(t) - f(s) \\
\nonumber & \leq 
 \sum_{j=1}^N \! \big(\|\xi(t_j)\|(t_j{-}t_{j-1})
             - \frac\lambda2 |t_j{-}t_{j-1}|^2 \big) .   
\end{align}
By a refined theory of the Riemann integral for $\rmL^1$ functions (see
\cite[Sec.\,4.4]{DaFrTo05QCGN} and \cite{Hahn15VRI} for the historic origin) it
can be shown that it is always possible to choose a sequence of partitions with
fineness tending to $0$ such that the limit of Riemann sums equals the Lebesgue
integral. (There $\xi$ is defined everywhere, and we can set $\xi(r)=0$ for $r
\not\in \Sigma$.) Hence, we conclude
\[
- \int_s^t \| \xi(r)\| \dd r \leq f(t)- f(s) \leq \int_s^t \| \xi(r)\| \dd r.
\]
Thus, we have established $| f(t) - f(s)| \leq \int_s^t \| \xi(r)\| \dd r$ for
all $s,t\in \Sigma\subset [0,T]$.  Because of $\xi \in \rmL^1([0,T];X^*)$ this
shows that there is a absolutely continuous function
$\wt f \in \rmW^{1,1}([0,T])\cap \rmC^0([0,T])$ satisfying $f(t)=\wt f(t)$ for
all $t \in \Sigma$.

\STEP{Step 2. $f=\wt f$ under arclength parametrization.} We continue under the
same conditions as in Step 1 and show $f(t)=\wt f(t)$ for all $t\in [0,T]$. 
By continuity of $t \mapsto
u(t)\in X$ and lsc of $\calF$ we know that $f$ is lower semicontinuous, which
implies $f(t)\leq \wt f(t)$ for all $t \in [0,T]$. 

To show the opposite inequality we restrict to $t\in [0,2T/3]$ and define for
$r \in {]0,T/3[}$ the averages 
\[
f_r(t)= \frac1r \int_t^{t+r} f(s) \dd s = \frac1r \int_t^{t+r} \wt f(s) \dd s \
\to \ \wt f(t) \quad \text{for } r \to 0^+.
\]
(For $t\in [T/3,T]$ one can proceed analogously by taking backward averages
$f_r(t)= \frac1r \int_{t-r}^t f(s) \dd s$.) Here $f_r$ is well defined, because
$f$ is bounded by assumption and lsc, hence (Borel) measurable and integrable.
Thus, it suffices to show $f(t) \geq \limsup_{r\to 0^+} f_r(t) = \wt f(t)$. For
this we proceed as above and obtain
\begin{align*}
&f(t){-} f_r(t)= \frac1r \int_t^{t+r} \!\! \big( f(t)-\wt f(s)\big) \dd s \geq 
\frac1r \int_t^{t+r}\!\! \big( {-}\|\xi(s)\|\,|t{-}s| -\frac{|\lambda|}2 \,|t{-}s|^2
\big) \dd s \big) \\ 
& \geq \frac1r \int_t^{t+r}\!\! \big( {-}\|\xi(s)\| \:r - \frac{|\lambda|}2\:r^2
\big) \dd s  = - \int_t^{t+r}\!\! \|\xi(s)\| \dd s - \frac{|\lambda|}2\:r^2 
 \to  0 \ \text{ for }r\to 0^+.
\end{align*} 
Thus, we conclude that $t \mapsto f(t)=\calF(u(t))$ is equal to the
continuous representative $\wt f$, and the desired
absolute continuity of $f$ is shown under the assumption $\| \dot u(t)\|=1$ a.e.

\STEP{Step 3: Reparametrization.}
For the general case with $\dot u \in \rmL^1([0,T];X)$ we follow
\cite[Lem.\,1.1.4]{AmGiSa05GFMS} and consider the
reparametrization
\[
\sigma(t)= \int_0^t \| \dot u(r)\| \dd r  \ \text{ giving } \sigma :[0,T] \to
[0,\ell], 
\]
where $\ell=\sigma(T)= \int_0^T \| \dot u(r)\| \dd r$. Clearly, $\sigma \in
\rmW^{1,1}([0,T])\subset \rmC^0([0,T])$ and $\sigma'(t)\geq 0$. We define the inverse 
\[
\tau: \left\{ \ba{ccc} [0,\ell] &\to & [0,T], \\
 s & \mapsto & \min\bigset{t\in [0,T]}{ \sigma(t)=s}, \ea \right.
\]
which is increasing and continuous from the left such that
$\sigma(\tau(s))=s$ for all $s \in [0,\ell]$. Moreover, we have 
\begin{equation}
  \label{eq:u.tau.sig}
  \tau(\sigma(t)) \leq t \qANDq u(\tau(\sigma(t)))=u(t) \quad \text{for all } t
\in [0,T].
\end{equation}
For the second relation note that on intervals ${]t_0,t_1[}$ where
$t_0=\tau(\sigma(t_0))=\tau(\sigma(t))<t$  we have $\dot u(t)=0$ giving 
$u(t)=u(t_0)$. 

With this we define $\wh u(s)=u(\tau(s))$, and for $0\leq s_0<s_1\leq \ell$ we have 
\[
\|\wh u(s_1){-}\wh u(s_0) \| = \| u(\tau(s_1)) - u(\tau(s_0))\| \leq
\int_{\tau(s_0)}^{\tau(s_1)} \| \dot u(r)\| \dd r = \sigma (\tau(s_1))- \sigma(
\tau(s_0)) = s_1 - s_0.
\] 
Thus, $\wh u$ is 1-Lipschitz. Moreover, the reflexivity of $X$ gives
$\rmC^\mafo{Lip}([0,T];X)= \rmW^{1,\infty}([0,T];X)$, such that the derivative 
$\wh u'(s)$ exists and $\| \wh u'(s)\|\leq 1$ a.e.\ in $[0,\ell]$.  

Moreover, for $0\leq t_0< t_1 \leq T$ we find 
\[
\|u(t_1){-}u(t_0) \| = \| \wh u(\sigma(t_1)) - \wh u(\sigma(t_0))\| \leq
\int_{\sigma(t_0)}^{\sigma(t_1)} \| \wh u'(\rho)\| \dd \rho = 
\int_{t_0}^{t_1}   \| \wh u'(\sigma(r))\| \dot \sigma(r) \dd r. 
\] 
This implies $\|\dot u(t)\| \leq \|\wh u'(\sigma(t))\| \dot \sigma(t) $ for
a.a.\ $t\in [0,T]$. Using $\dot \sigma(t)=\|\dot u(t)\|$ and 
$\| \wh u'(s)\|\leq 1$ from above we find $\| \wh u'(s)\| = 1$ a.e.\ in
$[0,\ell]$. 

\STEP{Step 4. Absolute continuity of $f$ in the general case.}
We now apply the reparametrization from the previous step also to $f$ and $\xi$
by setting $\wh f(s)= f(\tau(s))$ and $\wh\xi(s)=\xi(\tau(s))$ and obtain 
$f(t)=\wh f(\sigma(t)) = \calF(u(\tau(\sigma(t))))$ 
by using $u(\tau(\sigma(t)))=u(t)$ from 
\eqref{eq:u.tau.sig}. Moreover, we have 
\[
\int_0^\ell \!\! \| \wh \xi(s) \| \dd s \overset{\rmt\rmr}= \int_0^T \!\! \|
\wh \xi(\sigma(t))\| \dot\sigma(t) \dd t = \int_0^T \!\!
 \| \xi(\tau(\sigma(t)))\|\,
\|\dot u(t)\| \dd t \overset*= \int_0^T  \!\! \| \xi(t)\|\,
\|\dot u(t)\| \dd t  < \infty.
\]
(For a justification of the transformation rule in ``$\overset{\rmt\rmr}=$''
with $s=\sigma(t)$ we refer to \cite[Thm.\,5.8.30]{Boga07MT} and note that the
absolute continuous function $\sigma$ satisfies the Lusin property (N).) 
The last bound is simply the assumption, whereas in $\overset*=$ we use that 
$\dot u(t)=0$ whenever $\tau(\sigma(t)) \neq t$, see the comments after
\eqref{eq:u.tau.sig}. 
Hence we have $\wh\xi \in \rmL^1([0,\ell])$ as well as $\|\wh u'(s)\|=1$ a.e.\
in $[0,\ell]$. Thus, we can apply Step 2 and find $|\wh f(s_1){-}\wh f(s_0)|
\leq \int_{s_0}^{s_1} \| \wh\xi(s)\| \dd s $ and conclude via
\begin{align*}
|f(t_1)-f(t_0)| &= | \wh f(\sigma(t_1)) - \wh f(\sigma(t_0))| \leq 
\int_{\sigma(t_0)}^{\sigma(t_1)} \| \wh\xi(s)\| \dd s \\
&= 
\int_{t_0}^{t_1} \| \wh\xi(\sigma(t))\| \dot \sigma(t) \dd t = 
\int_{t_0}^{t_1}\| \xi(t)\|\, \|\dot u(t)\|
\dd t,
\end{align*}
which is the desired absolute integrability of $f: t \mapsto f(t)=\calF(u(t))$
as $t \mapsto \|\dot u\|\,\| \xi\|$ lies in $\rmL^1([0,T])$. 

\STEP{Step 5: Identification of the derivative.} As $f$ is differentiable a.e.\
in $[0,T] $ the set $\bbT\subset {]0,T[}$ on which $u:[0,T] \to H$ is
differentiable, $f:[0,T]\to \R$ is differentiable, and $\plF\calF(u(t))$ is
nonempty is of full measure. Now take $t \in \bbT$ and choose an arbitrary
$\eta \in \plF\calF(u(t))$. Then, for all $ h\in [-t,T{-}t]$ we have 
\[
f(t{+}h)- f(t) \geq \langle \eta, u(t{+}h){-}u(t)\rangle + \frac\lambda2 
\big\| u(t{+}h) - u(t)\big\|^2.
\]
Dividing by $h>0$ and taking the limit $h\to 0^+$ we find $\dot f(t) \geq
\langle \eta, \dot u(t)\rangle$. Dividing by $h<0$ and taking the limit $h \to
0^-$ gives the opposite estimate. Hence we have shown $\frac\rmd{\rmd t}
\calF(u(t)) = \dot f(t)= \langle \eta, \dot u(t)\rangle$ for all $t\in \bbT$, and the
chain rule is established. 

\STEP{Step 6: Abstract chain rule.} Starting from $\int_0^T\big(\calR(u,\dot u)
+ \calR^*(u,{-}\xi)\big) \dd t < \infty $ and the quantitative Young estimate
\eqref{eq:QYE} we obtain 
\[
\int_0^T \|\dot u\|_X\| \xi\|_{X^*} \dd t \leq \frac1{c_\rmY} \Big( \int_0^T\big(\calR(u,\dot u)
+ \calR^*(u,{-}\xi)\big) \dd t + C_\rmY \Big) \ < \ \infty.
\]
Thus the above results are applicable and we obtain \eqref{eq:ACRcond}.  
\end{proof}

\subsection{Existence theory via time-incremental minimization}
\label{su:BanachExist}

Our existence theory indeed follows very similar steps as in the Hilbert-space
setting. The difference is that we are now using the energy-dissipation
principle, i.e.\ we do not work with the evolutionary equation directly. We
rather exploit the favorable structure of the energy-dissipation inequality,
which allows us to pass to the limit by arguments of the calculus of
variations. 

We emphasize that uniqueness of solutions cannot be expected in this general
setting. Even if we are able to obtain uniqueness of the incremental minimizers
$u^\tau_k$ we cannot expect the continuous solutions to be unique because of
the doubly nonlinear structure. 

We start by collecting a set of sufficient condition that allow us to study a
large class of generalized gradient systems. However, the assumptions are
restricted for didactic reasons, and we will comment on possible extensions
and generalizations in the next subsection. 

For the following list of conditions we recall the sublevels
$S^\calF_E=\set{u\in X}{\calF(u)\leq E} $.%
\begin{subequations}%
\label{eq:Banach.Ass}%
\begin{align}
\label{eq:Banach.Ass.a}
& X \text{ is a separable reflexive Banach space}.
\\[0.4em]
\label{eq:Banach.Ass.b}
&\calF:X \to \Rinfty \text{ is semiconvex and has compact sublevels}. 
\\[0.4em]
\label{eq:Banach.Ass.c}
&\left.
\begin{aligned}
  &\calR:X\ti X\to [0,\infty]
  \AND \calR^*:X\ti X^*\to [0,\infty] \text{ are } \\
  &\text{uniformly superlinear on sublevels, i.e.\ } \forall\, E \in \R \\
  &\exists\,\text{ increasing, convex, superlinear } \psi_E:{[0,\infty[}
    \to{[0,\infty[} \text{ such that } \\ 
  &\forall\, (u,v,\xi)\in S^\calF_E\ti X \ti X^*: \ \ \calR(u,v)\geq \psi_E(\|v\|) 
   \AND  \calR^*(u,\xi) \geq \psi_E(\|\xi\|).  
\end{aligned}\right\}
\\[0.4em]
\label{eq:Banach.Ass.d}
&(X,\calF,\calR) \text{ satisfies the abstract chain rule condition
  \eqref{eq:ACRcond}}. 
\\[0.4em]
\label{eq:Banach.Ass.e}
&\left.\begin{aligned}
&\text{For each level $E\in \R$ there exists a modulus of continuity }
\omega^\calR_E \ \ 
\\
&\text{such that } \forall\, u_0,u_1\in S^\calF_E \ \forall\, v\in X:
\\
& \big| \calR(u_1,v)- \calR(u_0,v)\big| \ \leq \ \omega^\calR_E\big( \|
u_1{-}u_0\|\big) \,\big( 1 + \calR(u_0,v)\big). 
\end{aligned}\right\}
\end{align}
\end{subequations}

We emphasize that the semiconvexity condition in \eqref{eq:Banach.Ass.b} is
rather strong: First, it allows us to derive the discrete approximation of the
energy-dissipation inequality. Secondly, by our results in Section
\ref{su:Hilbert.ClosSubdiff} it implies the important condition of closedness
of the Fr\'echet subdifferential $\plF\calF:X\tosets X^*$.  Thirdly, it is a
very helpful condition of establishing the chain rule, see Theorem
\ref{thm:BanachChainRule}.

The upcoming existence result is now based on the time-incremental minimization
scheme \eqref{eq:Banach:TIM}, the four associated interpolants 
$\wh u_\tau$, $\ul u_\tau$, $\ol u_\tau$, and $\ol\xi_\tau$ (see
\eqref{eq:BanachInterpol}), and the discrete EDI \eqref{eq:Banach.DiscrEDI}.
The proof follows similar steps as the existence proof in the Hilbert-space
setting, but now in the last step we exploit the Energy-Dissipation Principle,
where the quantitative Young estimate \eqref{eq:Banach.Ass.d} is needed to
provide the abstract chain rule.   

\begin{theorem}[Existence for $(X,\calF,\calR)$]
\label{th:BanachExist}
Consider a generalized GS $(X,\calF,\calR)$ satisfying the assumptions
\eqref{eq:Banach.Ass}. Then, for all $u_0\in X$ with $\calF(u^0)<\infty$ there
exists a solution $(u,\xi)\in \rmW^{1,1}([0,T];X)\ti \rmL^1([0,T];X^*)$
satisfying $u(0)=u^0$, the gradient-flow equation 
\begin{equation}
  \label{eq:Banach.GFE4}
  0 \in \xi(t) + \pl\calR(u(t),\dot u(t)) \qANDq \xi(t) \in \plF\calF(u(t))
  \quad \text{ for a.a. } t \in [0,T],
\end{equation}
 and  the energy-dissipation balance 
$\calF(u(t)) + \int_s^t\big( \calR(u,\dot u)
{+}\calR^*(u,{-} \xi )\big) \dd r =  \calF(u(s)) $ for $0\leq s < t\leq T$.
\end{theorem}

Before starting the full proof, we provide a few auxiliary results that are
useful but are also of independent interest. First we recall that the Legendre
transformation $\mathfrak L$ is antimonotone, which implies that lower (upper)
bounds for $\calR$ imply upper (lower) bounds for $\calR^*$ and vice versa.

The lower bounds for $\calR$ and $\calR^*$ in \eqref{eq:Banach.Ass.e} hence imply
the upper bounds 
\[
\calR(u,v) \leq \psi_E^*(\|v\|) \qANDq \calR^*(u,\xi) \leq \psi_E^*(\|\xi\|),
\]
where $\psi^*(\zeta) = \sup\bigset{z\zeta - \psi(z)}{z\geq0}$. As $\psi$ is
finite everywhere, $\psi^*$ is again increasing, convex, and superlinear. 
As examples we can keep in mind $\psi(z)= cz^p-C$ for $p>1$ giving
$\psi^*(\zeta) = \wt c \zeta^{p^*}+C$ or $\psi(z)= (z{+}1)\log(z{+}1) -C$
giving $\psi^*(\zeta) = C +\ee^{\zeta-1} -\ee^{-1}$. Since $\calR$ and
$\calR^*$ are upper and lower bounded on each ball $B_R(0)$, they are even
Lipschitz continuous with bounded subdifferentials.

The continuity of $\calR$ in \eqref{eq:Banach.Ass.e} also provides upper and
lower bounds of $\calR(u_1,\cdot)$ in terms of $\calR(u_0,\cdot)$,
namely 
\begin{equation}
  \label{eq:Ru1.Ru0}
  -\omega_{1,0} + (1{-}\omega_{1,0})\calR(u_0,v) \leq \calR(u_1,v) \leq 
\omega_{1,0} + (1{+}\omega_{1,0}) \calR(u_0,v), 
\end{equation}
where $\omega_{1,0}= \omega^\calR_E\big( \|u_1{-}u_0\|\big) $. The upper bound
for $\calR(u_1,\cdot)$ transforms into a lower bound for $\calR^*(u_1,\cdot)$,
namely
\begin{equation}
  \label{eq:R*u1.R*u0}
  \calR^*(u_1,\xi) \geq   -\omega_{1,0} + (1{+}\omega_{1,0})\,
 \calR\big(u_0,\frac1{1{+}\omega_{1,0}}\,\xi \big).
\end{equation}

If for $w \in \rmW^{1,1}([0,T];X)$ we have the superlinear bound
$B:=\int_0^T\psi\big(\| \dot w\| \big) \dd t < \infty$, we obtain an explicit
equicontinuity.
\begin{align*} 
  \| w(t){-} w(s)\| & \leq \frac1\mu \int_s^t
  \mu\|\dot w\| \dd t \leq \frac1\mu \int_s^t\big( \psi^*(\mu)+ \psi(\|\dot
  w\|)\big) \dd t
 %\\ & 
\leq \big((t{-}s) \psi^*(\mu) + B\big)/\mu. 
\end{align*}
Taking the infimum over $\mu>0$, we obtain the desired result, namely 
\begin{equation}
  \label{eq:Banach:Equicont}
\big\| w(t){-}w (s)\big\| \leq \omega^B_\psi\big(|t{-}s|\big)
\quad \text{where } \omega^B_\psi(r):=
\inf\bigset{\frac1\mu\big(r\psi^*(\mu)+B\big) }{
  \mu>0}.
\end{equation}
For every $B>0$ the function $\omega^B_\psi$ is a modulus of
continuity, i.e., $\omega^B_\psi(r)\to 0$ for $r\to 0^+$.\bigskip

\noindent
\begin{proof}[Proof of Theorem \ref{th:BanachExist}] The proof consists of the
  typical steps.

\STEP{Step 0: construction of approximations via time-incremental
  minimization.} We first show that scheme in \eqref{eq:Banach:TIM}  
has minimizer $u_k=u^\tau_k$ for all $k =1,\ldots,N$. For this we use that $\calF$
is lsc (because of closed sublevels) and that $u \mapsto \tau
\calR \big( u_{k-1}, \frac1\tau(u {-} u_{k-1}) \big)$ is continuous
and coercive. Hence, $\Phi^{\calF,\calR}_\tau(u_{k-1}; \cdot)$ is lsc and
coercive. Moreover, the sublevels are compact, as they are contained in a
sublevel of $\calF$. Hence, by the one-sided Weierstra\ss\ extremal principle a
minimizer $u^\tau_k$ exists, namely 
\begin{equation}
  \label{eq:TIM.minim}
  \forall\, w \in X: \quad \tau \calR \big( u_{k-1}, \frac1\tau( u_k{-}
  u_{k-1}) \big) + \calF(u_k) \leq \tau \calR \big( u_{k-1}, \frac1\tau( w{-}
  u_{k-1}) \big) + \calF(w).
\end{equation}

As $\calR(u,\cdot)$ is convex and continuous, the sum rule gives
$\plF \Phi(u_*;u) = \pl_v\calR \big( u_*,\frac1\tau(u{-}u_*) \big) + \plF
\calF(u)$ and we obtain the inclusion
$ 0 \in \pl_v\calR \big( u_{k-1},\frac1\tau(u_k{-}u_{k-1}) \big) + \plF
\calF(u_k)$ or equivalently
\begin{equation}
  \label{eq:DiffIncl2}
 -\xi_k \in \pl_v\calR \big(
u_{k-1},\frac1\tau(u_k{-}u_{k-1}) \big)  \qANDq \xi_k \in \plF\calF(u_k) ,
\quad k=1,\ldots,N.
\end{equation}

\STEP{Step 1: a priori estimates.} Testing \eqref{eq:TIM.minim} with
$w=u_{k-1}$  we immediately
conclude $\calF(u_k) \leq \calF(u_{k-1}) \leq \calF(u^0) =:F_0< \infty$. 
Hence, all $u_k$ lie in the compact sublevel $S^\calF_{F_0} \Subset X$. 
The same test of \eqref{eq:TIM.minim} also provides a bound on increments,
namely 
\begin{align}
\label{eq:TIM.rate1}
\tau \psi\big(\frac1\tau \|u_k{-}u_{k-1} \| \big) & \leq
\tau \calR \big(u_{k-1},\frac1\tau(u_k{-}u_{k-1} ) \big))  \leq \calF(u_k) -
\calF(u_{k-1}). 
\end{align}

To obtain a supremum bound we divide by $\tau$ and estimate the energies:
\begin{equation}
  \label{eq:TIM.rateLinfty}
  \psi\big(\frac1\tau \|u_k{-}u_{k-1} \| \big) \leq \frac1\tau \big( \calF(u^0) -
F_\mafo{min}\big), \quad \text{where } F_\mafo{min} := \min_X \calF
>-\infty. 
\end{equation}
For an integral bound we sum \eqref{eq:TIM.rate1} over $k=1,\ldots,N$ to obtain 
\begin{equation}
  \label{eq:TIM.IntegrBdd}
   \int_0^T \psi\big( \| \dot{\wh u}_\tau\| \big) \dd t = \sum_{k=1}^N \tau \psi
 \big( \frac1\tau \|u_k{-}u_{k-1}\| \big) \leq \Delta_\calF:=\calF(u^0) -
F_\mafo{min} < \infty,
\end{equation}
where $\wh u_\tau$ is the piecewise affine interpolant. Hence,
\eqref{eq:Banach:Equicont} gives $\| \wh u_\tau(t){-}\wh u_\tau(s)\|  \leq
\omega^{\Delta_\calF}_\psi(|t{-}s|)$ for all $t,s\in [0,T]$.  

We also observe that $\dot{\wh u}_\tau$ is piecewise constant such that $\tau
\dot{\wh u}_\tau(t)= u^\tau_k{-}u^\tau_{k-1}$ for $t \in
{]k\tau{-}\tau,k\tau[}$.  Thus, we find $\tau \| \dot{\wh
  u}_\tau\|_{\rmL^\infty([0,T];X)} = \max\bigset{\| \wh u_\tau(k\tau){-} \wh
  u_\tau(k\tau{-}\tau) \| }{ k=1,...,N } \leq \omega^{\Delta_\calF}_\psi(\tau)
\to 0$ for $\tau  \to 0$. Later we will need the following estimate:
\begin{equation}
  \label{eq:TIM.tauSquared}
\begin{aligned}
E(\tau)&:=\tau \int_0^T \|\dot{\wh u}_\tau \|^2 \dd t  \leq \tau \|\dot{\wh
  u}_\tau\|_{\rmL^\infty} \int_0^T 1\, \|\dot{\wh u}_\tau \| \dd t 
\\
&  \leq  \omega^{\Delta_\calF}_\psi(\tau)\; \int_0^T \!\! 
 \Big( \psi^*(1)+  \psi\big( \| \dot{\wh u}_\tau\|\big) \Big) \dd t 
  \leq  \omega^{\Delta_\calF}_\psi(\tau)\;\big(T\psi^*(1){+} \Delta_\calF\big) , 
\end{aligned}
\end{equation}
such that $E(\tau)  \to  0$ for $\tau \to 0^+$.

To obtain an a priori estimate on the dual variable $\xi_k$, we proceed as at
the end of Section \ref{su:GenGS} where we use again the interpolants $\ol
u_\tau$, $\ul u_\tau$, and $\ol \xi_\tau$ and obtained the discrete approximate
energy-dissipation inequality \eqref{eq:Banach.DiscrEDI}, namely 
\begin{equation}
  \label{eq:TIM.discrEDI}
    \calF(\wh u_\tau(T)) + \int_0^T \!\!\Big(\calR \big( \ul u_\tau, \dot{\wh
    u}_\tau\big) {+}  \calR^*\big( \ul u_\tau, {-}\ol\xi_\tau\big)  
\Big) \dd t \leq \calF(u^0) - \frac{\lambda}2\, E(\tau), \quad \ol\xi_\tau \in
\plF\calF(\ol u_\tau). 
\end{equation}
This immediately implies 
\begin{equation}
  \label{eq:TIM.udot.xi}
  \int_0^T \!\! \Big( \psi\big( \|\dot{\wh u}_\tau\| \big) + \psi\big(
      \|\ol\xi_\tau\| \big)\Big) \dd t \leq \Delta_\calF + \frac{|\lambda|}2 \,E(\tau)
\leq \Delta_\calF + 1 <\infty, 
\end{equation}
for $0<\tau \ll 1$. Of course, from $u^\tau_k\in S^\calF_{F_0}\Subset X$ we
also have the a priori estimates
\[
\| \wh u_\tau\|_{\rmL^\infty([0,T];X)} \leq R, \quad
\| \ol u_\tau\|_{\rmL^\infty([0,T];X)} \leq R, \quad 
\| \ul u_\tau\|_{\rmL^\infty([0,T];X)} \leq R.
\]

\STEP{Step 2: extraction of convergent subsequences.} As all $\wh u_\tau$
satisfy the uniform bound \eqref{eq:TIM.IntegrBdd} we obtain equi-continuity via
\eqref{eq:Banach:Equicont}: 
\[
\| \wh u_\tau(t)- \wh u_\tau(s)\| \leq \omega^B_\psi \big(|t{-}s|\big) \quad
\text{with } B= \Delta_\calF .
\]
As the interpolants $\wh u_\tau$, $\ol u_\tau$, and $\ul u_\tau$ coincide for
$t=k\tau$ we conclude 
\[
\wt\omega(\tau):= \| \wh u_\tau - \ol u_\tau \|_{\rmL^\infty([0,T];X)} + 
 \| \wh u_\tau - \ul u_\tau \|_{\rmL^\infty([0,T];X)} \leq 2 \omega^B_\psi
 (\tau ) \to 0 \quad \text{for }\tau\to 0^+. 
\]
Moreover, exploiting the compactness of the sublevel $S^\calF_{F_0}$ in $X$, we
can apply the Arzel\`a-Ascoli selection principle to $(\wh u_\tau)_\tau $ and
obtain a subsequence (not relabeled) and a limit function
$u \in \rmC^0([0,T];X)$ such that
\[
\wh u_\tau \to u, \quad \ol u_\tau \to u, \quad \ul u_\tau \to u \quad \text{in
} \rmC^0([0,T];X).
\]

Moreover, since $\psi$ in  \eqref{eq:TIM.udot.xi} is superlinear, the criterion
of de la Vall\'ee Poussin shows that $(\dot{\wh u}_\tau)_\tau$ and $(\ol
\xi_\tau)_\tau$ are uniformly equi-integrable families in $\rmL^1([0,T];X)$ and
$\rmL^1([0,T];X^*)$, respectively. Hence there exists a further subsequence
(again not relabeled) and limits $v\in \rmL^1([0,T];X)$ and $\xi
\in\rmL^1([0,T];X^*)$ 
such that 
\[
\dot{\wh u}_\tau \weak v \ \text{ in } \rmL^1([0,T];X)  \qANDq 
\ol\xi_\tau \weak \xi \ \text{ in } \rmL^1([0,T];X^*). 
\]
Choosing a test function $\eta \in \rmC^1_\rmc({]0,T[};X^*)$ we can pass to the
limit $\tau \to 0^+$ in the identity $\int_0^T \langle \eta, \dot{\wh
  u}_\tau\rangle \dd t =- \int_0^T \langle \dot \eta, \wh u_\tau\rangle \dd t$
and find $v= \dot u$. Thus, we have $\wh u_\tau \weak u $ in
$\rmW^{1,1}([0,T];X) $ (along the subsequence chosen above). 

\STEP{Step 3: derivation of (EDI).} We derive (EDI) by passing to the limit
$\tau \to 0^+$ in \eqref{eq:TIM.discrEDI}. 

(3.a)  Because of $E(\tau)\to 0$ (cf.\
\eqref{eq:TIM.tauSquared}) the right-hand side in \eqref{eq:TIM.discrEDI}
converges to the desired limit $\calF(u^0) =F_0$. 

On the left-hand side we treat the three terms separately and note that it is
sufficient to derive a liminf estimate. 

(3.b) By $\wh u_\tau(T)\to u(T)$ and lower semicontinuity $\calF(u(T))
\leq \liminf_{\tau\to 0} \calF( \wh u_\tau (T) )$. 

(3.c) For the rate term $\calR$, we use the lower bound \eqref{eq:Ru1.Ru0} 
with $u_0=u(t)$, $u_1= \ul u_\tau(t)$, and $v= \dot{\wh u}_\tau(t)$. 
Note that $\ul u_\tau(t)$ and $u(t)$ lie in the sublevel $S^\calF_{F_0}$, such
that we can apply \eqref{eq:Ru1.Ru0} with
$\omega_{1,0}=\omega^\calR_{F_0}\big(\|\ul
u_\tau(t){-}u(t)\|_{\rmL^\infty}\big) \leq
\wh\omega(\tau):=\omega^\calR_{F_0}\big(\|\ul 
u_\tau{-}u\|_{\rmL^\infty}\big) 
\to 0$ for $\tau \to 0$. 
With this we obtain the estimate
\begin{align*}
& \liminf_{\tau\to 0} \int_0^T \calR\big( \ul u_\tau(t), \dot{\wh u}_\tau(t) \big)
\dd t   \geq \liminf_{\tau\to 0} 
  \int_0^T\!\! \Big(  {-} \wh\omega(\tau) + \big(1{-}\wh\omega(\tau)\big)
   \calR\big(u(t), \dot{\wh u}_\tau(t) \big) \Big) \dd t
\\
& = \liminf_{\tau\to 0} 
  \int_0^T\!\! \Big(  {-} 0 + 1\:
   \calR\big(u(t), \dot{\wh u}_\tau(t) \big) \Big) \dd t \geq 
\int_0^T\! \calR\big(u(t), \dot u(t) \big) \dd t,
\end{align*}
where in the last estimate we used the weak lower semicontinuity following from
the convexity of the primal dissipation potential $ \calR(u, \cdot)$. 

(3.d) For the $\calR^*$ term we proceed analogously now relying on
\eqref{eq:R*u1.R*u0}: 
\begin{align*}
& \liminf_{\tau\to 0} \int_0^T \!\!\calR^*\big( \ul u_\tau(t), {-}\ol \xi_\tau(t) \big)
\dd t   \geq \liminf_{\tau\to 0} 
  \int_0^T\!\! \Big(  {-} \wh\omega(\tau) + \big(1{-}\wh\omega(\tau)\big)
   \calR^*\big(u(t), \frac{-1}{1{-}\wh\omega(\tau) }\ol\xi_\tau(t) \big) \Big) \dd t
\\
& = \liminf_{\tau\to 0} 
  \int_0^T\!\! \Big(  {-} 0 + 1\:
   \calR^*\big(u(t), \frac{-1}{1{-}\wh\omega(\tau)} \ol\xi_\tau(t) \big) \Big) \dd t \geq 
\int_0^T\! \calR^*\big(u(t), {-}\xi(t) \big) \dd t. 
\end{align*}

Combining the results of (3.a-d) we obtain the desired EDI
\[
\calF(u(T)) + \int_0^T\big( \calR(u,\dot u) + \calR^*(u,{-}\xi)\big) \dd t \leq
\calF(u^0). 
\]
Clearly, we still have $u(0)=u^0$ and it remains to identify $\xi$. We recall
that for all $\tau$ we have 
$\ol\xi_\tau (t) \in \plF\calF(\ol u_\tau(t))$ for a.a.\ $t\in [0,T]$. 
Since $\ol u_\tau\to u$ and $\ol\xi_\tau\weak \xi$, we can use the strong-weak
closedness of $\plF\calF$ and obtain $\xi(t)\in \plF\calF(u(t))$ for a.a.\ $ t
\in [0,T]$. 

We remark here that this we need a generalization of the approach in Step 3
(see page \pageref{eq:HilbertAss.frF}) 
the proof of Theorem \ref{th:HilbertExist}, which relies on the result in
Exercise \ref{exerc:EvolClosed}. Instead we can exploit the result of Exercise  
\ref{exerc:EvolClosed2}, which only needs $\ul u_\tau \to u $ in
$\rmL^1([0,T];X)$ (strongly) and $\ol\xi_\tau \weak \xi$ in $\rmL^1([0,T];X^*)$
(weakly).

\STEP{Step 4: derivation of the gradient-flow equation.} It remains to apply
the Energy-Dissipation Principle from Theorem \ref{th:Banach.EDP}, which can be
applied because our assumption \eqref{eq:Banach.Ass.d} enforces the abstract
chain rule condition \eqref{eq:ACRcond}. Thus, we conclude that the constructed
pair $(u,\xi)$ satisfies the gradient-flow equation \eqref{eq:Banach.GFE3} and
the energy-dissipation balance \eqref{eq:Banach.EDB}. 
\end{proof}

We emphasize that in this case we are not able to show uniqueness. Thus,
different choices of the subsequences may lead to different solutions. Hence,
we cannot define a ``gradient flow'' as in Section \ref{su:Hilbert.EVI}. 

From the proof we can even learn more by observing that we did several liminf
estimates to obtain (EDI). However, later we showed that in fact (EDB)
holds. This implies that the liminf estimates must have been ``attained'' at least
along the chosen subsequence. Thus we additionally conclude:
\begin{itemize}
\item for $0\leq s < t\leq T$ we have $\calF(u(t))+ \int_s^t\!\big(\calR(..) 
  {+}\calR^*(..) \big) \dd r = \calF(u(s))$. 
\item $ \forall\, t\in [0,T]: \quad \calF(\wh u_\tau(t)) \to \calF(u(t)).$ 
\item $\int_0^T \calR(u , \dot{\wh u}_\tau) \dd t \to 
 \int_0^T \calR(u, \dot u) \dd t  $.

\item $\int_0^T \calR(u , \ol\xi_\tau) \dd t \to \int_0^T \calR(u , \xi) \dd t
  $.
\item $\calR^*(u(t),{-}\xi(t)) = \inf\bigset{\calR^*(u(t),{-}\eta}{ \eta\in
    \plF\calF(u(t)) }$ for a.a.\ $[t\in [0,T]$  
\end{itemize}
Hence, under additional strict convexity assumptions on $\calR(u,\cdot)$ and
$\calR^*(u,\cdot)$ it is even possible to show that the strong convergences 
$ \dot{\wh u_\tau} \to \dot u$ in $\rmL^1([0,T];X)$ and $\xi_\tau \to \xi$ in
$\rmL^1([0,T]; X^*)$, see \cite[Prop.\,C.3.3]{MieRou15RIST} for Visintin's
argument from \cite{Visi84SCRR}.

\begin{example}[State-dependent dissipation]\slshape
\label{ex:Ban.StateDepend}
We consider the GS $(\rmL^q(\Omega), \calF, \calR)$ with
\[
\dom(\calF)=\rmW^{1,p}_0(\Omega),\ \ \calF(u)=\int_\Omega \big( \frac1p
  |\nabla u|^p + F(u)\big) \dd x \qANDq 
\calR(u,v)= \int_\Omega \frac{a(u)}q  |v|^q \dd x,
\] 
where $\Omega\subset \R^d$ is a bounded Lipschitz domain, $p\in {]d,\infty[}$,
and $q\in {]1,\infty[}$. The function $F:\R\to {[0,\infty[}$ is $\rmC^1$ and
semiconvex and $a:\R\to {]0,\infty[}$ is continuous. 

\begin{exercise}
Formulate the associated gradient-flow equation and check that the
assumptions of Theorem \ref{th:BanachExist} hold. 
\end{exercise}
\end{example}

\subsection{Extensions} 
\label{su:Extensions}

We discuss a few possible extensions that allow us to widen the applicability
of the theory. 

\subsubsection{Time dependent gradient systems} 
\label{suu:TimeDepend}

Often one is interested in the case of time-dependent functionals
$\calF:[0,T]\ti X \to \Rinfty$. A typical case is $\calF(t,u)=\calE(u)- \langle
\ell(t), u\rangle$ implying that $\plF\calF(t,u)= \plF \calE(u) - \ell(t)$, where
the convention is now that $\plF \calF(t,u)= \plF \big(
\calF(t,\cdot)\big)(u)$. The forcing 
$\ell$ appears in the associated gradient-flow equation as source term:
\[
0 \in \pl\calR(u,\dot u) + \plF \calF(t,u) = \pl\calR(u,\dot u) + \plF \calE(u)
- \ell(t). 
\]
The above theory can be carried through under suitable technical assumptions
such as 
\begin{subequations}
\label{eq:BanTimDepAss}
\begin{align}
\label{eq:BanTimDep.a}
&\mathsf D:=\dom(\calF(t,\cdot)) \text{ is independent of } t \in [0,T],
\\
\label{eq:BanTimDep.b}
& \forall\, u \in \mathsf D: \quad \calF(\cdot,u) \in \rmC^1([0,T]),
\\
\label{eq:BanTimDep.c}
& \exists\, c_\rmF,\ C_\rmF>0 \ \forall\, u \in  \mathsf D, \ t\in [0,T]: \quad 
|\pl_t \calF(t,u)| \leq c_\rmF \calF(t,u) + C_\rmF,
\\
\label{eq:BanTimDep.d}
& (t_n,u_n) \to (t,u) \AND \sup_{n\in \N} \calF(t_n,u_n)<\infty \quad 
\text{imply} \quad \pl_t\calF(t_n,u_n) \to \pl_t\calF(t,u). 
\end{align}
\end{subequations}
With this the chain rule needs to be generalized into 
\[
\frac\rmd{\rmd t} \,\calF(t,u(t)) = \langle \xi(t),\dot u(t)\rangle +
\pl_t\calF(t,u(t))
\]
and the energy-dissipation balance takes correspondingly the form
\[
\calF(t,u(t)) + \int_s^t \big( \calR(u,\dot u)+ \calR^*(u,{-}\xi)\big) \dd r =
\calF(s,u(s)) + \int_s^t \pl_r\calF(r,u(r))\dd r ,
\]
where the last term can be understood as the work of the time-dependent
external forces. 

Now the energy $\calF(t,u(t))$ is no longer decreasing, but
\eqref{eq:BanTimDep.c} provides the upper bound 
\[
 \calF(t,u(t))+ C_\rmF \leq \ee^{c_\rmF (t-s)} \big( \calF(s,u(s)) +
 C_\rmF\big) \quad \text{for } 0\leq s < t.
\]

The construction of solutions still follows the time-incremental minimization
scheme \eqref{eq:Banach:TIM}, namely
\[
u^\tau_k \ \text{ minimizes } \ u \mapsto \tau \calR\big(u^\tau_{k-1},
\frac1\tau(u{-} u_{k-1}) \big) + \calF(k \tau, u). 
\]
As the minimizer $u^\tau_k$ satisfies $\xi^\tau_k \in
\plF\calF(k\tau,u^\tau_k)$ and $-\xi^\tau_k \in \pl\calR(u^\tau_{k-1},
\frac1\tau(u^\tau_k{-} u_{k-1}) \big) $, we can proceed as for
\eqref{eq:Ban.DiscEDB.la}. Using  the Fenchel equivalences and the
$\lambda$-convexity of $\calF(t,\cdot)$ we find 
\begin{align*}
& \calF(k\tau, u_k)+ \tau\Big(\calR \big( u_{k-1},
\frac1\tau(u_k{-}u_{k-1})\big) + \calR^*( u_{k-1}, {-}\xi_k) \Big) 
 \leq  \calF(k\tau, u_{k-1})  - \frac\lambda2 \|u_k {-} u_{k-1} \|^2
\\
& = \calF(k\tau{-}\tau, u_{k-1}) +\int_{k\tau-\tau}^{k\tau}
\pl_t\calF(t,u_{k-1}) \dd t  - \frac\lambda2 \|u_k {-} u_{k-1} \|^2.
\end{align*}

Using the interpolants as introduced in \eqref{eq:BanachInterpol} we see that
the approximate EDI \eqref{eq:Banach.DiscrEDI} generalizes to 
\begin{equation}
  \label{eq:Ban.AEDI.TimDep}
\begin{aligned}
  \calF(T,\ol u_\tau(T)) &+ \int_0^T \!\!\Big(\calR \big( \ul u_\tau, \dot{\wh
    u}_\tau\big) {+}  \calR^*\big( \ul u_\tau, {-}\ol\xi_\tau\big)  
\Big) \dd t\\
 &\leq \calF(0,u^0) + \int_0^T \pl_t\calF(t,\ul u_\tau) \dd t+ \frac{\tau\lambda}2\!
  \int_0^T  \!\|\dot{\wh u}_\tau(t)\|^2 \dd t.  
\end{aligned}
\end{equation}
From this, suitable a priori estimates can be derived and the limit passage
works as before, where \eqref{eq:BanTimDep.d} is used for the term involving
$\pl_t\calF$.

\subsubsection{Weakly compact sublevels}
\label{suu:WeakClosedSublev} 

A similar theory can be developed if the sublevels of the energy are not
compact in the strong topology, but only in the weak topology. The major
difference needed then, is that the closedness of the subdifferential has to be
imposed in the weak-weak topology. But this is the case of the leading term in
the energy is quadratic. For instance consider the Allen-Cahn energy
$\calF_{\rmA\rmC}$ on $H=\rmH^1_0(\Omega)$ with bounded $\Omega \subset \R^d$ and
$d\in \{1,2,3\}$:
\[
\calF_{\rmA\rmC}(u) = \int_\Omega \big( \frac\alpha2\,|\nabla u|^2 +
\frac\beta4\, (u^2{-}1)^2 \big) \dd x \quad \text{for }u \in
\mafo{dom}(\calF_{\rmA\rmC})  = H=\rmH^1_0(\Omega).
\]
Then $(u_n,\xi_n) \weak (u,\xi) $ in
$H\ti H^*= \rmH^1_0(\Omega)\ti \rmH^{-1}(\Omega)$ and
$\xi_n= \rmD\calF_{\rmA\rmC}(u_n) =\{ -\alpha \Delta u_n +
\beta(u_n^3{-}u_n)\}$. Hence, the embedding
$\rmH^1_0(\Omega) \leq \rmL^6(\Omega)$ for $p \in [1,p]$ which is compact for
$p<6$, implies boundedness of $\beta (u_n^3{-}u_n)$ and strong convergence to
the desired limit $\beta(u^3{-}u)$ in $\rmL^q(\Omega)$ for all $q\in
{[1,2[}$. Thus, we have the desired closedness
$\xi = \rmD \calF_{\rmA\rmC}(u)$.

\subsubsection{Approaches without semiconvexity and variational interpolants}
\label{suu:BanachVarInterpol}

Semiconvexity of the functional $\calF$ has proved to be a very useful
condition, because it implies closedness of the subdifferential, it helps to
establish the abstract chain rule, and it provides a simple approach 
the discrete EDI. However, for many applications semiconvexity is too strong
and it is desirable to avoid this assumption. 

For instance, in \cite[Prop.\,A.1]{MieRos23BVSI} the chain rule is established
under a much weaker ``uniform Fr\'echet differentiability''. Also the
closedness of the subdifferential can be shown by advanced PDE methods, thus
avoiding semiconvexity. 

The major problem is the derivation of the approximate discrete EDI, which then
provides an a priori estimate for the forces $\ol \xi_\tau$. The main idea is
to avoid the linear interpolation in the piecewise affine interpolant $\wh
u_\tau$, which can only be useful, if the functional $\calF$ can be controlled
along straight lines. The main new idea is due to Ennio De Giorgi, but he has never
published it. It can be found in the works \cite{Ambr95MM, AmGiSa05GFMS} of his
PhD student Luigi Ambrosio in the context of metric gradient flows, see Section
\ref{su:MetrVarInterpol}. For Hilbert-space gradient systems without
$\lambda$-convexity this idea was developed first in \cite{RosSav06GFNC} and
for generalized gradient systems on Banach
spaces in \cite[Lem.\,6.1]{MiRoSa13NADN}.  
 
We construct $(u^\tau_k)_{k=1:N}$ by time-incremental minimization as
before and define the \emph{variational (De Giorgi) interpolant} $\wt u_\tau:[0,T]\to
X$ such that for all $k=0,\ldots,N{-}1$ and $\theta \in {]0,1[}$ we have  
\[
\wt u_\tau\big(k\tau+\theta\tau\big) \ \text{ minimizes } \ 
u \mapsto   \theta\tau\,\calR\big(u^\tau_{k}, \frac1{\theta\tau}(u{-}u^\tau_{k})
\big) + \calF(u) .
\]   
As $\wt u_\tau(t)$ for $t \in {]k\tau,k\tau {+}\tau]}$ is obtained as a
minimizer, there is a $\wt\xi_\tau(t) \in \plF\calF(\wt u_\tau(t))$ with
$-\wt\xi_\tau(t)\in \pl\calR(u_k^\tau,
\frac1{t{-}k\tau}(\wt u_\tau(t){-}u_k^\tau)\big)$. Under suitable assumptions,
it is then possible to show
\[
\calF(u^\tau_k) + \tau \calR\big(u^\tau_{k-1}, \frac1\tau 
  (u^\tau_k{-}u^\tau_{k-1}) \big) + \int_{k\tau-\tau}^{k\tau} \calR^* 
 \big(u^\tau_{k-1}, {-}\wt\xi_\tau(t)\big) \dd t \leq \calF(u^\tau_{k-1})
\]  
with $\wt \xi_\tau(t) \in \plF\calF(\wt u_\tau(t))$ a.e.,
which replaces the former discrete EDI \eqref{eq:Ban.DiscEDB.la}, which was
derived using $\lambda$-convexity. 

But now $\lambda$-convexity is no longer
needed for obtaining the discrete EDI. It remains to generalize the abstract
chain rule to cases without $\lambda$-convexity.

\section{Metric gradient systems}
\label{se:MetricGS}

In this section we generalize the previous theory from Banach spaces to much
more general metric spaces, where we mainly follow \cite[Ch.\,2--4]{AmGiSa05GFMS}.
It is surprising that the concept of gradient
systems can be generalized to spaces without a linear structure. The main
reason for this is the variational character encoded in the time-incremental
minimization scheme via the energy functional
$\calF$ and the dissipation potential $\calR$. The major idea is to replace the
time derivative $\dot u(t) = \lim_{h\to 0} \frac1h\big( u(t{+}h)- u(t)\big) \in
X$ and the forces $\xi \in \plF \calF(u) \subset X^*$ by appropriate quantities
that are still available in metric spaces. In particular, convexity methods are
no longer available. In generalizing to metric spaces, we will also drop the
assumption of  semiconvexity that was very helpful in the Hilbert and Banach
space setting. For this we exploit the variational interpolant as introduced by
De Giorgi, see Section \ref{su:MetrVarInterpol}.

\subsection{Minimizing movements for metric gradient systems}
\label{su:MetrMinMov}

Throughout Section \ref{se:MetricGS} we will work with a complete metric space
$(M,\DD)$, i.e.\  $\DD:M\ti M \to {[0,\infty[}$ is a metric satisfying
positivity, symmetry and the triangle inequality. Completeness of $(M,\DD)$
means that all Cauchy sequences have a limit in $M$. For simplicity, we will
always use the topology on $M$ that is induced by the metric, however the
general theory needs to be developed by a second weaker topology, where
convergence is often denoted by $\overset{\sigma}\weak$, see e.g.\
\cite[Cha.\,3]{AmGiSa05GFMS}. This is in analogy to Banach spaces where
convergence in $\DD$ corresponds to norm convergence, whereas
$\overset{\sigma}\weak$ indicates weak convergence. 

\begin{definition}[Metric gradient systems and minimizing movements]
\label{de:MetrGSMM}
A quadruple $(M,\calF,\DD,\psi)$ is called \emph{generalized metric gradient system} if

\textbullet\ $(M,\DD)$ is a complete metric space,

\textbullet\ $\calF:M\to \Rinfty$ is a proper, lsc functional, 

\textbullet\ $\psi:\R \to [0,\infty]$ is a dissipation potential.
\\[0.2em]
Standard \emph{metric gradient systems} are given by the special choice
$\psi=\psi_\mafo{quadr} : r \mapsto \frac12r^2$. One then shortly writes 
$(M,\calF,\DD):= (M,\calF,\DD,\psi_\mafo{quadr})$. 
\medskip

The associated \emph{minimizing movement scheme  (MMS)} is given by 
\[
u^\tau_k \quad \text{minimizes} \quad u \ \mapsto \  \tau_k \, 
\psi\big( \frac1{\tau_k} \DD(u^\tau_{k-1}, u )\big) + \calF(u),
\]
where $\tau_k>0$ is a possibly variable time step. 
A curve $u:{[0,\infty[} \to M$ is called \emph{minimizing movement} for the
metric GS  $(M,\calF,\DD,\psi)$ if it is the limit (pointwise in $t$) of the
piecewise constant interpolants $\ol u _\tau:{[0,\infty[} \to M$ of the MMS
even, when varying time steps are allowed. One then writes $u \in
\rmM\rmM(M,\calF,\DD,\psi)$. If $u$ is only the limit of some sequence of
partitions (with fineness tending to $0$), then $u$ is called a
\emph{generalized minimizing movement} and we write  $u \in
\rmG\rmM\rmM(M,\calF,\DD,\psi)$.
\end{definition}

Note that the MMS for the standard metric GS
$(M,\calF,\DD)=(M,\calF,\DD,\psi_\mafo{quadr})$ leads to the standard 
\emph{minimizing movement scheme  (MMS)} given by 
\[
\boxed{\quad\text{standard MMS:} \qquad  u^\tau_k \quad \text{minimizes} \quad
  u \ \mapsto \   \frac1{2\tau_k} \,\DD(u^\tau_{k-1}, u )^2  + \calF(u),\quad }
\]
which is a direct generalization of the time-incremental minimization scheme
\eqref{eq:HilbertTIMS} in the Hilbert-space setting.

The notion of (generalized) minimizing movements can be seen as a solution
concept for metric GS. However, these solutions are only defined as limit
(or accumulation) points, which is a situation that is not always
satisfactory. The point is that 
it is difficult to derive further properties of solutions, in particular a
continuous dependence on a parameter $\mu$. The latter relies on
interchanging the two limits $\tau\to 0$ and $\mu_k \to \mu$, which is
absolutely nontrivial. See also Example \ref{ex:MetrMM}. Thus, it is desirable
to find a formulation of solutions that replaces the gradient-flow and
allows a direct study of solution without referring to the limiting process
$\tau\to 0$. 

\begin{example}[Missing upper semicontinuity for minimizing movements]\slshape
\label{ex:MetrMM}
Consider $M=\R^2$ with $\DD(u,w)=|u{-}w|_\text{Eucl}$ and $\psi(r)=\frac12r^2$
such that we are in the Hilbert-space setting $\calR(u,v)=\frac12(v_1^2{+}v_2^2)$ of
Section \ref{se:GSHilbert}. We choose the energy functional
$\calF(u)= \frac23 u_1 (u_1^2{+}u_2^4)^{5/4}$ which is smooth on
$\R^2\setminus\{0\}$. The gradient-flow equation reads 
\begin{equation}
  \label{eq:GFE.MMnotUSC}
  \dot u_1 = -\frac{3u_1^2 + 2u_2^4}{2(u_1^2{+}u_2^4)^{3/4}}, \quad 
\dot u_2 = -\frac{u_1u_2^3}{(u_1^2{+}u_2^4)^{3/4}},
\end{equation}
and Figure \ref{fig:MMnotUSC} shows the solutions. The vector field is locally
Lipschitz continuous on $\R^2\setminus\{0\}$, and the $u_1$ axis is invariant
because $\calF$ is even in $u_2$.
\begin{figure}
\begin{minipage}{0.32\textwidth}
\includegraphics[width=0.8\textwidth]{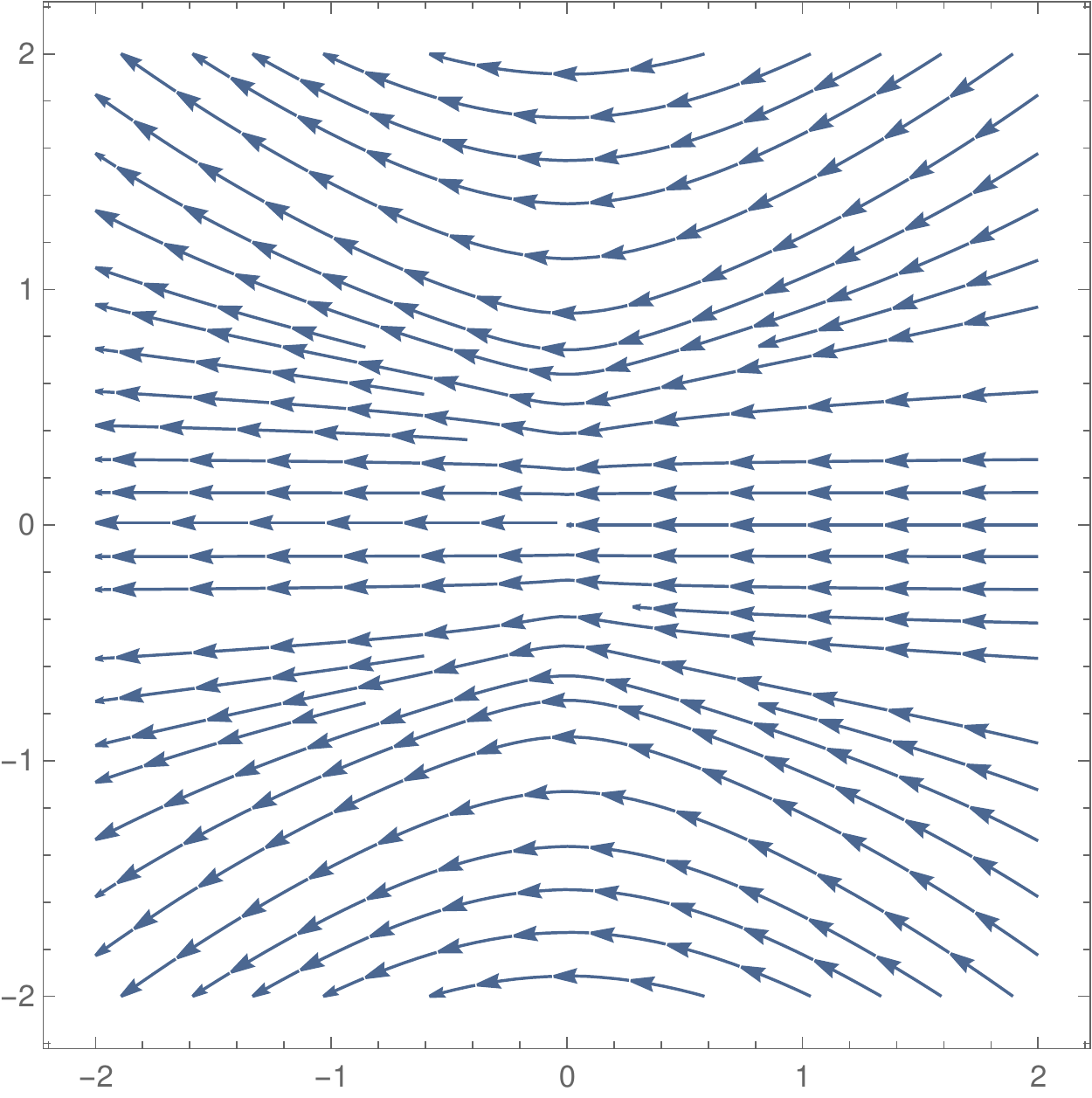}
\end{minipage}
\begin{minipage}{0.6\textwidth}
\caption{The figure shows the streamlines for the gradient-flow equations
  \eqref{eq:GFE.MMnotUSC}. All solutions satisfy $\dot u_1\leq 1$, and the
axis $u_2=0$ is invariant leading to the ODE $\dot u_1 = - |u_1|^{1/2}$ having
non-unique solutions. All other solutions stay away from $u_2=0$ and, hence,
are uniquely determined by their initial condition.} 
\label{fig:MMnotUSC}
\end{minipage}
\end{figure}

We consider the solutions starting at the initial points $u^0=(1,a)^\top$ for small
$a$ and denote these solutions by $t\mapsto U^a(t)$. We obviously have a
one-parameter family of solutions for $a=0$,  namely 
\[
U_{(\mu)}^0(t) = \begin{cases} \frac9{16}\big(\frac43{-}t\big)^2& \text{for }
  t\in [0,\frac43], \\  
0&\text{for } \frac43\leq t<\mu, \\ 
\frac9{16}(t{-}\mu)^2 & \text{for } t\geq \mu. 
\end{cases}
\] 
For $a\neq 0$ the solutions $U^a$ are unique, as they never hit the non-Lipschitz
point $u=(0,0)^\top$. To see this, observe that for $u_1(t)\leq 0$ we have $\dot u_2\geq
0$, while for $u_1(t)\in [0,1]$ and $u_2\geq 0$ we have $(u_1^2{+}u_2^4)^{3/4} \geq
\max\{u_1^{3/2},u_2^3\}$ which gives $\dot u_2 \geq - \min\{u_1,
u_2^3/u_1^{1/2}\}$ and bounds $u_2$ away from $0$. 
Hence, we see that taking the limit $0\neq a\to 0$ (from above or from below)
we see that   
\[
\forall \, t>0: \quad  U^a(t)\to U^0_{(4/3)}(t) \qquad \text{for } 0\neq a\to 0.
\]  

Below we will show that for this example, the MMS determines, for each initial
datum, a unique minimizing movement in the sense of Definition
\ref{de:MetrGSMM}.  The theory to be developed below will show that the MMS
always provides as least one solution (a GMM) for each initial condition
$u^0=(1,a)^\top$. If the solution $U^a$ of the gradient-flow equation is
unique, then the GMM is an MM and coincides with this solution.

This is the case for $a\neq 0$.  However, also for $a=0$ we obtain a unique
minimizing movement. For $a=0$ we start on the invariant line $u_2=0$ and the
following Euler-Lagrange equations show that we always stay there: $u=u^\tau_k$
has to satisfy
\[
\frac1\tau\big( u - u^\tau_{k-1}\big) + \frac1{2(u_1^2{+}u_2^4)^{3/4}}
\binom{3u_1^2+ 2 u_2^4}{ 2u_1 u_2^3}= \binom{0}0. 
\]
By induction over $k$ we see first that $u^\tau_{k,2}=0$ and then that
\[
\frac1\tau \big(u^\tau_{k,1}{-}u^\tau_{k-1,1}\big) + \frac32\sqrt{u^\tau_{k,1}}
=0  \quad 
\Longrightarrow \quad \sqrt{u^\tau_{k,1}}  = \sqrt{ \frac{9\tau^2}{16} +
  u^\tau_{k-1,1}} - \frac{3\tau}4  \ >\ 0.
\]
Since all $u^\tau_{k,1} $ are positive, we have convergence to the only solution
$U^0_{(\mu)}$ that is non-negative, i.e.\ $\mu = \infty$.

The important remark is now that considering the limit $0\neq a \to 0$ we see
that the limit of
$U^a= \mafo{MM} \big( \R^2,\calF, |\cdot|_\rmE, (1,a)^\top\big) $ is not in
$\mafo{GMM}\big(\R^2,\calF,|\cdot|_\rmE,(1,0)^\top \big)$, i.e.\ the solution
set is not upper semi-continuous. The notion of curves of maximal slope
encompasses this disadvantage of MM or GMM.

We refer to \cite{FleSav20RAGF} for a way to modify the MMS to obtain all
curves of maximal slope. 
\end{example}

\subsection{Curves of maximal slope}
\label{su:MetrCurvesMaxSlope}

The solution concept ``curves of maximal slopes for the GS
$(M,\calF,\DD,\psi)$'' will be tailored exactly to contain all $u \in
\rmG\rmM\rmM(M,\calF,\DD,\psi)$. Moreover, it is a direct generalization of the
solutions concept derived for Banach-space GS $(X,\calF,\calR)$. 

The major idea of generalizing the gradient-flow theory from Banach spaces to
metric spaces is obtained by looking at special classes in the Banach-space
setting. For this we consider generalized GS $(X,\calF,\calR)$ in a Banach
space $X$ with dissipation potentials $\calR(u,v)=\psi\big(\|v\|_X\big)$ where
$\psi:\R\to [0,\infty]$ is a scalar dissipation potential. The dual dissipation
potential reads $\calR^*(u,\xi)= \psi^*\big( \| \xi\|_{X^*}\big)$. Note that
this choice is a special instance of a generalized metric GS, where we choose
$M=X$ and $\calD(u,w)=\| w{-}u\|_X$. Moreover, the energy dissipation balance
(EDB), see e.g.\ \eqref{eq:Banach.EDB}, now takes the special form
\[
\calF(u(T)) + \int_0^T\!\! \Big(\psi\big(\|\dot u\|_X\big) + \psi^*\big( \|\rmD
\calF(u(t))\|_{X^*}\big) \Big) \dd t = \calF(u(0)),
\]  
where we assumed that $\plF\calF(u)$ is the singleton
$\big\{\rmD\calF(u)\big\}$.

The main observation is that within this special
class, we do not need the vector-valued quantities 
$\dot u(t)\in X$ and $\rmD\calF(u(t))\in X^*$, but it is enough to control the
real-valued quantities $\|\dot u(t)\|_X \in \R$ and
$\|\rmD\calF(u(t))\big\|_{X^*} \in \R$. We will see below that there are natural
generalizations of these two real-valued quantities in the metric setting,
where no linear structure is available. 

We first study absolutely continuous curves $\gamma:[0,T]\to M$ in the metric
space $(M,\DD)$. 

\begin{definition}[Absolutely continuous curves]
\label{de:MetrAC.Speed}
A curve $\gamma : [0,T] \to M$ is called \emph{absolutely continuous in
  $(M,\DD)$}, if there exists a function $g\in \rmL^1([0,T])$ such that
\begin{equation}
  \label{eq:MetrSpeedBd}
\DD\big(\gamma(t_1),\gamma(t_2)\big) \leq \int_{t_1}^{t_2} g(t) \dd t \quad
\text{for all } t_1,t_2 \in [0,T] \text{ with } t_1<t_2. 
\end{equation}
We then write $\gamma \in \AC\big([0,T];(M,\DD)\big)$ or shortly 
 $\gamma \in \AC\big([0,T];M\big)$ if $\DD$ is clear from the context. 
If additionally $g\in \rmL^p([0,T])$ for some $p\in [1,\infty]$, we write
$\gamma \in \AC^p\big([0,T];M\big)$. 
\end{definition}

As in the case of Banach spaces, we also have the embeddings in the Hölder
spaces $\AC^p([0,T];M) \subset \rmC^{1-1/p}([0,T];M)$, which follows via
H\"older's inequality:
\[
  \DD\big(\gamma(t_1),\gamma(t_2)\big) \leq \int_{t_1}^{t_2}1\cdot  g(t) \dd t
\leq \Big(\int_{t_1}^{t_2}1^{p^*} \dd t\Big)^{1/p^*} \Big(
\int_{t_1}^{t_2}g(t)^p \dd t \Big)^{1/p} \leq \big|t_2{-}t_1\big|^{1/p^*} \| g\|_{\rmL^p},
\]
where $p^*=p/(p{-}1)$. If $(M,\DD)$ is given by a reflexive Banach space
$(X,\|\cdot\|)$, then we have $\AC^p([0,T];X) = \rmW^{1,p}([0,T];X)$. However,
for general Banach spaces we only have the inclusion
$ \rmW^{1,p}([0,T];X) \subset \AC^p([0,T];X) $. As an example consider
$X=\rmL^1(\R)$ with the standard norm. Now consider the curve
$\wh\gamma:[0,T]\to \rmL^1(\R)$ with
\[
\wh\gamma(t)=\bm1_{[0,\cosh(t)]}:\ x \mapsto \left\{ \ba{cl}1 &\text{for } x \in
  [0,\cosh(t)],\\ 0 &\text{otherwise.} \ea\right. 
\]
Clearly, we have
$\| \wh\gamma(t_1)-\wh\gamma(t_2)\|_{\rmL^1} = |\cosh(t_2){-}\cosh(t_1)| $ such
that $\wh\gamma \in \AC^p([0,T];\rmL^1(\R))$ for all $p$ with function
$g: t\mapsto \sinh(t)$. However, $\wh\gamma$ does not lie in
$\rmW^{1,1}([0,T];\rmL^1(\R))$ because for $h>0$ the difference quotients
$\frac1h \big(\wh\gamma(t{+}h){-}\wh\gamma(t)\big) = \frac1h
\bm1_{{]\cosh(t),\cosh(t{+}h)]}}$ converge to $\sinh(t)\,\bfdelta_t(\cdot)$
(Dirac distribution at $x=t$) in the sense of measures, but do not converge in
$\rmL^1(\R)$, even though the difference quotients are bounded.

The following result from \cite[Thm.\,1.1.2]{AmGiSa05GFMS} shows that the
metric speed is well-defined a.e.\ along absolutely continuous curves.

\begin{theorem}[Metric speed]
\label{th:MetrSpeed}
For $p\in [1,\infty]$ assume $\gamma \in \AC^p\big([0,T];M\big)$. Then, the
metric speed $\SPE{\gamma}(t)$ defined via 
\[
\SPE \gamma (t) := \lim_{|h|\to 0} \frac1{|h|}\DD\big(\gamma(t),\gamma(t{+}h)\big)
\]
exists a.e.\ in $[0,T]$ and $\SPE \gamma(\cdot) \in \rmL^p([0,T])$. 

Moreover, for every $g$ satisfying \eqref{eq:MetrSpeedBd}, we have $\SPE \gamma
\leq g$ a.e.
\end{theorem}
\begin{proof} We choose a countable dense set $\bigset{s_n \in [0,T]}{ n \in
    \N}$ and define the auxiliary functions 
\[
\delta_n(t)= \DD\big(\gamma(s_n), \gamma(t)\big)  \quad \text{for } t \in
[0,T].
\]
By the inverse triangle inequality we find, for $0\leq t_1 <  t_2  \leq T$, 
\begin{equation}
  \label{eq:Metr.delta.n}
  |\delta_n(t_2)- \delta_n(t_1)| \leq
\DD(\gamma(t_1), \gamma(t_2)) \leq \int_{t_1}^{t_2} g (t)\dd t ,
\end{equation}
Thus, we conclude $\delta_n\in \AC^p([0,T])=\rmW^{1,p}([0,T])$, where we use that
$X=\R$ is a reflexive Banach space. Thus, $\delta_n$ is differentiable
a.e., more precisely $\dot\delta_n(t)=\lim_{h\to 0} \frac1h
\big(\delta_n(t{+}h)-\delta_n(t)\big) $ exists for $t\in [0,T]\setminus E_n$
with $\calL^1(E_n)=0$. Clearly, $|\dot\delta_n| \leq g$ a.e. 

We now define the function
\[
\mu(t)= \sup \bigset{ |\dot\delta_n(t)| }{ n\in \N } \ \text{ for } t\in
[0,T]\setminus E  \quad \text{and} \quad \mu(t)=0 \ \text{ for }t\in
E:=\mathop{\cup}\limits_{n\in \N} E_n 
\]
and observe $\mu \leq g \in \rmL^p([0,T])$.  Using \eqref{eq:Metr.delta.n} for
all $t\in [0,T]\setminus E$ we
find
\[
\liminf_{h\to 0} \frac1{|h|} \DD\big(\gamma(t),\gamma(t{+}h)\big) \geq
\sup_{n\in \N} \Big(
\liminf_{h\to 0}  \frac1{|h|}\big|\delta_n(t)-\delta_n(t{+}h)\big| \Big) = 
\sup_{n\in \N} \big| \dot\delta_n(t) \big| = \mu(t). 
\]

Moreover, if $s_{n_k} \to t_1$ then $\delta_{n_k}(t_2) -  \delta_{n_k}(t_1) 
\to \DD\big(\gamma(t_1),\gamma(t_2)\big)-0$.  Together with
\eqref{eq:Metr.delta.n} we observe, for $h>0$, 
\[
\DD\big(\gamma(t),\gamma(t{+}h)\big)  =
\sup_{ n\in \N} \big|\delta_n(t)-\delta_n(t{+}h)\big|
\leq \sup_{ n\in \N} \int_t^{t+h} |\dot\delta_n(r)| \dd r \leq \int_t^{t+h}
\mu(r) \dd r .
\]
Dividing by $h>0$ and doing the corresponding estimate for $h<0$ we arrive at
\[
\limsup_{h\to 0} \frac1{|h|} \DD\big(\gamma(t),\gamma(t{+}h)\big) \leq
\limsup_{h\to 0} \frac1{|h|} \Big| \int_t^{t+h} \mu(r) \dd r  \Big| \ = \ \mu(t),
\]
for a..a.\ $t\in [0,T]$, namely all right and left Lebesgue points of $\mu \in
\rmL^p([0,T])$. 

Together we have shown $\frac1{|h|} \DD\big(\gamma(t),\gamma(t{+}h)\big) \to
\mu(t) \leq g(t)$ a.e.
\end{proof}

We may return to the above example $\wh\gamma:[0,T]\to \rmL^1(\R)$ which does
not lie in $\rmW^{1,1}\big([0,T];\rmL^1(\R)\big)$. We can now easily verify
that the metric speed in $\rmL^1(\R)$ exists for all $t\in [0,T]$, namely 
$\BAR \dot{\wh\gamma}\BAR_{\rmL^1}(t) = \sinh(t)$. \bigskip 

The second important notion for metric gradient systems is a scalar notion for
the differential $\plF \calF:X \tosets X^*$ of the energy  functional
$\calF:X\to \Rinfty$.  In the
following definition of the metric slope $\SLO \calF$ we call a point $u\in M$
isolated if there exists a positive $r$ such that $B_r(u)\cap M = \{u\}$ and
use the notation $[  F ]_+:=\max\{F,0\}$ for the positive part. 

\begin{definition}[Metric slope]
\label{de:MetrSlope}
Given a metric GS $(M,\calF,\DD,\psi)$ we define the \emph{(local) metric slope  $\SLO
  \calF: M \to [0,\infty]$ of the functional $\calF$} via
\begin{equation}
  \label{eq:defMetrSlope}
  \SLO \calF (u):= \begin{cases} \infty & \text{for } u\not\in \dom(\calF), \\
         0& \text{for isolated } u \in \dom(\calF), \\
   \limsup\limits_{w\to u} \dfrac{\big[\calF(u)-\calF(w)\big]_+} {\DD(u,w)} &
\text{ for nonisolated }u \in \dom(\calF).  
  \end{cases}
\end{equation}
For $\lambda \in \R$ we also define the \emph{global metric $\lambda$-slope
  $\SLO{{}^\mafo{gl}_\lambda \calF}$ of $\calF$} via
\begin{equation}
  \label{eq:defGlobMetrSlope}
\SLO{{}^\mafo{gl}_\lambda \calF}(u):= \begin{cases} 
\infty & \text{for }u \not\in \dom(\calF), \\
\sup\limits_{w\in M\setminus\{u\}} \Big[ \dfrac{\calF(u){-}\calF(w)}{\DD(u,w)}
          + \dfrac\lambda2\, \DD(u,w)\Big]_+ 
& \text{for } u \in \dom(\calF).  
\end{cases}
\end{equation}
We say that $\calF$ \emph{has a $\lambda$-global metric slope} if
$\SLO{{}^\mafo{gl}_\lambda \calF} = \SLO\calF$, and we say that $\calF$ \emph{has
a semiglobal metric slope} if there exists $\lambda\in \R$ such that
$\SLO{{}^\mafo{gl}_\lambda \calF} = \SLO\calF$. 
\end{definition}

From the definitions we easily see that $\lambda_1 < \lambda_2$ implies 
\[
\SLO\calF(u) \leq \SLO{{}^\mafo{gl}_{\lambda_1} \calF}(u) \leq
\SLO{{}^\mafo{gl}_{\lambda_2} \calF}(u)  . 
\]
The important consequence of the $\lambda$-global slopes is that we have the
estimate
\begin{equation}
  \label{eq:MetricLamGlob}
  \forall\, u\in \dom(\calF)\ \forall \, w\in M:\quad 
\calF(w) \geq \calF(u) - \SLO\calF(u)\,\DD(u,w) + \frac\lambda2\,\DD(u,w)^2,
\end{equation}
which is a generalization of the characterization of the Fr\'echet 
subdifferential in Lemma~\ref{le:CharFrechSub}. In particular, $\SLO\calF(u)$
is the smallest number such that \eqref{eq:MetricLamGlob} for all $w\in M$.

The notion of semiglobal and $\lambda$-global metric slopes will play a similar
role as semiconvexity and $\lambda$-convexity of functionals in the
Banach-space setting. In Sections \ref{se:GSHilbert} and \ref{se:GSBanach} we
used semiconvexity for three important steps, namely (i) showing strong-weak
closedness of the Fr\'echet subdifferential, (ii) establishing the chain rule,
and (iii) deriving a discrete energy-dissipation estimate from the
time-incremental minimization scheme. In the metric setting the notion of
``semiglobal slopes'' will be good enough to how (i') the lower semicontinuity
of the metric slope and (ii') a metric chain-rule estimate.

\begin{example}[Local and semiglobal slopes]
\label{ex:Slopes}\mbox{} 

(A) Consider $(M,\DD)=(\R,\DD_\mafo{Eucl})$ and $\calF(u) =a^\pm u$ for $\pm
u\geq 0$. For $u >0$ we obviously have $\SLO\calF(u)=|a^+|$, while
$\SLO\calF(u)=|a^-|$ for $u<0$. The case $u=0$ is special and we obtain
$\SLO\calF(u)=\max\{ 0, a^-,{-}a^+\}$.\medskip 

(B) Consider $(M,\DD)=(\R,\DD_\mafo{Eucl})$ and $\calF(u)=\big||u|{-}1\big|$. 
We easily find the local slope $\SLO\calF(u)=1$ for $u\neq \pm1$ and
$\SLO\calF(\pm1)=0$, i.e.\ $\SLO\calF$ is lsc but not continuous. 

For the $\lambda$-global slope we obtain
$\SLO{{}^\mafo{gl}_\lambda\calF}=\SLO\calF$ for all $\lambda\leq 0$, i.e.\ 
$\SLO\calF$ is a semiglobal slope. 

For $\lambda>0$ we obtain larger values, e.g.\ for $u>$ we have 
$\SLO{{}^\mafo{gl}_\lambda\calF}(u) = 1+\frac\lambda2(u{-}1)$ if $\lambda\in
{]0,\frac2{u{+}1}]}$ and $\SLO{{}^\mafo{gl}_\lambda\calF}(u)=\frac{u-1}{u+1} +
  \frac\lambda2(u{+}1)$ for $\lambda \geq \frac2{u{+}1}$.\medskip

(C)If $(M,\DD)$ is given by a Banach space $(X;\|\cdot\|)$ and $\calF:X\to
\Rinfty$ is lsc, then for $u\in \dom(\calF)$ we have 
\[
  \SLO\calF (u)= \bigset{\| \xi\|_{X^*}}{ \xi \in \plF\calF(u)}.
\]
Moreover, if $\calF$ is $\mu$-convex, then $\SLO\calF$ is a
$\lambda$-global slope for all $\lambda\geq \mu$. 
\end{example}

\begin{exercise}[Slopes]
\label{exer:Slopes} (a) $(M,\DD)=(\R,\DD_\mafo{Eucl})$ consider
$\calF(u)=\min\{u,0\}$. Calculate $\SLO\calF$ explicitly and show that it is
not lsc. Moreover, calculate $\SLO{{}^\mafo{gl}_\lambda\calF}$ for all $\lambda
\in \R$. 

(b) Establish the claims in part (C) of Example \ref{ex:Slopes}.  
\end{exercise}

Note that the proof of the following result is very similar to the
corresponding closedness of the Fr\'echet subdifferential for semiconvex
functionals, see Proposition \ref{pr:ClosedFrechet}. Part (A) in Example
\ref{ex:Slopes} shows that $\SLO\calF$ is not lsc in general, see Exercise
\ref{ex:Slopes} 

\begin{proposition}[Lsc of semiglobal metric slopes]
\label{pr:Metric.LSC} 
If $\calF$ is lsc and has a semiglobal slope $\SLO\calF$ on the metric space
$(M,\DD)$, then $\SLO\calF:M\to [0,\infty]$ is lower semicontinuous.
\end{proposition}
\begin{proof} For a sequence $u_k\to u$ in $(M,\DD)$,  we have to show
$\sigma:=\liminf_{k\to \infty} \SLO\calF(u_k) \geq \SLO\calF(u)$. Obviously,
the case $\alpha=\infty$ is trivial. 

Hence we assume $\sigma<\infty$ which implies $u_k\in \dom(\calF)$. Thus we
have \eqref{eq:MetricLamGlob} for $u=u_k$ for all $k$. Using the lsc of $\calF$
and $\DD(u_k,w)\to \DD(u,w)$ we immediately find 
\[
\calF(w)\geq \calF(u)- \sigma  \,\DD(u,w) + \frac\lambda2\, \DD(u,w)^2
\quad \text{for all } w\in M.
\]
But this implies $ \SLO\calF(u) \leq \sigma $, which is the desired estimate. 
\end{proof}

We have now all the ingredients to define the metric version of the generalized
gradient-flow equation. 

\begin{definition}[Curves of maximal slope]
\label{de:CurvMaxSlope}
Given a generalized metric GS $(M,\calF,\DD,\psi)$ we call a curve $u:[0,T]\to
M$ a \emph{$\psi$-curve of maximal slope} if $u \in \AC\big([0,T];(M,\DD)\big)$
and for all $t_1,t_2\in [0,T]$ with $t_1<t_2$ we have
\begin{equation}
  \label{eq:def.CurMaxSlo}
  \calF(u(t_2)) + \int_{t_1}^{t_2} \!\Big( \psi\big( \SPE u(t)\big)+
  \psi^*\big( \SLO\calF (u(t))\big) \Big) \dd t = \calF(u(t_1)). 
\end{equation}
If $\psi(r)=\frac1p r^p$ for $p \in {]1,\infty[}$ we shortly say that $u$ is a
\emph{$p$-curve of maximal slope of} $(M,\calF,\DD)$. 
If $\psi=\psi_\mafo{quadr}: r \mapsto \frac12 r^2$, then $u$ is simply called a
\emph{curve of maximal slope} for the standard metric GS $(M,\calF,\DD)$. 
\end{definition}

As we have learned in Sections \ref{se:GSHilbert} and \ref{se:GSBanach}, we
know that the above formulations are enough to characterize the solutions of
the corresponding gradient-flow equations, if we are in the special case
$\calR(u,v)= \psi(\|v\|_X)$. Of course, the notion is much more general as will
become clear by the following examples. 

\begin{example}[Different instances of curves of maximal slope]
\label{ex:CurMaxSlo} \mbox{}
\begin{enumerate}[label={(\Alph*)}, leftmargin=2em]

\item {\bfseries Nonuniqueness.} We consider $(\R,\calF,\DD_\mafo{Eucl},
  \psi_\mafo{quadr})$ with $F(u)=\frac12u^2-|u|$. 

This system can also be treated as a Hilbert-space GS but, then the
subdifferential $\plF\calF$ is not closed: $\plF\calF(u)=\{u-1\}$ for $u>0$ and
$\plF\calF(0)=\emptyset$. 

Treating it as a metric GS leads to the metric slope $\SLO\calF(u)=\big|
1{-}|u|\big|$ which is even semiglobal with $\lambda =0$. 

We now show that there are two solutions starting at $u^0=0$, namely
$u(t)=\pm(1{-}\ee^{-t})$. To show that these two solutions are curves
of maximal slope, we can simply check that 
\[
\frac\rmd{\rmd t} \calF(u(t)) = - \frac12 \big(\SPE u(t)\big)^2 - \frac12 
\big(\SLO\calF(u(t)))^2 
\]
by inserting the explicit solutions.
 
\item {\bfseries Riemannian manifold.} We consider a Riemannian manifold
  $(M,\bbG)$ with a smooth functional $\calF\in \rmC^1(M)$. For the smooth GS
  $(M,\calF,\bbG)$ we have the associated GFE $\dot u = - \mafo{grad}_\bbG\calF(u)=-
  \bbG(u)^{-1} \rmD \calF(u)$. 

We now want to switch to the metric picture. For this we define the metric
distance
\[
\DD_\bbG(u_0,u_1):= \inf \Bigset{\int_0^1  \| \dot\gamma\|_\bbG \dd
  t}{ \gamma \in \rmC^1([0,T];M),\ \gamma(0)=u_0, \ \gamma(1)= u_1 }, 
\] 
where $\|\dot\gamma \|_\bbG^2 = \langle \bbG(\gamma)\dot\gamma, \dot\gamma
\rangle$. 

Doing some classical calculations in local charts one finds
$\AC^p \big( [0,T];(M,\DD_\bbG)\big) = \rmW^{1,p} ([0,T];M)$ and
$\SPE u(t) = \|\dot \gamma (t)\|_\bbG$.

Similarly, the metric slope takes the form $\SLO\calF (u) = \|
\rmD\calF(u)\|_{\bbG^{-1}} $. 

With this the condition for curves of maximal slope takes the form 
\begin{align*}
0&= \frac\rmd{\rmd t} \calF(u(t)) + \frac12 \SPE u(t)^2+ \frac12\SLO\calF
(u(t))^2
\\
&= \langle \rmD\calF(u),\dot u\rangle + \frac12\langle \bbG(u)\dot u,\dot
u\rangle + \langle \rmD\calF(u),\bbG(u)^{-1} \rmD\calF(u)\rangle \\
& = \frac12\big\langle \bbG(u)\big(\dot u{-}\bbG(u)^{-1}\rmD\calF(u)\big) , 
\dot u{-}\bbG(u)^{-1}\rmD\calF(u)\big\rangle = \frac12\big\| \dot
u{-}\bbG(u)^{-1}\rmD\calF(u) \big\|_\bbG^2.  
\end{align*}
Thus, we see that for this nice case the metric formulation is equivalent to
the classical gradient-flow equation. 

\item {\bfseries Wasserstein space and Otto diffusion.} We consider a bounded
  open set $\Omega\subset \R^d$ and denote by $\mafo{Prob}(\ol\Omega)$ the
  space of probability measures, which is a closed convex subset of the signed
  measures $\rmS\rmM(\ol\Omega) = \big( \rmC(\ol\Omega)\big)^*$. On this set
  the Kantorovich-Wasserstein distances $\rmW_p$ (cf.\ \cite{AmGiSa05GFMS,
    Vill09OTON}) are defined via
\[
\rmW_p(\mu_0,\mu_1)^p := \inf\Bigset{ \int_{\ol\Omega\ti \ol\Omega} |x{-}y|^p
  \Pi(\rmd x,\rmd y) }{ \Pi \in \calC(\mu_0,\mu_1) },  \quad \text{where}
\]
$\calC(\mu_0,\mu_1):=\bigset{ \Pi {\in} \mafo{Prob} (\ol\Omega\ti\ol\Omega)}{
  \forall\,\text{meas.}\, A \subset 
  \ol\Omega : \ \Pi(A\ti \ol\Omega)=\mu_0(A),\ \Pi(\ol\Omega\ti A)=\mu_1(A)} .$ 
For all $p\in {[1,\infty[}$, the pair $(\mafo{Prob}(\ol\Omega), \rmW_p)$
defines a complete metric space and the convergence is equal to the weak*
convergence. 

For $p=2 $ the Wasserstein space $(\mafo{Prob}(\ol\Omega), \rmW_2)$ is even a
geodesic metric space (see Definition \ref{de:GeodMetrSpaces}) which has many
similarities with a Riemannian manifold with nonsmooth boundaries. In a series
of papers around 2000, the corresponding metric theory was developed and
summarized in \cite{AmGiSa05GFMS}. The metric speed of a curve
$\mu \in \AC\big( [0,T]; (\mafo{Prob}(\ol\Omega), \rmW_2) \big)$ can be defined
as follows. For every such function, there exists  a vector field
$V \in \rmL^1([0,T] \ti \ol\Omega;\R^d)$ such that the continuity equation
\[
\dot \mu + \DIV\big( V \mu\big) = 0 \quad \text{holds in }
\big( \rmC^\infty_\rmc({]0,T[}\ti \ol\Omega)\big)^* 
\]
(where ``$_\rmc$'' stands for compactly contained support),
and the metric speed takes the form $\SPEED{\mu}{\rmW_2}(t) = \big(
\int_{\ol\Omega} |V(t,x)|^2 \mu(t,\rmd x) \big)^{1/2}$ for a.a.\ $t\in [0,T]$. 

Similarly one can derive a formula for the metric slope of certain
functionals. Choosing a lsc, convex, and superlinear functional $E:{[0,\infty[}\to
[0,\infty]$ and $\varphi\in \rmC^1(\ol\Omega)$ on can define 
\[
\calF(\mu) = \begin{cases} \int_\Omega \big( E(\rho(x)) + \varphi(x) \rho(x)\big) \dd
  x & \text{for } \mu = \rho \dd x \text{ with }\rho\in \rmL^1_{\geq}(\Omega),
  \\ 
\infty & \text{otherwise},
\end{cases}  
\]
where $\rmL^1_\geq(\Omega)$ denotes the non-negative functions in
$\rmL^1(\Omega)$. 
Then, $\calF$ is lsc on $(\mafo{Prob}(\ol\Omega), \rmW_2)$ and the metric slope
is given by 
\[
 \SLOPE{\calF}{\rmW_2}(\rho\rmd x) = \Big( \int_\Omega \big| \nabla
 (E'(\rho){+}\varphi)\big|^2 \rho \dd x \Big) ^{1/2},
\]
see \cite{AmGiSa05GFMS} for more precise statements and the justification of
these relations.

Under the assumption that a curve of maximal slope for $( 
\mafo{Prob}(\ol\Omega), \calF, \rmW_2)$ has the form $\mu(t) = \rho(t,\cdot)
\dd x$ with $\rho$ sufficiently smooth and bounded from below, one can show
that $\rho$ satisfies a drift-diffusion equation. We argue as in Example (B): 
\begin{align*}
0&\ \ = \ \  \frac\rmd{\rmd t} \calF(\mu(t)) + \frac12 \SPEED{\mu}{\rmW_2}(t)^2 
+ \frac12 \SLOPE{\calF}{\rmW_2}(\mu(t))^2 \\
&\overset{\mu=\rho\rmd x} = \int_\Omega\Big((E'(\rho){+} \varphi)\dot\rho + \frac12
\rho|V|^2 + \frac12 \rho\big|\nabla(E'(\rho){+}\varphi)\big|^2 \Big) \dd x \\
&\ \ \overset{*}= \ \ \int_\Omega \frac\rho2 \big| V{+} \nabla
(E'(\rho){+}\varphi) \big|^2 \dd x ,   
\end{align*}
where for the last identity we inserted the continuity equation and integrated
by parts, thus finding a complete square. 

This shows that for curves $\mu=\rho\dd x$ of maximal slope (with $\rho$
sufficiently smooth and positive), the velocity field $V$ in the continuity
equation can be identified as $-\nabla (E'(\rho)+\varphi)$. This leads to the
nonlinear drift-diffusion equation
\[
\dot \rho = - \DIV(\rho V) = \div\big( \rho\nabla(E'(\rho){+}\varphi)\big) =
\DIV\big( \rho E''(\rho) \nabla \rho + \rho \nabla \varphi\big).
\]
In particular, one may consider the Boltzmann entropy with $E(\rho)=\rho \log
\rho - \rho +1$. Then $E''(\rho)=1/\rho$ and we left with the linear
Fokker-Planck equation as the associated gradient-flow equation
\[
\dot \rho = \DIV\big( \nabla \rho + \rho \nabla \varphi \Big). 
\]
This link between Wasserstein distance and the linear Fokker-Planck equation
was first observed in \cite{Otto96DDDE,JoKiOt98VFFP}. For that reason the
Minimizing Movement Scheme in the case of $\DD=\rmW_2$ is nowadays called the
JKO scheme for ``Jordan-Kinderlehrer-Otto''. 

For the entropies $E(\rho)=\big(\rho^m-m\rho +m-1\big)/(m^2{-}m) $  we have
$E''(\rho)=\rho^{m-2}$ and in the case $\varphi= 0$ the associated
gradient-flow equation is the \emph{porous medium equation}
\[
\dot \rho= \DIV\big( \rho^{m-1}\nabla \rho\big) = \frac1m \Delta \rho^m. 
\]

\end{enumerate}
\end{example}

\begin{exercise}[Nontrivial metric space]
\label{exer:NontrivMetric}
We consider $M=\R^k$ with the nontrivial metric
$\DD_\mafo{sq}(u,w)=\sqrt{|u{-}w|_\mafo{Eucl}}$ that is topologically equivalent to the
Euclidean one. 

(a) Show that $\AC\big([0,T];(\R^k,\DD_\mafo{sq})\big)$ is trivial in the
sense that it only contains constant functions. 

(b) For a smooth function $\calF\in \rmC^1(\R^k)$ calculate $\SLO\calF$ with
$\DD=\DD_\mafo{sq}$. 

(c) Characterize all curves of maximal slope for $(M,\calF,\DD_\mafo{sq},\psi)$
in terms of their initial condition $u(0)=u^0$.  
\end{exercise}

\subsection{The metric chain-rule inequality}
\label{su:MetrCRIneq}

To prove existence of curves of maximal slopes we will use the minimization
scheme in a similar way as for in the Banach-space setting. In this subsection
we develop the corresponding replacement of the chain rule formula
$\frac\rmd{\rmd t} \calF(u(t)) = \langle \xi(t), \dot u(t)\rangle$ with $\xi(t)
\in \plF\calF(u(t))$. As we do not have any linear structure in the metric space
$(M,\DD)$, we use the fact that in the existence proof for the GFE in Banach
spaces we do not really need the above chain-rule identity. 
From the Fenchel-Young inequality we already have
an inequality such that it would be
sufficient to have the lower estimate $\frac\rmd{\rmd t} \calF(u(t)) \geq
\langle \xi(t), \dot u(t)\rangle$. It turns out that in the metric setting a
corresponding chain-rule inequality can be established. 

\begin{definition}[Abstract metric chain-rule inequality]
\label{de:AbstrMetrCRI}
We say that the generalized metric GS $(M,\calF,\DD,\psi)$ satisfies the
\emph{abstract metric chain-rule inequality} if the following holds. 
\begin{quote}
If $u\in \AC([0,T];M)$ satisfies $\sup_{t\in [0,T]}\calF(u(t)) <\infty$\\[0.3em]
\hspace*{2em}
and $\int_0^T \big( \psi(\SPE u(t)) + \psi^*\big(\SLO\calF(u(t))\big)\big) \dd t
<\infty$,\\[0.3em]
then $ t \mapsto \calF(u(t))$ is absolutely continuous and 
\end{quote}
\begin{equation}
  \label{eq:MetrCRIneq}
   \frac\rmd{\rmd t} \calF(u(t)) \geq 
   -\SLO\calF(u(t))\,\SPE u(t)\quad \text{a.e.\ in } [0,T]. 
\end{equation}
\end{definition}

We will see that this inequality is enough for completing the
existence proof for curves of maximal slope. The next result demonstrates
that the condition that $\SLO\calF$ is a semiglobal slope is sufficient 
for showing that the abstract metric chain-rule inequality holds. 
Moreover, the proof is almost identical to the
corresponding Theorem \ref{thm:BanachChainRule} in Banach spaces. In fact, the
origin of the proof of the latter result is \cite[Thm.\,1.2.5]{AmGiSa05GFMS},
which is almost identical to our next result. Hence, a full proof for the case
$\SLO\calF = \SLO{{}^\mafo{gl}_0\calF}$ can be found there. Here we only give a
sketch of the proof, by referring back to our proof of Theorem
\ref{thm:BanachChainRule}. 

\begin{proposition}[Metric chain-rule inequality]
\label{pr:MetrCRIneq}
Consider a generalized metric GS $(M,\calF,\DD,\psi)$ such that $\SLO\calF$ is
a semiglobal slope. Then, the following holds:
\begin{quote}
If $u\in \AC([0,T];M)$ satisfies $\sup_{t\in [0,T]}\calF(u(t)) <\infty$\\
\hspace*{2em}
and $\int_0^T \big( \psi(\SPE u(t)) + \psi^*\big(\SLO\calF(u(t))\big)\big) \dd t
<\infty$,\\[0.2em]
then $ t \mapsto \calF(u(t))$ is absolutely continuous and
\end{quote}
\begin{equation}
  \label{eq:MetrCREst}
   \big|\frac\rmd{\rmd t}
\calF(u(t))\big|  \leq \SLO\calF(u(t))\,\SPE u(t) \quad \text{a.e.\ in } [0,T]. 
\end{equation}
In particular, for all dissipation potentials $\psi:\R\to {[0,\infty[}$ the
abstract metric chain-rule inequality of Definition \ref{de:AbstrMetrCRI} holds.  
\end{proposition}
\begin{proof}
[Sketch of proof]
We follow the proof of Theorem \ref{thm:BanachChainRule} and set
$f(t)=\calF(u(t))$ and $\sigma(t)=\SLO\calF(u(t))$. Using that $\calF$ has a
semiglobal slope, i.e.\ there exists $\lambda \in \R$ such that $\SLO\calF=
\SLO{{}^\mafo{gl}_\lambda \calF}$, and choosing any partition of $[s,t]$ lying in
$\Sigma:=\bigset{t\in [0,T]}{ \sigma(t)<\infty}$, we obtain the estimates 
\begin{align*}
&\sum_{j=1}^N \Big({-}\sigma(t_{j-1})\DD\big(u(t_{j-1}),u(t_j)\big) + \frac\lambda2
\DD\big(u(t_{j-1}),u(t_j)\big)^2 \Big)\\
&\quad \leq  \calF(u(t))- \calF(u(s))= f(t)-f(s)
\\
&\qquad\leq \sum_{j=1}^N \Big({-}\sigma(t_{j})\DD\big(u(t_{j-1}),u(t_j)\big) + 
\frac\lambda2 \DD\big(u(t_{j-1}),u(t_j)\big)^2\Big),
\end{align*}   
which correspond to \eqref{eq:Ba.LowUppEst} in the case of arclength
parametrization, i.e.\ $\DD(u(s_1),u(s_2))=|s_2{-}s_1|$ for all $s_1,s_2\in
[0,T]$. 

As before we can pass to the limit and find the desired estimate 
\[
\big| \calF(u(t)) - \calF(u(s))\big| \leq \int_s^t \sigma(r)\: \SPE u(r) \dd r,
\]
which provides the absolute continuity as well as the desired estimate.  

Finally, the abstract metric chain-rule inequality follows by applying the
Fenchel-Young inequality to the scalar dissipation potential $\psi$, namely 
\[
\int_0^T \sigma(r)\: \SPE u(r) \dd r \leq \int_0^T \!\! 
\Big(\psi\big(\sigma(r)\big) + \psi^*\big( \SPE u(r)\big) \Big) \dd r <
\infty. 
\]

As \eqref{eq:MetrCREst} implies \eqref{eq:MetrCRIneq}, Proposition
\ref{pr:MetrCRIneq} is established.  
\end{proof}

In \cite{AmGiSa05GFMS} the names ``chain rule'' and ``metric chain-rule
inequality'' are not used as here. There, the same notion is encoded in the
term ``\emph{strong upper gradient}''. For instance,
\cite[Thm.\,1.2.5]{AmGiSa05GFMS} states that ``if $\calF$ is $\calD$-lsc, then
$\SLO{{}^\mafo{gl}_0\calF}$ is a strong upper gradient for $\calF$'', which
means that the metric chain-rule inequality holds, if we replace
$\SLO\calF$ by $\SLO{{}^\mafo{gl}_0\calF}$ in \eqref{eq:MetrCREst}. 

As for generalized GS $(X,\calF,\calR) $ in Banach spaces we again have an
Energy-Dissipation Principle in the following form. 

\begin{proposition}[Metric energy-dissipation principle]
\label{pr:MetrEDP}
Consider a generalized metric GS $(M,\calF,\DD,\psi)$ that satisfies the
abstract metric chain-rule inequality of Definition \ref{de:AbstrMetrCRI}.  
Then, for $u\in \AC([0,T];M)$ the following two statements are equivalent:

(A) $u$ satisfies the $\mafo{EDI}$ given by  $\calF(u(T)) + \int_0^T \big( \psi(\SPE u)+
\psi^*(\SLO\calF(u)) \big) \dd t \leq \calF(u(0))$. 

(B) $u$ is a $\psi$-curve of maximal slope, i.e.\ $\mafo{(EDB)}_{[t_0,t_1]}$ holds for
$0\leq t_0<t_1\leq T$.  
\end{proposition}
\begin{proof} \STEP{$(B)\ \Longrightarrow (A)$.} This direction is trivial. 

\STEP{$(A)\ \Longrightarrow (B)$.} We set 
\[
 f(t)= \calF(u(t)), \quad v(t)=\SPE u(t), \quad \text{ and }
 \sigma(t)=\SLO\calF(u(t))
\]
and observe that the chain-rule inequality implies $\dot f + \sigma v\geq 0$ a.e. 
Hence, we obtain
\[
0\leq \int_0^T\!\big(\dot f + \sigma v\big) \dd t 
  \overset{\text{FenYou}}\leq 
\int_0^T \!\big(\dot f + \psi(v)+ \psi^*(\sigma)\big) \dd t 
\overset{\text{(EDI)}}\leq 0. 
\]
Thus, all inequalities ``$\leq$'' must be equalities ``$=$''. Moreover the
nonnegative integrand  $\dot f + \psi(v)+ \psi^*(\sigma) \geq \dot f + \sigma
v\geq 0$ must vanish a.e.\ in $[0,T]$. However, integrating $\dot f + \psi(v)+
\psi^*(\sigma) =0$ a.e.\ over $t\in [t_0,t_1]$ given (EDB) on $ [t_0,t_1]$.
\end{proof}

\subsection{De Giorgi's variational interpolant}
\label{su:MetrVarInterpol}

%% Start 12. Januar 2022 

In the subsequent analysis we will use the  following assumptions on $\psi$:
\begin{equation}
  \label{eq:Ass.psi}
  \psi:\R \to {[0,\infty[} \text{ is a strictly convex dissipation potential
    and } \psi\in \rmC^1\big({[0,\infty[}\big). 
\end{equation}

The MMS gives global minimizers $u_k=u^\tau_k$, namely 
\begin{equation}
  \label{eq:MMS.varInter}
  \forall \, w \in M: \quad \tau \psi\big( \frac1\tau \DD(u_{k-1}, u_k)\big)  +
\calF(u_k) \leq \tau \psi\big( \frac1\tau\DD(u_{k-1},w)\big) + \calF(w).
\end{equation}
From this we can derive a first slope estimate.

\begin{proposition}[Metric slope estimate]
\label{pr:MetrSlopeEstim}
Assume that $\psi$ satisfies \eqref{eq:Ass.psi} and let $(u_k)_{k=1,..,N}$ be
the sequence obtained via the MMS for the generalized metric GS
$(M,\calF,\DD,\psi)$, then
\[
\SLO\calF(u_k) \leq \psi'\big( \frac1\tau\DD(u_{k-1},u_k)\big) \quad \text{for
} k=1,\ldots, N.
\]
\end{proposition} 
\begin{proof} Rearranging the terms in \eqref{eq:MMS.varInter} for $w\neq u$ we have 
\[
\frac{\calF(u_k) - \calF(w) }{ \DD (u_k,w) } \leq 
\frac{\psi\big(\frac1\tau \DD(u_{k-1},w) \big) - \psi\big(\frac1\tau
  \DD(u_{k-1},u_k) \big)} {\frac1\tau \DD(u_k,w)}.
\]
Using the triangle inequality and the monotonicity of $\psi$ we have $ 
\psi\big(\frac1\tau \DD(u_{k-1},w) \big) \leq \psi\big(\frac1\tau
\DD(u_{k-1},u_k) + \frac1\tau \DD(u_k,w) \big)$. Now taking the limit $w\to
u_k$ we find 
\begin{align*}
&\limsup_{w\to u_k} \frac{\calF(u_k) {-} \calF(w) }{ \DD (u_k,w) } 
\leq \limsup_{w\to u_k} \frac{\psi\big(\frac1\tau
\DD(u_{k-1},u_k) {+} \frac1\tau \DD(u_k,w) \big) - \psi\big(\frac1\tau
\DD(u_{k-1},u_k) \big) } {\frac1\tau \DD(u_k,w)}
\\
&= \limsup_{H\to 0^+}  \frac{\psi\big(\frac1\tau
\DD(u_{k-1},u_k) + H \big) - \psi\big(\frac1\tau
\DD(u_{k-1},u_k) \big) } {H} \ = \ \psi'\big(\frac1\tau
\DD(u_{k-1},u_k) \big) .
\end{align*} 
As the right-hand side is non-negative (as $\psi$ is a dissipation potential),
we can take the positive part on both sides and obtain the desired result.  
\end{proof}

In some sense, the last result can be seen as a generalization of the
Euler-Lagrange equations $0 \in \pl\calR\big(\frac1\tau(u_k{-}u_{k-1})\big) +
\plF\calF(u_k)$ in the Banach-space setting. There we used Fenchel's
equivalence and $\lambda$-convexity of $\calF$ to derive the discrete EDI 
(with $\xi_k \in \plF\calF(u_k)$) 
\[
\tau\Big( \calR\big( \frac1\tau(u_k{-}u_{k_1})\big) + \calR^*( {-} \xi_k)
\Big) \leq -\langle \xi_k,u_k{-}u_{k-1}\rangle \leq \calF(u_{k-1}) - \calF(u_k)
- \frac\lambda2\| u_k{-}u_{k-1}\|^2.
\]
The importance of this inequality is the telescoping structure with respect to
the energies $\calF(u_j)$.

In the metric setting we can also apply Fenchel's equivalence to the scalar
relation $\sigma_k:=\psi'(v_k)$ where
$v_k:=\frac1\tau\DD(u_{k-1},u_k)$. Exploiting the slope estimate in Proposition
\ref{pr:MetrSlopeEstim} we obtain
\begin{align*}
& \tau\Big(\psi\big( \frac1\tau \DD(u_{k-1},u_k)\big) {+}\psi^*\big(
\SLO\calF(u_k)\big) \Big)  \overset{\text{Prop.}}\leq 
\tau\big(\psi( v_k) {+}\psi^*\big(\psi'(v_k)\big) \big)\\
&= \tau\big(\psi( v_k) {+}\psi^*(\sigma_k) \big) \overset{\text{Fenchel}}=
\tau \sigma_k v_k = \psi'\big(\frac1\tau\DD(u_{k-1},u_k)\big)\, 
\DD(u_{k-1},u_k)\\
& \overset{??}\leq \ \ \SLO\calF(u_k)\:\DD(u_{k-1},u_k).  
\end{align*}
The last estimate $\overset{??}\leq$ would be necessary to exploit the
$\lambda$-global slope in \eqref{eq:MetricLamGlob} for generating again a
discrete energy estimate with a telescoping structure. However, this would mean
that one has to show equality in the slope estimate of Proposition
\ref{pr:MetrSlopeEstim}, which is false in general metric spaces, e.g.\ in the
simple case $M=R$, $\calF(u)=\frac\alpha u^2/2$, and $\DD(u,w)=\arctan(|u{-}w|)$. 

De Giorgi's variational interpolant will be a way around this problem and, much
more importantly, paves the way to solve problems without semiconvexity in 
Banach spaces or semiglobal slopes in metric spaces. The definition of the
interpolant is based on minimization only. We refer to \cite[Lem.\,2.5]{Ambr95MM} and
\cite[Def.\,3.2.1]{AmGiSa05GFMS} for the first occurrences of the variational
interpolant.  

\begin{definition}[De Giorgi's variational interpolant]
\label{de:DeGiorgiVarInter} For a generalized metric GS $(M,\calF,\DD,\psi)$,
a starting point $u^0\in M$, and a time step $\tau=T/N$ we consider a discrete
approximant $(u^\tau_k)_{k=0,..,N}$ obtained via the MMS for $u^\tau_0=u^0$. 
Then, the variational interpolants $\wt u_\tau:[0,T]\to M$ are defined via $\wt
u_\tau(j\tau)=u^\tau_j$ for $j=0,1,\ldots,N$ and 
\[
\wt u_\tau(k\tau{+}r) \text{ minimizes } \ w \mapsto
\Phi_r(u_{k}^\tau , w ) := r\psi\big(\frac1r\DD(u^\tau_{k},w) \big) + \calF(w). 
\]
for $k=0,\ldots,N{-}1$ and $r \in {]0,\tau[}$. 
\end{definition}

In general, the variational interpolant will not be continuous in $t$, but
nevertheless it has good properties, because it is created by the intrinsic
building blocks of the metric GS. In particular, we will not need the
interpolant $\wt u_\tau$ so often, but can rely on the so-called \emph{value function} 
\[
\phi (r,u_{k-1}^\tau ) :=  \Phi_r\big(u_{k-1}^\tau , \wt u_\tau(k \tau {+}r)
\big). 
\]
It will turn out that $r \mapsto \phi(r,u_{k-1}^\tau ) $ is absolutely
continuous and that the derivative can be expressed by the derivative of
$ r \mapsto \Phi_r(u_{k-1}^\tau, w)$.  For showing this, we introduce the auxiliary
function
\[
\Psi:{[0,\infty[}^2 \to [0,\infty]; \quad \Psi(r,a)= \begin{cases} 
0 &\text{for } a=0, \\ r\psi(a/r) & \text{for } r>0, \\ 
\infty& \text{for } r=0 \text{ and } a>0.
\end{cases} 
\]

The following gives a series of properties of $\Psi$ that will be used in the
upcoming analysis. We leave the elementary proof to the reader. 

\begin{lemma}[Properties of $\Psi$]
\label{le:Metric.Psi} 
Assume that the dissipation potential $\psi:\R \to {[0,\infty[}$ satisfies
\eqref{eq:Ass.psi}, then $\Psi:{[0,\infty[}^2 \to [0,\infty]$ is lsc and
satisfies  the following properties:
\begin{enumerate}
[label={\upshape(\roman*)}, leftmargin=4em, rightmargin=2em]
%[label={\upshape(\roman*)}, leftmargin=4em, rightmargin=2em]

\item For all $a\geq 0$ the function $r\mapsto \Psi(r,a)$ is decreasing with 
      $\pl_r \Psi(r,a)= - \psi^*\big( \psi'(a/r)\big)$ for all $r>0$. 
\item For all $a\geq 0$ the function $r\mapsto \Psi(r,a)$ is strictly convex.
\item For $0<r_1<r_2$ the function $a \mapsto \Psi(r_1,a)- \Psi(r_2,a)$ is
      strictly increasing. 
\end{enumerate} 
\end{lemma}

With this we are able to provide some first results concerning the value
function. For this we introduce a few simplifying notations. We fix a state
$\UU \in M$ and define, for $r>0$, 
\begin{align*}
\phi(r,\UU) &:= \inf\bigset{\Phi_r(u_*, w)}{ w\in M} ,
\\
A(r,\UU)&:= \mafo{Arg\!\;min}\bigset{\Phi_r(\ol u, w)}{ w\in M}:= 
 \bigset{u_r \in M}{ \Phi_r(\ol u,u_r)= \phi(r,\UU)},
\\
d^+(r,\UU)&:=\sup\bigset{ \DD(\ol u,u_r)}{ u_r \in A(r,\UU)}, \quad 
  d^-(r,\UU):=\inf\bigset{ \DD(\ol u,u_r)}{ u_r \in A(r,\UU)}.
\end{align*}

\begin{proposition}[Value function and distances]
\label{pr:ValueFctDist} Consider a generalized metric GS $(M,\calF,\DD,\psi)$
such that $\calF$ has compact sublevels and $\psi$ satisfies
\eqref{eq:Ass.psi}. Then for all $\UU \in M$ and $r_2>r_1>0$ the functions
$\phi$, $A$, and $d^\pm$ are well defined and satisfy 
\begin{enumerate}
[label={\upshape(\alph*)}, leftmargin=4em, rightmargin=2em]

\item\label{EN:phi.monotone} 
$\phi(r_2,\UU) \leq \phi(r_1,\UU) \leq \calF(\UU)$;

\item\label{EN:d+.d-} 
  $d^+(r_2,\UU) \geq d^-(r_2,\UU) \geq d^+(r_1,\UU)$; 

\item\label{EN:phi.to.calF}
 $\phi(r,\UU) \to \calF(\UU)$ for $r \to 0^+$;

\item\label{EN:d+to0} 
If $\UU \in \ol{\dom(\calF)}$, then $d^+(r,\UU)\to 0$ for $r\to 0^+$. 
\end{enumerate} 
Property \ref{EN:d+.d-} implies that
$d^\pm(\cdot,\UU):{]0,\infty[} \to \R$ are increasing and continuous for
$t\in [0,\tau]\setminus J$, where $J$ is at most countable. Moreover,
$d^+(r,\UU)=d^-(r,\UU)$ for all $\in {]0,\infty[} \setminus J$.%
\end{proposition}
\begin{proof}
\STEP{Step 1: Wellposedness and attainment.} We first observe that the
  properties that $\calF$ is proper and has compact sublevels guarantee that
  $A(r,\UU)$ is nonempty and compact.  Hence, the infimum in the definition of
  $\phi$ is attained. Moreover, by Weierstra\ss' extreme-value principle the
  continuous function $w \mapsto \DD(\UU,w)$ attains its minimum and maximum on
  $A(t,\UU)$.

\STEP{Step 2: Monotonicity \ref{EN:phi.monotone}.}  Clearly
$\Phi_r(\UU,w)=\Psi(r,\DD(\UU,w)) + \calF(w)$ is decreasing in $r$. Thus, 
choosing any $u_j \in A(r_j,\UU)$ we
have 
\[
\calF(\UU)=\Phi_{r_1}(\UU,\UU)\geq 
\phi(r_1,\UU)=\Phi_{r_1}(\UU,u_1) \geq \Phi_{r_2}(\UU,u_1) \geq
\Phi_{r_2}(\UU,u_2)= \phi(r_2,\UU),
\]
which is the desired monotonicity.

\STEP{Step 3. Intertwining property.} We trivially have $d^+\geq d^-$ because
$\sup \geq \inf$. However, it is absolutely nontrivial that $d^+(r_1,\UU)$ can
be estimated from above by $d^-(r_2,\UU)$ for all $r_2>r_1$. To see this, we
have to exploit the special structure of $\Phi_r(\UU,w)= \Psi(r,\DD(\UU,w)) +
\calF(w)$. Again choose arbitrary $u_j \in A(r_j,\UU)$ and set
$\DD_j:=\DD(\UU,u_j)$, then we have
\begin{align*}
\Psi(r_1,\DD_1)&+ \calF(u_1)= \Phi_{r_1}(\UU,u_1)=\phi(r_1,\UU) \leq
\Phi_{r_1}(\UU,u_2)= \Psi(r_1,\DD_2)+ \calF(u_2) \\
& = \Phi_{r_2}(\UU,u_2) + \Psi(r_1,\DD_2) - \Psi(r_2,\DD_2) 
\leq \Phi_{r_2}(\UU,u_1) + \Psi(r_1,\DD_2) - \Psi(r_2,\DD_2) \\
& = \Psi(r_2,\DD_1) + \calF(u_1)  + \Psi(r_1,\DD_2) - \Psi(r_2,\DD_2). 
\end{align*}
We observe that $\calF(u_1)$ can be eliminated on both ends, and rearranging
gives
\[
\Psi(r_1,\DD_1)-\Psi(r_2,\DD_1) \leq \Psi(r_1,\DD_2)- \Psi(r_2,\DD_2). 
\]
Now we can exploit Lemma \ref{le:Metric.Psi}(iii) and conclude $\DD(\UU,u_1) =
\DD_1 \leq \DD_2 = \DD(\UU,u_2)$. As $u_j \in A(r_j)$ were arbitrary, we can
take the supremum over $u_1$ and the infimum over $u_2$ and obtain
$d^+(r_1,\UU)\leq d^-(r_2,\UU)$ as desired. 

\STEP{Step 4: $d^+=d^-$ whenever one is continuous.} By the last step we know
that $d^+$ and $d^-$ are increasing functions. Hence, they are continuous for
all $t$ except for an at most countable jump set $J^+$ or $J^-$, respectively. 
However, if $d^+$ is continuous at $r_*>0$, then for $0<\eps_n\to 0$ we have 
\begin{align*}
d^+(r_*,\UU) \leftarrow 
d^+(r_*{-}\eps_n,\UU) &\leq d^-(r^*{-}\eps_n/2,\UU) \leq d^+(r_*,\UU)
\\
&
\leq d^-(r_*{+}\eps_n/2,\UU) \leq d^+(r_*{+}\eps_n,\UU) \to
d^+(r_*,\UU). 
\end{align*}
As $\eps_n\to 0$ was arbitrary, we conclude that $d^-$ is continuous at
$r=r_*$, i.e.\ $J^-\subset J^+$, as well as $d^+(r_*,\UU)=d^-(r_*,\UU)$. 
Interchanging ``+'' and ``$-$'' we obtain $J^+=J^-=:J$, and the final assertion
is established.

\STEP{Step 5: $\phi(r,\UU)\to \calF(\UU)$.} From Step 1 we have $\phi(r,\UU)
\leq \calF(\UU)$. Hence, by the monotonicity we have $\phi(r,\UU)\to \phi_*
\leq \calF(\UU)$ for $r \to 0^+$. 

Choose $d^+(r,\UU)\to d_*>0$ for $r\to 0$, then $\phi(r,\UU)=\Psi(r,\DD(\UU,u_r))+
\calF(u_r) \geq \Psi(r,d^+(r,\UU))+\calF_\text{min} \to \infty$. This means
$\phi_*=\calF(\UU)=\infty$ and the assertion holds. 

If $d^+(r,\UU)\to 0$ for $r\to 0^+$, then there exists $u_r \in A(t,\UU)$ such
that $\DD(\UU,u_r)\leq d^+(r,\UU)\to 0$, and the lsc of $\calF$ implies
$\liminf_{r\to 0^+} \calF(u_r)\geq \calF(\UU)$. Because of $\Psi\geq 0$ we find
\[
\phi(r,\UU) = \Phi_r(\UU,u_r) = \Psi(r,\DD(\UU,u_r))+\calF(u_r) \geq
\calF(u_r),
\]
which now implies $\phi_*\geq \calF(\UU)$. Thus $\phi_*=\calF(\UU)$ is
established. 

\STEP{Step 6: $d^+(r,\UU)\to 0$.} For arbitrary $w\in \dom(\calF)$ and $u_r\in
A(r,\UU)$ we have 
\[
\phi(r,\UU) = r\psi\big( \frac1r \DD(\UU,u_r) \big) + \calF(u_r) 
\leq  r\psi\big( \frac1r \DD(\UU,w) \big) + \calF(w) .
\]
Solving for $\DD(u,u_r)$ we use the strict monotonicity of $\psi $ and find 
\[
\DD(u,u_r) \leq r \psi^{-1}\Big(\psi\big(\frac1r\DD(u,w)\big) + \frac1\tau
\big( \calF(w) - \calF(u_r)\big) \Big) .
\]
As $\psi$ is convex with $\psi(0)=0$, the inverse $\psi^{-1}$ is concave with
$\psi^{-1}(0)=0$, and thus subadditive. Hence, we have 
\begin{equation}
  \label{eq:DD.VarInterpol}
  \DD(u,u_r) \leq \underbrace{r \psi^{-1}\Big(\psi\big(\frac1r\DD(u,w)\big)
  \Big)}_{=\DD(u,w)}  + 
r \psi^{-1}\Big( \frac1\tau
\big[\calF(w) - \calF(u_r)\big]_+ \Big) .
\end{equation}
As $\Psi$ is superlinear, we have $\psi^{-1}(a) = o(a)_{a\to \infty}$ such that
the last term tends to $0$ for $r\to 0^+$. This implies $\lim_{r\to 0^+}
d^+(r,\UU) \leq \DD(\UU,w)$. Since $w \in \dom(\calF)$ was arbitrary, and
$\UU\in \ol{\dom(\calF)}$ we obtain $\lim_{r\to 0^+}
d^+(r,\UU) =0$. 
\end{proof}

We are now ready to prove the following discrete energy-dissipation estimate,
which first appears in \cite[Lem.\,2.5]{Ambr95MM} and in a slightly more
elaborate version in \cite[Thm.\,3.1.4]{AmGiSa05GFMS}. According to several
oral presentations of these authors, the following result should be
called ``\emph{De Giorgi's lemma}'', as it was inspired by his personal
communication. Our version is slightly more general, as we treat arbitrary
dissipation potentials $\psi$. We can now show that the value function
$r\mapsto \phi(r,\UU)$ is differentiable and satisfies
\[
\frac\rmd{\rmd r} \phi(r,\UU) = - \psi^*\big( \psi'\big(\frac1r\,
d^+(r,\UU)\big)\big) \quad \text{ a.e.\ in } {]0,\tau]}.
\]

\begin{theorem}[De Giorgi's lemma]
\label{th:DeGiorgisLemma}
Consider a generalized metric GS $(M,\calF,\DD,\psi)$ where $\calF$ has compact
sublevels and $\psi$ satisfies \eqref{eq:Ass.psi}. Fix $\UU \in \dom(\calF)$ and 
$\tau>0$ and define $\phi$, $d^+$, and the variational interpolant $\wt
u: [0,\tau] \to M$ as above. Then, we have 
\begin{equation}
\label{eq:IdentityDeGiorgi}
\phi(\tau,\UU) + \int_0^\tau \psi^*\big( \psi'\big(\frac1r
\,d^+(r,\UU)\big)\big) \dd r = \calF(\UU). 
\end{equation}
If additionally the function $r \mapsto \wt u_\tau(r) \in M$  is
measurable, then 
\begin{equation}
  \label{eq:EstimDeGiorgi}
  \tau \psi\big( \frac1\tau \DD(\UU,\wt u_\tau(\tau))\big) + \int_0^\tau 
\psi^*\big( \SLO\calF (\wt u_\tau(r))\big) \dd r \leq \calF(\UU).  
\end{equation}
\end{theorem}
\begin{proof}
\STEP{Step 1: \eqref{eq:IdentityDeGiorgi} implies \eqref{eq:EstimDeGiorgi}.} 
We exploit the metric slope estimate in Proposition \ref{pr:MetrSlopeEstim} and
the monotonicity of $\psi^*$ giving
\begin{equation}
  \label{eq:Metr333}
  \psi^*\big( \SLO\calF (\wt u_\tau(r))\big) \leq 
\psi^*\big( \psi'\big(\frac{\DD(\UU,\wt u_\tau(r))}r  \big) \big) 
 \leq \psi^*\big( \psi'\big(\frac{d^+(r,\UU)}r  \big) \big) \ 
\text{ for } r\in {]0,\tau]}.
\end{equation}
As $r \mapsto \wt u_\tau(r) \in M$ is measurable, and
$u \mapsto \psi^*\big(\SLO\calF (u) \big) \in [0,\infty]$ is Borel measurable
(as a composition of a continuous and a lsc map), we see that
$r \mapsto \psi^*\big( \psi'\big(\frac1r \,d^+(r,\UU)\big)\big)\geq 0$ is
integrable and we obtain the desired estimate \eqref{eq:EstimDeGiorgi} by
integrating \eqref{eq:Metr333} and exploiting \eqref{eq:IdentityDeGiorgi}. 

\STEP{Step 2: Local Lipschitz continuity of ${]0,\tau]}\ni r \mapsto
    \phi(r,\UU)$.} 
For $0<r<s$ and all $u_s \in A(s,\UU)$ and $u_r\in A(r,\UU)$ we have
\begin{align*}
\Phi_s(\UU,u_r)- \Phi_r(\UU,u_r) &\geq 
\Phi_s(\UU,u_s)- \Phi_r(\UU,u_r) = \phi(s,\UU)- \phi(r,\UU) 
\\  &
\geq \Phi_s(\UU,u_s)- \Phi_r(\UU,u_s). 
\end{align*}
In the first and last term the appearance of $\calF$ cancels and we are left
with the estimate 
\[
\Psi(s,\DD(\UU,u_r))- \Psi(r,\DD(\UU,u_r)) \geq 
 \phi(s,\UU)- \phi(r,\UU) \geq \Psi(s,\DD(\UU,u_s))- \Psi(r,\DD(\UU,u_s)). 
\]
By Lemma \ref{le:Metric.Psi}(iii) the mapping $a \mapsto  \Psi(s,a)- \Psi(r,a)$
is decreasing hence, we may maximize for $u_r\in A(r,\UU)$ and minimize for
$u_s \in A(s,\UU)$ to obtain 
\begin{align}
  \label{eq:ValueFcnEnclosed}
&\Psi(s,d^+(r,\UU))- \Psi(r,d^+(r,\UU)) \\
\nonumber
& \geq \phi(s,\UU)- \phi(r,\UU) \geq \Psi(s,d^-(s,\UU))- \Psi(r,d^-(s,\UU)). 
\end{align}
From this we can derive Lipschitz continuity by assuming $0<r_*\leq r<s\leq
\tau$, namely
\begin{align*}
0&\geq \phi(s,\UU)- \phi(r,\UU) \geq \int_r^s \pl_t\Psi\big(t, d^-(s,\UU)\big)
\dd t \\
&\overset{\text{(i)}}= -\int_r^s \psi^*\big( \psi'\big(\frac1t d^-(t,\UU) 
\big)\big) \dd t   \ \geq \ - (s{-}r) \,\psi^*\big( \psi'\big(\frac1{r_*}
d^+(\tau,\UU) \big)\big) =: -(s{-}r) K_*,
\end{align*}
where we used the monotonicity of $ \psi^*\circ \psi'$ in the last step  and 
$\overset{\text{(i)}}=$ indicates the identity derived in Lemma
\ref{le:Metric.Psi}(i). Thus, we have Lipschitz continuity with Lipschitz
constant $K_*$ on $[r_*,\tau]$. 

\STEP{Step 3: Identification of the derivative.} Because of local Lipschitz
continuity, we have differentiability a.e.\ in ${]0,\tau]}$. To identify the
derivative we divide \eqref{eq:ValueFcnEnclosed} by $s-r>0$ and obtain, again
using Lemma \ref{le:Metric.Psi}(i),
\begin{align} 
\nonumber 
-\psi^*\big( \psi'\big( \frac{d^+(r,\UU)}r\big) \big) &= 
\lim_{s\to r^+} \! \frac{\Psi(s,d^+(r,\UU)){-} \Psi(r,d^+(r,\UU))}{s\ - \ r} 
\\ 
\label{eq:ValFcnUppBdd} 
&\geq
\limsup_{s\to r^+}\frac{\phi(s,\UU) {-} \phi(r,\UU))}{s\ - \ r} \qquad
\text{and} 
\\
\nonumber
-\psi^*\big( \psi'\big( \frac{d^-(s,\UU)}s\big) \big) &= 
\lim_{r\to s^-}\! \frac{\Psi(s,d^-(s,\UU)){-} \Psi(r,d^-(s,\UU))}{s\ - \ r}
\\
\label{eq:ValFcnLowBdd}
& \leq
\liminf_{r\to s^-}\frac{\phi(s,\UU) {-} \phi(r,\UU))}{s\ - \ r}.
\end{align}
Denote by  $\bbT\subset {]0,\tau]}$ the set of points where $r\mapsto \phi(r,\UU)$ is
differentiable and where $d^+$ and $d^-$ are continuous. Together with
Propositions \ref{pr:ValueFctDist} we know that $\bbT$ is a set of full
measure and that $d^+(t,\UU)=d^-(t,\UU)$ on $\bbT$.  
Taking $r=t$ in \eqref{eq:ValFcnUppBdd} and $s=t$ in \eqref{eq:ValFcnLowBdd} we
obtain 
\[
\frac\rmd{\rmd
    t} \phi(t,\UU) =  -\psi^*\big( \psi'\big( \frac{d^\pm(t,\UU)}t\big)
  \big) \quad \text{for all } t \in \bbT. 
\]

\STEP{Step 4: Integral formula on $[0,\tau]$.} Step 3 implies, for all $r\in
{]0,\tau[}$, the relation
\[
\phi(\tau,\UU)+ \int_r^\tau \psi^*\big( \psi'\big( \frac1s \,d^+(s,\UU)\big)
\big) \dd s = \phi(r,\UU). 
\]
By Proposition \ref{pr:ValueFctDist}\ref{EN:phi.to.calF} we have
$\phi(r,\UU)\to \calF(\UU)$ for $r\to 0^+$, i.e.\ convergence on the right-hand
side. The convergence for $r\to 0^+$ on the left-hand side follows from Beppo
Levi's monotone convergence theorem as the integrand is nonnegative. Thus, identity
\eqref{eq:IdentityDeGiorgi} is established.  
\end{proof}

\subsection{Existence of curves of maximal slopes via MMS}
\label{su:MetrExist}

We are now ready to show the existence of $\psi$-curves of maximal slope. Of
course, the construction is based on the MMS and it will follow closely the
proof of Theorem \ref{th:BanachExist} for Banach-space gradient systems. 
The major difference is that we do no longer assume any type of
$\lambda$-convexity (of $\lambda$-global slopes) and exploit De Giorgi's
variational interpolant instead.

\begin{theorem}[Existence of $\psi$-curves of maximal slope]
\label{th:MetricExistence}
Consider a generalized metric gradient system $(M,\calF,\DD,\psi)$ that
additionally satisfies
\begin{subequations}
\label{eq:MetrAssExist}
\begin{align}
\label{eq:MetrAssExi.a}
& \calF \text{ has compact sublevels } S^\calF_E\subset M; 
\\
\label{eq:MetrAssExi.b}
& \SLO\calF :M\to [0,\infty] \text{ is lower semicontinuous}; 
\\
\label{eq:MetrAssExi.c}
& \psi \in \rmC^1({[0,\infty[}) \text{ and is strictly convex};
\\ 
\label{eq:MetrAssExi.d}
& (M,\calF,\DD,\psi) \text{ satisfies the metric chain-rule inequality
  \eqref{eq:MetrCRIneq}}. 
\end{align}
Then, for all $u^0\in \dom(\calF)$ there exists a $\psi$-curve of maximal slope
$u:{[0,\infty[} \to M$ satisfying $u(0)=u^0$.  
\end{subequations}
\end{theorem}
\begin{proof}
We fix a time $T>0$ and construct solutions on $[0,T]$ at first. For $N\in \N$
we define the time step $\tau>0$.  

\STEP{Step 0: Construction of approximants.} Because of 
the compact sublevels of $\calF$ (see \eqref{eq:MetrAssExi.a}) we know that
$\calF(u)\geq \calF_\mafo{min}$ 
for all $u\in M$. Moreover, using $\calF(u^0)<\infty$ we know that the MMS
produces solutions $(u^\tau_k)_{k=0,..,N}$ lying in the compact sublevel
$S^\calF_{\calF(u^0)}$.  Moreover, we can construct De Giorgi's variational
interpolant $\wt u_\tau:[0,T]\to M$ and apply De Giorgi's lemma (i.e.\
Theorem \ref{th:DeGiorgisLemma}) on each time interval $[k\tau{-}\tau, k\tau]$
and obtain 
\begin{equation}
  \label{eq:Metr.22}
  \calF(\wt u_\tau(k\tau)) +  \int_{k\tau{-}\tau}^{k\tau}
 \Big(\psi\big(S_\tau(r)\big)+ 
\psi^*\big(G_\tau(r)\big) \Big) \dd r = \calF(\wt u_\tau(k\tau{-}\tau))
\end{equation}
for $k =1,\ldots,N=T/\tau$, 
where we introduced the functions $S_\tau$ and $G_\tau$ as follows: 
\begin{align*}
S_\tau(t) &= \frac1\tau \DD(\wt
u_\tau(k\tau{-}\tau), \wt u_\tau(k\tau) )  &\text{for }& t \in
{]k\tau{-}\tau, k\tau]}, 
\\
G_\tau(t) &= \psi'\big( \frac1r\,d^+(t,\wt u_\tau(k\tau{-}\tau))\big) 
&\text{for }& t=k\tau{-}\tau+r  \in
{]k\tau{-}\tau, k\tau]}.
\end{align*} 

We note that it is tempting to replace $G_\tau(t)$ by the smaller value
$\SLO\calF(\wt u_\tau(t))$ (cf.\
the slope estimate in Proposition \ref{pr:MetrSlopeEstim}), however
we refrain from doing so because then we would need to show measurability
(which is possible but technical). It is better to keep $G_\tau$ as defined,
which is automatically measurable and apply the slope estimate later (see Step 3). 

\STEP{Step  1: A priori estimates.} Clearly, summing \eqref{eq:Metr.22}  over
$k=1,\ldots, N$ leads to a telescope sum and we find 
\begin{equation}
  \label{eq:Apri.Stau.Gtau}
  \int_0^T \!\!\big( \psi(S_\tau(t)) + \psi^*(G_\tau(t))\big)\dd t  
 = \calF(\wt u_\tau(0)) {-} \calF(\wt u_\tau(T)) \leq \calF(u^0) {-} 
  \calF_\mafo{min}=:\Delta_\calF <\infty.
\end{equation}
This provides superlinear a priori estimates for $S_\tau$ and $G_\tau$. 

We also want to derive a ``kind of equi-continuity'' of the sequence $(\wt
u_\tau)_\tau$. Of course, we cannot expect the individual $\wt u_\tau$ for
fixed $\tau=T/N$ to be continuous but it should be close to a continuous
function. We will show that there exists a modulus of continuity $\wt\omega$
such that 
\begin{equation}
  \label{eq:wtu.tau.equicont}
  \DD\big( \wt u_\tau(s), \wt u_\tau(t)\big) \leq \wt\omega\big(\tau+
|t{-}s|\big) \quad \text{for all }s,t\in [0,T] \text{ and all }\tau=T/N.
\end{equation}

For this we first quantify the convergence $d^+(r,u_*)\to 0$ in Proposition
\ref{pr:ValueFctDist}\ref{EN:d+to0}, i.e.\ we show that variational
interpolants $\wt u_\tau$ are close to the nodal points $\wt u_\tau(k\tau)$. 
Setting $u=w=\wt u_\tau(k\tau)$ in \eqref{eq:DD.VarInterpol}, for $k=0,\ldots,
N{-}1$ and $r\in {]0,\tau[}$ we find 
\begin{align*}
\DD\big(\wt u_\tau(k\tau), \wt u_\tau(k\tau{+}r) \big) &\leq r\, \psi^{-1}\Big(
\frac1r\big(\calF (\wt u_\tau(k\tau))- \calF(\wt u_\tau(k\tau{+}r))\big)\Big) 
\\
& \leq  r\, \psi^{-1}\big( \frac1r\,\Delta_\calF\big) \ =:\ \wh\omega(r) =
o(1)_{r\to 0^+}.
\end{align*}
Here we used that $\psi^{-1}$ is increasing and growing less than linear,
because $\psi$ is superlinear. Hence $\wh \omega$ is an modulus of continuity.

We define the function $\ul t_\tau:[0,T]\to [0,T]$ via $\ul t_\tau(s) =
\max\set{ k\tau}{ k\tau\leq s} = \tau\lfloor s/\tau\rfloor$. With this, we
obtain, for $0\leq r< s\leq T$, the estimate 
\begin{align}
\nonumber
\DD\big(\wt u_\tau(r),\wt u_\tau(s)\big)&\leq 
\DD\big(\wt u_\tau(r),\wt u_\tau(\ul t_\tau(r))\big) +
\DD\big(\wt u_\tau(\ul t_\tau(r)) ,\wt u_\tau(\ul t_\tau(s)) \big)+
\DD\big(\wt u_\tau(\ul t_\tau(s)),\wt u_\tau(s)\big)
\\
\nonumber
&\leq \wh\omega\big(r{-}\ul t_\tau(r)\big) + 
 \sum_{k=\lfloor r/\tau\rfloor}^{\lfloor s/\tau\rfloor-1} \tau
\:\frac1\tau \, \DD\big(\wt u_\tau(k\tau), \wt u_\tau (k\tau {+}\tau) \big)  +
\wh\omega\big(s{-} \ul t_\tau(s)\big) 
\\
\label{eq:wtu.tau.equi22}
&\leq \wh\omega(\tau) + \int_{\ul t_\tau(r)}^{\ul t_\tau(s)}
S_\tau(t) \dd t +  \wh\omega(\tau).
\end{align}
We proceed as in the Banach-space case (cf.\ Section \ref{su:BanachExist}) by
estimating $S_\tau \leq \frac1\mu \:\mu S_\tau\leq \frac1\mu
\big(\psi(S_\tau){+}\psi^*(\mu)\big) $ and obtain 
\begin{align*}
\DD\big(\wt u_\tau(r),\wt u_\tau(s)\big)&\leq 
 2\wh\omega(\tau) + \int_{[r-\tau]_+}^s S_\tau(t) \dd t \leq 2\wh\omega(\tau) +
  \omega^{\Delta_\calF}_\psi\big( s{-}r + \tau\big),
\end{align*}
where $\omega^B_\psi$ is defined in \eqref{eq:Banach:Equicont} and
$\Delta_\calF$ in \eqref{eq:Apri.Stau.Gtau}. Hence, \eqref{eq:wtu.tau.equicont}
is established with $\wt\omega = 2\wh\omega + \omega^{\Delta_\calF}_\psi$.

\STEP{Step  2: Extraction of converging subsequences.}
Since $\psi$ and $\psi^*$ are superlinear, the a priori estimate
\eqref{eq:Apri.Stau.Gtau} and the criterion of de la
Vall\'ee-Poussin guarantee that the sequences $(S_\tau)_\tau$ and
$(G_\tau)_\tau$ are equi-integrable and there exists a subsequence (not
relabeled) such that 
\[
S_\tau \weak S_0 \qAND G_\tau \weak G_0 \quad \text{in } \rmL^1([0,T]).
\]

Moreover, the equi-continuity \eqref{eq:wtu.tau.equicont} allows us to employ
the generalized Arzel\`a-Ascoli theorem, 
such that along a further subsequence (not relabeled) we have pointwise
convergence to a continuous limit function $u:[0,T]\to M$, namely
\[
\forall\, t\in [0,T] : \quad \wt u_\tau(t) \to u(t) \ \text{ as } \tau \to 0^+. 
\]
Because of $\wt u_\tau(0)=u^0$, we also have $u(0)=u^0$. 
By passing to the limit  in \eqref{eq:wtu.tau.equi22} we obtain
\begin{equation}
  \label{eq:SpeedEstim}
  \forall\, s,t\in [0,T] \text{ with } s<t:\quad 
\DD(u(s),u(t)) \leq \int_s^t S_0(t) \dd t,
\end{equation}
which shows $ u \in \AC([0,T];M)$. 

\STEP{Step 3: Derivation of (EDI).}
We return to \eqref{eq:Apri.Stau.Gtau} in the form 
\begin{equation*}
 \calF(\wt u_\tau(T))+ \int_0^T \! \psi(S_\tau(t))\dd t  + \int_0^T \!
 \psi^*(G_\tau(t))\dd t    = \calF(\wt u_\tau(0)),
\end{equation*}
and calculate the liminf for $\tau \to 0^+$ for the three terms on the
left-hand side. 

From $\wt u_\tau(T) \to u(T)$ and the lsc of $\calF$ we have $\liminf_{\tau \to
  0^+} \calF(\wt u_\tau(T)) \geq \calF(u(T))$. 

For the second term we observe that the mapping $\alpha \mapsto
\int_0^T\psi(\alpha(t)) \dd t$ is convex and strongly lsc on
$\rmL^1([0,T])$. Hence, the mapping is also weakly lsc and $S_\tau \weak S_0$
implies $\liminf_{\tau \to 0^+} \int_0^T \! \psi(S_\tau(t))\dd t \geq \int_0^T
\! \psi(S_0 (t))\dd t \geq \int_0^T \psi\big(\SPE u(t)\big) \dd t$. For the
last estimate we used that $\psi: {[0,\infty[} \to {[0,\infty[}$ is increasing
and the characterization of the metric speed in Theorem \ref{th:MetrSpeed},
i.e.\ $\SPE u \leq S_0$ because of \eqref{eq:SpeedEstim}. 

For the third term we fix $t\in [0,T]$ and exploit the slope estimate in
Proposition \ref{pr:MetrSlopeEstim} as well as the lsc of the slope
$\SLO\calF$, see assumption \eqref{eq:MetrAssExi.b}. Using that
$\psi^*:{[0,\infty[} \to {[0,\infty[}$ is continuous and increasing and that
$\wt u_\tau(t) \to u(t)$ we have
\[
\psi^*\big(\SLO\calF(u(t))\big)  \leq \liminf_{\tau \to 0^+} \psi^*\big(
\SLO\calF(\wt u_\tau(t))\big)  \leq \liminf_{\tau \to 0^+} \psi^*\big(G_\tau(t)\big) .
\]
Thus, Fatou's lemma yields $\liminf_{\tau \to 0^+} \int_0^T
\psi^*\big(G_\tau(t)\big) \dd t \geq \int_0^T \psi^*\big(\SLO\calF(u(t))\big)
\dd t$. 

In summary, we find the EDI 
\[
\calF(u(T)) + \int_0^T\!\! \Big( \psi\big( \SPE u(t)\big) + \psi^*\big(
\SLO\calF(u(t))\big) \Big) \dd t \leq \calF(u(0)).
\]

\STEP{Step 4: Derivation of (EDB).} As we have assumed the abstract metric chain-rule
inequality in \eqref{eq:MetrAssExi.d} we can apply the metric
energy-dissipation principle from Proposition \ref{pr:MetrEDP}. Hence, $u$ is a
$\psi$-curve of maximal slope. 
\end{proof}

As in Section \ref{se:GSBanach} one can infer additional convergences (along the
chosen subsequence), if we assume strict convexity of $\psi$ and $\psi^*$:
\begin{align*} 
 \forall\; t\in [0,T]: \quad &\wt u_\tau(t) \to u(t) \AND \calF(\wt
u_\tau(t)) \to \calF(u(t)). 
\\
\forall_\text{a.a.} t\in [0,T]:\quad & \frac1\tau \DD \big(
\wt u_\tau(\ul t_\tau(t)),\wt u_\tau(\ul t_\tau(t){+}\tau)\big)  \to \SPE
u(t)\\
& 
\AND \SLO\calF(\wt u_\tau(t)) \to \SLO\calF(u(t)). 
\end{align*}

We emphasize that there is no easy way of showing uniqueness in
this general setting. Example \ref{ex:CurMaxSlo}(A) provides a case where all
assumption of the above existence theorem are satisfied, but uniqueness fails. 
Moreover, the following example shows that one may have even uncountably many
solutions for a given initial point $u^0$. 

\begin{example}[Non-uniqueness for curves of maximal slope] \slshape
Consider the gradient system $(\R^2,\calF,|\cdot|_1, \psi_\mafo{quadr})$
with $\calF(u)=u_1+u_2$ and $|(v_1,v_2)|_1=|v_1|+|v_2|$.  
A curve $u:[0,T]\to \R^2$ is a curve of maximal slope if and only if 
$u\in \rmW^{1,\infty}([0,T];\R^2)$ with 
\[
\dot u_1(t), \dot u_2(t) \in [-1,0] \ \text{ and } \ \dot u_1(t)+\dot u_2(t)=-1
\quad \text{ a.e.\ in } [0,T].
\]
Thus, all the curves $u(t)=u^0-t(1{-}\theta,\theta)+ g\sin(\omega t)(1,-1)$
with $\theta\in [0,1]$  and $|g\omega| \leq \min\{\theta,1{-}\theta\}$ are
curves of maximal slope starting at $u^0$. 
\end{example}

\subsection{Metric evolutionary variational inequalities (EVI)}
\label{su:MetrEVI}

We recall that in the case of Hilbert spaces (see Section \ref{su:Hilbert.EVI})
the evolutionary variational inequality (EVI)$_\lambda$ did only use the norms
$\|u{-}w\|$ and no time derivatives $\dot u$  or sub\-differentials
$\plF\calF(u)$ appear. Hence, we can easily define the corresponding EVI notion
for metric GS. We emphasize that this theory is restricted to the quadratic
dissipation function $\psi=\psi_\text{quadr}:\delta \mapsto \delta^2/2$, thus
use the short-hand $(M,\calF,\DD):=(M,\calF,\DD,\psi_\text{quadr})$.  

\begin{definition}[Metric EVI$_\lambda$ solutions]
\label{de:MetrEVI.sol} We consider a metric GS $(M,\calF,\DD)$. Then, we call
$u:[0,T]\to M$ an (EVI)$_\lambda$ solution, if
\begin{align*}
&\forall\,s,t\in [0,T] \text{ with } s<t\ \forall\, w\in \dom(\calF):
\\
&\frac12\DD( u(t) , w)^2 \leq \frac12 \ee^{-\lambda(t-s)} \DD(u(s),w)^2 +
M_\lambda(t{-}s) \big( \calF(w)-\calF(u(t))\big),
\end{align*} 
where $M_\lambda (r)=\int_0^r \ee^{-\lambda(r-s)}\dd s$. 
\end{definition} 

We will see below that it is possible to derive uniqueness for EVI solutions,
however it is very difficult to establish existence. Except for the
Hilbert-space case discussed in Section \ref{se:GSHilbert}, there is no
direct way of showing that curves of maximal slope (with
$\psi=\psi_\text{quadr}$) are also EVI solutions if $\calF$ satisfies a
suitable $\lambda$-convexity condition.  

Instead, there is an independent existence theory for EVI solutions based on
rather strong assumptions on the metric space $(M,\DD)$ and on the functional
$\calF$. We refer to \cite[Cha.\,4]{AmGiSa05GFMS} and \cite{Sava07GFDS,
  DanSav14LNGF, MurSav22?GFEV} because the general existence theory is ongoing
research. 

The major new assumption is that of the existence of geodesic curves. 

\begin{definition}[Geodesic metric spaces]
\label{de:GeodMetrSpaces}  
In a metric space $(M,\DD)$ a curve $\gamma:[0,1]\to M$ is called a \emph{(constant
speed) geodesic} if 
\[
\forall\, r,s\in [0,1]: \quad \DD(\gamma(r),\gamma(s)) =|s{-}r|\,
\DD(\gamma(0),\gamma(1)). 
\] 
In this case we say that the geodesic $\gamma$ connects the points $\gamma(0)$
and $\gamma(1)$ and write $\mafo{Geod}\big(\gamma(0),\gamma(1)\big)$ for the
set of all such geodesics.  

The metric space $(M,\DD)$ is called a \emph{geodesic space}, if for all 
$u_0,u_1\in M$ there exists a geodesic connecting $u_0$ and $u_1$. 

A function $\calF:M\to \Rinfty$ is called \emph{geodesically $\lambda$-convex}
if
\begin{align*}
&\forall\,u_0,u_1\in \dom(\calF)\ \exists\,\gamma \in \mafo{Geod}(u_0,u_1)\ \forall\, s\in
[0,1]:
\\
& \calF(\gamma(s)) \leq (1{-}s) \calF(\gamma(0)) + s \calF(\gamma(1)) -
\frac\lambda 2 \, s(1{-}s) \,\DD(\gamma(0),\gamma(1))^2. 
\end{align*}

\end{definition}

With these conditions we are able to state the following simplified version of
the existence result in \cite[Thm.\,4.0.4]{AmGiSa05GFMS}. Again, the
construction uses the MMS and, because of uniqueness, the constructed solutions
are minimizing movements in the sense of Definition \ref{de:MetrGSMM}. In this
case we also have a true gradient flow $(S_t)_{t\geq 0}$ on $\ol{\dom(\calF)}$,
similar to Theorem \ref{th:CompleteGF} for Hilbert spaces.

\begin{theorem}[Existence of EVI solutions]
\label{th:ExistEVIsol}
Consider the metric GS $(M,\calF,\DD)$ (with $\psi=\psi_\text{quadr}$) with
the following properties
\begin{subequations}
\label{eq:metrEVIass}
\begin{align} 
&\label{eq:metrEVIass.a} 
 (M,\DD) \text{ is a geodesic space},
\\
& \label{eq:metrEVIass.b}
 \forall\, u_*\in M:\quad u\mapsto \frac12\DD(u_*,u)^2 \text{ is geodesically
   $1$-convex}, 
\\
&\label{eq:metrEVIass.c}
\exists\, \lambda\in \R:\quad  \calF;M\to \Rinfty \text{ is geodesically
  $\lambda$-convex}.  
\end{align}
\end{subequations}
Then, for all $u^0\in \scrD:=\ol{\dom(\calF)}$ there exists a unique (EVI)$_\lambda$
solution $u:{[0,\infty[}\to M$ which satisfies $u(0)=u^0$ and $u\in
\mafo{MM}(M,\calF,\DD)$. 

Moreover,  the mapping $S_t:\scrD \to \scrD$ defined by the unique solutions
via  $S_t(u(0)):=u(t)$ is a
$\lambda$-contractive, continuous semigroup, namely
\\[0.2em]
(S1)\quad $S_t: \scrD \to \scrD$, \quad $S_0=\mafo{id}_\scrD$, \quad $S_t\circ
S_r= S_{t+r}$ for  all $r,t\geq 0$.
\\[0.2em]
(S2)\quad For all $u^0$ the function ${[0,\infty[} \ni t\mapsto S_t(u^0)$ is
continuous.
\\[0.2em]
(S3)\quad For all $u_0, u_1\in \scrD$ we have $\DD\big(S_t(u_0),S_t(u_1)\big) \leq
\ee^{-\lambda t} \DD(u_0,u_1)$.  
\end{theorem}

The critical condition in the above theorem is
that of the geodesic $1$-convexity of $u\mapsto \frac12\DD(u_*,u)^2$ in
\eqref{eq:metrEVIass.b}. This condition is satisfied in Hilbert spaces, but it
does not hold for many geodesic spaces.  In particular, it does not hold for
the Wasserstein space $(\mafo{Prob}(\ol\Omega),\rmW_2)$ from Example
\ref{ex:CurMaxSlo}(C). Thus, in \cite[Thm.\,4.0.4]{AmGiSa05GFMS} condition
\eqref{eq:metrEVIass.b} is replaced by a weaker one.

In \cite[Ch.\,3+4]{MurSav20GFEV} the question is addressed how EVI solutions
and curves of maximal slope are related. From
\cite[Thm.\,3.5,\;cf.\,(3.17)]{MurSav20GFEV} one easily sees that every EVI
solution is a curve of maximal slope. The reverse statement that a curve of
maximal slope is also an EVI solution (and hence unique) is more desirable, but
it is known only under strong additional conditions, see
\cite[Thm.\,4.2]{MurSav20GFEV}. In particular, one needs an independent
existence result for EVI solutions.\medskip

For the proof of the above existence result we refer to
\cite[Cha.\,4]{AmGiSa05GFMS}.  Here we provide the analysis that is  necessary
for establishing the $\lambda$-contractivity. For this we derive a few
properties (P.$n$) for all EVI solutions $u$.\smallskip

\STEP{(P.1) Finite energy: For all $t \gneqq 0$ we have $\calF(u(t)) < \infty$.}

We insert $s=0$ and $w \in \dom(\calF) \neq \emptyset$ into (EVI)$_\lambda$ and
obtain after dropping $\frac12\DD(u(t),w)^2$ the estimate
\[
\calF(u(t)) \leq  \calF(w) + \frac{\ee^{-\lambda t} } {2M_\lambda(t)}  \,\DD(
u(0), w)^2 < \infty.
\] 

\STEP{(P.2): $t\mapsto \calF(u(t))$ is decreasing.}

For $0<s<t$  we insert $w=u(s)$ into (EVI)$_\lambda$ and
obtain 
\[
\calF(u(t)) \leq  \calF(u(s)) + \frac{\ee^{-\lambda (t-s)} } {2M_\lambda(t)}
\big(\; 0\; - \frac12\DD(u(t),u(s))^2\big) \leq  \calF(u(s)). 
\] 
If $u(0)\in \dom(\calF)$ we can also do this for $s=0$, whereas in the case
$\calF(u(0))=\infty$ we have $\infty = \calF(u(0))>\calF(u(s)) \geq
\calF(u(t))$ for $0< s < t$. 

\STEP{(P.3) Local H\"older continuity: $u \in
  C^{1/2}_\mafo{loc}({]0,\infty[};M)$.} 

Choose $[t_0, T]\Subset {]0,\infty[}$ (compactly contained), then for
$t_0\leq s < t\leq T$ and $w=u(s)$ in (EVI)$_\lambda$ we find
\[
\DD(u(s),u(t))^2 \leq 2M_\lambda(t{-}s) \big(\calF(u(s))-\calF(u(t))\big) \leq
C_{t_0,T,\lambda} \,|t{-}s|\,\big(\calF(u(t_0))-\calF(u(T))\big) .
\]
This implies $\DD(u(s),u(t)) \leq \wt C_{t_0,T,\lambda} \,|t{-}s|^{1/2}$ as
desired. 

\STEP{(P.4) Local absolute continuity: $u \in
  \AC^2_\mafo{loc}({]0,\infty[};M)$.}

For $[t_0, T]\Subset {]0,\infty[}$ as above and $N\in N$ we define
$\tau_N=(T{-}t_0)/N$ and the partition $t^N_k=t_0 + k\tau_N$ for
$k=0,1,...,N$. Now (EVI)$_\lambda$ gives 
\[
\calF(u(t^N_k)) + \frac1{2M_\lambda(\tau_N)} 
   \DD\big( u(t^N_{k-1}),u(t^N_k) \big)^2 \leq \calF(u(t^N_{k-1} )). 
\]
When adding over $k=1,...,N$ we can exploit the telescope sum and obtain 
\[
\frac{\tau_N}{2M_\lambda(\tau_N)} 
\sum_{k=1}^N  \tau_N \Big(\frac1{\tau_N} \DD\big( u(t^N_{k-1}) , u(t^N_k) \big)
\Big)^2 \leq \calF(u(t_0 ))  - \calF(u(T))=:\Delta. 
\]
Defining the piecewise constant function $S^N$ via $S^N(t)=\frac1{\tau_N} \DD \big(
u(t^N_{k-1}) , u(t^N_k) \big) $ for $t \in {] t^N_{k-1},t^N_k ]}$, we have the
$\rmL^2$ bound $\int_{t_0}^T S^N(t)^2 \dd t \leq \Delta$. Thus, after
extracting a subsequence (not relabeled) we may assume $S^N\weak S_0$ in
$\rmL^2([t_0,T])$. 

For arbitrary $r,s\in [t_0,T]$ with $r<s$ we choose $l(N),
m(N)\in\{0,1,...,N\}$ such that $\wt r_N:=t^N_{l(N)} \to r$ and $\wt
s_N:=t^N_{m(N)} \to s $. Using the triangle inequality we obtain
\begin{align*}
\DD(u(r),u(s))& \leq \DD(u(r),u(\wt r_N)) + \Big(\sum_{k=l(N)+1}^{m(N)} \DD\big(
u(t^N_{k-1}), u(t^N_k)\big)  \Big) + \DD(u(\wt s_N),u(s))
\\
& \leq C\big| r -\wt r_N  \big|^{1/2}  + \int_{t_0}^T \bm1_{[\wt r_N, \wt
  s_N]}(t) \,S^N(t) \dd t +  C\big| s -\wt s_N  \big|^{1/2} ,
\end{align*} 
where we used the H\"older continuity (P.3) and the definition of $S^N$. We can
now pass to the limit $N\to \infty$ on the right-hand side and arrive at
$ \DD(u(r),u(s)) \leq \int_r^s S_0(t) \dd t$ which implies $u \in
\AC^2([t_0,T];M)$ with $\SPE u \leq S_0 \in \rmL^2([t_0,T])$ a.e.\ in
$[t_0,T]$, see Theorem \ref{th:MetrSpeed}.  

\begin{proposition}[$\lambda$-contractivity for solutions of (EVI)$_\lambda$]
\label{pr:lambdaContr}
Consider two (EVI)$_\lambda$ solutions $u,\wt u:{[0,\infty[} \to M$  for 
$(M,\calF,\DD)$. Then, we have 
\begin{equation}
  \label{eq:lambda.Contr}
  \DD\big(u(t),\wt u(t)\big) \leq \ee^{-\lambda(t-s)} \,
  \DD\big( u(s), \wt u(s)\big)  \quad\text{for } 0\leq s < t. 
\end{equation} 
\end{proposition}
\begin{proof}
  \STEP{Step 1: First two applications of EVI.} We insert $w=\wt u(t)$ into the
  (EVI)$_\lambda$ for $u$ and $\wt w= u(t)$ into the (EVI)$_\lambda$ for
  $\wt u$. Adding the two inequalities we see that all terms involving $\calF$
  cancel, and we obtain
\[
\DD\big(u(t),\wt u(t)\big)^2 = \big(\frac12+\frac12\big) \DD\big(u(t),\wt
u(t)\big)^2  \leq  \ee^{-\lambda(t-s)}\Big(\frac12 \DD\big(u(s),\wt u(t)\big)^2  + \frac12
\DD\big(\wt  u(s),u(t)\big)^2 \Big) .
\]
Note that on the right-hand side the four different points $u(s),\: u(t),\: \wt
u(s)$, and $\wt u(t)$ appear. 

\STEP{Step 2: Third and fourth application of EVI.} We again use
(EVI)$_\lambda$ for $u$ but now with $w= \wt u(s)$ and (EVI)$_\lambda$ for $\wt
u$ with $\wt w = u(s)$. Thus we can estimate the terms on the right-hand side
and arrive at 
\begin{align}
  \label{eq:Metr55}
\DD\big(u(t),\wt u(t)\big)^2& \leq  \ee^{-\lambda(t-s)}\Big(
\ee^{-\lambda(t-s)}\big(\frac12+\frac12\big) \DD\big(u(s),\wt u(s)\big)^2
\\ \nonumber &
\qquad \qquad + M_\tau(t{-}s) \big(\calF(u(s)){-}\calF(u(t)) 
  + \calF(\wt u(s)) - \calF(\wt u(t)) \big) \Big).
\end{align} 

\STEP{Step 3: Absolute continuity of $[t_0,T]\ni t\to \delta(t)= \DD\big(u(t),\wt
  u(t)\big)$.} For $r,s\in [t_0,T]$ the triangle inequality gives 
\begin{align*}
\big|\delta(r)- \delta(s)\big| & 
 = \big| \DD\big(u(r),\wt u(r)\big) - \DD\big(u(s),\wt u(s)\big) \big| \\
&
\leq \big| \DD\big(u(r),\wt u(r)\big) - \DD\big(u(s),\wt u(r)\big) \big|
   + \big| \DD\big(u(s),\wt u(r)\big) - \DD\big(u(s),\wt u(s)\big) \big|
\\ &
\leq  \DD\big(u(r),u(s)\big) + \DD\big(\wt u(r),\wt u(s)\big) 
\leq \int_r^s \big( \SPE u(t) + \SPE{\wt u}(t)\big) \dd t .
\end{align*}
Hence, $u,\wt u\in \AC^2([t_0,T];M)$ implies $\delta \in
\AC^2([t_0,T];\R)=\rmW^{1,2} ([t_0,T])$. 

\STEP{Step 4: Conclusion.} We set $\rho(t)= \ee^{2\lambda t} \delta(t)^2$, then
the product rule and Step 3 give $\rho \in \rmW^{1,2} ([t_0,T])$. Moreover, by
the definition of $\rho$, the estimate  
\eqref{eq:Metr55} turns into 
\[
\rho(t) - \rho(s) \leq \ee^{\lambda(t+s)} M_\lambda(t{-}s)  \big(\calF(u(s)){-}\calF(u(t)) 
  + \calF(\wt u(s)) - \calF(\wt u(t)) \big),
\]
for $t_0\leq s < t\leq T$.

Now assume that $s=s_*$ is a point of differentiability of $\rho$, which is
true on a set of full measure. Then, dividing by $t-s_*>0$ and taking the
limit $t\to s_*^+$ gives 
\begin{align*}
\dot \rho(s_*) &= \lim_{t\to s_*^+} \frac{\rho(t)-\rho(s_*)}{t - s_*} 
\\ &
\leq \limsup_{t\to s_*^+} \Big(B_\lambda(t,s_*) \,\big(\calF(u(s_*)){-}\calF(u(t)) 
  + \calF(\wt u(s_*)) - \calF(\wt u(t)) \big) \Big) 
\end{align*}
with $B_\lambda(t,s_*) =\ee^{\lambda(t+s_*)} M_\lambda(t{-}s_*)/(t{-}s_*) \to
\ee^{2\lambda s_*}$ for $t\to s_*$. 

Using the H\"older continuity (P.3) and lsc of $\calF$ we have 
\[
\limsup_{t\to s_*} \big(\calF(u(s_*))- \calF(u(t)) \big) =  
\calF(u(s_*)) - \liminf_{t\to s_*} \calF(u(t)) \leq  \calF(u(s_*))
-\calF(u(s_*)) =0,
\]
and similarly for $\wt u$. Hence, we conclude $\dot\rho(s_*) \leq 0$. Because
$\rho$ is absolutely continuous, we have the monotonicity $\rho(t)\leq \rho(s)$
for $s<t$, which is the desired estimate \eqref{eq:lambda.Contr} when recalling
the definition $\rho(t) = \ee^{2\lambda t} \DD\big(u(t),\wt u(t)\big)^2$.
\end{proof}

\section[Evolutionary $\bm\Gamma$-convergence for gradient systems]
{Evolutionary $\Gamma$-convergence for gradient systems}
\label{se:EGC}

In this section we study families of gradient systems $(X,\calF_\eps,
\calR_\eps)$ or $(M,\calF_\eps, \DD_\eps, \psi_\eps)$ where $\eps\in
{[0,1]}$. The typical question one is interested are the following:

\begin{itemize}
\item[Q1] Assume we have solutions $u_\eps:[0,T]\to X$ for
  $(X,\calF_\eps,\calR_\eps)$ with $u_\eps(0)\leadsto u^0$. Is it possible to
  find a subsequence (not relabeled) and a limit function $u:[0,T] \to X$ 
  such that $u_\eps(t) \leadsto u_0(t)$ for all $t\in [0,T]$. 

\item[Q2] Is their a notion of convergence for the energies $\calF_\eps
  \overset{\mafo{energ}}\leadsto \calF_0$ and for dissipation potentials $\calR_\eps
  \overset{\text{diss}}\leadsto \calR_0$ such that $u_0$ is a solution of
  the \textbf{effective gradient system} $(X,\calF_0, \calR_0)$. 

\item[Q3] There are cases, where limits $\calF_0$ and $\calR_0$ as in Q2
  exists, but they produce the wrong solutions! Is there a direct way 
  to construct the correct effective GS $(X,\calF_\eff,\calR_\eff)$ is the
  sense that $(\calF_\eps,\calR_\eps) \overset{\text{GS}}\leadsto
  (\calF_\eff,\calR_\eff)$. 

\end{itemize}

In light of our examples in Section \ref{s:Intro} question Q3
cannot be answered by studying the solutions $u_\eps$ of the gradient-flow
equations $0 \in \pl\calR_\eps(u,\dot u)+ \plF\calF_\eps(u)$ and then showing
that the limits $u_0$ of sequences $u_\eps$ solve the \emph{effective}
evolution equation $ \dot u= \bfV_\eff(u)$.  
Of course, it is always a major achievement to find the effective evolution
equation, but it does not answer the question whether the effective equation
has a gradient structure. Moreover, if it has a gradient structure it may have
many of them. Hence, it is of independent interest, in particular in the
sense of physical modeling, to show how the gradient structure passes to the
limit. 

Of course, we are not so interested to study the case of ``continuous
dependence on parameters'' as is studied in the theory of ODEs. If
$V:[0,1]\ti [0,T]\ti \R^n \to \R^n$ is continuous and globally Lipschitz in $u
\in \R^n$., then the unique solution $u_\eps:[0,T]\to \R^n$ of $\dot u_\eps(t) =
V(\eps,t,u_\eps(t))$ depends continuously
on $\eps \in [0,1]$ and $t\in [0,T]$.
If we follow this approach in the setting of classical gradient systems $(M,\calF_\eps,
\bbG_\eps)$ on a finite-dimensional manifold $M$, then we need assumptions on
the energy $\calF:[0,1]\to M\to \R$ as well as on the Riemannian
tensor $\bbG_\eps(u):\rmT_uM\to \rmT^*_uM$ may depend on $\eps$. The
gradient-flow equation reads 
\[
\dot u= - \bbG_\eps(u)\rmD \calF_\eps(u) =:V(\eps,u).
\]
Thus, to apply the above-mentioned continuous dependence result for ODEs, we need 
$\bbG \in \rmC^0\big([0,1];\rmC^\mafo{Lip}(M;F_2(M))\big)$ and  $\calF \in
\rmC^0\big([0,1];\rmC^{1,\mafo{Lip}}( M) \big)$.

Such results are not relevant for PDEs because the vector fields are not
not smooth and only defined on dense subsets. There the question of
``singular limits'' is studied (cf.\ \cite{FeiNov09SLTV}), for instance PDEs of the form
\begin{align*}
\dot u_\eps&= \DIV\big(A(\frac1\eps x) \nabla u \big) - b(\frac1\eps
x)u_\eps , \quad x \in \Omega, \quad u_\eps|_{\pl\Omega}=0.
\\
\lambda_\eps \dot w_\eps &= \eps \pl_x^2(w_\eps) + \frac1\eps \big( w_\eps -
w_\eps^3) , \quad x \in \Omega, \quad w_\eps|_{\pl\Omega}=1.
\end{align*}

We refer to \cite{SanSer04GCGF, Serf11GCGF, Brai14LMVE, MiMoPe21EFED, Miel16EGCG,
  MurSav22?GFEV} for general approaches in evolutionary $\Gamma$-convergence.

\subsection{$\Gamma$-convergence for (static)  functionals}
\label{su:staticGCvg}

To study limits of functionals we define a notion of convergence in the spirit
of question Q3 above, but now in the static case. If the ``problem'' associated
with a GS $(X,\calF,\calR)$ is the solution of the gradient-flow equation, then
the ``problem'' associated with a static functional $\calJ$ is to find its
minimizer. Of course, we have seen that these problems are strongly linked
by the time-incremental minimization sometimes also called minimizing movement
scheme. Thus, for a family $(\calJ_\eps)_{\eps>0}$ of functionals $\calJ_\eps
: M \to \Rinfty$, we ask the (static) question:
\begin{description}
\item[Question:] What is a good notion of convergence $\calJ_\eps \leadsto \calJ_0$
  such that any limit $u_0$ of (a subsequence of) minimizers $u_\eps$ of
  $\calJ_\eps$ is automatically a minimizer of $\calJ_0$. 
\end{description}
Again, we are not so much interested in the case $\calJ_\eps \to \calJ_0$ in
$\rmC^1_\mafo{loc}(X)$, which is of course sufficient to show convergence in
the associated Euler-Lagrange equations.

We consider a complete metric space $(M,\DD)$ and functionals 
$\calJ_\eps: X \to \R_\infty$. In a metric space ``$u_k \to u$'' will always
denote convergence in the
metric; if $M$ is a Banach space $X$ then $u_k \to u$ and
$v_k\weak v$ denote strong and weak convergence, respectively. 
We first introduce more classical notions
of convergence of functionals, namely the \emph{pointwise convergence} 
$\calJ_\eps \xrightarrow{\text{pw}} \calJ_0$ and 
\emph{continuous convergence} (also weak in Banach spaces) defined via
\begin{subequations}\label{eq:ContConv}
\begin{align}
  \label{eq:pwCvg}
\calJ_\eps \overset{\mafo{pw}}\to \calJ_0,&\quad \text{ if } \
\calJ_\eps(u) \to \calJ_0(u) \  \text{ for all }u \in M; \\
  \label{eq:contCvg}
\calJ_\eps \Cto \calJ_0,& \quad \text{ if } \ u_\eps
\to u\ \Longrightarrow \ \calJ_\eps(u_\eps) \to \calJ_0(u).
\end{align}
\end{subequations}

In the context of minimization of functionals, the concept of
\emph{$\Gamma$-convergence} is more natural, see Theorem
\ref{th:GammaCvg}. This convergence was originally called \emph{variational
  convergence}  or \emph{epi-graph convergence} (cf.\
\cite{DegFra75TCV,Degi77GCGC,Atto84VCFO}), but nowadays the term
$\Gamma$-convergence is more common and we refer to
\cite{Dalm93IGC,Brai02GCB,Brai06HGC,Brai14LMVE} for further details.

\begin{definition}[$\Gamma$ and Mosco convergence]
\label{de:GammaCvg}
Let $(M,\DD)$ be a complete metric space.  We say that $\calJ_\eps$ \
\emph{$\Gamma$-converges to} $\calJ_0$ and write $\calJ_\eps \Gweak \calJ_0$ or
$\calJ_0 = \Glim\limits_{\eps\to 0} \calJ_\eps$, if $(\Gamma.\mafo{inf})$ and
$(\Gamma.\mafo{sup})$ hold:
\\[0.2em]
\hspace*{1.2ex}\ $(\Gamma.\mafo{inf})$\quad
$u_\eps \to\, u \ \Longrightarrow\ \calJ_0(u) \leq \liminf\limits_{\eps\to 0}
\calJ_\eps(u_\eps)$ \hfill {\small (liminf estimate)}
\\[0.2em]
\hspace*{1.2ex}\ $(\Gamma.\mafo{sup})$\quad $\forall\, \wh u\ \exists\, (\wh
u_\eps)_\eps{:}\ \wh u_\eps \to\, \wh u \ \text{ and } \calJ_0(\wh
u)=\limsup\limits_{\eps\to 0} \calJ_\eps(\wh u_\eps)$ \hfill {\small (limsup
estimate)}%
\\[0.4em]
If $(M,\DD)$ is a Banach space $(X;\|\cdot\|)$ we say that $\calJ_\eps$
(sequentially) \emph{weakly $\Gamma$-converges to} $\calJ_0$ and write
$\calJ_\eps \Gweak \calJ_0$ and $\calJ_0= \Gwlim\limits_{\eps\to 0 } \calJ_\eps$,
if $(\Gamma.\mafo{inf})$ and $(\Gamma.\mafo{sup})$ when ``$\to$'' is replaced
by ``$\weak$''.  If $\calJ_\eps \Gto \calJ_0$ and $\calJ_\eps \Gweak \calJ_0$
hold, then we say that $\calJ_\eps$ \emph{Mosco-converges to} $\calJ_0$ and
write $\calJ_\eps \Mto \calJ_0$ or
$\calJ_0 = \Mlim\limits_{\eps\to 0} \calJ_\eps$. In this case, it suffices to
show $(\Gamma.\mafo{inf})$ for weak convergence and $(\Gamma.\mafo{sup})$ for
strong convergence.
\end{definition}
 
We will see in Lemma \ref{le:Average} that
there are simple quadratic functionals for which weak and strong
$\Gamma$-limits exist, but they are different. 

The conditions $(\Gamma.\mafo{sup})$ is often replaced by the so-called
\emph{existence of recovery sequences}:
\\[0.2em]
\hspace*{1.2ex}\ $(\Gamma.\mafo{rec})$\quad
$\forall\, \wh u\ \exists\, (\wh u_\eps)_\eps{:}\ \wh u_\eps \to\, \wh u \
\text{ and } \calJ_\eps(\wh u_\eps) \to \calJ_0(\wh u)$. \hfill {\small
  (recovery sequence)}%
\\[0.4em]
Of course, $(\Gamma.\mafo{rec})$ implies $(\Gamma.\mafo{sup})$. Moreover,
assuming that $(\Gamma.\mafo{inf})$ holds, $(\Gamma.\mafo{rec})$ follows from
$(\Gamma.\mafo{sup})$.  The sequence $(\wh u_\eps)_\eps$ is called recovery
sequence as it recovers the correct energy $\calJ_0(\wh u)$. Moreover, one sees
in several examples that $\wh u_\eps$ has to ``recover'' the correct
microscopic structure which makes the energy $\calJ_\eps(\wh u_\eps)$ small
enough to reach (recover) the lowest possible value for $\calJ_0(\wh
u)$. \medskip

We emphasize that the definition of $\Gamma$-convergence is asymmetric and
fits to ``minimization''. For ``liminf'' we impose a condition for \emph{all}
sequences, while for ``limsup'' we only need \emph{one} sequence. This way we lose
the linearity for $\Gamma$-convergence. If $\Glim \calJ_\eps$, $\Glim
\calG_\eps$, and $\Glim(\calG_\eps{+}\calJ_\eps)$ exist we do \emph{not} have 
$\Glim(\calG_\eps{+}\calJ_\eps) =\big( \Glim \calG_\eps \big)  + \big(\Glim
\calJ_\eps \big)$ in general. 

\begin{example}
\label{ex:GammaLimits} 
(A) Consider $X=\R^1$ and $\calJ_\eps(u)= \frac12 u^2 -\cos(u/\eps)$. We claim
\[
\calJ_\eps \Gto \calJ_0 \quad \text{with } \calJ_0(u)=\frac12 u^2 -1.
\]  
To show $(\Gamma.\mafo{inf})$ we use $\cos \alpha \leq 1$ and obtain
$\calJ_\eps(u)\geq \calJ_0(u)$ for all $u$. As $\calJ_0$ is continuous, the
result follows. To show $(\Gamma.\mafo{sup})$ we start from an arbitrary $\wh
u\in \R$ and look for a close-by $\wh u_\eps$ such that $\calJ_\eps(\wh u_\eps)$
is close to $\calJ_0(\wh u)$. This means that we want to have $\cos(\wh u_\eps)$
close to $1$. Thus, we choose $\wh u_\eps = 2\pi\eps \lfloor \wh
u/(2\pi\eps)\rfloor$, where the floor function $\lfloor\cdot\rfloor$ rounds down
to the nearest integer. Obviously, we have $\wh u_\eps \to \wh u$ and
$\calJ_\eps(\wh u_\eps) =\calJ_0(\wh u_\eps) \to \calJ_0(\wh u)$ as desired. 

(B) For an arbitrary $\lambda \in \R$, we set  $\calG_\eps = \lambda
\calJ_\eps$. With an analogous argument we obtain 
\[
\calG_\eps \Gto \calG_0\quad 
\text{with }\calG_0(0)=\frac\lambda 2 u^2 -|\lambda| 
\]
In the case $\lambda<0$ one chooses  $\wh u_\eps = \pi\eps \big(2\lfloor \wh
u/(2\pi\eps)\rfloor+1\big)$ to find $\cos (\wh u_\eps/\eps)=-1$. 

(C) We see that linearity is destroyed, in particular we have 
\[
0 = \Glim_{\eps\to 0}(\calJ_\eps-\calJ_\eps) \neq 
\big( \Glim_{\eps\to 0}\calJ_\eps \big) + \Glim_{\eps\to 0}\big( {-}\calJ_\eps
\big) = -1 + (-1) = -2.
\] 
\end{example}

The following lemma presents a simple quadratic example in which the
weak and the strong $\Gamma$-limits exist but they are different. We define
\[
\calF_\eps(w) = \int_\Omega \frac12 w(x) \cdot \bbA\big(\frac1\eps x\big)
w(x)\dd x  \quad\text{for } w \in X =\rmL^2(\Omega;\R^m), 
\]
where $\Omega \subset \R^d$ is a bounded Lipschitz domain and $\bbA
\in \rmL^\infty(\R^d;\R^{m\ti m}_\text{sym})$ is  1-periodic, i.e.\
$\bbA(y{+}n)=\bbA(y)$ for all $y\in \R^d$ and all $n\in
\Z^d$. Moreover, we assume that $\bbA$ is uniformly positive definite,
i.e.\ $\underline{a} |w|^2 \leq w\cdot \bbA(y)w\leq \ol{a} |w|^2$ for
$\ol a >\underline a >0$.  The main tools is the Riemann lemma stating that 
the sequence $A_\eps: x \mapsto \bbA(\frac1\eps x)$ satisfies $A_\eps \overset*\weak
\bbA_\text{arith}$, see \cite[Exa.\,6.6]{Dalm93IGC} for more general results of this
type. 

\begin{lemma} \label{le:Average}
Define the arithmetic and harmonic mean of $\bbA$ via
\[
\bbA_\text{arith} := \int_{[0,1]^d} \bbA(y) \dd y 
\quad \text{ and } \quad 
\bbA_\text{harm} := \Big(\int_{[0,1]^d} \bbA(y)^{-1} \dd y\Big)^{-1}
\]
and the two functionals 
\[
\calF_\text{arith} (w) = \int_\Omega \frac12 w(x)\cdot
\bbA_\text{arith} w(x) \dd x 
\ \text{ and } \ 
\calF_\text{harm} (w) = \int_\Omega \frac12 w(x)\cdot
\bbA_\text{harm} w(x) \dd x . 
\]
In $X=\rmL^2(\Omega;\R^m)$ we have $\calF_\eps \Gweak
\calF_\text{harm}$ and $\calF_\eps \Cto \calF_\text{arith}$, which
implies $\calF_\eps \Gto \calF_\text{arith}$. 
\end{lemma}
\begin{proof}
We first prove $\calF_\eps \Gweak \calF_\text{harm}$. 
For the liminf estimate assume $w_\eps \weak w$ in
$\rmL^2(\Omega)$. Writing $\bbA_\eps(x)=\bbA(\frac1\eps x)$ we have
\begin{align}
\label{eq:calF-decomp}
 \calF_\eps(w_\eps)&= \frac12\int_\Omega w_\eps\cdot \bbA_\eps
w_\eps\dd x =\\ \nonumber
&=\frac12\int_\Omega
\underbrace{(w_\eps{-}\bbA_\eps^{-1}\bbA_\text{harm}w){\cdot} \bbA_\eps 
(w_\eps{-}\bbA_\eps^{-1}\bbA_\text{harm}w)}_{\geq 0} \\
&\qquad\qquad
+2\underbrace{w_\eps}_{\weak w} {\cdot}\bbA_\text{harm} w 
- \bbA_\text{harm} w{\cdot}
\underbrace{ \bbA_\eps^{-1}}_{\weaks\bbA_\text{harm}^{-1}} \bbA_\text{harm} w \dd x.
\end{align}
Dropping the nonnegative term and taking the limit $\eps\to 0$ give
the desired lower estimate  $\liminf_\eps \calF_\eps(w_\eps)\geq
\frac12\int_\Omega 0 +2w{\cdot}\bbA_\text{harm}w -
w{\cdot}\bbA_\text{harm}w \dd x =  \calF_0(w)$.  

For the limsup-estimate we use the same reformulation of $\calF_\eps$
as in \eqref{eq:calF-decomp}. For a
given $\wh w$ we choose $\wh w_\eps = \bbA_\eps^{-1}\bbA_\text{harm} \wh
w$. Since by construction the first term in the integral is $0$ we
find $\calF_\eps(\wh w_\eps) =  \frac12\int_\Omega  0  + 
2 \bbA_\eps^{-1}\bbA_\text{harm} \wh w {\cdot}\bbA_\text{harm} \wh w  
- \bbA_\text{harm} \wh w {\cdot} \bbA_\eps^{-1}\bbA_\text{harm} \wh w \dd x  \to 
\calF_\text{harm}(\wh w)$.
\smallskip

For strong continuous convergence take any  $w_\eps \to w$
in $\rmL^2(\Omega)$ and write
\begin{align}
\calF_\eps(w_\eps)&=\frac12\int_\Omega w{\cdot}
 \underbrace{\bbA_\eps w}_{\weak \bbA_\text{arith}w} -  2 w{\cdot}
\bbA_\eps\underbrace{(w{-}w_\eps)}_{\to 0} +
\underbrace{(w{-}w_\eps)}_{\to 0} {\cdot} \bbA_\eps(w{-}w_\eps) \dd x \\
& \to \ \calF_\text{arith}(w).
\end{align}
This proves the strong continuous and hence the strong $\Gamma$-convergence.
\end{proof}

Clearly, continuous convergence is much stronger than
$\Gamma$-convergence. We have the following relations.

\begin{lemma}[Properties of $\Gamma$-limits]
\label{le:Cto} On the complete metric space $(M,\DD)$ consider
  functionals $\calJ_\eps,\,\calK_\eps:M\to \Rinfty$. 

 (a) $\calJ_\eps \Gto \calJ_0 \
  \Longrightarrow \ \calJ_0:M\to \Rinfty$ is lsc. 

 (b)\  $\calJ_\eps \Gto \calJ_0$ and $\calK_\eps \Cto \calK_0 \
  \Longrightarrow \ \calJ_\eps{+}\calK_\eps \Gto \calJ_0{+} \calK_0$

 (c) \  $\calJ_\eps \Cto \calJ_0 \
  \Longrightarrow \ \calJ_\eps \Gto \calJ_0$

\end{lemma}
\begin{proof}
\STEP{Part (a):} We use an argument that is standardly used for constructing
recovery sequences. For $u_n\to u$ we have to show $\calJ_0(u)\leq
\liminf_{n\to \infty} \calJ_0(u_n)$.  As $\calJ_0= \Glim\calJ_\eps$ we find
$(\wh u^n_\eps)_\eps$ with $\wh u^n_\eps \to u_n$ and $\calJ_\eps( \wh
u^n_\eps) \to \calJ_0(u_n)$. Thus for each $n$ we can find $\eps_n>0$ such that 
\[
\eps_n\in
{]0,1/n[}, \quad \DD(\wh u^n_{\eps_n}, u_n) \leq 1/n, \quad \calJ_{\eps_n}(\wh
u^n_{\eps_n} ) \leq \calJ_0(u_n) + 1/n.
\]
Setting $\wt u_{\eps_n}:= \wh u^n_{\eps_n}$ we have $\eps_n \to 0$ and $\DD(\wt
u_{\eps_n},u) \leq \DD(\wt u_{\eps_n},u_n)+ \DD(u_n,u)\to 0$. Setting $\wt
u_\eps=u$ for $\eps \not\in \set{\eps_n}{n\in \N}$, we have $\wt u_\eps \to u$
and obtain
\[
\liminf_{n\to \infty}\calJ_0(u_n) \geq  \liminf_{n\to \infty}
\calJ_{\eps_n}(\wt  u_{\eps_n}) \geq \liminf_{\eps \to 0^+} \calJ_\eps(\wt
u_\eps) \geq \calJ_0(u),
\]
where the last estimate follows from $(\Gamma.\inf)$ and $\wt u_\eps \to u$. 

\STEP{Part (b):} This follows easily as convergent sequences can be chosen as
needed for $\calJ_\eps$, and continuous convergence for $\calK_\eps$ gives the
result. 

\STEP{Part (c):} This is trivial because the liminf estimate is a limit. As
recovery sequence one can take any convergent sequence, e.g.\ the constant
sequence with $\wh u_\eps = \wh u$. 
\end{proof}

The following properties of sequences of functionals will be useful in the
formulation of the following results. Recall the sublevels
$S^\calF_E:=\bigset{u \in M}{\calF(u)\leq E}$. 

\begin{definition}[Uniform properties]
\label{def:Equi.calJeps}
On a complete metric space $(M,\DD)$ consider a family $(\calJ_\eps)_\eps$ of
functionals $\calJ_\eps:M\to \Rinfty$. 

(i) The family is called \emph{equi-coercive}, if 
\[ 
  \forall\, E\in \R\ \exists\, R>0,\ u_*\in M\ \forall\, \eps: \quad
  S^{\calJ_\eps}_E \subset B_R(u_*).
\]

(ii) The family is called \emph{equi-compact}, if (where ``$\Subset$'' means
compactly contained)  
\[ 
  \forall\, E\in \R\ \exists\, K\Subset M \  \forall\, \eps: \quad
  S^{\calJ_\eps}_E \subset K. 
\]

(iii) If $(M,\DD)$ is a Banach space $(X,\|\cdot\|)$, we call the family
\emph{equi-superlinear}, if there exists a superlinear function
$\varphi:{[0,\infty[}\to \R$ such that
\[
  \forall\, u \in X\ \forall\, \eps: \quad \calJ_\eps(u) \geq \varphi\big(
  \|u\|\big). 
\]
\end{definition}

\textbf{Warning:} In many papers and textbooks our notion of
``equi-compactness'' is simply called ``equi-coercivity''.  
We distinguish these two concepts, which is quite useful for gradient
systems where different functionals like $\calF_\eps$ and $\calR_\eps$ or
$\DD_\eps$ are considered on the same space. Moreover, it allows us to avoid
switching between weak and strong topologies in Banach spaces, where
equi-coercivity implies weak equi-compactness.

\noindent The origin for the definition of $\Gamma$-convergence, which is
clearer in the original name ``variational convergence'', is the
following convergence of minimizers, see
\cite{Dalm93IGC, Brai02GCB}.

\begin{theorem}[Convergence of minimizers] \label{th:GammaCvg}
In a complete metric space $(M,\DD)$ assume $\calJ_\eps \Gto \calJ_0$ with $\inf \calJ_0
=:\alpha_0 \in \R$. 

(a) If $u_\eps \to u_0$ and $\liminf_{\eps \to 0^+}\calJ_\eps(u_\eps) =
\alpha_0$, then $u_0$ is a minimizer of $\calJ_0$.

(b) If the family $(\calJ_\eps)_\eps$ is equi-compact, then $\alpha_\eps =
\inf\calJ_\eps$ satisfies $\alpha_\eps \to \alpha_0$. Moreover, every sequence
$(u_\eps)_{\eps>0}$ with $\calJ_\eps(u_\eps)\to \alpha_0$ has a convergent
subsequence $u_{\eps_k}\to u_0$ and each such limit $u_0$ is a minimizer of
$\calJ_0$. In particular, if $(u_{\eps_k})$ is a sequence of minimizers for
$\calJ_{\eps_k}$, then all accumulation points $u_0$ of this sequence are
minimizers of $\calJ_0$.
\end{theorem} 
\begin{proof} \STEP{Part (a).} By the $(\Gamma.\inf)$ we have $\calJ_0(u_0)\leq
  \liminf_{\eps\to 0} \calJ_\eps(u_\eps) = \alpha_0$. However, with $\alpha_0=
  \inf \calJ_0\leq \calJ_0(u_0)$ we conclude $\alpha_0=\calJ_0(u_0)$, i.e.\
  $u_0$ is a minimizer. 

\STEP{Part (b).} By Lemma \ref{le:Cto} we know that $\calJ_0$ is lsc and the
equi-compactness implies that the sublevels of $\calJ_0$ are compact. Hence
$\calJ_0$ has a minimizer $u_0$ with $\calJ_0(u_0)=\alpha_0$. 

By $(\Gamma.\sup)$ there exists a recovery sequence $\wh u_\eps \to u_0$ with
$\calJ_\eps (\wh u_\eps)\to \calJ_0(u_0)=\alpha_0$. Using
$\alpha_\eps =\inf \calJ_\eps \leq \calJ_\eps(\wh u_\eps)$ we find
$\limsup_{\eps\to 0} \alpha_\eps \leq \alpha_0$. Moreover, we can choose a
subsequence $\eps_k\to 0$ such that
$\liminf_{\eps \to 0} \alpha_\eps =\lim_{k\to \infty} \alpha_{\eps_k}$. In
addition, there exist $u_k$ with
$\calJ_{\eps_k}(u_k)\leq \alpha_{\eps_k} + \eps_k$, and the equi-compactness
guarantees the existence of a convergent subsequence $u_{k(l)} \to u_*$ for
$l \to \infty$. Now $(\Gamma.\inf)$ implies
\[
\alpha_0 \leq \calJ_0(u_*) \leq \liminf_{l \to \infty}
\calJ_{\eps_{k(l)}}(u_{k(l)}) \leq \liminf_{l\to \infty}
(\alpha_{\eps_{k(l)}}{+}\eps_{k(l)}) =\liminf_{\eps\to 0} \alpha_\eps \leq 
\limsup_{\eps\to 0} \alpha_\eps \leq \alpha_0.  
\]
Hence, $\alpha_\eps \to \alpha_0$ is established. 

If a sequence $u_\eps$ satisfies $\calJ_\eps(u_\eps)\to \alpha_0 \in \R$, it
lies in a compact set, because of equicompactness. By (a) all accumulation
points are minimizers. 

The last statement is a consequence of the previous assertion and the
convergence $\alpha_\eps \to \alpha_0$. 
\end{proof}

\begin{example} For $(M,\DD)=(\R,\DD_\mafo{Eucl})$ the sequence
  $\calJ_\eps(u)=\cos(\eps u)$ satisfies $\calJ_\eps\Gto \calJ_0$ with
  $\calJ_0\equiv 1$. Indeed, for $u$ in the compact interval $[-R,R]$ we have
  $|\eps u|\leq \eps R$ and find $0\leq 1-\calJ_\eps(u) \leq \frac12\eps^2R^2$,
  which gives uniform convergence on compact sets to $\calJ_0:u\mapsto 1$. 

However, for all $\eps>0$ we have $\alpha_\eps=\inf_\R \calJ_\eps  =-1$, whereas
for $\eps=0$ we have $\alpha_0 \inf_\R \calJ_0=1$. 
\end{example}

The main result of the above theorem is that solving a minimization problem for
$\calJ_\eps$ can be interchanged with passing to the limit $\eps\to 0$. This
can be depicted by the following commuting diagram:\medskip

\centerline{
{\centering
\begin{tikzpicture}
\draw [color=gray!10,fill=gray!10] (-6,-1.8) rectangle (6,1.8);
\draw [fill=green!20, very thick] (-1.85,0.6) rectangle (-1.15,1.4);
\draw [fill=green!20, very thick] (1.85,0.6) rectangle (1.15,1.4);
\node at (-1.5,1){$\calJ_\eps$};
\draw[ ->] (-0.9,1)-- node[above, pos=0.5]{$\Gamma$} (0.9,1); 
\node at (1.5,1){$\calJ_0$};

\node at (0,0) {{\footnotesize $\eps \longrightarrow 0$}}; 

\node[left] at (-2.5,0) {minimizing $\calJ_\eps$};
\node[right] at (2.5,0) {minimizing $\calJ_0$};

\draw[->, thick] (-1.5,0.5)--(-1.5,-0.5);
\draw[->, thick] (1.5,0.5)--(1.5,-0.5);

\node at (-2.5,-1){$\mafo{Argmin}\,\calJ_\eps \ni u_\eps$}; 
\node at (2.5,-1){$ u_0 \in \mafo{Argmin}\,\calJ_0$}; 
\draw[ ->] (-0.9,-1)--(0.9, -1); 

\end{tikzpicture}\par
}}

We make this more explicit in the Banach space setting by considering an
equi-superlinear family $\calF_\eps$ with $\calF_\eps\Gweak \calF_0$. Then, for
all $\ell\in X^*$ we set $\calJ^\ell_\eps = \calF_\eps
-\langle\ell,\cdot\rangle$ and observe $\calJ_\eps^\ell \Gto \calJ_0^\ell$,
cf.\ Lemma \ref{le:Cto}(b). We define the $\mafo{LimSup}$ for a family
$\big(A_\eps\big)_\eps$ of sets $A_\eps\subset X$ via 
\[
\mathop{\mafo{LimSup}}_{\eps\to 0} A_\eps:= \bigset{u \in X}{ \exists\,
  (\eps_k,u_k)_{k\in \N}: \ \ 0<\eps_k\to 0, \ u_k\in A_{\eps_k}, \ u_k \to u} 
\]
Thus, the theory of $\Gamma$-convergence leads to the following result on
the upper semicontinuity of minimizers.

\begin{corollary}[Upper semicontinuity of the sets of minimizers]
\label{co:UppSemicontMinim} 
If $\big(\calF_\eps\big)_\eps$ is equi-superlinear on a Banach space $X$. Then,
we have 
\[
\calF_\eps \Gweak \calF_0 \quad \Longrightarrow \quad 
\forall\: \ell \in X^*: \quad \mathop{\mafo{LimSup}}_{\eps\to 0}
\mafo{Argmin}\big(\calF_\eps{-}\ell\big) \ \subset\
\mafo{Argmin}\big(\calF_0{-}\ell\big).  
\]
\end{corollary}

The following useful result seems to be folklore, but it is not easy to locate
a specific reference. Hence, we give a full proof. 

\begin{proposition}[$\Gamma$-convergence versus Mosco convergence] 
\label{pr:G-Mosco}
Assume that $X$ and $\bfZ$ are reflexive Banach spaces such that
$\bfZ$ is compactly embedded in $X$, written $\bfZ\Subset
X$. Moreover, assume that the functionals $\calJ_\eps$ are
equi-coercive in $\bfZ$, i.e.\ 
\begin{align}
  \label{eq:EquiCoerJ}
\forall\,J>0\ \exists\, R>0\ \forall\, \eps>0,\ u\in X:\quad 
\calJ_\eps(u)\leq J\ \Rightarrow\ \| u\|_\bfZ\leq R:
\end{align}
Then, $\calJ_\eps \Mto \calJ_0$ in $X$ is equivalent to $\calJ_\eps
\Gweak \calJ_0$ in $\bfZ$. 
\end{proposition}
\begin{proof} The equi-coercivity is meant such that all $\calJ_\eps$ take the
  value $+\infty$ on $X\setminus\bfZ$.

  ``$\Rightarrow$''\ We start from $\calJ_\eps \Mto \calJ_0$ in $X$.
  If $u_\eps \weak u$ in $\bfZ$, then this also holds in
  $X$. Hence, the liminf estimate follows. To construct a recovery
  sequence $\wh u_\eps \weak \wh u$ in $\bfZ$ for arbitrary $\wh u\in
  \bfZ$, we first assume $\calJ_0(\wh u)<\infty$. We choose the
  recovery sequence $\wh u_\eps$ guaranteed by $\calJ_\eps \Mto
  \calJ_0$ in $X$, i.e.\ we know $\wh u_\eps \to \wh u$ in
  $X$. The equi-coercivity \eqref{eq:EquiCoerJ} and $\calJ_\eps(\wh
  u_\eps)\to \calJ_0(\wh u)<\infty$ imply $\|\wh u_\eps\|_\bfZ\leq
  R$. Hence, $\wh u_\eps \weak \wh u$ in $\bfZ$ by reflexivity of
  $\bfZ$.  If $\calJ_0(\wh u)=\infty$, we choose $\wh u_\eps = \wh u$
  giving $\wh u_\eps \to \wh u$ in $\bfZ$. Hence, the liminf estimate
  yields $\infty=\calJ_0(\wh u)\leq \liminf_{\eps \to 0} \calJ_\eps(\wh
  u) $, which shows that we have a recovery sequence in $\bfZ$.\medskip

  ``$\Leftarrow$''\ Given $\calJ_\eps \Gweak \calJ_0$ in $\bfZ$, we
  take any sequence $u_\eps \weak u$ in $X$. If we have $\liminf_{\eps\to
    0}\|u_\eps\|_\bfZ =\infty$, then the equi-coercivity implies
  $\calJ_\eps(u_\eps)\to \infty$ and the liminf estimate holds.  If
  for some subsequence $\|u_{\eps_k}\|_\bfZ\leq C$, then $u_{\eps_k}
  \weak u$ in $\bfZ$, and the liminf estimate follows from that of
  $\calJ_\eps \Gweak \calJ_0$ in $\bfZ$.  For the construction of
  recovery sequences, we can choose $\wh u_\eps = \wh u$ if $\wh u\in
  X\setminus \bfZ$.  If $\wh u\in \bfZ$ we choose a recovery
  sequence $\wh u_\eps \weak \wh u$ in $\bfZ$. By the compact
  embedding we have $\wh u_\eps \to \wh u$ in $X$ and the proof is
  finished.
\end{proof}

The following result will be very useful for studying the evolutionary
$\Gamma$-convergence for gradient systems $(X,\calF_\eps,\Psi_\eps)$ in Banach
spaces, because there we need $\Gamma$-convergence for $\Psi_\eps$ and for
$\Psi^*_\eps$. The connection between Legendre transform and
$\Gamma$-convergence is nontrivial because it involves the duality product
$X\ti X^* \ni (v,\xi) \mapsto \langle \xi,v\rangle$ which is only weak-strongly
or strong-weakly continuous and moreover it is order reversing because of
$-\Psi_\eps$, hence ``$\inf$'' and ``$\sup$'' are interchanged.

\begin{theorem}[{\cite[pp.\,271]{Atto84VCFO}}]
\label{th:Attouch}
Let $X$ be a separable, reflexive Banach space and assume that 
all $\Psi_\eps:X\to [0,\infty]$ are dissipation potentials (namely lsc, convex 
and $\Psi_\eps(0)=0$).  Then,
\[
\Psi_\eps \ \Gweak\  \Psi \qquad \Longleftrightarrow \qquad \Psi_\eps^* \
\Gto \ \Psi^*\:.
\]
\end{theorem}

The proof uses techniques from \cite{Mosc71CYFT}, where the following
equivalence was shown:  
\begin{align}
  \label{eq:MoscoEquiv}
\Psi_\eps \Mto \Psi \quad  \Longleftrightarrow \quad \Psi_\eps^* \Mto \Psi^*,
\end{align}
which is a direct consequence of the above theorem, but holds under
weaker assumptions. 

\noindent
\begin{proof}[Sketch of proof] 
The following four implications imply the desired result.  

(1)  $(\Gamma_\rmw.\inf)$ for $\Psi_\eps \quad 
\Longrightarrow\quad (\Gamma_\rms.\sup)$  for $\Psi^*_\eps$
 
(2)  $(\Gamma_\rmw.\sup)$ for $\Psi_\eps \ \  
\Longrightarrow\quad (\Gamma_\rms.\inf)$  for $\Psi^*_\eps$

(3)  $(\Gamma_\rms.\inf)$ for $\Psi^*_\eps \quad 
\Longrightarrow\quad (\Gamma_\rmw.\sup)$  for $\Psi_\eps$

(4)  $(\Gamma_\rms.\sup)$ for $\Psi^*_\eps \ \ 
\Longrightarrow\quad (\Gamma_\rmw.\inf)$  for $\Psi_\eps$

The simpler directions are from ``$\sup$'' to ``$\inf$'', because we don't have
to show existence of a converging sequence. We give the proof of (2) and
observe that (4) is analogous. 

\STEP{Part (2):} We consider an arbitrary sequence $\xi_\eps \to \xi$.

For $\delta >0$ we find $\wh v_0$ such that $\Psi^*(\xi)\leq \delta +\langle
\xi, \wh v_0\rangle - \Psi(\wh v_0)$. By $(\Gamma_\rmw.\sup)$ we find a
recovery sequence $\wh v_\eps \weak \wh v_0$ and $\Psi(\wh v_0)\geq
\limsup_{\eps\to 0} \Psi_\eps(\wh v_\eps)$. With this, we have
\begin{align*}
\liminf_{\eps\to 0} &\Psi^*_\eps(\xi_\eps) \overset{\text{Legr}}\geq
\liminf_{\eps\to 0} \big( \langle \xi_\eps, \wh v_\eps\rangle 
  -\Psi_\eps(\wh v_\eps)\big) 
\\
&\overset{*}= \langle \xi,\wh v_0\rangle - \limsup_{\eps\to 0} \Psi_\eps(\wh
v_\eps) \geq  \langle \xi,\wh v_0\rangle- \Psi(\wh v_0) \geq
\Psi^*(\xi)-\delta. 
\end{align*} 
In $\overset*=$ we use the weak-strong continuity of
$(v,\xi) \mapsto \langle \xi,v\rangle$. As $\delta>0$ was arbitrary, we have
$\liminf_{\eps\to 0} \Psi^*_\eps(\xi_\eps) \geq \Psi^*(\xi)$ as desired for
$(\Gamma_\rms.\inf)$.

\STEP{Part (1):} We show this under the additional assumption that the family 
$\big(\Psi_\eps\big)_\eps$ is equi-superlinear. 
In this case, the constant recovery sequence $\wh\xi_\eps =
\wh\xi$ always works. 

We first observe $v \mapsto \Psi_\eps(v)-\langle \xi,v\rangle$ has at least one
minimizer $v_\eps$ because $\Psi_\eps$ is superlinear, lsc, and convex. By
equi-superlinearity we find $\|v_\eps\|\leq C<\infty$ for all $\eps>0$. 
We first choose a subsequence $\eps_k\to 0$ such that $\Psi_{\eps_k}(\xi)\to
\limsup_{\eps \to 0} \Psi_\eps(\xi)$. Next, we can extract a further
subsequence such that $v_{\eps_{k(l)}} \weak v_*$ for $l\to \infty$ and conclude
\begin{align*}
&-\Psi^*(\xi) \overset{\text{Legr}}= \inf_{v\in X} \big(  \Psi(v)-\langle
\xi,v\rangle \big)  \leq \Psi(v_*)-\langle \xi,v_*\rangle  
\overset{(\Gamma_\rmw.\inf)}\leq  \liminf_{l\to \infty} \big(  \Psi(v_{\eps_{k(l)}}) -
\langle \xi,v_{\eps_{k(l)}}\rangle \big)\\
& = \lim_{l\to \infty}
\big({-}\Psi^*_{\eps_{k(l)}}(\xi) \big) = - \limsup_{\eps \to 0} \Psi_{\eps}(\xi),
\end{align*}
which is the desired estimate of $(\Gamma_\rms.\sup)$. 

\STEP{Part (3):} This is much more difficult and we refer the reader to
\cite[pp.\,271]{Atto84VCFO}.
\end{proof}

Lemma \ref{le:Average} provides an interesting example for the
application of Theorem \ref{th:Attouch}.  In fact, we have $\calF_\eps^*(\xi) =
\frac12\int_\Omega \xi{\cdot}\bbA_\eps^{-1} \xi \dd x$. Thus, the
strong convergence for $\calF_\eps^*$ leads to an effective matrix
$\text{arith}(\bbA^{-1}) = \text{harm}(\bbA)^{-1} $.\medskip

Another important tool of convex analysis is the weak-strong
closedness of the graphs of the subdifferentials $\plF\calF_\eps:X
\rightrightarrows X^*$ in the limit $\eps\to 0$. The following result is a
variant of \cite[Thm.\,3.66]{Atto84VCFO}, and it again relies strongly on 
semi-convexity.

\begin{proposition}[Strong-weak closedness for subdifferentials for $\Gamma$-limits]
\label{pr:ClosednessGammaLim}
Assume that all $\calF_\eps:X\to \Rinfty$ are proper and lsc and $\calF_\eps
\Gto \calF_0$ in the reflexive Banach space $X$. Moreover, assume that
$\big(\calF_\eps\big)_\eps$ is equi-semiconvex, i.e.\ there exists $\lambda \in
\R$ such that all $\calF_\eps$ are $\lambda$-convex. Then, we have 
\begin{align}
  \label{eq:Closedn}
\left. \begin{aligned}
 &u_\eps \to u \text{ in } X, \quad \xi_\eps \weak \xi \text{ in } X^*\\
 &\forall\, \eps >0: \ \xi_\eps \in \plF\calF_\eps(u_\eps) 
\end{aligned} \right\}\quad \Longrightarrow
\quad 
\calF_\eps(u_\eps) \to \calF_0(u) \text{ and }\xi \in \plF\calF_0(u).
\end{align}
\end{proposition} 
\begin{proof} 
The $\lambda$-convexity of $\calF_\eps$ gives 
\begin{equation}
  \label{eq:calFeps.laCvx} \calF_\eps(w) \geq \calF_\eps(u_\eps) + \langle
\xi_\eps, w{-}u_\eps\rangle + \frac\lambda 2 \| w{-}u_\eps\|^2 \quad \text{ for
all } w \in X.
\end{equation}
Choosing an arbitrary $\wh w\in X$ the limsup condition $(\Gamma.\sup)$
provides a recovery sequence $\wh w_\eps \to \wh w$ with
$\calF_\eps(\wh w_\eps) \to \calF_0(\wh w)$. Inserting $w=\wh w_\eps$ into
\eqref{eq:calFeps.laCvx} and passing to the limit $\eps\to 0$ we can exploit
the strong convergence $\wh w_\eps- u_\eps \to \wh w-u$ and the weak
convergence $\xi_\eps\weak \xi$. Setting $\ol F_0=\limsup_{\eps\to 0}
\calF_\eps(u_\eps)$ we find
\begin{equation}
  \label{eq:calFeps.lowerBdd} 
\calF_0(\wh w) \geq \ol F_0 + \langle \xi, \wh w{-}u\rangle +
\frac\lambda2\| \wh w{-} u\|^2 \quad \text{for all } \wh w \in X.
\end{equation}

Using $(\Gamma.\inf)$ we have $\calF_0(u)\leq \liminf_{\eps\to 0} \calF_\eps(u_\eps)$, while
\eqref{eq:calFeps.lowerBdd} with $\wh w=u$ gives $\calF_0(u)\geq \ol
F_0=\limsup \calF_\eps(u_\eps)$, which
provides the desired convergence $\calF_\eps(u_\eps)\to \calF_0(u)$. 

Moreover, replacing $\ol F_0$ in \eqref{eq:calFeps.lowerBdd} by
$\calF_0(u)$ we conclude $\xi \in \plF\calF_0(u)$ as desired. 
\end{proof}

\subsection{Evolutionary $\bm\Gamma$-convergence via EVI}
\label{su:EvolGCvg.EVI}

We consider metric GS $(M,\calF_\eps,\DD_\eps)$ (recall that this notation
implies $\psi=\psi_\mafo{quadr}$) and the associated EVI formulation which is
ideal to pass to the limit $\eps\to 0$ because the formulation only contains
the functionals $\calF_\eps$ and $\DD_\eps$, but no derivatives like $\BAR \dot
u_\eps\BAR_{\DD_\eps}$ or $\BAR \pl\calF_\eps\BAR_{\DD_\eps}$ appear. 
 
The following result is a variant of \cite[Thm.\,2.17]{DanSav14LNGF},  where the
more restrictive case $\DD_\eps=\DD$ is treated, see also  \cite{MurSav22?GFEV}. 

\begin{theorem}[Evolutionary $\Gamma$-convergence via EVI]
\label{th:EvolGCvg.EVI} 
Consider a complete metric space $(M,\DD)$ and the metric GS
$(M,\calF_\eps,\DD_\eps)$ with the following properties:
\begin{subequations}
\label{eq:EVIcvgAss}
\begin{align}
\label{eq:EVIcvgAss.a}
&\exists\, C\geq 1\ \forall\, \eps>0:\quad \DD\leq \DD_\eps \leq C\DD;
\\
\label{eq:EVIcvgAss.b}
&  \DD_\eps \Cto \DD_0 \ \text{  in } (M,\DD);
\\
\label{eq:EVIcvgAss.c}
& \big(\calF_\eps\big)_\eps \ \text{ is equi-compact in } (M,\DD);
\\
\label{eq:EVIcvgAss.d}
&   \calF_\eps \Gto \calF_0 \ \text{ in } (M,\DD), \text{where } \calF_0 \text{
  is proper}; 
\\
\nonumber
& \exists\,\lambda \in \R\ \forall \,\eps>0: \ \ 
  (M,\calF_\eps,\DD_\eps) \text{ has an (EVI)}_\lambda \text{ semiflow } \\
\label{eq:EVIcvgAss.e}
&\hspace*{8.5em} 
   S^\eps_t:\scrD_\eps   \to \scrD_\eps:=\ol{\dom(\calF_\eps)}. 
\end{align}
\end{subequations}
Then, (EVI)$_\lambda$ for $(M,\calF_0,\DD_0)$ has for each $u^0_0\in \scrD_0$ a
unique solution $t \mapsto u_0(t)=: S^0_t(u^0_0)$. Moreover, $S^0_t:\scrD_0\to
\scrD_0$ is a $\lambda$-contractive semiflow, and we have convergence of
solutions as follows 
\[
\scrD_\eps \ni u^0_\eps \to u^0_0\in \scrD_0 \quad \Longrightarrow \quad
\forall\, t>0: \ \begin{cases} S^\eps_t(u^0_\eps) \to S^0_t(u^0_0) \ \ \text{ and} \\
\calF_\eps\big( S^\eps_t(u^0_\eps)\big) \to \calF_0\big( S^0_t(u^0_0)\big).
\end{cases}
\] 
\end{theorem}
\begin{proof}
The proof follows closely the general strategy of the existence proofs.

\STEP{Step 0: Approximating sequences.} Here $u_\eps:[0,T]\to M$ are given as EVI
solutions. 

\STEP{Step 1: A priori estimate for finite energies.}  We start with
$u^0_0\in \dom(\calF_0)$ and use $(\Gamma.\sup)$ to construct a recovery
sequence $u^0_\eps \to u^0_0$ with $\calF_\eps(u^0_\eps) \to
\calF_0(u^0_0)$. Moreover, $\calF_0$ is lsc and has compact sublevels, hence by
equi-compactness \eqref{eq:EVIcvgAss.c} $\inf \calF_0 =:\alpha_0\in \R$ and
$ \inf \calF_\eps =:\alpha_\eps \to \alpha_0$, see Theorem
\ref{th:GammaCvg}. Thus, for $\eps \in {]0,\eps_*[}$ we have
\[
\calF_\eps(u^0_\eps) \leq \calF_0(u^0_0)+1, \quad 
 \alpha_\eps  \geq \alpha_0-1, \quad 
  \calF_\eps(u^0_\eps) - \alpha_\eps \leq \calF_0(u^0_0) - \alpha_0 +2
=:\Delta_\calF.
\]

Our a priori estimates for EVI solutions in Section \ref{su:MetrEVI} 
provide 
\[
\int_0^T \frac12 \BAR \dot u_\eps\BAR_{\DD}(t)^2 \dd t \leq \int_0^T \frac12
\BAR \dot u_\eps\BAR_{\DD_\eps}(t)^2 \dd t \leq \calF_\eps(u^0_\eps) -
\calF_\eps(u_\eps(T)) \leq \Delta_\calF, 
\] 
where we used the first estimate in \eqref{eq:EVIcvgAss.a}, which implies $\SPE u
 \leq \BAR \dot u\BAR_{\DD_\eps}$ a.e.  Moreover, using $\calF_\eps(u_\eps(t))
 \leq \calF_\eps(u^0_\eps) \leq \calF_0(u^0_0)+1$ and the equi-compactness
 \eqref{eq:EVIcvgAss.c} show that there is a compact set $K \Subset M$ such
 that $u_\eps(t) \in K$   for all $t\in [0,T]$ and all $\eps \in
 {]0,\eps_*]}$. 

\STEP{Step 2: Extraction of converging subsequences.} With the results from
Step 1 we have the equi-continuity $\DD\big(u_\eps(s),u_\eps(t)\big) \leq
C|t{-}s|^{1/2}$ and we can apply the Arzel\`a-Ascoli theorem to obtain a
uniformly converging subsequence (not relabeled) in $[0,T]$, where $T>0$ was
arbitrary, hence we have
\[
\forall\, t \geq 0: \quad u_\eps(t) \to u(t) \ \text{ in }(M,\DD). 
\]

\STEP{Step 3: Limit passage $\eps\to 0$.} For all $\eps>0$ we have
(EVI)$_\lambda$:
\begin{equation}
  \label{eq:EVI.eps.22}
\begin{aligned}
 & \forall\, w_\eps \in \dom(\calF_\eps) \ \forall \, 0\leq s < t: \\
&\quad
\frac12 \DD_\eps \big(w_\eps,u_\eps(t)\big)^2 \leq 
\frac{\ee^{-\lambda(t-s)}}2 \DD_\eps \big( w_\eps,u_\eps(s)\big)^2 + M_\lambda
(t{-}s) \big(\calF_\eps(w_\eps) - \calF_\eps(u_\eps(t))\big).
\end{aligned}
\end{equation}
We emphasize here that $\lambda$ is independent of $\eps$. 

For given $w\in \dom(\calF_0)$  $(\Gamma.sup)$ from \eqref{eq:EVIcvgAss.d}
provides a recovery sequence $\wh w_\eps\to w$ with $\calF_\eps(\wh w_\eps) \to
\calF_0(w)$. Inserting $w_\eps = \wh w_\eps$ into
\eqref{eq:EVI.eps.22} we can pass to the limit $\eps\to 0^+$, where we use
$M_\lambda(t{-}s)>0$, the continuous convergence \eqref{eq:EVIcvgAss.b} for the
distance, and $(\Gamma.\inf)$ for $\calF_\eps(u_\eps(t))$:
\begin{align*}
&  \forall\, w \in \dom(\calF_0) \ \forall \, 0\leq s < t: 
\\
&\quad
  \frac12  \DD_0 \big( w, u(t)\big)^2 \leq 
\frac{\ee^{-\lambda(t-s)}}2 \DD_0 \big( w , u(s)\big)^2 + M_\lambda
(t{-}s) \big(\calF_0 (w) - \calF_0(u(t))\big).
\end{align*}
Thus, $u:{[0,\infty[}\to M$ solves (EVI)$_\lambda$ for the GS
$(M,\calF_0,\DD_0)$. 

As the EVI solutions are unique, we conclude that the convergence does hold for
the whole family, i.e.\ without the extraction of a subsequence.

\STEP{Step 4: Convergence of general initial data.}  From Section
\ref{su:MetrEVI} we know that the induced semigroups $(S^\eps_t)_{t\geq 0}$
are $\lambda$-contractions in $(M,\DD_\eps)$ wherever they are defined. In
particular, we can extend the domain to its closure
$\scrD_\eps:= \ol{\dom(\calF_\eps)}$. This also holds for the case $\eps = 0$.
Assume now
\[
\scrD_\eps \ni u^0_\eps \to u^0_0 \in \scrD_0 \quad \text{and} \quad u_\eps(t)
= S^\eps_t(u^0_\eps).
\]
For arbitrary $\delta>0$ we choose $\wh u^0_0\in \dom(\calF_0)$ with
$\DD(u^0_0, \wh u^0_0) <\delta $ and a recovery sequence $\wh u^0_\eps \to \wh u^0_0$
with $\calF_\eps( \wh u^0_\eps ) \to \calF_0(\wh u^0_0)$. Then,
\[
\DD( u^0_\eps, \wh u^0_\eps) \leq \DD(u^0_\eps, u^0_0) + \DD( u^0_0, \wh
u^0_0 ) + \DD( \wh u^0_0, \wh u^0_\eps) < 2\delta  \text{ for } \eps \in
{]0,\eps_1[}.
\]
With this and setting $\wh\Delta_\eps^\delta (t) := \DD( S^\eps_t(\wh u_\eps^0),
S^0_t(\wh u_0^0)\big)$, we can estimate for all $t\geq 0$ as follows:
\begin{align*}
 &\DD \big(u_\eps(t),u_0(t)\big) \  \leq \ \DD\big(u_\eps(t),S^\eps_t(\wh
 u_\eps^0) \big) +\wh \Delta_\eps^\delta(t) + \DD(S^0_t(\wh u_0^0), u_0(t)\big) 
\\
& \overset{\text{\eqref{eq:EVIcvgAss.a}}}\leq 
 \wh\Delta_\eps^\delta(t) + C \,\ee^{\lambda t} \,\DD\big(u_\eps^0,\wh u_\eps^0\big) +
C\, \ee^{-\lambda t}\, \DD(\wh u_0^0, u_0^0 \big)
\  \leq  \  \wh\Delta_\eps^\delta(t) + C\,\ee^{-\lambda t}\,\big( 2\delta + \delta\big).
\end{align*}
Because Step 2 shows the uniform convergence of  $\wh\Delta_\eps
^\delta \to 0$  on all $[0,T]$ for $\delta>0$ fixed and $\eps \to 0$, we obtain
uniform convergence of $u_\eps\to u=u_0$ on $[0,T]$ by first making $\delta$
small and then $\eps$. 

\STEP{Step 5: Energy convergence.} We refer to Step 4 in the proof of
\cite[Thm.\,2.17]{DanSav14LNGF}.
\end{proof}

As in the static case we have a commuting diagram. Passing to the ``right limit'' in
$(\calF_\eps, \DD_\eps, u^0_\eps)$  for $\eps\to 0$ (horizontal direction) can
be interchanged by solving (EVI)$_\lambda$ (vertical direction). 
\begin{figure}[h]
{\centering
\begin{tikzpicture}
\draw [color=gray!10,fill=gray!10] (-7.5,-2) rectangle (6,3);
\draw [fill=green!20, very thick] (-1.85,0.6) rectangle (-1.15,2.4);
\draw [fill=green!20, very thick] (1.85,0.6) rectangle (1.15,2.4);
\node at (-1.5,2){$\calF_\eps$};
\node at (-1.5,1.5){$\DD_\eps$};
\node at (-1.5,1){$u^0_\eps$};
\draw[ ->] (-1,2)-- node[above, pos=0.5]{$\Gamma$} (1 , 2); 
\draw[ ->] (-1,1.5)-- node[above, pos=0.5]{$\mafo{cc}$}(1 , 1.5); 
\draw[ ->] (-1,1)--(1 , 1); 
\node at (1.5,2){$\calF_0$};
\node at (1.5,1.5){$\DD_0$};
\node at (1.5,1){$u^0_0$};

\node at (0,0) {{\footnotesize $\eps \longrightarrow 0$}}; 

\node[left] at (-2.5,0) {solving EVI$^\eps_\lambda$};
\node[right] at (2.5,0) {solving EVI$^0_\lambda$};

\draw[->, thick] (-1.5,0.5)--(-1.5,-0.5);
\draw[->, thick] (1.5,0.5)--(1.5,-0.5);

\node at (-6,-1){for all $t>0$: };
\node at (-2.5,-1){$u_\eps(t)=S^\eps_t(u^0_\eps)$}; 
\node at (2.5,-1){$u_0(t)=S^0_t(u^0_0)$}; 
\draw[ ->] (-1,-1)--(1 , -1); 

\end{tikzpicture}\par
}
\caption{Commuting diagram for $\Gamma$-convergence of EVI solutions.}\label{fig.EVI}
\end{figure}
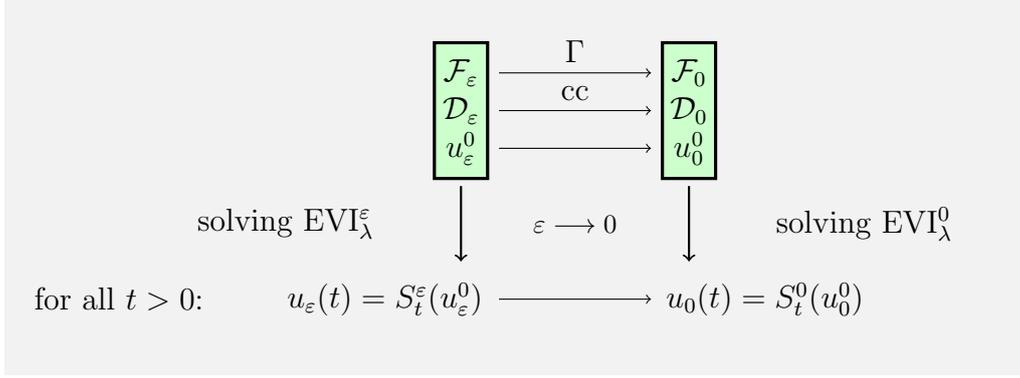

We consider two applications, where the first treats a linear parabolic
equation and shows that homogenization can be treated with the above result. 
In the second example we treat a simple ODE in $M=H=\R^1$ and show that
solutions do not converge, because of the solutions of the gradient-flow
equation cannot be EVI solutions. 

\begin{example}[Homogenization of an Allen-Cahn equation]
\label{exa:Homogenization}
Consider the Hilbert space $M=H=\rmL^2(\Omega)$ with $\Omega = {]0,\ell[}
\subset \R^1$. For 1-periodic functions $a,\,b,\, B,\,c \in \rmL^\infty(\R)$
satisfying $a(y)\geq \ul a>0$, $B(y)\geq \ul B>0$ and $c(y)\geq
\ul c>0$ for all $y\in \R$ we set $a_\eps(x) = a(x/\eps)$ and similarly
$b_\eps$, $B_\eps$, and $c_\eps$. With this, we define the energy functional 
\[
\calF_\eps(u)=\begin{cases} \ds \int_\Omega  \big( \frac{a_\eps}2\,u_x^2 + \frac{b_\eps}2\,
  u^2 + \frac{B_\eps}4 \,u^4\big)  \dd x & \text{for } u\in \rmH^1(\Omega), \\ \infty &
  \text{otherwise}. 
\end{cases}
\]
and the distances $\DD_\eps$ and $\DD$ via 
\[
\DD_\eps(u,w)^2= \!\int_\Omega \! c_\eps|u{-}w|^2\dd x, \quad 
\DD_0(u,w)^2= \! \int_\Omega\! c_\mafo{arith}|u{-}w|^2\dd x,
\DD(u,w)^2 = \!\int_\Omega\! \underline c\,|u{-}w|^2\dd x,
\]
where the subscripts ``$\mafo{arith}$'' and ``$\mafo{harm}$'' denote the
arithmetic and harmonic mean as in Lemma \ref{le:Average}.  Clearly, the
assumptions \eqref{eq:EVIcvgAss.a} and \eqref{eq:EVIcvgAss.b} for $\DD_\eps$
and $\calD$ are satisfied with $C=\|c\|/\ul{c}$.

Using $\ul a,\,\ul B>0$ we obtain equi-coercivity of
$\big(\calF_\eps\big)_{\eps>0}$ in $\rmH^1(\Omega)$, which implies
equi-compactness in $H=\rmL^2(\Omega)$, i.e.\ \eqref{eq:EVIcvgAss.c} is also
satisfied. Moreover, using the results from the Section \ref{su:staticGCvg} 
it is not difficult to show that we have the $\Gamma$-convergence
\eqref{eq:EVIcvgAss.d}, namely
\[
\calF_\eps \Gto \calF_0: u \mapsto \begin{cases} 
\ds \int_\Omega  \big( \frac{a_\mafo{harm}}2\,u_x^2 + \frac{b_\mafo{arith}}2\,
  u^2 + \frac{B_\mafo{arith}}4 \,u^4\big)  \dd x & \text{for } u\in
  \rmH^1(\Omega), \\ \infty & 
  \text{otherwise}. 
\end{cases} 
\]

Finally, we can use the existence results of Section \ref{su:Hilbert.EVI} to
show that EVI solutions exist. The importance is of course that we find one
$\lambda \in \R$ that works for all $\eps>0$. In our Hilbert-space case the
only non-convexity of $\calF_\eps$ can arise from the quadratic term $b_\eps
u^2/2$ which can be negative if $\mafo{ess\!\;inf} b<0$. Indeed, choosing 
$\lambda =\mafo{ess\!\;inf}\bigset{b(y)/c(y)}{ y \in \R} $ we see that 
$u \mapsto \calF_\eps(u) - \frac\lambda 2 \DD_\eps(0,u)^2$ is convex. Hence, 
we obtain an EVI$_\lambda$ semiflow for $ \big( \rmL^2(\Omega),
\calF_\eps,\DD_\eps \big) $. 

Because we have $\dom(\calF_\eps)=\rmH^1(\Omega)$ for all $\eps\geq 0$, we have
$\scrD_\eps = \rmL^2(\Omega)$, and the convergence Theorem
\ref{th:EvolGCvg.EVI} shows that for all sequence $u_\eps^0\to u_0^0$ the
solutions $u_\eps:{[0,\infty[}\to \rmL^2(\Omega)$  of the Allen-Cahn equation 
\[
c_\eps \dot u_\eps = \pl_x\big( a_\eps \pl_x u_\eps\big) -b_\eps u_\eps -B_\eps
u_\eps^3 \text{ in }\Omega, \quad \pl_x u_\eps(t,x)=0\text{ on }\pl\Omega,
\quad u_\eps(0,\cdot)=u_\eps^0
\]
converge to the unique solution $u_0:{[0,\infty[}\to \rmL^2(\Omega)$ of the
homogenized Allen-Cahn equation 
\[
c_\mafo{arith} \dot u_0 = \pl_x\big( a_\mafo{harm} \pl_x u_0\big)
-b_\mafo{arith} u_0 - B_\mafo{arith}
u_0^3 \text{ in }\Omega, \quad \pl_x u_0(t,x)=0\text{ on }\pl\Omega,
\quad u_0(0,\cdot)=u_0^0
\]
in the sense that $u_\eps(t) \to u_0(t)$ in $\rmL^2(\Omega)$ locally uniformly
in ${[0,\infty[}$. 
\end{example}  

The next example is the opposite, because it describes a situation where the
interchanging of limiting process $\eps \to 0$ and solving the gradient-flow
equation does not work.

\begin{example}[The wiggly-energy problem]\slshape
\label{ex:WigglyEnergy} 
The following model was introduced by \cite{Jame96HPT, AbChJa96KMWE}, but it
goes back to much earlier \cite{Pran28GKTF, Toml29MTF} explaining the emergence
of dry friction from a molecular origin. A treatment of this problem using
EDP-convergence, as is discussed in the following section, can be found in
\cite{DoFrMi19GSWE}.

We consider the Hilbert-space gradient system 
\[
M=H=\R^1, \quad \calF_\eps(u) = \frac12\,u^2 + \eps^\alpha \cos(u/\eps) \text{
  with }\alpha>0, \quad
\calR_\eps(v) = \frac12\, v^2.
\]
We see that the dissipation does not depend on $\eps$ at all and is given by
the Euclidean distance $\DD_\eps = \DD_\mafo{Eucl}$, in particular we have $\DD_\eps =
\DD_\mafo{Eucl} \Cto \DD_\mafo{Eucl}$. For the energy, the condition $\alpha>0$
gives 
\[
\calF_\eps \Cto \calF_0:u \mapsto \frac12\, u^2 \quad \Longrightarrow \quad
\calF_\eps \Mto \calF_0.
\]

In this simple case, we can study the gradient-flow equation directly:
\[
\dot u_\eps = - \rmD\calF_\eps(u_\eps) = - u_\eps + \eps^{\alpha-1}\,\sin\big(
u_\eps/\eps\big) , \quad u_\eps(0)=u_\eps^0.
\]
For $\alpha>1$ the right-hand side in the ODE convergences uniformly to $-u$
and hence, we can expect $u_\eps(t)\to u_0^0\,\ee^{-t}$ for all $t \geq 0$, if
$u_\eps^0 \to u_0^0$. 

For $\alpha \in {]0,1[}$ the situation is different. We
see that $u \mapsto -\rmD\calF_\eps(u)$ has many zeros, indeed their spacing
around $u_\eps^0$ is roughly $\pi \eps$, if $\eps^{\alpha-1} \gg
|u_\eps^0|$. Thus, the solutions $t\mapsto u_\eps(t)$ get stuck between two 
zeros. Indeed if $\rmD\calF_\eps( \ul u_\eps)=0=\rmD\calF_\eps(\ol u_\eps) $
and $u_\eps^0 \in [\ul u_\eps , \ol u_\eps] $, then we have $u_\eps(t) \in
[\ul u_\eps , \ol u_\eps] $ for all $t\geq 0$.  By the fast oscillations of $u
 \mapsto \sin(u/\eps)$ we can always find  $\ul u_\eps$ and $\ol u_\eps$
 with $\ol u_\eps - \ul u_\eps \leq 4 \eps \pi$. Thus, we  conclude 
\[
\alpha \in {]0,1[} \ \text{ and } \ u_\eps^0\to u_0^0 \quad 
\Longrightarrow \quad u_\eps(t) \to u_0^0 \text{ for all } t\geq 0. 
\]
The constant limits $u_0$ of the solutions $u_\eps$ are certainly not the
solutions of the limiting gradient system $(\R^1,\calF_0,\DD_\mafo{Eucl})$.  

For $\alpha=1$ a similar problem occurs: solutions $u_\eps$ starting with $u_\eps^0\to
u_0^0 \in [-1,1]$ get stuck and satisfy $u_\eps(t) \to u_0^0$. For $u_0^0>1$,
first decay and reach $u=+1$ in finite time, namely $u_\eps(t) \to u_0(t)= \cosh\big(
\max\{\mafo{Arcosh}(u_0^0){-}t,0\})$. 

One can easily check that all assumptions in \eqref{eq:EVIcvgAss} are satisfied
except for \eqref{eq:EVIcvgAss.e}. This implies that for the case $\alpha\in
{]0,1]}$ there is no $\lambda \in \R$, such that the evolutionary variational
inequality (EVI)$_\lambda$ has a solution for all $\eps \in {]0,1]}$. Indeed,
from the general existence result in Theorem \ref{th:ExistEVIsol}, we
know that geodesic $\lambda$-convexity is a sufficient condition for
existence. In this simple example we have $\DD=\DD_\mafo{Eucl}$, hence geodesic
$\lambda$-convexity is Hilbert-space $\lambda$-convexity, which means $\rmD^2
\calF_\eps(u) = 1 - \eps^{\alpha-2} \cos(u/\eps) \geq \lambda$ for all $u \in
\R$. Clearly, equi-semiconvexity only holds for $\alpha\geq 2$.
\end{example}

\subsection{Evolutionary $\Gamma$-convergence using the energy-dissipation
  balance} 
\label{su:EvolGCvgEDB}

The approach to evolutionary $\Gamma$-convergence in the previous section is
restrictive because of two major assumptions, namely (i) it applies only to
classical gradient systems, i.e.\ $\psi=\psi_\mafo{quadr}$ (but of course it
allows metric gradient systems), and (ii) it needs equi-$\lambda$-convexity. 

The following result uses the energy-dissipation balance and hence is more
flexible. Of course, the result is weaker which is seen in two aspects. First,
we will not have uniqueness of solutions and hence can only establish
convergence along subsequences. Nevertheless, one can show that all
accumulation points of families of solutions solve the limiting
problem. Second, we have to impose a stronger condition on the convergence of
the initial condition, i.e.\ they need to be \emph{well prepared}: 
\[
\text{well-preparedness of initial conditions: } \ u_\eps^0 \to u_0^0 \text{ and
} \calF_\eps(u_\eps^0) \to \calF_0(u_0^0) <\infty.
\]
Thus, we need the sequence of initial conditions $(u_\eps^0)_{\eps>0}$ to be a
recovery sequence for $u_0^0 \in \dom(\calF_0)$. While the restriction to
recovery sequences is not too severe, the restriction to finite energy is
significant as we see in the Allen-Cahn equation where
$\dom(\calF_\eps)=\rmH^1(\Omega)$ is significantly smaller than the whole space
$H = \rmL^2(\Omega) $.\medskip

The proof of following convergence result is only a small variant of the
existence result provided in Theorem \ref{th:BanachExist}, but now we can start
directly from the EDB for $\eps>0$ and pass to the limit in the four terms.
Results of this type were originally developed in
\cite[Thm.\,4.8]{MiRoSa13NADN}, where still the stronger condition
$\calR_\eps \Mto \calR_0$ was imposed. Only in \cite{LieRei18HCHT} it was shown
that the weaker condition $\calR_\eps \Gto \calR_0$ is sufficient.

\begin{theorem}[Evolutionary $\Gamma$-convergence using EDB]
\label{th:EvolGCvgEDB} 
On a reflexive Banach space we consider a family
$(X,\calF_\eps,\calR_\eps)_{\eps\geq 0}$ of gradient systems. If we assume
\begin{subequations}
  \label{eq:Ass.EGC.EDB}
\begin{align}
  &\label{eq:Ass.EGC.EDB.a}
\big(\calR_\eps\big)_{\eps\geq 0} \text{ and } \big(\calR^*_\eps)_{\eps\geq 0}
\text{ are state-independent  and equi-superlinear}; 
\\
& \label{eq:Ass.EGC.EDB.b}
\calR_\eps \Gto \calR_0 \quad (\text{or equivalently } \ \calR_\eps^*\Gweak \calR_0^*);
\\
& \label{eq:Ass.EGC.EDB.c}
\big(\calF_\eps\big)_{\eps\geq 0} \ \text{ is equi-compact};
\\
& \label{eq:Ass.EGC.EDB.d}
\calF_\eps \Gto \calF_0, \ \text{ where } \calF_0:X \to \Rinfty \text{ is proper}; 
\\
&\nonumber
\text{the subdifferentials } \ \big( \plF\calF_\eps(\cdot)\big)_{\eps\geq 0} \ \text{ are  ``closed for $\eps
  \to 0$'', i.e.}
\\&  \label{eq:Ass.EGC.EDB.e}
\qquad \left. \ba{c} u_\eps \to u_0, \ \xi_\eps \weak \xi_0, \\
  \forall\, \eps>0: \ \xi_\eps \in \plF\calF_\eps(u_\eps)  \ea \right\}  \quad
\Longrightarrow \quad \xi_0 \in \plF\calF_0(u_0).
\\
&  \label{eq:Ass.EGC.EDB.f}
 (X,\calF_0, \calR_0) \ \text{ satisfies the abstract chain rule
  \eqref{eq:ACRcond}};
\end{align}
\end{subequations} 
then the following holds. If $\big( u_\eps\big)_{\eps>0}$ is a family of EDB
solutions $u_\eps:[0,T] \to X$ for $(X, \calF_\eps, \calR_\eps)$  with 
\begin{equation}
  \label{eq:EGC.EDB.WellPrep}
  \text{well-prepared initial conditions, i.e. } u_\eps(0)\to u_0^0 \text{ and }
  \calF_\eps( u_\eps(0)) \to \calF_0(u_0^0),
\end{equation}
then there exists a subsequence $\eps_k\to 0$ and an EDB solution $u_0:[0,T]\to
X$ for $(X, \calF_0 , \calR_0)$ with $u_0(0)=u_0^0$ such that
\begin{subequations}
  \label{eq:EGC.EDB.cvg}
\begin{align}
 \label{eq:EGC.EDB.cvg.a}
 u_{\eps_k}(t) &\to u_0(t) \quad \text{ in } X \quad \text{ for all } t \in [0,T];
\\
 \label{eq:EGC.EDB.cvg.b}
\dot u_{\eps_k} &\weak \dot u_0 \qquad  \; \text{ in } \rmL^1([0,T];X);
\\
 \label{eq:EGC.EDB.cvg.c}
 \calF_{\eps_k} (u_{\eps_k}(t)) &\to \calF_0(u_0(t)) \ \text{ for all } t \in [0,T].
\end{align}
\end{subequations}
\end{theorem}

In the proof we will also show the convergences 
\[
\int_0^T \calR_{\eps_k} \big(\dot u_{\eps_k}(t) \big) \dd t \to \int_0^T
\calR_0 \big( \dot u_0(t)\big) \dd t \quad \text{and} \quad 
\int_0^T \calR^*_{\eps_k} \big({-} \xi_{\eps_k}(t) \big) \dd t \to \int_0^T
\calR^*_0 \big( {-}\xi_0(t)\big) \dd t, 
\]
which may be used to improve the convergences of $\dot u_\eps$ and $\xi_\eps$. 

Before giving the proof of the above theorem we provide
two auxiliary results. We leave the proof of the first lemma as an exercise.

\begin{lemma}[$\Gamma$-convergence of integral functionals]
\label{le:GammaCvgIntegral}
Consider an equi-superlinear family $\big(\calG_\eps\big)_{\eps\geq 0}$ of proper lsc
functionals $\calG_\eps:Y\to \Rinfty$ on a reflexive Banach space $Y$ satisfying
$ \calG_\eps \Gweak \calG_0$.  On $Z=\rmL^1([0,T];Y)$ define
$\calJ_\eps(u(\cdot)) := \int_0^T \calG_\eps(u(t)) \dd t$ for $T>0$ and all
$\eps \in [0,1]$. Then, we have $\calJ_\eps \Gweak \calJ_0$. 
\end{lemma}

The next result should be seen as a simple generalization of the closedness
result derived in Proposition \ref{pr:ClosedFrechet}. It shows that equi-semiconvexity
for the family $\big( \calF_\eps)_{\eps\geq 0}$ is sufficient to establish the condition 
``closed for $\eps\to 0$'' imposed abstractly in
\eqref{eq:Ass.EGC.EDB.e}. However, equi-semiconvexity is not necessary which is
easily seen in Example \ref{ex:WigglyEnergy}. There the wiggly energy
$\calF_\eps(u)= \frac12 u^2 + \eps^\alpha \cos( u/\eps)$ is equi-semiconvex for
$\alpha \geq 2$, but the subdifferentials $\rmD \calF_\eps(u)= u -
\eps^{\alpha-1} \sin(u/\eps)$ are ``closed for $\eps\to 0$'' whenever
$\alpha>1$.

\begin{proposition}[Closedness for $\eps\to 0$]
\label{pr:Closedness.eps0}
On a reflexive Banach space $X$ we consider a family
$\big( \calF_\eps)_{\eps\geq 0}$ that is equi-semiconvex, i.e.
\[
\exists\, \lambda \in \R \ \forall\, \eps \in [0,1]: \quad \calF_\eps \text{ is
  $\lambda$-convex on } X.
\]
If $\calF_\eps \Gto \calF_0$, then we have the following closedness for $\eps
\to 0$:
\begin{equation}
  \label{eq:ClosednessEnergy}
  \left. \ba{c} u_\eps \to u_0, \ \xi_\eps \weak \xi_0, \\
  \forall\, \eps>0: \ \xi_\eps \in \plF\calF_\eps(u_\eps)  \ea \right\}  \quad
\Longrightarrow \quad \calF_\eps(u_\eps) \to \calF_0(u_0) \ \text{ and } \ 
 \xi_0 \in \plF\calF_0(u_0).
\end{equation}
\end{proposition}
\begin{proof} We follow the proof of Proposition  \ref{pr:ClosedFrechet} but
  need to use the $\Gamma$-convergence $\calF_\eps \Gto \calF_0$. 

Using the global characterization of
the Fr\'echet subdifferential in Lemma \ref{le:CharFrechSub} we have
\begin{equation}
  \label{eq:EGC.EDB.plFeps}
\forall\, \eps \in [0,1]\ \forall\, w_\eps \in W:\quad 
  \calF_\eps ( w_\eps)  \geq \calF_\eps (u_\eps) + \langle \xi_\eps, w_\eps{-}
  u_\eps\rangle + \frac\lambda 2 \, \| w_\eps{-}u_\eps \|^2. 
\end{equation}
For all $\wh w\in X$ \ $(\Gamma.\sup)$ provides a recovery sequence $\wh w_\eps
\to \wh w$ with $\calF_\eps( \wh w_\eps) \to \calF_0( \wh w)$. Setting
$F^*=\limsup_{\eps \to 0} \calF_\eps(u_\eps)$ and inserting $w_\eps = \wh
w_\eps$ we can pass to the limit in \eqref{eq:EGC.EDB.plFeps} and arrive at
\[
\calF_0(\wh w) \geq F^* + \langle \xi_0, \wh w{-} u_0\rangle + \frac\lambda2
\,\| \wh w {-} u_0\| ^2. 
\]
Choosing $\wh w = u_0$ we find $\limsup_{\eps \to 0} \calF_\eps(u_\eps) = F^*
\leq \calF_0(u_0)$. By $(\Gamma.\inf)$ we also have $\calF_0(u_0) \leq
\liminf_{\eps \to 0} \calF_\eps( u_\eps)$, which implies $\calF_\eps( u_\eps)
\to \calF_0( u_0)$. Replacing $F^*$ by $\calF_0(u_0)$ is the last displayed
formula gives $\xi_0 \in \plF\calF_0(u_0)$ as desired. 
\end{proof} 

\noindent\begin{proof}[Proof of Theorem \ref{th:EvolGCvgEDB}]
\mbox{}

\STEP{Step 0: approximating sequences.} Here the given solutions $u_\eps: [0,T] \to
X$ serve as the approximations for the desired limiting solution $u_0:[0,T] \to
X$.

\STEP{Step 1: a priori estimates.} For $\eps>0$ we have EDB solutions, i.e.
\[
\calF_\eps(u_\eps(T)) + \int_0^T \Big( \calR_\eps\big( \dot u_\eps(t)\big) +
\calR_\eps^*\big( {-}\xi_\eps(t)\big) \Big) \dd t = \calF_\eps(u_\eps(0)). 
\]
As in Step 1 of the proof of Theorem \ref{th:EvolGCvg.EVI} we have 
\begin{equation}
  \label{eq:EDB.eps}
  \calF_\eps(u_\eps(0)) \leq \calF_0(u_0^0){+} 1 \ \text{ and } \ 
\calF_\eps(u_\eps(0)) - \calF_\eps(u_\eps(T)) \leq \calF_0(u_0^0) - \inf
\calF_0 +2 =:\Delta_\calF. 
\end{equation}
According to assumption \eqref{eq:Ass.EGC.EDB.a} there exists a superlinear
function $\psi:\R\to \R$ such that 
\[
\int_0^T \psi\big( \| \dot u_\eps(t)\|_X\big) \dd t \leq \Delta_\calF
\quad \text{and}  \quad 
\int_0^T \psi\big( \| \xi_\eps(t)\|_{X^*}\big) \dd t \leq \Delta_\calF.
\]
As in Step 1 of the proof of Theorem \ref{th:BanachExist} we obtain the
equi-continuity 
\[
\big\| u_\eps(t) -u_\eps(s) \big\|_X \ \leq \ \omega^{\Delta_\calF}_\psi \big(
|t{-}s|\big) \quad \text{for all } t,s\in [0,T] \text{ and } \eps >0. 
\]
Finally, using $\calF_\eps(u_\eps(t)) \leq \calF_\eps(u_\eps(0)) \leq
\calF_0(u_0^0)+1$ and the equi-compactness assumed in \eqref{eq:Ass.EGC.EDB.c},
we find a compact set 
\[
 K \Subset X \quad \text{such that} \quad  u_\eps(t) \in K   \quad \text{for
   all } t,s\in [0,T] \text{ and } \eps >0.  
\]

\STEP{Step 2: extraction of convergent subsequences.} 
According to Step 1 we can apply Arzel\`a-Ascoli's selection principle and find
a subsequence $\eps_k \to 0$ and a continuous limit function $u_0:[0,T] \to X$
such that $u_{\eps_k} \to u_0 $ in $\rmC^0([0,T];X)$. Moreover, using the
superlinear bounds in the reflexive Banach spaces $X$ and $X^*$ we can choose
a further subsequence (not relabeled) such that 
\[
\dot u_{\eps_k} \weak \dot u_0 \  \text{ in } \rmL^1([0,T];X) 
\qANDq 
\xi_{\eps_k} \weak \xi_0 \  \text{ in } \rmL^1([0,T];X^*) .
\]

\STEP{Step 3: limit passage in (EDB)$_\eps$, derivation of (EDI).} 
For passing to the limit in \eqref{eq:EDB.eps} we will first derive an
energy-dissipation inequality (EDI), i.e.\ it will be enough to derive liminf
estimates on the left-hand side, but we need a limsup estimate for the
right-hand side. 

We first consider the two energy
terms. For the right-hand side we simply use the well-preparedness
\eqref{eq:EGC.EDB.WellPrep} to obtain the desired convergence. For the first
term on the left-hand side we use $(\Gamma.\inf)$ from $\calF_\eps \Gto
\calF_0$ and the pointwise convergence from Step 2 and conclude $\calF_0(T))
\leq \liminf_{\eps \to 0} \calF_\eps(u_\eps(T))$. 

The two dissipation terms involving $\calR_\eps$ and $\calR_\eps^*$,
respectively, can be treated separately. Using $\xi_{\eps_k} \weak \xi_0$ in
$\rmL^1([0,T]; X^*)$ and $\calR_\eps^* \Gweak \calR_0^*$ from
\eqref{eq:Ass.EGC.EDB.b} 
we can apply Lemma \ref{le:GammaCvgIntegral} and find
\[
\int_0^T \calR_0^* \big({-}\xi_0(t)\big) \dd t \ \leq \ \liminf_{k\to \infty}
\int_0^T \calR_{\eps_k}^* \big( {-}\xi_{\eps_k}(t) \big) \dd t .
\]

For the remaining term the convergence $\dot u_{\eps_k} \weak \dot u_0$ is not
enough, because we only have $\calR_\eps \Gto \calR_0$ which does not allow for
weak convergence. To compensate for that we do a time discretization via
$\tau=T/N$ and $N \in \N$. Defining $\wh u_\tau:[0,T]\to X$ to be the piecewise
affine interpolant of $u_0:[0,T]\to X$ satisfying $\wh u_\tau(j\tau)=u_0(j\tau)$
for $j=0,...,N$, we obtain  
\begin{align*}
&\int_0^T \calR_0\big(\dot u_0(t) \big) \dd t = \sum_{j=1}^N \int_{k\tau-\tau}^{k\tau}
\calR_0(\dot u_0(t)) \dd t 
%  \\ &
\overset{\text{Jensen}}\geq 
 \sum_{j=1}^N 
\tau\, \calR_0\Big(\frac1\tau {\ts \int_{j\tau-\tau}^{j\tau} \dot  u_0(t) \dd t} \Big)
\\
&  =\sum_{j=1}^N  \tau \,\calR_0\Big( \frac1\tau\big(u_0(j\tau) -
u_0(j\tau{-}\tau) \big) \Big)
%\\ &
 = \sum_{j=1}^N 
\tau\, \calR_0\Big(\frac1\tau \ts \int_{j\tau-\tau}^{j\tau} \dot{\wh u}_\tau(t) \dd t \Big)
=\ds \int_0^T\calR_0\big( \dot{\wh u}_\tau (t) \big) \dd t .
\end{align*}

Doing the same time discretization for $\eps>0$ we obtain the lower estimate 
\begin{align*}
& \liminf_{k\to \infty} \int_0^T \calR_{\eps_k} \big(\dot u_{\eps_k} (t)\big) \dd t 
\geq 
 \liminf_{k\to \infty} \sum_{j=1}^N  \tau \, \calR_{\eps_k}\Big( \frac1\tau 
  \big(u_{\eps_k}(j\tau) - u_{\eps_k} (j\tau{-}\tau)\big) \Big)
\\ &
 \overset*\geq  \sum_{j=1}^N 
\tau\, \calR_0\Big( \frac1\tau\big(u_0(j\tau) -
u_0(j\tau{-}\tau) \big) \Big) 
=\ds \int_0^T\calR_0\big( \dot{\wh u}_\tau (t) \big) \dd t ,
\end{align*}
where $\overset*\geq $ uses the liminf estimate of $\calR_\eps \Gto \calR_0$
and the strong convergence $u_{\eps_k}(t) \to u_0(t)$ established in Step 2. 

In the last estimate we can now perform the limit $\tau= T/N \to 0$ and use
$\dot{\wh u}_\tau \to \dot u_0$ strongly in $\rmL^1([0,T];X)$, which implies,
after extracting a subsequence $\tau_n\to 0$, the convergence
$\dot{\wh u}_{\tau_n}(t) \to \dot u_0(t)$ a.e.\ in $[0,T]$. Thus, using Fatou's
lemma and the lsc of $\calR_0$ we have
\[
\liminf_{\tau_n\to 0} \int_0^T \calR_0 \big(\dot{\wh u}_{\tau_n}(t) \big)\dd t
\ \overset{\text{Fatou}}\geq \  \int_0^T \liminf_{\tau_n\to 0}  \calR_0 \big( 
 \dot{\wh u}_{\tau_n}(t)  \big)\dd t \ \overset{\text{lsc}}\geq \ 
 \int_0^T  \calR_0 \big( \dot u_0(t)  \big)\dd t .
\]
With this we have shown $ \liminf_{k\to \infty} \int_0^T \calR_{\eps_k}
\big(\dot u_{\eps_k} (t)\big) \dd t  \geq \int_0^T  \calR_0\big(\dot 
  u_0(t) \big)\dd t $, and the energy-dissipation inequality 
\[
\calF_0(u_0(T)) + \int_0^T \Big( \calR_0 \big(\dot u_0 (t) \big) + \calR_0^*\big(
{-}\xi_0(t)\big)\Big) \dd t \ \leq \ \calF_0(u_0(0))
\] 
is established. 

It remains to identify $\xi_0$, for this we use the closedness condition
\eqref{eq:Ass.EGC.EDB.e} and argue as in Exercise \ref{exerc:EvolClosed2} and
obtain $\xi_0(t) \in \plF\calF_0(u_0(t))$ a.e.\ in $[0,T]$. 

\STEP{Step 4: Derivation of (EDB) for $\eps=0$.} With the abstract
chain rule assumed to hold in \eqref{eq:Ass.EGC.EDB.f} we can apply the
energy-dissipation principle from Theorem \ref{th:Banach.EDP} and conclude that
$u_0$ is indeed a EDB solution. 

Moreover, the chain rule implies that the liminf estimates in Step 3 were
indeed limits such that $\calF_{\eps_k}(u_{\eps_k}(T)) \to \calF_0(u_0(T))$
holds. However, we could have performed the liminf estimates on any subinterval
$[0,t_*]$ with $t_*\in {]0,T[}$, from which we obtain
$\calF_{\eps_k}(u_{\eps_k}(t_*)) \to \calF_0(u_0(t_*))$ for all $t_*\in [0,T]$. 
\end{proof}

\subsection{EDP-convergence for gradient systems}
\label{su:EDPcvg}

In the above two section we studied the convergence of (a subsequence of) the
solutions $u_\eps$ of a gradient system $(X,\calF_\eps,\calR_\eps)$ to a
solution $u_0$ of the limiting gradient system $(X,\calF_0,\calR_\eff)$ where
$\calF_0= \Glim \calF_\eps$ and $\calR_\eff= \Glim \calR_\eps$ in suitable
topologies. It is important to note here that the two $\Gamma$-limits are not
independent, because both have to be considered in the same Banach space $X$. The
choice of $X$ is dictated by the family of dissipation potentials
$\big(\calR_\eps \big)_\eps$. Then, the family $\big(\calF_\eps\big)_\eps$ has
to be considered in the same space $X$, and not in a so-called ``energy space''
which is often constructed as the smallest space in which 
$\big(\calF_\eps\big)_\eps$ is weakly equi-compact.

However, there are situations in which the interaction between energy and
dissipation is even stronger. Recovery sequences for the energy may not be
compatible with recovery sequences for the dissipation.  In such cases, the
effective dissipation $\calR_\eff$ cannot be obtained by looking at the family
$\big(\calR_\eps \big)_\eps$ alone, but one needs to consider the family of
pairs $\big((\calF_\eps,\calR_\eps)\big)_\eps$. Such a definition is the
so-called \emph{EDP-convergence} which was first defined in \cite{LMPR17MOGG}
and made more precise in \cite{MiMoPe21EFED}. We also refer to
\cite{DoFrMi19GSWE} for a treatment of the wiggly-energy model of Example
\ref{ex:WigglyEnergy}, to \cite{MieSte20CGED,MiPeSt21EDPC} for applications in
reaction systems with slow and fast reactions, and to
\cite{Fren19DEGS,FreMie21?DKRF,Step21CGED,FreLie21EDTS,PelSch22?CGST} for
reaction-diffusion systems.

The name of EDP-convergence derives from \emph{convergence in the sense of the
  energy-dissipation principle}, because this notion of convergence is strongly
linked to the EDP as formulated in Theorem \ref{th:Banach.EDP} or Proposition
\ref{pr:MetrEDP}. We give the main ideas and a few examples by using the
Banach space formulation, but a similar theory can be obtained in the metric
setting. 

Considering the family $(X,\calF_\eps,\calR_\eps)_{\eps>0}$ of gradient systems
and a time horizon $T>0$, we define the dissipation functionals
\begin{equation}
  \label{eq:def.mfD}
  \mfD_\eps(u) := \int_0^T \Big( \calR_\eps(u(t),\dot u(t)) +
  \calR^*_\eps \big(u(t),{-}\rmD\calF_\eps(u(t))\big) \Big) \dd t .
\end{equation}
Here we wrote the so-called ``slope term of the dissipation'' in terms of
$\calR^*_\eps$ and a single-valued single-valued subdifferential
$\rmD\calF_\eps$. However, in general one can replace this term by the more
correct definition 
\begin{align*}
&\calS_\eps(u):= \mafo{lsc}\big(\wt\calS_\eps\big): u \mapsto
\inf\bigset{\inf_{u_k\to u} \wt\calS_\eps(u_k)}{ u_k \to u} ,
\text{ \ where} 
\\
&\wt\calS_\eps(u):= \begin{cases}\inf\set{\calR^*_\eps(u,{-}\xi)}{ \xi\in \plF
    \calF(u)} & \text{for } \plF\calF(u)\neq \emptyset, \\
\infty& \text{otherwise}.   \end{cases} 
\end{align*}
With this, the proper definition of $\mfD_\eps $ is given by
$\mfD_\eps(u)=\int_0^T \! \big( \calR_\eps(u,\dot u) {+} \calS_\eps(u)\big) \dd t$,
however we will continue to use the $(\calR,\calR^*)$ form to emphasize the
special duality character encoded into $\mfD_\eps$. 

The main point of the definition of $\mfD_\eps$ is that it is a functional on curves
$u:[0,T] \to  X$, unlike to $\calF_\eps$, which are functionals on the state space $X$. 
The idea is now to use classical $\Gamma$-convergence for the functionals $
\mfD_\eps $ as well, but now on a space of curves, let us say
$\rmL^2([0,T];X)$. To reflect the idea of gradient flows with well-prepared
initial conditions we adapt the topology of $\rmL^2([0,T];X)$ by asking the
families of functions $(u_\eps)_{\eps>0}$ additionally have uniformly bounded
energy. 

\begin{definition}[Energy-bounded $\Gamma$-convergence of $\mfD_\eps$]
\label{de:EnergyBddGam.mfD}
Given a Banach space $X$, a family $(\calF_\eps)_{\eps>0}$ of energies
$\calF_\eps:X \to \Rinfty$, and a family $(\mfD_\eps)_{\eps\geq 0}$ of
functionals $\mfD_\eps : \rmL^2([0,T];X) \to \Rinfty$ we say that \emph{$\mfD_\eps$
$\Gamma$-converges to $\mfD_0$ with bounded energies}, 
and shortly write  $\mfD_\eps \GEto \mfD_0$ or $\mfD_0 = \GElim_{\eps\to 0}
\mfD_\eps$, if the following holds:
\begin{subequations}
 \label{eq:def.GE.conv}
\begin{align}
\label{eq:def.GE.liminf}
&\hspace{-1em}\text{Energy-bounded liminf estimate:}\\
\nonumber
&\left. \ba{c}  u_\eps \to u_0 \text{ in } \rmL^2([0,T];X) \ \text{ and }\\ 
  \sup_{\eps\geq 0,\; t\in [0,T]} \calF_\eps(u_\eps(t)) \leq F_*<\infty \ea \right\} 
\quad \Longrightarrow \quad \liminf_{\eps\to 0^+} \mfD_\eps(u_\eps) \geq \mfD_0(u_0),
\\
&\hspace{-1em}\text{Energy-bounded limsup estimate:}\label{eq:def.GE.limsup} 
\\
&\nonumber\quad
\forall\, \wh u_0 \in \rmL^2([0,T];X) \text{ with } \sup\nolimits_{t\in [0,T]}
\calF_0(\wh u_0(t)) \leq F_0<\infty \\
&\nonumber\qquad \exists\: F_* \in \R\ \exists\: 
\big( \wh u_\eps\big)_{\eps>0} \text{ with }\sup\nolimits_{\eps>0,\; t\in [0,T]}
\calF_\eps(\wh u_\eps(t)) \leq F_*:
\\
&\nonumber\hspace{6em} \wh u_\eps\to \wh u_0 \text{ in } \rmL^2([0,T]; X) \text{ and } 
\limsup_{\eps \to 0^+} \mfD_\eps( \wh u_\eps) \leq \mfD_0(\wh u_0). 
\end{align}
\end{subequations}
\end{definition}

In particular applications, the choice $\rmL^2([0,T];X)$ for the space of
curves can be replaced by
other function spaces and the condition of energy boundedness can be dropped or
amended by other conditions. The choice of a good notion of
$\Gamma$-convergence should be seen as a problem-specific task or a modeling
issue.
 
Using the above notion we can now define the simplest notion of
EDP-convergence, and we refer to \cite{DoFrMi19GSWE, MiMoPe21EFED} for the more
advance notions of ``EDP-convergence with tilting'' (in short tilt-EDP
convergence) and ``contact EDP-convergence with tilting'' (in short
``contact-EDP convergence).   

\begin{definition}[EDP-convergence of gradient system]
\label{de:EDPcvg.GS}
A family $\big((X,\calF_\eps,\calR_\eps)\big)_{\eps>0}$ of gradient systems is
said to \emph{converge in the sense of the energy-dissipation
  principle} (in short ``to EDP-converge''), if there exists an effective gradient
system $(X,\calF_\eff,\calR_\eff)$ such that for all $T>0$ the following holds:
\[
\calF_\eps \Gto \calF_\eff \text{ in } X \qANDq
\mfD_\eps \GEto \mfD_0 \text{ in } \rmL^2([0,T];X)
\]
where $\mfD_\eps$ is defined in \eqref{eq:def.mfD} and $\mfD_0$ has the form 
\[
\mfD_0(u) = \int_0^T \!\Big( \calR_\eff\big(u(t),\dot u(t)\big) + \calR^*_\eff
\big( u(t) , {-}\rmD\calF_\eff(u(t))\big) \Big) \dd t.  
\]
We then shortly write $ (X,\calF_\eps,\calR_\eps) \EDPto
(X,\calF_\eff,\calR_\eff)$ for $\eps\to 0^+$. 
\end{definition}

We observe that EDP-convergence of gradient systems has similar properties as
$\Gamma$-convergence of functionals:
\begin{enumerate}[label={(\Roman*)}]
\item
 The notion is independent of the concept of ``solution'', which in the case
of classical functionals means minimizer (after adding a linear loading
$-\langle \ell, \cdot\rangle$)  and in the case of gradient systems systems
means solutions of the gradient-flow equation (after adding an initial
condition $u(0)=u^0$. 

\item Nevertheless, under suitable technical assumptions EDP-convergence of
  gradient systems implies the convergence of solutions if the initial
  conditions are well-prepared, see Proposition \ref{pr:EDPcvg.Sol.cvg}. 

\item The EDP limit $(X,\calE_\eff,\calR_\eff)$ is uniquely determined by the
  family $(X,\calF_\eps, \calR_\eps)$, see Remark \ref{re:EDPlimUnique}. Asking
  only the liminf estimate for $\mfD_0$ will be enough to find some gradient
  structure that produces the correct gradient-flow equation, but the
  uniqueness of the gradient structure is lost, see Remark
  \ref{ex:NonuniqueGStr}. 

\item The involvement of general curves $u\in \rmL^2([0,T];X)$ in the
  definition of EDP-convergence can be understood in the sense of fluctuation
  theory and the associated large-deviation principle, which provide a
  thermodynamical justification of the theory of gradient systems, see e.g.\
  the discussion in \cite[Chap.\,4]{Pele14VMEG} and \cite{ADPZ11LDPW,
    MiPeRe14RGFL, MPPR17NETP}.  
\end{enumerate}

The next result shows that EDP-converge implies convergence of the solutions if
suitable conditions are met. This result corresponds to Theorem
\ref{th:GammaCvg} and Corollary \ref{co:UppSemicontMinim} for the case of
(static) $\Gamma$-convergence of functionals. 

\newcommand{\epsk}{{\eps_k}}%
\begin{proposition}[EDP-convergence implies convergence of solutions]
\label{pr:EDPcvg.Sol.cvg}
Assume\linebreak[4] $ (X,\calF_\eps,\calR_\eps) \EDPto
(X,\calF_\eff,\calR_\eff)$ for $\eps\to 0^+$ and that
$(X,\calF_\eff,\calR_\eff)$ satisfies the abstract chain rule
\eqref{eq:ACRcond}. Moreover, assume that for a sequence  $(\epsk)_{k\in \N}$
with $\epsk\to 0^+$ there are
EDB solutions $u_\epsk:[0,T]\to X$ for $ (X,\calF_\epsk,\calR_\epsk)$ satisfying
\begin{equation}
  \label{eq:Ass.u.epsk}
  u_\epsk \to u \ \text{ in } \rmL^2([0,T];X), \quad \forall\: t\in [0,T]{:}\ 
u_\epsk(t)\to u(t) \ \text{ in } X, \ \text{ and } \calF_\epsk(u_\epsk(0)) \to
\calF_0(u(0)).
\end{equation}
If additionally $u\in \AC([0,T];X)$, then it is an EDB solution for
$(X,\calF_\eff,\calR_\eff)$, and for $\epsk \to 0^+$ we have 
\[
\calF_\epsk(u_\epsk(t)) \to \calF_\eff(u(t)) \text{ for all }t\in [0,T]
\quad\text{and} \quad \mfD_\epsk(u_\epsk)\to \mfD_0 (u). 
\]
\end{proposition}
\begin{proof} To simplify notation, we write $\eps$ in place of $\epsk$. 

The argument uses the lsc property of the energy-dissipation
balance as in previous sections. As $u_\eps$ is an EDB solution we have 
\begin{equation*}
%\label{eq:EDB.mfDeps}
\calF_\eps(u_\eps(T)) {+} \mfD_\eps(u_\eps)= \calF_\eps(u_\eps(T))+\!
\int_0^T\!\!\big( \calR_\eps(u_\eps,\dot u_\eps) {+} \calR^*_\eps ( u_\eps, 
 {-}\rmD\calF_\eps(u_\eps)) \big) \dd t = \calF_\eps(u_\eps(0)). 
\end{equation*}
We pass to the limit $\eps \to 0^+$ in this relation. 
By the assumption of the well-preparedness of the initial conditions
$u_\eps(0)$ we have convergence on the right-hand side. 

On the left-hand side we use $\calF_\eps \Gto \calF_\eff$ and pointwise
convergence $u_\eps(T)\to u(T)$ to obtain $\calF_\eff (u(T))\leq
\liminf_{\eps\to 0^+} \calF_\eps(u_\eps(T))$. 

To treat the term $\mfD_\eps(u_\eps)$ we observe that by the well-preparedness
we have $\calF_\eps(u_\eps(0))\leq \calF_\eff(u(0))+1=:E_*$ for sufficiently small
$\eps$. Hence, the EDB solutions $u_\eps$ satisfy $\calF_\eps(u_\eps(t))\leq
\calF_\eps(u_\eps(0)) \leq E_*$. Thus, we can use the energy-bounded liminf
estimate and obtain $\mfD_0(u)\leq \liminf_{\eps\to 0^+}
\mfD_\eps(u_\eps)$. 

Using the duality structure of $\mfD_0$ in terms of
$\calF_\eff$ and $\calR_\eff$ we see that $u$ satisfies the energy-dissipation
inequality 
\[
\calF_\eff(u(T)) + \int_0^T \! \big( \calR_\eff(u,\dot u) + \calR^*_\eff(u,
{-}\rmD \calF_\eff(u))\big) \dd t \leq \calF_\eff(u(0)). 
\]
Since $u$ is absolutely continuous and $(X,\calF_\eff,\calR_\eff)$ satisfies
the abstract chain rule, we can apply the energy-dissipation principle  from
Theorem \eqref{th:Banach.EDP} and conclude that $u$ is an EDB solution. 

As $u$ is an EDB solution we know that EDI is in fact an EDB which implies that
the liminf estimates are indeed limits providing an equality. This proves
$\calF_\eps(u_\eps(T)) \to \calF_\eff(u(T))$ and $\mfD_\eps(u_\eps)\to
\mfD_0(u)$. Since $T$ can be replaced by any $T'\in {]0,T[}$ the assertion is
established. 
\end{proof}

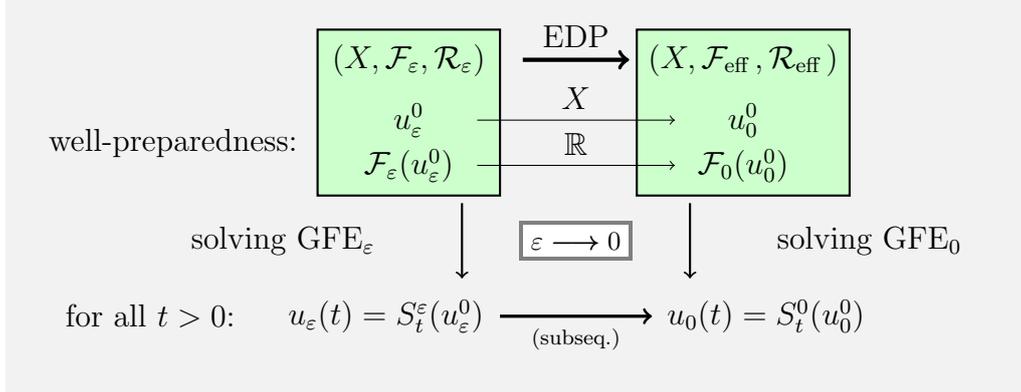
\begin{figure}[h]
{\centering
\begin{tikzpicture}
\draw [color=gray!10,fill=gray!10] (-7.5,-2) rectangle (6,3.2);
\draw [fill=green!20,  thick] (-3.4,0.6) rectangle (-1,2.8);
\draw [fill=green!20,  thick] (0.8,0.6) rectangle (3.6,2.8);
\node at (-2.2,2.4){$(X,\calF_\eps,\calR_\eps)$};
\node at (-2.2,1.6){$u^0_\eps$};
\node at (-2.2,1){$\calF_\eps(u^0_\eps)$};
\draw[ultra thick, ->] (-0.7,2.4)-- node[above, pos=0.5]{EDP} (0.7 , 2.4); 
\draw[ ->] (-1.3,1.6)-- node[above, pos=0.5]{$X$}(1.3 , 1.6); 
\draw[ ->] (-1.3,1)--node[above, pos=0.5]{$\R $}(1.3 , 1); 
\node at (2.2,2.4){$(X,\calF_\eff,\calR_\eff)$};
\node at (2.2,1.6){$u_0^0$};
\node at (2.2,1){$\calF_0(u^0_0)$};

\node at (0,0) {{\footnotesize \fcolorbox{gray}{white}{$\eps \bm\longrightarrow 0$}}}; 

\node[left] at (-2.5,0) {solving GFE$_\eps$};
\node[right] at (2.5,0) {solving GFE$_0$};

\draw[->, thick] (-1.5,0.5)--(-1.5,-0.5);
\draw[->, thick] (1.5,0.5)--(1.5,-0.5);

\node at (-5.6,-1){for all $t>0$: };
\node at (-5.3,1.3){well-preparedness:};
\node at (-2.5,-1){$u_\eps(t)=S^\eps_t(u^0_\eps)$}; 
\node at (2.5,-1){$u_0(t)=S^0_t(u^0_0)$}; 
\draw[very thick, ->] (-1,-1)--node[pos=0.5, below]{\scriptsize (subseq.)} (1 , -1); 

\end{tikzpicture}\par
}
\caption{Commuting diagram for EDP-convergence and EDB solutions under suitable
  technical conditions.}\label{fig.EDP.cvg}
\end{figure}
Figure \eqref{fig.EDP.cvg} shows the corresponding commuting diagram that can
be established if we have enough compactness on the solutions
$(u_\eps)_{\eps>0}$ to extract subsequences satisfying the assumptions in
\eqref{eq:Ass.u.epsk}.  

\begin{exercise}
Discuss what additional conditions are needed such that the evolutionary
$\Gamma$-convergence in Theorem \ref{th:EvolGCvgEDB} can be turned into a
result on EDP-convergence. 
\end{exercise}

\begin{remark}[On the uniqueness of EDP-limits]
\label{re:EDPlimUnique}\slshape
Assuming that there are two gradient
structures we first observe that $\calE_\eff$ and $\mfD_0$ as $\Gamma$-limits
are uniquely determined. Hence, if there exist two effective dissipation potentials
$\calR_\eff$ and $\ol\calR_\eff$ 
generating $\mfD_0$ we must have 
\begin{equation}
  \label{eq:Two.calReff}
  \calR_\eff(u,\dot u)+ \calR^*_\eff\big(u,{-}\rmD\calF_\eff(u)\big) = 
\ol\calR_\eff(u,\dot u)+ \ol\calR^*_\eff\big(u,{-}\rmD\calF_\eff(u)\big)
=:\calM(u,\dot u).  
\end{equation}  
Setting $\dot u=0$ we have $\calR_\eff(u,0)=0=\ol\calR_\eff(u,0)$ and find immediately
$\calM(u,0)=\calR^*_\eff\big(u,{-}\rmD\calF_\eff(u)\big) =
\ol\calR^*_\eff\big(u,{-}\rmD\calF_\eff(u)\big)$. Subtracting this identity
from \eqref{eq:Two.calReff} and assuming $\calM(u,0)< \infty$, we obtain
$\calR_\eff(u,\dot u)=\ol\calR_\eff(u,\dot u)$, which is the desired uniqueness. 
\end{remark}

The main advantage in the definition of EDP convergence is that it can be
applied in degenerate cases, where $\calR_\eps$ and $\calR_\eps^*$ are not
uniformly coercive, but may degenerate for $\eps\to 0^+$. Moreover, keeping the
two terms $\calR_\eps(u_\eps,\dot u_\eps)$ and
$\calR_\eps^*(u_\eps,{-}\rmD\calF_\eps(u_\eps))$ together we allow for the
option that ``microscopic information of $\calF_\eps$ may move into
$\calR_\eff$''. We may define 
\[
\mfD^\text{rate}_\eps(u)= \int_0^T \calR_\eps(u,\dot u)\dd t  \qANDq 
\mfD^\text{slope}_\eps(u) = \int_0^T \calR^*_\eps(u, {-}\rmD
\calF_\eps(u)) \dd t,
\]
such that $\mfD_\eps = \mfD^\text{rate}_\eps+ \mfD^\text{slope}_\eps$.

Keeping the sum $\mfD^\text{rate}_\eps+ \mfD^\text{slope}_\eps$ together 
is the main difference to the theory developed in \cite{SanSer04GCGF,
  Serf11GCGF} where along EDB solutions $u_\eps \to u$ 
the two independent liminf estimates 
\begin{equation}
  \label{eq:TwoSanSerEstim}
\int_0^T \calR_\eff(u,\dot u)\dd t \leq \liminf_{\eps\to 0^+}
\mfD^\text{rate}_\eps(u_\eps) \qANDq 
\int_0^T \calR^*_\eff(u,{-}\rmD \calF_\eff(u))\dd t  \leq \liminf_{\eps\to 0^+}
\mfD^\text{slope}_\eps(u_\eps)
\end{equation}
are supposed. We emphasize that here the estimates are on EDB solutions and not
on general curves. For a discussion of this concept we refer to
\cite[Sec.\,11.2]{Brai14LMVE} and \cite[Sec.\,3.3.3]{Miel16EGCG}.
Note also that our assumption for Theorem \ref{th:EvolGCvgEDB} are such that in
Step 3 of the proof we can establish the two estimates in
\eqref{eq:TwoSanSerEstim}.

We discuss now three simple ODE examples of EDP-convergence and refer to
\cite{Fren19DEGS, MieSte20CGED, MiPeSt21EDPC, FreMie21?DKRF, Step21CGED,
  FreLie21EDTS, PelSch22?CGST} for further applications including PDEs. 
 
\begin{example}[Two binary reactions generate one ternary reaction]
\label{ex:TwoBinaryOneTernary} 
In \cite{Miel23?NESS} as reaction system with four species with density vector 
$\bfc=(c_1,c_2,c_3,c_4)\in \bfC:={[0,\infty[}^4$ is considered that react by two
binary reaction pairs $X_1+X_2 \rightleftharpoons X_4$
and $X_1+ X_4 \rightleftharpoons X_3$. The point is that $X_4$ is considered as
an intermediate product that exists only with a much lower equilibrium density
$c_4^*(\eps)=\eps^2 w^*$, while the other equilibrium densities $c_i^*$ for
$i=1,2,3$ are independent of $\eps$. 

The gradient system
$(\bfC,\calF_\eps,\calR_\eps^*)$ is given by 
\begin{align*}
&\calF_\eps(\bfc)= \LB\big(\frac{c_4}{\eps^2 w^*}\big) \eps^2 w^* + \sum_{i=1}^3
\LB\big(\frac{c_i}{c^*_i}\big) c^*_i \qANDq 
\\
&\calR^*_\eps(\bfc;\bfxi) = 
\frac{\ol\kappa_1}{\eps} (c_1c_2c_4)^{1/2} \sfC^*\big(\xi_1{+}\xi_2{-}\xi_4\big)+ 
\frac{\ol\kappa_2}{\eps} (c_1c_3c_4)^{1/2} \sfC^*\big(\xi_1{-}\xi_3{+}\xi_4\big).
\end{align*}
The associated gradient-flow equation is the following reaction-rate equation
{\small\renewcommand{\arraystretch}{0.83}\normalsize %
\[
\dot\bfc = \frac{\ol\kappa_1}{\eps} \Big( 
   \frac{A_1}{\eps}\, c_4 - \frac{\eps}{A_1}\,c_1c_2\Big) 
   \bma{c} 1\\ 1\\ 0 \\ \!\!-1\!\! \ema + 
 \frac{\ol\kappa_2}{\eps}  \Big( 
      \frac{\eps}{A_2}\, c_3 - 
      \frac{A_2}{\eps} \, c_1 c_4 \Big)
   \bma{c} 1\\ 0\\ \!\!-1\!\! \\ 1 \ema ,
\]}%
where $A_1:= (c_1^*c_2^*/w^*)^{1/2}$ and $A_2=
\big(c_3^*/(c_1^*w^*)\big)^{1/2}$. We see that that setting $c_4 = \eps^2 w$
leads to a right-hand side that is independent of $\eps$, but then we have
$\eps^2 \dot w$ on the left-hand side. 

Doing the formal limit $\eps\to 0^+$ (which can be justified rigorously, see
\cite{Both03ILRC}) we arrive at 
{\small\renewcommand{\arraystretch}{0.83}\normalsize %
\[
\bma{c} \dot c_1\\ \dot c_2\\ \dot c_3\\ 0\ema = \ol\kappa_1 \Big( 
    A_1 w  - \frac{1}{A_1}\,c_1c_2\Big) 
   \bma{c} 1\\ 1\\ 0 \\ \!\!-1\!\! \ema + 
  \ol\kappa_2  \Big( \frac{1}{A_2}\, c_3 - A_2 \, c_1 w \Big)
   \bma{c} 1\\ 0\\ \!\!-1\!\! \\ 1 \ema .
\]}%
From the last equation we can calculate $w$ explicitly as via $(\ol\kappa_1
A_1{+}\ol\kappa_2 A_2) w = \frac{\ol\kappa_1}{A_1} c_1c_2 +
\frac{\ol\kappa_2}{A_2} c_3$. Note that the relation for $w$ guarantees that
the two terms in front of the stoichiometric vectors must be equal, such that we
are left with one reaction only having the form 
{\small\renewcommand{\arraystretch}{0.83}\normalsize %
\[
\bma{c} \dot c_1\\ \dot c_2\\ \dot c_3\ema = \ol\kappa_\eff(c_1) \Big(
\frac{A_1}{A_2}\, c_3 - \frac{A_2}{A_1} \, c_1^2 c_2\Big) 
   \bma{c} 2\\ 1\\  \!\!-1\!\! \ema \quad \text{with }\ol\kappa_\eff(c_1) =
   \frac{\ol\kappa_1\,\ol\kappa_1}{\ol\kappa_1A_1 + \ol\kappa_2 A_2 c_1}. 
\]}%
Thus, the effective reaction for $\eps\to 0$ is the ternary reaction pair $2X_1+ X_2
\rightleftharpoons X_3$. 

So far, the analysis was on the gradient-flow equation only. The EDP-limit
$(\bfC,\calF_\eff, \calR_\eff)$ is
shown to exist in \cite{Miel23?NESS}, where  
\begin{align*}
  &\calF_\eps\Gto \calF_\eff :\bfc \mapsto \begin{cases}
    \sum_{i=1}^3 \LB(c_i/c_i^*) c_i^*& \text{for } c_4=0, \\
    \infty&\text{for } c_4>0,\end{cases} \qANDq
  \\ 
& \calR_\eff^*(\bfc;\bfxi) = \ol\kappa_\eff(c_1)\big( c_1^2c_2c_3\big)^{1/2}
\sfC^*\big( 2\xi_1{+}\xi_2 {-}\xi_3\big).
\end{align*}
\end{example}

In the next example we return to the wiggly-energy model that was already
discussed in Example \ref{ex:WigglyEnergy}. We now follow the analysis in
\cite{DoFrMi19GSWE, MiMoPe21EFED} where \emph{contact EDP-convergence with
  tilting} to the gradient system $(\R,\calF_\eff,\calR_\eff)$ was
established (cf.\ \cite[Def.\,2.14]{MiMoPe21EFED}). Here we establish the
weaker notion of EDP-convergence to the gradient system
$(\R,\calF_\eff,\ol\calR_\eff)$.  

\begin{figure}
\centerline{
\begin{tikzpicture}
\node at (0,0){\includegraphics[width=0.5\textwidth, trim=0 0 0 0, clip=true]{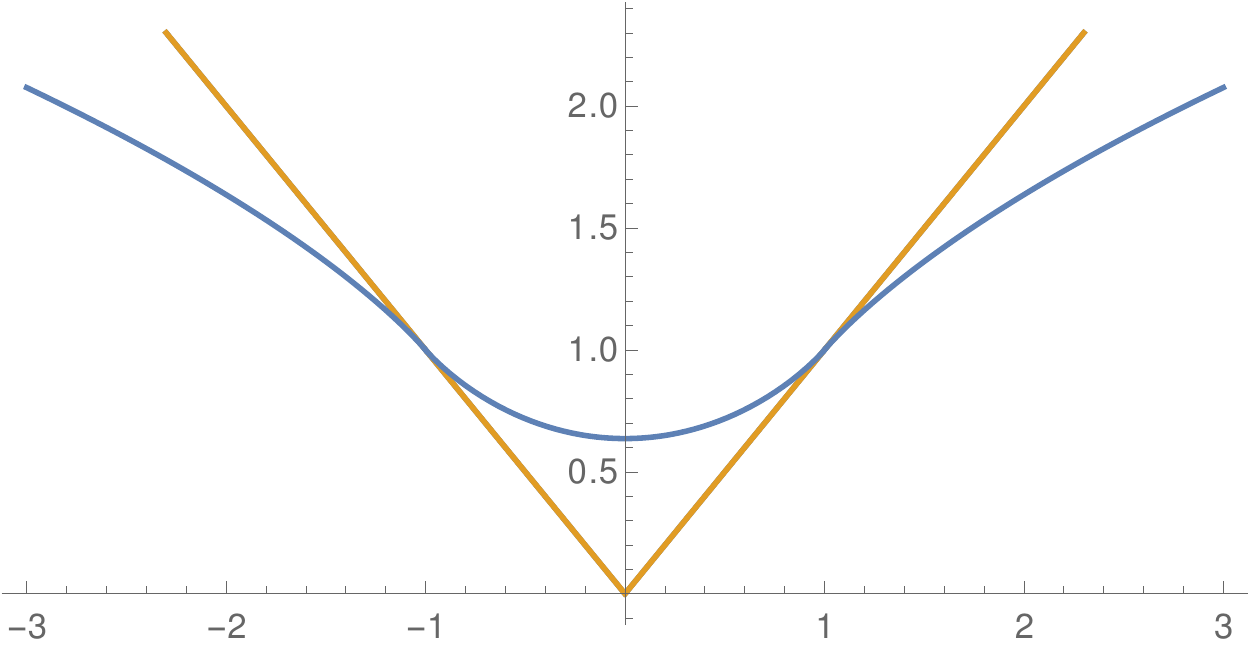}};
\node [color=blue!70!gray] at (3,0.5) {$M_1(\xi)$};
\node [color=orange!35!brown] at (1.7,1.2){$|\xi|$};
\end{tikzpicture}
}
\caption{The function $\xi_1 \mapsto M_1(\xi)$ (blue) from
  \eqref{eq:calM.v.approx0} is plotted for the case $A=1$ together with
  $\xi \mapsto |x|$ (orange). }
\label{fig:calM1}
\end{figure}

\begin{example}[EDP-convergence for the wiggly-energy model]
\label{ex:EDPWigglyEnergy}
We consider a variant of the wiggly-energy problem studied in Example
\ref{ex:WigglyEnergy}, namely $(\R,\calF_\eps,\calR_\eps)$ with 
\[
\calF_\eps(u) = \phi(u)  - \frac{A \eps}{2\pi} \, \cos(2\pi u/\eps) 
\qANDq \calR_\eps(v)= \frac12 \,v^2,
\]
where $A$ is a positive constant.  Obviously, we have
\[
\calF_\eps \to \calF_\eff: \; u\mapsto \phi(u)  \qANDq \calR_\eps \to
\calR_0: v \mapsto \frac12\,v^2.
\]
However, the $\Gamma$-limit $\mfD_0$ of 
\[
\mfD_\eps: \ u \mapsto \int_0^T \!\! \Big( \frac12 \,\dot u{}^2 +
\frac12 \big(\phi'(u)+ A\sin(2\pi u/\eps)\big)^2 \Big)\dd t 
\]
is nontrivial, see \cite{DoFrMi19GSWE}, and has the form $ \mfD_0(u)=\int_0^T
M(\dot u, \phi'(u))  \dd t$ with 
\begin{align*}
 M(v,\xi)&= 
  \inf\Bigset{\int_0^1\!\!\Big( \frac{v^2}2 z'(s)^2 +\frac12\big(\xi {+}A
    \sin(2\pi z(s))\big)^2\Big) \dd s }{ z \in \rmH^1({]0,1[}),\ z(1)=z(0){+}1 } .
\end{align*}
From the definitions we easily see the symmetries
$M(-v,\xi)=M(v,\xi)=M(v,-\xi)$. 

Moreover, \cite[Lem.\,4.3]{DoFrMi19GSWE} provides the following expansion for
$v\approx 0$:
\begin{equation}
  \label{eq:calM.v.approx0}
  \begin{aligned}
   &M(v,\xi)= M_0(\xi) + M_1(\xi)|v| + O(|v|^{3/2}) \quad\text{with }
  M_0(\xi)= \frac12\min\{ |\xi|{-} A, 0\}^2
\\
&\qquad  \AND M_1(\xi)= \int_0^1 \! \Big( \big( \xi{+} A \sin(2\pi y)\big)^2 -
2M_0(\xi) \Big)^{1/2} \dd y .  
\end{aligned}
\end{equation}
Here $M_1$ can be evaluated explicitly (see also Figure \ref{fig:calM1}) giving
\[
M_1(\xi) = \begin{cases} 
\frac2\pi \big( \sqrt{A^2{-}\xi^2} + \xi \arcsin(\xi/A)\big) &\text{for } |\xi|\leq A, \\ 
\frac2\pi \big( \sqrt{|\xi|{-}A} + |\xi|  \arcsin\big(\sqrt{A/|\xi|}\big) \big)
&
\text{for } |\xi|\geq A. 
\end{cases}
\]
We first observe that we have the estimate $M(v,\xi)\geq \xi v$ for all
$v,\xi\in \R$, which is a remainder of the Fenchel-Young inequality. 
To see this, we observe 
\begin{align}
\label{eq:M(v,xi)estim} 
\int_0^1&\!\!\Big( \frac{v^2}2 z'(s)^2 +\frac12\big(\xi {+}A
    \sin(2\pi z(s)\big)^2\Big) \dd s \geq \int_0^1\!\! v z'(s)\big(\xi {+}A
    \sin(2\pi z(s)\big) \dd s \\
\nonumber
&=  v\xi \int_0^1 \! z'(s)\dd s + vA \int_0^1 z'(s) \sin(2\pi z(s))\dd s\\
\nonumber & =
  v\xi \big(z(1){-}z(0)\big) + \frac{vA}{2\pi}\big( \cos(2\pi z(0))-\cos(2\pi
  z(1))\big) = v\xi, 
\end{align}
where we used the boundary condition $z(1)=z(0){+}1$ for the last
identity. Taking the infimum over $z$ gives $M(v,\xi)\geq \xi v$ as desired. 

Moreover, we can discuss the equality $M(v,\xi)=\xi v$ explicitly. For $v=0$ we
have $M(0,\xi)=0$ if and only if $\xi\in [-A,A]$ by the form of $M_0$. For
$v> 0$, we see that the equality $M(v,\xi)=\xi v$ implies equality a.e.\ for
the integrand in \eqref{eq:M(v,xi)estim}, i.e.\ $vz'(s) = \xi + A\sin(2\pi
z(s))>0$ and hence $\xi>A$. With this we find 
\[
1 = \int_{s=0}^1 \dd s   = \int_{s=0}^1 \frac{v z'(s)\:\dd s}{\xi{+}A\sin(2\pi z(s)) }
 = \int_{z=z(0)}^{z(0)+1} \frac{v\:\dd z}{\xi + A\sin(2\pi z)} = \frac{ v
 }{\sqrt{\xi^2{-}A^2}} .
\]
With the similar argument for $v<0$,  we obtain $0\neq |v|=
\sqrt{\xi^2{-}A^2}$. As a result we have shown that 
\begin{equation}
  \label{eq:M.v.xi-vxi}
  \forall\, \xi \in \R: \quad \min\bigset{M(v,\xi) - \xi v }{v\in \R}  =0. 
\end{equation}

We now define the effective dissipation potential $\ol\calR_\eff$ via
\[
\ol\calR_\eff(u,v):= M(v,\phi'(u)) - M(0,\phi'(u)).
\]
By definition we have $\ol\calR_\eff(u,0)=0$, and the results in
\cite[Prop.\,4.11]{DoFrMi19GSWE} show
$\ol\calR_\eff(u,v)=\ol\calR_\eff(u,-v)\geq 0$ and the convexity of
$\ol\calR_\eff(u,\,\cdot\,)$.

It remains to show the representation  
\begin{equation}
  \label{eq:WE.MRR*}
  M(v,\phi'(u)) = \ol\calR_\eff(u,v) +
\ol\calR_\eff^*\big(u,{-}\rmD\calF_\eff(u)\big). 
\end{equation}
Using $\rmD\calF_\eff(u)= \phi'(u)$ we obtain 
\begin{align*}
\ol\calR_\eff^*\big(u,-\phi'(u))&=\sup_{v\in \R} \big({-}\phi'(u)v -
\ol\calR_\eff(u,v) \big)  =  \sup_{v\in \R} \big( {-}\phi'(u)v {-}M(v,\phi'(u))
     {+} M(0,\phi'(u)) \big) \\
&=\sup_{v\in \R} \big( {-}\phi'(u)v {-}M(v,{-}\phi'(u))\big)\  + M(0,\phi'(u))
=  0 + M(u,\phi'(u)).  
\end{align*}
Using the definition of $\ol\calR_\eff$ this implies \eqref{eq:WE.MRR*}, and
the desired EDP-convergence for the wiggly-energy model  is established, i.e.\
we have 
$(\R,\calF_\eps, \calR)\EDPto (\R,\calF_\eff,\ol\calR_\eff)$. 

However, following the argumentation in \cite{DoFrMi19GSWE, MiMoPe21EFED} the
derived effective dissipation potential $\ol\calR_\eff$ is somehow
artificial, because $\ol\calR_\eff$ depends on the force $\xi=\phi'(u)$ via 
$\ol\calR_\eff(u,v)= M(v,\phi'(u))-M(0,\phi'(u))$. The notion of contact
EDP-convergence appears tilting (see \cite[Def.\,2.14]{MiMoPe21EFED}) is more
more natural and leads to the effective dissipation potential $\calR_\eff$ with 
\[
\calR_\eff(v) =A^2\mathsf R(v/A) \ \text{ with }\mathsf R(w)= \frac12\Big(
|w|\sqrt{1{+}w^2} + \log\big(|w|{+}\sqrt{1{+}w^2}\big)\Big),
\]
which is independent of $u$ and hence of the force $\phi'(u)$. But the limit
$\mfD_0$ of the dissipation integrals $\mfD_\eps$ coincides with $\int_0^T \!
\big( \calR_\eff(\dot u){+}\calR_\eff^*(-\phi'(u))\big) \dd t$ only along
solutions $u$ of the effective gradient-flow equation $\dot u=
\pl_\xi\ol\calR^*_\eff(u,- \rmD\calF_\eff(u))= \pl_\xi\calR^*_\eff({-}
\rmD\calF_\eff(u))$. 
\end{example}

\begin{remark}[Nonuniqueness when using liminf only]
\label{ex:NonuniqueGStr} Often it is argued that for obtaining the effective
gradient-flow equation it is not necessary to establish any limsup estimate for
$\mfD_0$. In particular, in the Sandier-Serfaty theory \cite{SanSer04GCGF,
  Serf11GCGF} only the liminf estimates \eqref{eq:TwoSanSerEstim} are
requested. This is indeed true, but one has to be aware that by this approach
we lose the uniqueness of the gradient structure. If we only impose the liminf
estimates we may have two different gradient structures, which both generate
the same effective equation.  

As an example consider the wiggly-energy model consider in the previous
example. We claim that we can find $\wt\calR$ such that
$\wt\calR(u,v)+\wt\calR^*(u,-\rmD\calF_\eff(u))\lneqq M(u,v)$. Clearly, then
the liminf estimate holds trivially, but the limsup estimate is false. To find
an example the ideas is to make $\calR$ smaller in some region where it does
not increase the slope term $\calR^*(u,-\rmD\calF_\eff(u))$. To be more
precise, we choose
$\theta \in \rmC^0(\R;[0,1])$ with $\theta(\xi)=1$ for $|\xi|\geq A$ and
$\theta(\xi) \in {]0,1[}$ for $|\xi|<A$ and set 
\[
\wt\calR(u,v) = \theta(\phi'(u))\ol\calR_\eff(u,v) +
\big(1{-}\theta(\phi'(u))\big) |\phi'(u)|\: |v| \ \geq |\phi'(u)|\: |v|.
\]
For the last estimate we used $ \ol\calR_\eff(u,v)\geq M_1(\phi'(u))|v|$ and
$M_1(\xi) \geq |\xi|$  with $M_1$ defined in \eqref{eq:calM.v.approx0}, see also  Figure
\ref{fig:calM1}. 

Using the convexity of $\ol\calR_\eff(u,\cdot)$ we have $\wt\calR(u,v)\lneqq
\ol\calR_\eff(u,v)$. Because $\ol\calR_\eff(u,\cdot)=\wt \calR(u,\cdot)$ for
$|\phi'(u)|\geq A$ we also have $\wt\calR^*(u,-\rmD\calF_\eff(u))=
\ol\calR^*_\eff(u,-\rmD\calF_\eff(u)) $ in that range. However, for 
$|\phi'(u)|<A $ we easily obtain $\wt\calR^*(u,-\rmD\calF_\eff(u))=0=
\ol\calR^*_\eff(u,-\rmD\calF_\eff(u)) $. 

Thus, we see that $\mfD_0$ generated by $\ol\calR_\eff{\oplus}\ol\calR^*_\eff$ as well as
$\wt\mfD$ generated by $\wt\calR {\oplus} \wt\calR^*$ satisfy the liminf
estimate for the family $\mfD_\eps$. 
\end{remark}

\section{Rate-independent systems}
\label{se:RateIndepSyst}

\subsection{Introduction to rate independence}
\label{su:IntroRateIndep}

A very special case of gradient systems is obtained in the so-called rate-independent
case. This is a very degenerate model class where
$\calR(u,\cdot):X\to [0,\infty]$ is positively homogeneous of degree 1
(shortly: one-homogeneous), i.e.\
\[
\forall \, \lambda >0 \ \forall \, u,v\in X: \quad \calR(u,\lambda v) = \lambda
\calR(u,v). 
\]
This case is only interesting if the energy depends on $t\in
[0,T]$, i.e.\ we consider $\calF:[0,T] \ti X \to \Rinfty$ where the dependence
$t \mapsto \calF(t,u)$ for fixed $u$ describes an external loading like in a
Banach space with $\calF(t,u)=\calE(u)-\langle \ell(t),u\rangle$. 
The gradient-flow equation reads 
\begin{equation}
  \label{eq:RI.DiffForm}
  0 \in \pl\calR(u(t),\dot u(t)) + \plF \calF(t,u(t))  \ \in X^*.
\end{equation}
The term ``\emph{rate independence}'' stems from the fact that for a (smooth
and) strictly increasing transformation $\phi:[0,S]\to [0,T]$ of the loading in 
the form $\wt \calF(s,u)= \calF(\phi(s),u)$ a solution $u:[0,T]\to X$ for
$(X,\calF,\calR)$ transforms to a solution $\wt u: s \mapsto u(\phi(s))$ for
$(X,\wt\calF,\calR)$, and vice versa. The reason for this is that $v \mapsto
\pl\calR(u,v)$ is positively $0$-homogeneous, i.e.\ $\pl\calR(u,\lambda v) =
\pl\calR(u,v)$. Indeed the following result shows that the subdifferential of a
one-homogeneous function has very special properties. 

\begin{lemma}[Subdifferential of one-homogeneous functionals]
\label{le:SubdiffOnehom}
Consider a lsc, positively one-homogeneous functional $\Psi:X\to \Rinfty$, then
the subdifferential $\pl\Psi$ satisfies 
\[
\forall \:v\in X:\quad \pl\Psi(v) = \bigset{\xi \in \pl\Psi(0)\subset X^*}{
  \langle \xi, v\rangle= \Psi(v)}. 
\]
\end{lemma} 

This formula shows that rate independence of $\pl\Psi(v)$ in the sense that the
subdifferential does not depend on the length of $v$ but only on the
direction. 

Of course, we see that our existence theory developed in previous sections does
not apply, because $v \mapsto \calR(u,v)$ is not superlinear. Hence, a special
theory needs to be develop but nevertheless many similarities to the superlinear
case remain.  In the metric setting the rate-independent case corresponds to
the choice $\psi = \psi_\mafo{id}: r \mapsto r$.

We refer to the surveys \cite{Miel05ERIS, Miel11DEMF} and the monograph
\cite{MieRou15RIST} for the full theory which was developed in parallel in the
works starting with \cite{MieThe99MMRI, MiThLe02VFRI, MieThe04RIHM} using the
name ``rate-independent systems'' and the works \cite{FraMar98RBFE,
  DalToa10QCGE, FraLar03ECQS, DaFrTo05QCGN} using the name ``quasistatic
evolution''. In the following we give a very short introduction into the theory
with the single goal to show the connections of this theory with the general
theory of gradient systems.

As a simple example we consider the case $M=X=\R^1$ with the energy
$\calF(t,u)= \frac a2 u^2 - u\lambda t$, where $a>0$ and $\lambda\in R$, and
the dissipation potential $\calR(u,v)= 2|v|+v$ satisfying
$\pl\calR(u,0)=[-1,3]$. The differential form \eqref{eq:RI.DiffForm} of the
system takes the form
\begin{equation}
  \label{eq:RI.Toy.Exa}
  0 \in 2\,\mafo{Sign}(\dot u) +\dot u + a u -\lambda t,
\end{equation}
where $v\mapsto \mafo{Sign}(v) \subset \R$ is the set-valued signum function
obtained as subdifferential of $v \mapsto |v|$. E.g.\ starting with $u(0)=0$ we
obtain the solution
\[
u(t) = \begin{cases} \max\big\{ 0, (\lambda t{-}3)/a\big\} & \text{ for
  }\lambda \geq 0,
\\
\min\big\{ 0, (\lambda t{+}1)/a\big\} & \text{ for
  }\lambda \leq 0.
\end{cases}
\]

\subsection{Energetic solutions}
\label{su:EnergeticSol}

The concept of energetic solutions plays the role of curves of maximal slope in
the metric setting, but there are two major differences. First, the solutions
are no longer absolutely continuous, i.e.\ they are allowed to have jumps with
respect to the time variable $t\in [0,T]$. Second, we can allow the dissipation
distance $\DD:M\ti M\to
[0,\infty]$  to be an extended quasi-distance, i.e.\ $\DD$ doesn't have to be
symmetric and it may take the value $\infty$. Hence, we have to be careful
about the order of arguments when writing the triangle inequality for $\DD$. We
emphasize that in the following we will always use the order
``$\DD(u_\text{old},u_\text{new})$'', where `old' and `new' refer to the
ordering of the time
variable $t \in [0,T]$, because $\DD$ is considered to be a dissipation
distance which associates with an arrow of time.  

To simplify our exposition here, we assume that there is another true metric
$D:M\ti M\to {[0,\infty[}$ satisfying $D(u,w)\leq \DD(u,w)$.

\begin{definition}[Energetic rate-independent system]
A triple $(M,\calF,\DD)$ is called an \emph{energetic rate-independent system (ERIS)}
with metric $D:M\ti M\to {[0,\infty[}$, if
\begin{enumerate}[label={\upshape (E.\arabic*)}]\itemsep-0.1em
\item \label{ERIS.1} $(M,D)$ is a complete metric space; 
\item $\calF:[0,T]\ti M\to \Rinfty$ is lsc on $(M,D)$ with
  domain $\dom \calF = [0,T] \ti F_{\dom}\neq \emptyset$;
\item\label{eq:calF.Gronw} $\exists\: C_\rmE, c_\rmE>0\ \forall \:u\in F_{\dom}:\ \calF(\cdot,u)\in
  \rmC^1([0,T])$  \text{ and }\\
 \hspace*{11em}  $|\pl_t\calF(t,u)| \leq C_\rmE\big(
  \calF(t,u){+} c_\rmE\big)$ for all $t\in [0,T]$;
\item $\DD:M\ti M\to [0,\infty]$  is lsc on  on $(M,D)$ and $D(u,w)\leq \DD(u,w)$ for all
  $u,w\in M$;

\item \label{ERIS.5} $\forall\:u_1,u_2,u_3\in M: \quad \DD(u_1,u_3) \leq \DD(u_1,u_2) +
  \DD(u_2,u_3)$ and $\calD(u_1,u_1)=0$. 
\end{enumerate}
\end{definition}

Below we will define \emph{energetic solutions} (also called quasistatic
evolutions) as natural limit of the time-incremental minimization scheme. We
emphasize that the rate-independent case associates with the scalar dissipation
function $\psi_\text{ri}(r)=r$, whence the metric construction 
\[
\tau \,\psi_\text{ri}\big( \frac1\tau \DD(u_{k-1},u)\big) \ = \ \DD(u_{k-1},u)
\]
in Definition \ref{de:MetrGSMM} simplifies considerably. In particular, the time step
$\tau$ disappears completely, which can be seen again as a manifestation of
rate independence.   Thus, defining a partition $0=t_0< t_1 < \cdots
 < t_{N-1}<t_N=T$ we obtain \medskip

\noindent%
\fbox{\begin{minipage}{0.98\textwidth}
\centerline{\textbf{rate-independent time-incremental minimization scheme} (TIMS)}
\begin{equation}
  \label{eq:RI.TIMS}
  u_k \quad \text{ minimizes } \ \ u \ \mapsto \ \DD(u_{k-1},u)+ \calF(t_k,u). \vspace{0.3em}
\end{equation}
\end{minipage}}\medskip

We again emphasize that the time step $\tau_k=t_k{-}t_{k-1}$ does not show up
because of rate independence. This fact can be used in material modeling for
the study of microstructures in nonlinear plasticity \cite{OrtRep99NEMD,
  CaHaMi02NCPM, ConThe05SSEM}, in shape memory alloys \cite{MiThLe02VFRI,
  BarKru11EANS, DesKru13DPHP}, or in crack propagation \cite{DalToa02MQSG,
  DaFrTo05QCGN, DalZan07QSCG, DRST21?VESM}.

The following result shows that one easily obtains useful information from this
minimization scheme even without having a subdifferentials. 

\begin{proposition}[TIMS for ERIS]
\label{pr:TIMS.ERIS}
Assume that $(u_k)_{k=1,..,N}$ solve the TIMS for the ERIS $(M,\calF,\DD)$,
then we have, for all $k\in \{1,\ldots,N\}$, \smallskip

{\upshape (i)} \ \; $\calF(t_k,u_k) + \DD(u_{k-1},u_k) \leq \calF(t_k,
u_{k-1})=\calF(t_{k-1},u_{k-1}) + \int_{t_{k-1}}^{t_k} \pl_s \calF(s,u_{k-1})
\dd s $.\smallskip

{\upshape (ii)} \ \ $\calF(T,u_N) + \sum_{m=1}^N\DD(u_{m-1},u_m) \leq \calF(0,u_0) 
+ \int_0^T  \pl_s \calF(s,\ul u(s)) \dd s $.\smallskip

{\upshape (iii)} \ $u_{k}$ minimizes the functional $w \mapsto
\DD(u_{k},w)+\calF(t_k,w)$.\smallskip

{\upshape (iv)} \ \ $ \calF(t_k,u_k) + \sum_{j=1}^k \DD(u_{j-1}, u_j) \leq \ee^{C_\rmE t_k}
\,\big(\calF(0,u_0) + c_\rmE\big) - c_\rmE$. 
\\[0.2em]
Assertion (ii) uses the right-continuous interpolant $\ul u:[0,T]\to M$, see
\eqref{eq:BanachInterpol}. 
\end{proposition}
\begin{proof}
(i) is a simple consequence of \eqref{eq:RI.TIMS} when comparing with $u=u_k$
and $u=u_{k-1}$.

(ii) then follows by summing over $k=1$ to $N$.

To obtain (iii) we use the triangle inequality for $\DD$ in
$\overset\triangle\leq$ and obtain 
\begin{align*}
\DD(u_k,u_k)+\calF(t_k,u_k) &\leq 0+\calF(t_k,u_k)+ \DD(u_{k-1},u_k)-
\DD(u_{k-1},u_k)
\\
&\overset{\text{\eqref{eq:RI.TIMS}}}\leq \calF(t_k,w)+ \DD(u_{k-1},w)-
\DD(u_{k-1},u_k) \overset\triangle\leq \DD(u_k,w) + \calF(t_k,w),
\end{align*}
which is the desirable result. 

For (iv) we abbreviate $f_k=\calF(t_k,u_k)+c_\rmE$ and $d_k= \DD(u_{k-1},u_k)$ and
find 
\begin{align}
\label{eq:fk.dk}
f_k+d_k &\overset{\text{(i)}}\leq f_{k-1} +\int_{t_{k-1}}^{t_k} \!\!
C_\rmE \big(\calF(s,u_{k-1}){+}c_\rmE\big) \dd s \\
\nonumber
& \overset{**}\leq  
f_{k-1} + \int_{t_{k-1}}^{t_k}\!\! C_\rmE \ee^{C_\rmE(s{-}t_{k-1})} f_{k-1}\dd s
 = \ee^{C_\rmE(t_k{-}t_{k-1})}f_{k-1} ,
\end{align}
where $\overset{**}\leq$ exploits that \ref{eq:calF.Gronw} combined with 
Grönwall's estimate gives $\calF(s,u_{k-1}){+}c_\rmE \leq \ee^{C_\rmE|t{-}s|}
(\calF(t,u_{k-1}){+}c_\rmE)$.  Using $d_k\geq 0$ we first obtain $ f_k \leq
\ee^{C_\rmE t_k} f_0$. 

With this we return to \eqref{eq:fk.dk} and estimate as follows:
\begin{align*}
f_N+ \sum_{j=1}^Nd_N &= \sum_{j=1}^N(f_k{+}d_k) -  \sum_{k=1}^{N-1}f_k
\leq   \sum_{k=1}^Nf_{k-1}  \,\ee^{C_\rmE(t_k{-}t_{k-1})}  -  \sum_{k=1}^{N-1}f_k
\\
&= f_0 \, \ee^{C_\rmE(t_1{-}t_{0})} +\sum_{k=2}^Nf_{k-1}
\big(\ee^{C_\rmE(t_k{-}t_{k-1})}-1\big)\\
&\leq f_0  \,\ee^{C_\rmE t_1} +\sum_{k=2}^Nf_0  \,\ee^{C_\rmE t_{k-1}} 
\big(\ee^{C_\rmE(t_k{-}t_{k-1})}-1\big) = f_0 \,\ee^{C_\rmE t_N}.
\end{align*}
Noting that $N$ can be replaced by any $k\in \{1,..,N\}$ assertion (iv) is established. 
\end{proof}

In the above we recognize that (ii) is a discrete energy balance in the spirit
of \eqref{eq:Banach.DiscrEDI} or \eqref{eq:Metr.22}; however, it is unclear
whether a term involving $\calR^*$ or $\psi^*$ is missing. We will see that
this is not the case, because of the special structure of
$\psi=\psi_\text{id}$, leading to the dual function $\psi^*_\text{id}(\zeta)=0$
for $\zeta\in [0,1]$ and $\psi^*_\text{id}(\zeta)=\infty$ for $\zeta>1$.   

An important observation is the so-called \emph{global stability} satisfied by
$u_k$ as is shown in (iii).  We define the \emph{set of globally stable states}
\[
\calS(t):= \bigset{u\in M }{ \calF(t,u)<\infty \text{ and } 
 \forall\:w\in M: \ \calF(t,u)\leq  \calF(t,w)+\DD(u,w) }
\]
and call its elements the \emph{(globally) stable states}. This stability has the
simple interpretation that it is energetically not favorable to move from $u$ to another
point $w$ if the dissipated energy $\DD(u,w)$ is taken into account. 
In the toy example  \eqref{eq:RI.Toy.Exa} we have $\calS(t)= \big[(\lambda
t{-}3)/a, (\lambda t{+}1)/a\big]$.

To compare this concept with the metric theory we recall the notion of global
metric slope \eqref{eq:defGlobMetrSlope} from the classical metric theory 
and introduce the same object also for the 
extended quasi-metric $\DD$, where we have to be careful about the order of the
arguments: 
\begin{equation}
  \label{eq:RI.GlobSlope}
  \SLO{{}^\mafo{gl}_0 \calF(t,\cdot)}(u) :=\begin{cases} \infty& \text{for } u
    \not\in \dom(\calF(t,\cdot)), \\
  \sup\Bigset{\dfrac{\big[\calF(t,u){-}\calF(t,w)\big]_+}{\DD(u,w)} }{ w \in M}  &
\text{for } u
    \not\in \dom(\calF(t,\cdot)).
  \end{cases}
\end{equation}
By simply comparing the definitions we clearly obtain the equivalence 
\begin{equation}
\label{eq:RI.EquivStabSlope}
u\in \calS(t) \quad \Longleftrightarrow \quad \SLO{{}^\mafo{gl}_0
  \calF(t,\cdot)}(u) \leq 1.  
\end{equation}

For the dissipated energy we also need an adaptation as follows. For arbitrary
curves $u : [0,T]\to M$ defined pointwise but assuming no continuity or
measurability, we define for all $s,t\in [0,T]$ with $s<t$ the \emph{variation
  dissipation} 
\[
\mafo{Var}_\DD(u,[s,t]) := \sup\Bigset{
\sum_{k=1}^N \DD \big( u(t_{k-1}),u(t_k)\big) }{ N\in \N, \ 
s\leq t_0<t_1<\cdots < t_N\leq t} .
\]
By our assumption $D\leq \DD$ every curve with $\mafo{Var}_\DD(u,[0,T])<\infty$
also satisfies $\mafo{Var}_D(u,[0,T])<\infty$ in the complete metric space
$(M,D)$. This implies that such a $u$ can have at most countably many jump
points and that left and right limits 
\[
u(t^-):= \lim_{h\to 0^+} u(t{-}h) \quad \text{and} \quad 
u(t^+):= \lim_{h\to 0^+} u(t{+}h)
\]
exist for all $t \in [0,T]$ (by definition one sets $u(0^-)=u(0)$ and $u(T^+)=u(T)$). 

We are now ready to give a precise definition of a suitable notion of solutions
for ERIS.

\begin{definition}[Energetic solutions {\cite[Def.\,3.1]{Miel05ERIS}}] 
\label{de:EnergSol} 
A curve $u:[0,T]\to M$ is called an \emph{energetic
  solution} for the ERIS $(M,\calF,\DD)$ if the global
stability (S) and the energy equality (E) hold:
\begin{align*}
\textbf{\upshape(S)}\quad & u(t) \in \calS(t) \ \text{ for all } t \in [0,T],
\\
\textbf{\upshape(E)}\quad & \calF(T;u(T)) + \mafo{Var}_\DD(u,[0,T]) = \calF(0,u(0)) +
\int_0^T\!\! \pl_s\calF(s,u(s)) \dd s. 
\end{align*}
\end{definition}

We emphasize that the solutions are defined pointwise and that the condition of
global stability is asked for \emph{all} $t\in [0,T]$. Moreover, the energy
balance (E) is posed only for the whole time interval $[0,T]$. However, using
the chain rule from below it follows that it is valid on all subintervals,
i.e.\ for all $r,t\in [0,T]$ with $r<t$ we have 
\[
 \calF(t;u(t)) + \mafo{Var}_\DD(u,[r,t]) = \calF(r,u(r)) +
\int_r^t\!\! \pl_s\calF(s,u(s)) \dd s. 
\]
It is even possible to consider the limits $r\nearrow s$ and $t\searrow s$ to
obtain the jump conditions 
\[
\calF(s,u(s^+)) + \DD(u(s),u(s^+))=\calF(s,u(s)) \ \text{ and } \ 
\calF(s,u(s)) + \DD(u(s^-),u(s))=\calF(s,u(s^-)).
\]
Recall that it is possible that the three states $u(s^-)$, $u(s)$, and $u(s^+)$
may be mutually different. 

Finally, we remark that it is tempting to rewrite (S) and (E) in  
$\calR{\oplus}\calR^*$ form: 
\[
\calF(T;u(T)) + \int_0^T\!\!\Big( \psi_\mafo{id}\big(\SPE u(t)\big) +
\psi^*_\mafo{id}\big( \SLO{{}^\mafo{gl}_0 \calF(t,\cdot)}(u) \big) \Big) \dd t
= \calF(0,u(0)) + \int_0^T\!\! \pl_s\calF(s,u(s)) \dd s.
\]
Since $\psi^*_\mafo{id}$ only takes the value $0$ and $\infty$, the finiteness
of the left integral encodes the condition (S) at least almost
everywhere. However, the major difficulty is to define the metric speed at jump
points taking care of the possibly three different values $u(s^-)$, $u(s)$, and
$u(s^+)$. Hence, it turns out that it is much easier and truly necessary to
use the exact and pointwise formulation (S)\&(E) from Definition
\ref{de:EnergSol}.

\subsection{Existence of energetic solutions}
\label{suExistEnergSol} 

The following existence result follows exactly along the lines of the existence
theory for curves of maximal slope. We will repeat the main arguments to show the
analogies as well as the differences. The first major difference is that we
cannot appeal to the Arzel\'a-Ascoli theorem because of the missing
superlinearity. However, a metric version of Helly's selection theorem as
derived in \cite[Thm.\,3.2]{MaiMie05EREM}.

A second difference is more formal than mathematical. It was already observed
in \cite[Thm.\,2.5]{MiThLe02VFRI} that the global stability (S) implies a ``lower
energy estimate'' which is the corresponding version of the metric chain-rule inequality,
see \eqref{eq:MetrCRIneq}. We will see that the
proof is considerably simpler than that of Proposition \ref{pr:MetrCRIneq},
because the stability condition is equivalent to the property that the global
slope is bounded by $1$, see \eqref{eq:RI.EquivStabSlope}. 

The essential new condition is the so-called ``\emph{closedness of the stable
  sets} in \eqref{eq:RI.ClosStabSet}, which can be seen as a replacement of the
lower semicontinuity of the (global) slope. This condition is nontrivial here
because we allow $\DD$ to be non-continuous and take the value $+\infty$, see
the discussion in Section \ref{su:ClosedStabSets}.

\begin{theorem}[Existence of energetic solutions] 
\label{th:ExiEnergSol} 
Let the ERIS $(M,\calF,\DD)$ satisfy the conditions
\ref{ERIS.1}--\ref{ERIS.5}. Moreover, assume the following properties:
\begin{subequations}
\begin{align}
&\nonumber\hspace*{-1em}\text{compactness of sublevels:}\\
&\label{eq:RI.SublCmpt}
\forall\: E>0\ \forall\:t\in [0,T]:\quad S^{\calF(t,\cdot)}_E=\bigset{u\in M}{
  \calF(t,u)\leq E} \text{ is compact},
\\ 
&\nonumber\hspace*{-1em}\text{closedness of the stable sets:}
\\
\label{eq:RI.ClosStabSet}
&t_i\to t,\ u_i \to u,\ u_i\in \calS(t_i) \quad \Longrightarrow \quad u\in \calS(t),
\\
&\nonumber \hspace*{-1em}\text{conditional continuity of the power
$\pl_t\calF$:}
\\
&  \label{eq:RI.ContPower}
  t_i\to t,\ u_i\to u, \ \sup_{i\in \N}\calF(t_i,u_i)< \infty \quad 
\Longrightarrow \quad \pl_t\calF(t_i,u_i)\to \pl_t\calF(t,u).
\end{align}
\end{subequations}
Then, for all $u_0\in \calS(0)$ there exists an energetic solution $u:[0,T] \to
M$ for the ERIS $(M,\calF,\DD)$ with $u(0)=u_0$. In particular, every
accumulation point in the sense of pointwise convergence of a sequence of
piecewise interpolants for the time-incremental minimization scheme
\eqref{eq:RI.TIMS} is an energetic solution.
\end{theorem}
 
Before going into the proof of the existence theorem, we will shortly discuss
the version of the metric chain-rule inequality for ERIS. An important point is
now that the solutions are not continuous, hence we can only derive an
integrated version. Moreover, we need to generalize the theory to
time-dependent energies. To see the analogy we observe that integrating the
differential metric chain-rule inequality \eqref{eq:MetrCRIneq} over $t\in
[r,s]$ we find 
\[
\calF(u(s)) + \int_r^s \SPE u (t) \,\SLO \calF (u(t))  \,\dd t \geq \calF(u(r)) 
\]
For stable states  we have $\SLO \calF (u(t))\leq \SLO {{}_\mafo{gl}\calF}
(u(t)) \leq 1$, such that   $\int_r^s \SPE u (t)\dd t = \mafo{Var}_\DD(u,[r,s])$
remains, where the last identity holds for absolutely continuous curves. Thus, the
chain-rule inequality \eqref{eq:RI.ChRuIneq} appears naturally in the context
of ERIS. Because of the global slope condition  the
proof is considerably simpler than that of Proposition \ref{pr:MetrCRIneq}.

\begin{proposition}[Rate-indep.\ chain-rule inequality]
\label{pr:RI.ChainRuleIneq}  
Consider the ERIS $(M,\calF,\DD)$ satisfying the conditions
\ref{ERIS.1}--\ref{ERIS.5} as well as \eqref{eq:RI.ContPower}.  If the curve
$u:[0,T]\to M$ satisfies $u(t)\in \calS(t)$ for all $t \in {[r,s[}$ and
$\sup_{t\in [r,s]} \calF(t,u(t)) < \infty$, then we have the chain-rule
inequality
\begin{equation}
  \label{eq:RI.ChRuIneq}
  \calF(s,u(s) ) + \mafo{Var}_\DD(u,[r,s]) \geq \calF(r,u(r)) + \int_r^s
  \pl_t\calF(t,u(t)) \dd t. 
\end{equation}
\end{proposition}
\begin{proof} By assumption $t\to \calF(t,u(t))$ is bounded. Using
  \ref{eq:calF.Gronw} also the power $\pl_t\calF(t,u(t))$ is bounded such that
  the right-hand side in \eqref{eq:RI.ChRuIneq} is finite. Hence, the assertion
  holds if $\mafo{Var}_\DD(u,[r,s])=\infty$. Thus, we can assume
  $\mafo{Var}_\DD(u,[r,s]) <\infty$ from now on.

We choose an arbitrary partition $r=t_0< t_1 < \cdots < t_N=s$ and set
$u_j=u(t_j)$, $f_j=\calF(t_j,u_j)$ and $d_j=\DD(u_{j-1},u_{j})$. For
$j=0,...,N{-}1$, we have $u_j\in \calS(t_j)$ which implies $f_j \leq
\calF(t_j,u_{j+1}) + d_{j+1}$. Hence, we have 
\[
f_{j+1} + d_{j+1} - f_j = f_{j+1} - \calF(t_j,u_{j+1})= \int_{t_j}^{t_{j+1}}
\pl_t \calF(t,u_{j+1}) \dd t \ \text{ for } j=0,1,...,N{-}1.
\]
Summing of these $j$ and using the left-continuous interpolant $\ol u$
(cf.\ \eqref{eq:BanachInterpol}) we find
\begin{equation}
  \label{eq:RI.EstimPartit}
  \calF(s,u(s))+\mafo{Var}_\DD(u,[r,s])-\calF(r,u(r)) \geq f_N+\sum_{j=0}^{j-1}
d_{j+1} - f_0 \geq \int_r^s \pl_t \calF(t,\ol u(t)) \dd t.
\end{equation}

Finally we choose the sequence of partitions by setting $\tau_N=(s{-}r)/N$ and
$t^N_j=r{+}j\tau_N $. This gives the piecewise constant interpolants $\ol u_N:
[0,T] \to M$. As $\mafo{Var}_\DD(u,[r,s]) <\infty$, we have $\ol u_N(t) \to
u(t)$ for all $t\in [r,s]$ except for the jump points of $u$, which are at most
countable. Moreover, \ref{eq:calF.Gronw} and the boundedness of $t\mapsto
\calF(t,u(t))$ implies $ |\pl\calF(t,\ol u_N(t)) |\leq C$. Together with the assumed
continuity of the power \eqref{eq:RI.ContPower} we can pass to the limit in the
right-hand side of \eqref{eq:RI.EstimPartit} and obtain the desired lower
energy estimate.  
\end{proof}

\noindent
\begin{proof}[Proof of Theorem \ref{th:ExiEnergSol}]
We follow the same five steps as in the existence proof for curves of maximal
slope, see Theorem \ref{th:MetricExistence}. 

\STEP{Step 0: Construction of approximants.} We choose an arbitrary sequence of
partitions $0=t_0^N< t_1^N < \cdots < t^N_{N_1}<t_N^N=T$ whose fineness
$\phi_N:= \max\bigset{ t_j^N{-}t^N_{j-1}} {j=1,...,N}$ tends to $0$ for $N\to \infty$. 

The time-incremental minimization problem \eqref{eq:RI.TIMS} is solvable in
each step, because $\DD(u^N_{k-1},\cdot)$ and $\calF(t_k,\cdot)$ are lsc
on $M$ and $\calF(t_k,\cdot)$ has compact sublevels by \eqref{eq:RI.SublCmpt}.  
By Proposition \ref{pr:TIMS.ERIS} the right-continuous interpolants $\ul
u^N:[0,T]\to M $ satisfy the discrete a priori estimate 
\begin{equation}
  \label{eq:RI.DiscrEDI}
  \calF(T,\ul u^N(T)) + \mafo{Var}_\DD(\ul u^N,[0,T]) \leq \calF(0, u_0) +
\int_0^T\pl_t \calF(t, \ul u_N (t)) \dd t ,
\end{equation}
where we use the identity $\mafo{Var}_\DD(\ul u^N,[0,T])= \sum_{j=1}^N \DD \big( 
\ul u^N(t_{j-1}) , \ul u^N(t_{j})\big)  $ which holds for piecewise constant
interpolants.
 
\STEP{Step 1: A priori estimates.} Proposition \ref{pr:TIMS.ERIS} provides the
a priori estimates
\[
\forall\:N\in \N\ \forall\: t\in [0,T]{:} \ \  \calF(t,\ul u_N(t)) +
\mafo{Var}_\DD(\ul u^N,[0,T]) \leq \ee^{C_\rmE T}\big(\calF(0,u_0) 
{+} c_\rmE\big) - c_\rmE=:C_*.
\]
Using \ref{eq:calF.Gronw} we obtain $\calF(0,\ul u_N(t)){+}c_\rmE \leq
\ee^{C_\rmE t} \big(\calF(t,\ul u_N(t)){+}c_\rmE\big) \leq \ee^{C_\rmE T} C_* +
c_\rmE=C_{**}$. Thus, we have
\[
\forall\:N\in \N\ \forall\: t\in [0,T]{:} \ \ \ul u_N(t)\in
S^{\calF(0,\cdot)}_{C_{**}} \Subset M,
\]
where we used the compactness of sublevels from \eqref{eq:RI.SublCmpt}.  

\STEP{Step 2: Extraction of a converging subsequence.} The a priori estimates
from Step 1 allows us to apply the abstract version of Helly's selection
principle (see \cite[Thm.\,3.2]{MaiMie05EREM} or
\cite[Thm.\,B.5.13]{MieRou15RIST}). This implies that there exists a
subsequence $ \big( \ul u_{N_l} \big)_{l\in \N} $ and a limit function $u:[0,T]
\to M$ such that we have the
pointwise convergence 
\[
\forall\:t\in [0,T]: \quad \ul u_{N_l}(t) \to u(t) \quad \text{in } (M,D).
\]
In particular, from $\ul u_N(0)=u_0$ we conclude $u(0)=u_0$ as desired. 

\STEP{Step 3: Derivation of the upper energy estimate.} To pass to the limit
$N_l\to \infty$ in \eqref{eq:RI.DiscrEDI} we first observe that the lsc of
$\calF(T,\cdot)$ gives $\calF(T,u(T)) \leq \liminf_{l \to \infty} \calF(T,\ul
u_{N_l}(T))$. For the second term on the left-hand side we deduce lsc from the
lsc of $\DD$ as follows. 
follows. 

For arbitrary partitions $0= t_1< t_1 < \cdots < t_N=T$ we have 
\[
\sum_{j=1}^N \DD\big(u(t_{j-1}),u(t_j)\big) \overset{\DD\text{ lsc}}\leq 
\liminf_{l\to \infty}  \sum_{j=1}^N 
\DD\big(\ul u_{N_l}(t_{j-1}), \ul u_{N_l}(t_j)\big) \leq 
\liminf_{l\to \infty} \mafo{Var}_\DD(\ul u_{N_l},[0,T]) 
\leq C_*.
\] 
Taking now the supremum over all partitions on the left-hand side gives 
$\mafo{Var}_\DD(u,[0,T]) \leq  \liminf_{l\to \infty} \mafo{Var}_\DD(\ul
u_{N_l},[0,T]) $ as desired. 

For the power integral on the right-hand side in \eqref{eq:RI.DiscrEDI} we can
pass to the limit (not liminf) by the same arguments as at the end of the proof
of Proposition \ref{pr:RI.ChainRuleIneq}, i.e.\ we use \ref{eq:calF.Gronw} and
\eqref{eq:RI.ContPower} once again. In summary, we have shown that the limiting
curve $u:[0,T] \to M$ satisfies the upper energy estimate 
\begin{equation}
  \label{eq:RI.UpperEDI}
  \calF(T,u(T)) + \mafo{Var}_\DD(u,[0,T]) \leq \calF(0, u(0)) +
\int_0^T\pl_t \calF(t,  u(t)) \dd t . 
\end{equation}
   
\STEP{Step 4: Derivation of energetic solutions.} By Proposition
\ref{pr:TIMS.ERIS}(iii) we have the discrete global stability
$\ul u_N(t_j^N) \in \calS(t_j^N)$. Now fix a $t \in [0,T]$ such that
$\ul u_{N_l}(t) \to u(t)$. By the construction of the piecewise constant
interpolants we have $\ul u_{N}(t)=\ul u_N(t_{j_N(t)}^N)$ for
$t-\phi_{N}<t^{N}_{j_{N}(t)}\leq t$, where $\phi_{N}$ is the fineness of the
partition. Hence, $\wt t_l=t^{N_l}_{j_{N_l}(t)} \to t$, \ $\ul u_{N_l}(t) \to
u(t)$, and $\ul u_{N_l}(t) \in \calS(t)$, which implies $u(t)\in \calS(t)$ by
the closedness assumption \eqref{eq:RI.ClosStabSet}. Since $t\in [0,T]$ was
arbitrary, we have established the global stability condition \textbf{(S)}. 

Moreover, we have shown now all the conditions that are necessary for
Proposition \ref{pr:RI.ChainRuleIneq}, and we obtain the the lower energy
estimate \eqref{eq:RI.ChRuIneq}. Together with the upper estimate in
\eqref{eq:RI.UpperEDI}, we have established the energy balance \textbf{(E)},
and hence $u:[0,T]\to M$ is an energetic solution.   
\end{proof}

\subsection{Closedness of the stable sets}
\label{su:ClosedStabSets}

The crucial and nontrivial condition for showing existence of energetic
solutions is the closedness of the stable sets, namely condition
\eqref{eq:RI.ClosStabSet}. This difficulty is comparable to the difficulty of
showing closedness of the subdifferentials in 
rate-dependent gradient system in Banach spaces or to showing lsc of the metric
slope. 

The first case is the easiest case, namely when $\DD$ is continuous. 

\begin{lemma}[Closedness of $\calS$ via continuity]
Assume that the ERIS $(M,\calF,\DD)$ satisfies \ref{ERIS.1}--\ref{ERIS.5} and
that $\DD:M\ti M \to {[0,\infty[}$ is continuous, then the closedness
condition \eqref{eq:RI.ClosStabSet} holds. 
\end{lemma}
\begin{proof} From $u_i\in \calS(t_i)$ we have 
\[
\forall \:w\in M:\quad \calF(t_i,u_i) \leq \calF(t_i,w) + \DD(u_i,w). 
\]
We simply pass to the limit $i\to \infty$ using $t_i\to t$, \ $u_i \to u$, 
\ref{eq:calF.Gronw}, and the 
lsc of $\calF(t,\cdot)$. This we obtain 
\[
\calF(t,u)\leq \liminf_{i\to \infty} \calF(t_i,u_i) \leq \lim_{i \to \infty}
\big( \calF(t_i,w) + \DD(u_i,w)\big) = \calF(t,u)+ \DD(u,w),
\]
which is the desired result. 
\end{proof}

A typical application of this theory are models used for hysteresis in
ferromagnetic materials, see \cite[Sec.\,4.4]{MieRou15RIST}. A simplistic
version is given by 
\[
M=\rmL^1(\Omega;\R^d), \quad \DD (u,w)=\rho\|u{-}w\|_{\rmL^1}, \quad \calF(t,u) =
\int_\Omega \!\!\big( \frac\kappa2|\nabla u|^2 {+}F(u){-}H(t){\cdot} u\big) \dd x,  
\]
where $u:\Omega\to \R^d$ plays the role of the magnetization and $H(t):\Omega
\to \R^d$ is a time-dependent, applied field. 
\medskip

However, in many applications the continuity of $\DD$ is too strong. In some
cases a unidirectionality condition is desirable, which leads to 
\[
\DD_\mafo{unidir}(u,w) = \begin{cases} 
\int_\Omega\big( w(x)-u(x)\big) \dd x & \text{if } w\geq u \text{ a.e.\ in
}\Omega,
\\ \infty & \text{else}. 
\end{cases}
\]
Typical applications of this idea are in damage processes (\cite{Thom10PhD, KneSch12GSPE,
  KnRoZa13VVAR}) or crack propagation \cite{FraLar03ECQS, DaFrTo05QCGN,
  DalZan07QSCG, DalToa10QCGE}, not allowing for any healing. 

In such cases the theory of ``\emph{mutual recovery sequences}'' can be
helpful. The MRS condition (introduced in  \cite{MiRoSt08GLRR} as JRS) reads as
follows:
\begin{align}
\nonumber
&\text{for } (t_j,u_j)\to (t_*,u_*)  \text{ with }u_j\in \calS(t_j) \text{ and
} \wh u \in M \\
\label{eq:MRSeq}
&\text{there exists } \big( \wh u_j\big)_{j\in \N} \text{ such that } \wh u_j
\to \wh u \ \text{ and } \ \\
&\hspace{3em} \nonumber
\limsup_{j\to \infty}\big(\calF(t_j,\wh u_j){+}\DD(u_j,\wh u_j){-}\calF(t_j,u_j) \big)\leq 
\calF(t_*,\wh u){+}\DD(u_*,\wh u){-}\calF(t_*,u_*). 
\end{align}
In the theory of crack propagation this condition is established via the
so-called ``jump transfer lemma'', see \cite{FraLar03ECQS, DaFrTo05QCGN}. 

\begin{lemma}[Closedness of $\calS$ via MRS]
Assume that the ERIS $(M,\calF,\DD)$ satisfies \ref{ERIS.1}--\ref{ERIS.5} and
\eqref{eq:MRSeq}, then the closedness
condition \eqref{eq:RI.ClosStabSet} holds. 
\end{lemma}
\begin{proof} We consider $t_j, t_*, u_j$, and $u_*$ as in
  \eqref{eq:RI.ClosStabSet}. The closedness is established if we can show
  $u_*\in \calS(t_*)$. 

For an arbitrary test state $\wh u$ we choose $\big( \wh u_j\big)_{j\in \N}$ as
provided in \eqref{eq:RI.ClosStabSet}. Then, we have 
\begin{align*} 
\calF(t_*,\wh u){+}\DD(u_*,\wh u){-}\calF(t_*,u_*) \geq \limsup_{j\to
 \infty}\big(\calF(t_j,\wh u_j){+}\DD(u_j,\wh u_j){-}\calF(t_j,u_j) \big) \geq
0
\end{align*}
where the last estimate follows via $u_i\in \calS(t_i)$. Rearranging the terms
gives $u_*\in \calS(t_*)$. 
\end{proof}

The usefulness of this condition is already seen in classical linearized
elastoplasticity, where we have 
\[
\DD(u,w)= \big\| u{-}w\|_{\rmL^1(\Omega)} \quad \text{and} \quad 
\calF(t,u) = \frac12 \langle \bbA u , u\rangle_{\rmL^2(\Omega)} - \langle
\ell(t), u\rangle. 
\]
Here $\bbA=\bbA^*$ is bounded and positive definite operator on
$\rmL^2(\Omega)$. Since $\rmL^2(\Omega)$ does not compactly embed into
$\rmL^1(\Omega)$ the construction of solutions has to be based on the weak
topology, in $\rmL^2(\Omega)$, but $\DD$ is only lsc but not continuous.  

Nevertheless, the construction of a recovery sequence works because we can use
cancellations in the terms appearing in the limsup condition in
\eqref{eq:MRSeq}. 
For a sequence $u_j\weak u$ in $\rmL^2$ and a fixed $\wh u\in \rmL^2$ we define 
\[
\wh u_j = \wh u + u_j - u.
\]
Clearly, we have $\DD(u_j,\wh u_j)=\DD(u,\wh u)$, i.e.\ the two weakly
converging sequences cancel each other. Similarly, using the quadratic
structure of $\calF(t,\cdot)$ we have 
\begin{align*}
\calF(t_j,\wh u_j)-\calF(t_j,u_j)&=\frac12\langle \bbA (\wh u{-}u), \wh u {+}
2u_j{-}u\rangle - \langle \ell(t_j), \wh u{-}u\rangle \\
& \to \  
 \frac12\langle \bbA (\wh u{-}u), \wh u {+}u\rangle - \langle \ell(t_*),
 \wh u{-}u\rangle = \calF(t_*,\wh u) - \calF(t_*,u). 
\end{align*}
This shows that the construction of mutual recovery sequences in the sense of 
\eqref{eq:MRSeq} works for this case.

  \paragraph*{Acknowledgments.} The author is grateful to Moritz Gau and
  Jia-Jie Zhu for several critical and constructive remarks that helped to
  improve these lecture notes. Of course, this work benefited greatly from
  fruitful discussion with many collaborators, in particular Thomas Frenzel,
  Matthias Liero, Mark Peletier, Riccarda Rossi, Giuseppe Savar\'e, and Artur
  Stephan.

\footnotesize 
%\bibliographystyle{alpha_AMs}
%\bibliography{alex_pub,bib_alex}
\newcommand{\etalchar}[1]{$^{#1}$}
\def\cprime{$'$}
\providecommand{\bysame}{\leavevmode\hbox to3em{\hrulefill}\thinspace}
\providecommand{\MR}{}

\end{document}